\documentclass[Times_New_Roman,12pt, a4paper, oneside]{Report} 

\graphicspath{{./Pictures/}} 
\usepackage{times}
\usepackage[comma, sort&compress]{natbib}
\usepackage{doi,hyperref}
\usepackage{multicol}
\usepackage{aas_macros}
\usepackage{isotope}
\usepackage[utf8]{inputenc}
\usepackage[version=3]{mhchem}
\usepackage[titletoc]{appendix}
\usepackage{pdfpages}
\usepackage[utf8]{inputenc}
\usepackage{textcomp}

\DeclareUnicodeCharacter{2010}{-}
\hypersetup{urlcolor=blue, colorlinks=true} 

\begin{document}
\frontmatter 

\setstretch{1.3} 

\fancyhead{} 
\rhead{\thepage} 
\lhead{} 

\pagestyle{fancy} 

\newcommand{\HRule}{\rule{\linewidth}{0.5mm}} 
\hypersetup{pdftitle={\ttitle}}
\hypersetup{pdfsubject=\subjectname}
\hypersetup{pdfauthor=\authornames}
\hypersetup{pdfkeywords=\keywordnames}

\begin{titlepage}


\begin{center}
\begin{figure}
	\centering
	\begin{subfigure}[b]{0.2\textwidth}
		\centering
		\includegraphics[width=1.1\textwidth]{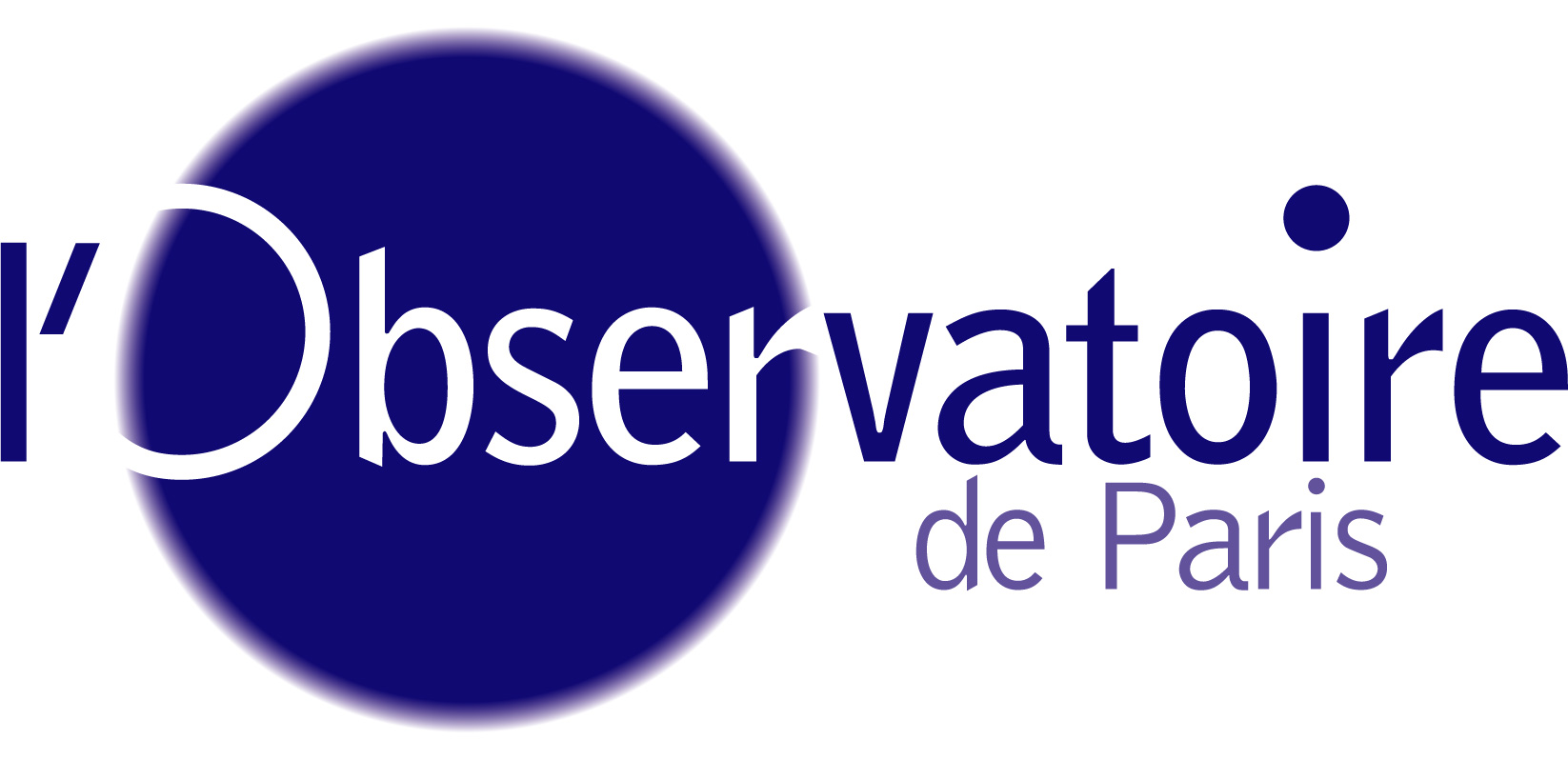}
	\end{subfigure}
	\hfill
	\begin{subfigure}[b]{0.2\textwidth}
		\centering
		\includegraphics[width = 0.7\linewidth]{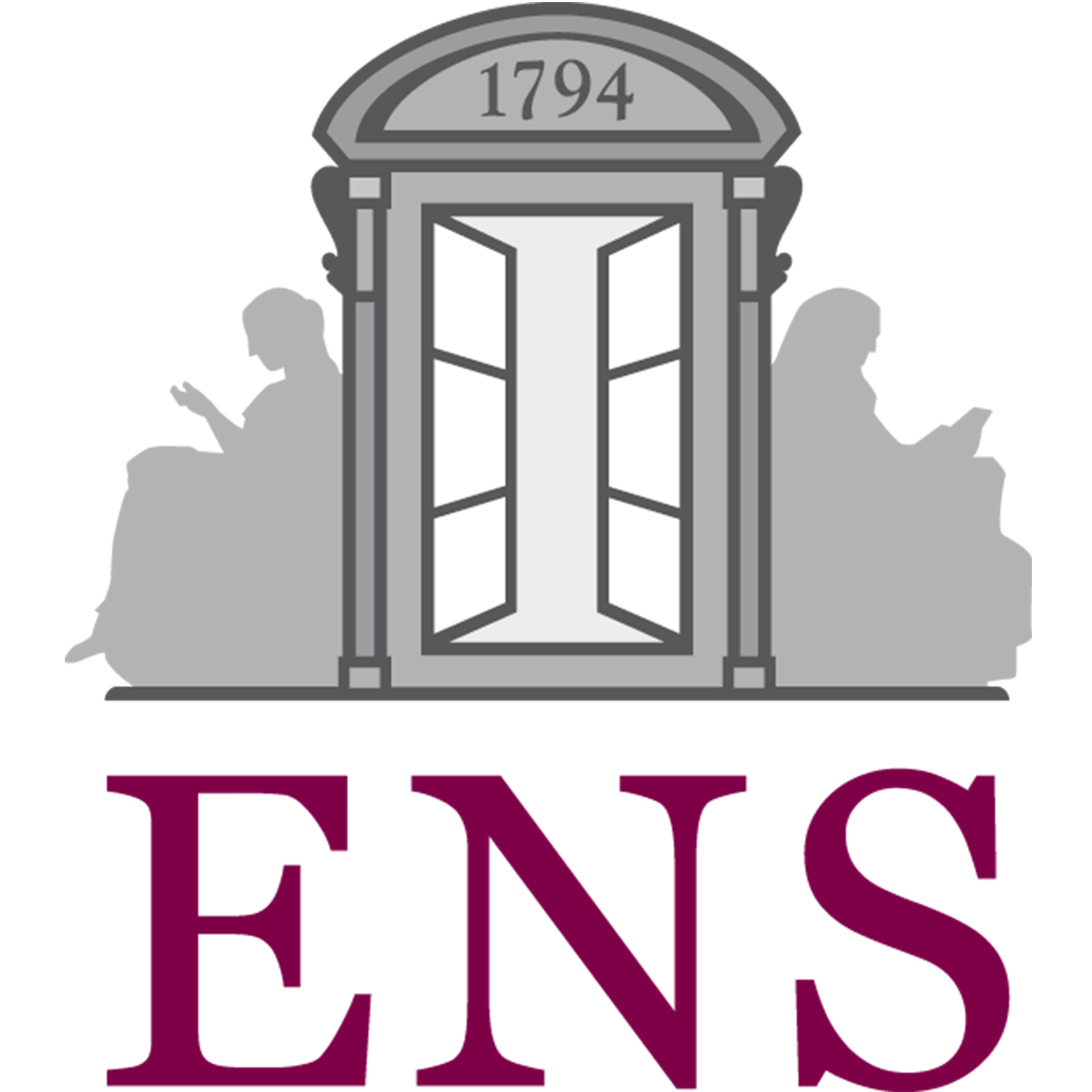}
	\end{subfigure}
	\hfill
	\begin{subfigure}[b]{0.2\textwidth}
		\centering
		\includegraphics[width = 1.1\linewidth]{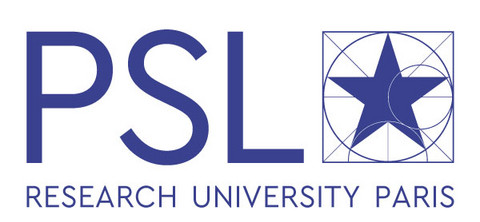}
	\end{subfigure}
	\hfill
	\begin{subfigure}[b]{0.2\textwidth}
		\centering
		\includegraphics[width =1.1\linewidth]{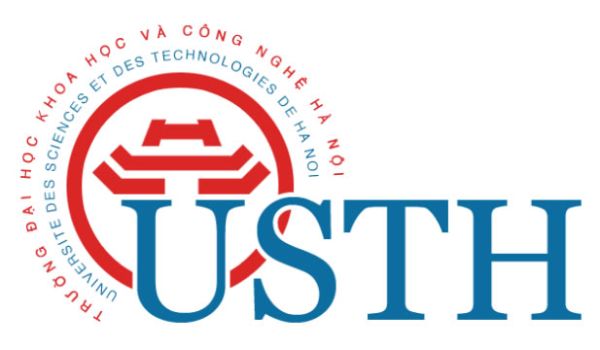}
	\end{subfigure}
\end{figure}
\textsc{\small \univname}\\[0.9cm] 
\HRule \\[0.5cm] 
{\LARGE \bfseries \ttitlecover}\\[0.5cm] 
\HRule \\[1.3cm] 
\textsc{\LARGE THESIS}\\[0.5cm] 
 \textit{submitted and publicly defended on March 28$^{th}$, 2018\\
  in fulfilment of the requirements for the} \\[0.5cm] 
 \textsc{\LARGE doctorat de l'Observatoire de Paris}\\ [0.5cm]
 \textsf{by}\\[0.5cm] 
 \textbf{\large \authornames}\\[1.3cm] 

\begin{frame}
\centering Jury Members \\
Pr\'esident : Pierre ENCRENAZ\par\medskip
\begin{tabular}[t]{@{}l@{\hspace{3pt}}p{.32\textwidth}@{}}
Rapporteurs: & David FLOWER \\
& Leen DECIN\\
Examinateurs: &  St\'ephane GUILLOTEAU\\
&  Eva VILLAVER
\end{tabular}    
\begin{tabular}[t]{@{}l@{\hspace{3pt}}p{.3\textwidth}@{}}
Directeur: & Pierre LESAFFRE \\
Co-directrice: &  Sylvie CABRIT \\
Co-directrice: & Pham Thi Tuyet NHUNG
\end{tabular}%
\end{frame}

%


\end{center}
\end{titlepage}

\Declaration{

\addtocontents{toc}{\vspace{1em}} 

I, \authornames, declare that this thesis titled, '\ttitle' and the work presented in it are my own. I confirm that:

\begin{itemize} 
\item[\tiny{$\blacksquare$}] This work was done wholly or mainly while in candidature for a doctoral degree at Observatoire de Paris.
\item[\tiny{$\blacksquare$}] Where any part of this thesis has previously been submitted for a degree or any other qualification at Observatoire de Paris or any other institution, this has been clearly stated.
\item[\tiny{$\blacksquare$}] Where I have consulted the published work of others, this is always clearly attributed.
\item[\tiny{$\blacksquare$}] Where I have quoted from the work of others, the source is always given. With the exception of such quotations, this thesis is entirely my own work.
\item[\tiny{$\blacksquare$}] I have acknowledged all main sources of help.
\item[\tiny{$\blacksquare$}] Where the thesis is based on work done by myself jointly with others, I have made clear exactly what was done by others and what I have contributed myself. \\
\end{itemize}
 
Signed:\\
\rule[1em]{25em}{0.5pt} 
 
Date:\\
\rule[1em]{25em}{0.5pt} 
}

\clearpage 


\pagestyle{empty} 

\null\vfill 

\textit{For my mother MA Thi Kieu, my younger sister LE Thi Ngoc Tram and my family ...}


\vfill\vfill\vfill\vfill\vfill\vfill\null 

\clearpage 
\setstretch{1.3} 
\acknowledgements{\addtocontents{toc}{\vspace{1em}} 

I honorably thank Pierre Lesaffre for his enthusiastic supervision. He always was patient in the way of helping me solve practical problems that I have faced during my thesis. I am grateful to have been given the opportunity to learn science over these years. Furthermore, his scientific abilities to figure out the point and his attitudes at work were a model for me.

I also wish to thank Sylvie Cabrit for the deep intensity of her supervision of this thesis. Over these three years, I benefited from her general vision of the Interstellar physics, as well as from her excellent ideas that made my thesis logical and constituted a magnificent information for a young researcher like me.

I am obviously thankful to both of them for training me at using the shock code that they have been updating, improving for a long time. I personally could not find any words to express my gratitude. This gave me a general picture how the research should be, and what we understood about shock in the insterstellar medium.

I thank my collaborators A.Gusdorf, David Neufeld, Thibeau Le Bertre, Jan-Martin Winters, P.Tuan-Anh and P.T Nhung for working with me. They do not only give beautiful data, useful materials and enlightening suggestions, but they also willingly respond to me for any questions. It is really interesting and valuable whenever I discuss with them, and I always learn something special after all.

Here is the place where I want to thank Antoine Gusdorf, who gave me beautiful observational data and explained to me the observational techniques that I did not really understand, as well as gave me a chance to be a part in his collaborations. And to Benjamin Godard, who is developing the shock code for helping me deeply understand the algorithm of the code.

Indispensably, I am thankful to the Vietnam International Education Development (VIED) for mainly funding this thesis and the ANR SILAMPA (ANR-12-BS09-0025) for partly supporting me. I am also thankful the University of Science and Technology of Hanoi (USTH), who accepted me to be a candidate studying abroad.

I am glad to mention my godfather St\'ephane Jacquemoud here, who always stood by me and encouraged me for these three years and in the future for sure. He helped me set up the life in Paris and solve any social problems which could happen. He also taught me French, as well as the history, the cultures and the arts of France.   
  
Of course, I am thankful to my family for constantly supporting me for these years, and the way they help me becoming what I am now. I would like to send a special thank to my mother who worked hard and traded off here life and health to make my dream come true. I also want to mention my uncles and my aunts who covered up my family when we were in trouble. 

Last but not least, I want to thank my friends, who were always beside me when I needed it, who gave me some advises, experiences for some specific cases.
}
\clearpage 

\clearpage 
\addtotoc{English summary} 
\SummaryENG{\addtocontents{toc}{\vspace{1em}} 
\setstretch{1.1} 

Stars are bad neighbors: they often disturb their surroundings. They sometimes travel very fast through the interstellar medium (ISM). They frequently undergo violent ejection events which leave an imprint on their neighborhood (jets, winds, supernovae). These supersonic flows generate shocks both in the ejected material and in the stellar environment. The study of these shocks constitute the subject of this thesis, and we model them with the Paris-Durham planar shock code, which incorporates a wealth of micro-physics and chemical processes relevant to the magnetized ISM.

  First, we use this code to model 3D magnetized axisymmetric bow shocks with arbitrary shapes, thanks to a formalism which links mathematically the shape of shocks to an equivalent statistical distribution of 1D shocks. For the first time, we examine systematically the effect of the geometry, age, and various other parameters on the H$_2$ excitation diagram and emission line profiles. For example, we unveil a geometrical effect which shows that 1D planar shocks emission fits to 3D bow shocks are biased towards small velocities. We also apply our models to spatially integrated H$_2$ observations of bow-shocks in Orion BN-KL and BHR71 where a much better match is obtained with only a limited number of additional parameters compared to former planar models. We illustrate on the Herbig-Haro object HH54 how spectrally resolved H$_2$ line emission profiles can be used to extract a wealth of dynamical information.
  
  Second, we include in the Paris-Durham shock code a minimum set of processes necessary to describe asymptotic giant branch (AGB) wind models: geometrical dilution, external interstellar radiation, radiative pressure on grains, gravity, heating from stellar radiation pumping, three-body reactions, and sonic-point crossing. With this tool, we started to examine the time-dependent chemistry of hydrogen in winds of hot and cool AGB stars. We suggest that the low abundance of HI inferred from observations is due to hydrogen locked in its molecular form, and we use our model to try and reproduce HI line observations lines in a hot AGB (Y CVn) and a cold AGB (CW Leo).
  
  Although we have mainly focused on atomic or molecular hydrogen in this study it would be straightforward to extend it to other molecules with optically thin transitions. These simplified tools to model chemistry for complex geometries and dynamics are proving very useful at a time when new instruments such as ALMA discover a wealth of spectral and spatial information for a multitude of chemical tracers, and also when the JWST will soon provide complementary data in the infrared H$_2$ and ionic lines with unprecedented resolution and sensitivity.
}
\clearpage
\addtotoc{Résumé français} 
\SummaryFREN{\addtocontents{toc}{\vspace{1em}} 
\setstretch{1.1} 
Les étoiles sont de très mauvaises voisines: elles perturbent souvent leur environnement. Parfois, elles se déplacent à grande vitesse dans le milieu interstellaire (MIS). Souvent, elles subissent des soubresauts violents qui laissent une empreinte dans leur voisinage (jets, vents, supernovae). Ces flots supersoniques génèrent des chocs à la fois dans le matériau éjecté par l'étoile et dans l'environnement stellaire. L'étude de ces chocs constituent le sujet de cette thèse, et nous les modélisons avec le code de chocs stationnaires plan parallèle Paris-Durham, qui incorpore une riche panoplie de processus microphysiques et chimiques adaptés au MIS magnétisé.

  Tout d'abord, nous utilisons ce code pour modéliser des chocs magnétisés 3D pour des formes arbitraires à symétrie axiale, grace à un formalisme qui lie mathématiquement la forme des chocs à une fonction de distribution de chocs 1D équivalente. Pour la première fois, nous examinons systématiquement l'effet de la géométrie, de l'âge, et de quelques autres paramètres sur le diagramme d'excitation de H$_2$ résultant et la forme des profils raies d'émission de H$_2$. Par exemple, nous dévoilons un effet géométrique qui montre que l'ajustement par des modèles 1D de l'émission de H$_2$ observée sur un choc 3D est sujette à un biais vers les basses vitesses. Nous appliquons aussi nos modèles à l'observation de H$_2$ spatialement intégrée de chocs d'étrave dans Orion BN-KL et BHR71 où nous obtenons un bien meilleur ajustement des observations avec un nombre à peine plus grand de paramètres comparé aux modèles précédents. Nous illustrons sur l'objet de Herbig-Haro HH54 la grande richesse d'information dynamique que renferme le profil des raies d'émission résolues de H$_2$.
  
  Ensuite, nous incluons dans le code de Paris-Durham un ensemble minimal de processus nécessaires pour décrire les modèles de vents d'étoiles de la branche asymptotique des géantes (AGB): la dilution géométrique, l'irradition externe, la pression de radiation sur les grains, la gravité, le chauffage dû au pompage radiatif par l'étoile, les réactions à trois corps et le passage du point sonique. Avec cet outil, nous commençons à examiner la cinétique chimique de l'hydrogène dans les vents d'étoiles AGB chaudes et froides. Nous suggérons que la faible abondance de HI déduite des observations s'explique par la forme principalement moléculaire que prend l'hydrogène. Nous générons le choc terminal dans le vent et nous essayons avec nos modèles de reproduire les observations de la raie HI dans une AGB chaude (Y CVn) et une froide (CW Leo).
  
  Bien que nous ayons principalement concentré notre attention sur l'hydrogène (atomique ou bien moléculaire) dans cette étude, l'extension de ce travail à des transitions optiquement minces d'autres molécules est assez directe. Ces modèles simplifiés pour modéliser la chimie dans des géométries et dynamiques néanmoins complexes se révèlent très utiles au moment où de nouveaux instruments comme ALMA dévoilent une grande richesse spectrale et spatiale pour une multitude de traceurs chimiques. Ceci alors que le JWST est sur le point d'apporter dans l'infra-rouge de l'information complémentaire sur les raies de H$_2$ et les raies ioniques avec une résolution et une sensibilité inégalées.
}
\clearpage

\pagestyle{fancy} 

\lhead{\emph{Contents}} 
\tableofcontents 
\lhead{\emph{List of Figures}} 
\listoffigures 
\lhead{\emph{SUBSYSTEM DESIGN}} 
\listoftables 

\mainmatter 

\pagestyle{fancy} 

%
%
\part{INTRODUCTION}
\setstretch{1.1} 
\chapter{INTERSTELLAR SHOCKS}
\lhead{\emph{INTRODUCTION}} 
\section{Introduction}

The gas in between stars is usually much colder than them.
As a result of the slow velocity of sound, flows can easily become
supersonic. The relative motions of stars
or gaseous clouds can be sufficient to trigger shocks. On top of that,
stars are subject to violent events during their life. 
Immediately after their birth, the gas which does not make it onto the surface can be
ejected and impact the surrounding interstellar medium (ISM) through
outflow cavities and protostellar jets. Later in their evolution, 
stars launch winds which can become supersonic very quickly with respect
to their cold environment. At the end of their lives, some
stars end up in a burst of supernovae ejecta which generate shocks
at extremely large velocities. Finally, large scale galaxy collisions
can also shake the gas supersonically and generate shocks.
The interstellar gas inside galaxies is thus continuously permeated
by traveling shock waves, which heat up and illuminate the gas. 
The emission of light, which can be observed by astronomers, 
gives us as many opportunities to
access informations on the galactic dynamics.

\begin{figure}
	\centering
		\includegraphics[width=1\linewidth]
    	{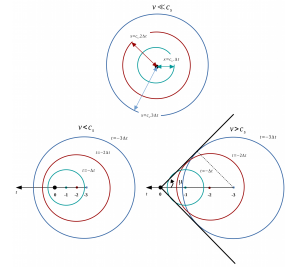}
	\caption[Wave, subsonic and supersonic motions]
	{Wave, subsonic and supersonic motions of a source point.}
	\label{fig:shock_wave}
\end{figure}

\section{Shock waves} \label{sec:shock_define}
A wave is a perturbance propagating in a fluid without changing. 
A shock wave is a pressure wave that moves faster than the speed of sound in that fluid.
In nature, shock waves or simply shocks are common phenomena. In principle, an object will deflect the gas molecules when it penetrates through it. If the speed of the object is much smaller than the speed of sound, the density of gas remains approximately constant (\autoref{fig:shock_wave}, \textit{top}). If it is comparable (but lower) to the speed of sound, its motion is always behind the sound wave launched from the previous position (\autoref{fig:shock_wave}, \textit{bottom left}), and the gas is swept away and its density is compressed by the object. This compressive gas flow is then nearly \textit{reversible} and its properties are well described by the \textit{isentropic condition}, with entropy remaining constant. When the object moves faster than the speed of sound, all the compressive waves sent ahead to sweep the gas are caught up by the object, and gathered in an abrupt structure: a shock wave is formed (\autoref{fig:shock_wave}, \textit{bottom right}) 
with an opening angle of the cone $\mu$. This angle allows us to estimate the speed of the supersonic motion through the \textit{Mach number} ($M$), defined as $M=v/c_s$ with $v$ and $c_{s}$ the speed of the object and the speed of sound. The opening angle of the cone satisfies
\begin{equation} \label{eq:Mach_number}
	\sin(\mu) = \frac{c_{s}}{v} = \frac{1}{M} \mbox{.}
\end{equation}

Unlike sound waves, shock waves are non linear waves and they largely change the gas properties. Across the shock wave, the pressure, the density, the temperature and the entropy of the gas abruptly jump. Downstream, the kinetic and thermal energy of the gas in the shock wave dissipate rapidly with respect to the distance: a shock wave is an \textit{irreversible} (or \textit{non-isentropic}) process which dissipates kinetic energy into heat and radiation. 
If not sustained, a shock wave loses its energy over some distance as it heats gas and it degenerates into a conventional sound wave. 

\begin{figure}
	\centering
		\includegraphics[width=1\linewidth, height=0.25\textheight]
    	{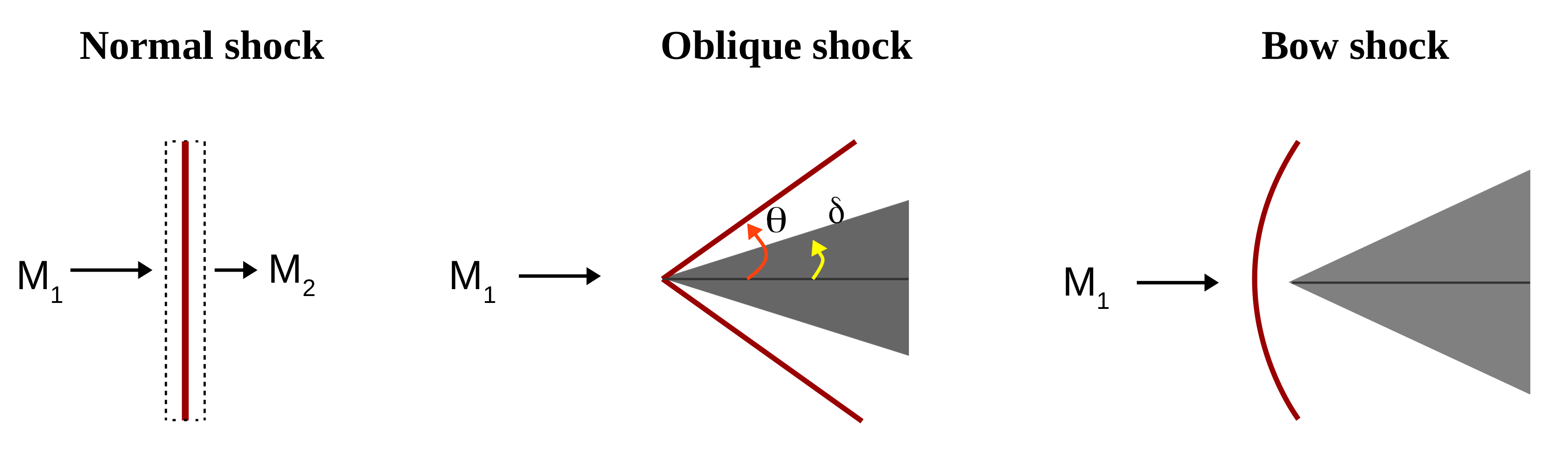}
	\caption[Sketch of shock wave divisions]
	{Sketch of shock wave geometries: normal shock, oblique shock, and bow shock.}
	\label{fig:shock_definition}
\end{figure}

\begin{figure}
	\centering
	\begin{minipage}[c]{0.8\textwidth}
		\includegraphics[width=1\linewidth]
    	{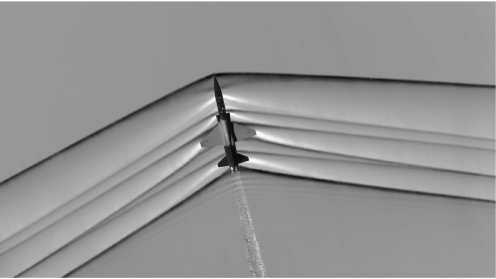}
	\end{minipage}
	\caption[Oblique shocks around a supersonic aircraft]
	{Oblique shocks around the T-38 Talon aircraft 
	in supersonic flight over the Mojave desert 
	(credit: NASA \& US Air Force, 2015)}.
	\label{fig:obliqueshock_air}
\end{figure}

\begin{figure}
	\centering
	\begin{minipage}[c]{0.8\textwidth}
		\includegraphics[width=1\linewidth]
    	{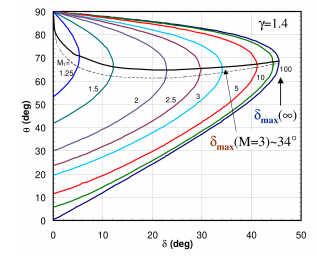}
	\end{minipage}
	\caption[Geometric condition of the supersonic object in order to form bow shock]
	{Geometric condition of the supersonic object in order to form a bow shock. 
	The bow shock is formed when the size-angle of the infinite-wedge object $\delta$ 
	exceeds its maximum value at a given Mach number. 
	$\theta$ is the angle of the oblique shock 
	(\autoref{fig:shock_definition}, \textit{middle panel}).}
	\label{fig:oblique_angle}
\end{figure}

\begin{figure}
	\centering
	\begin{minipage}[c]{0.7\textwidth}
    	\includegraphics[width=1\linewidth]
    	{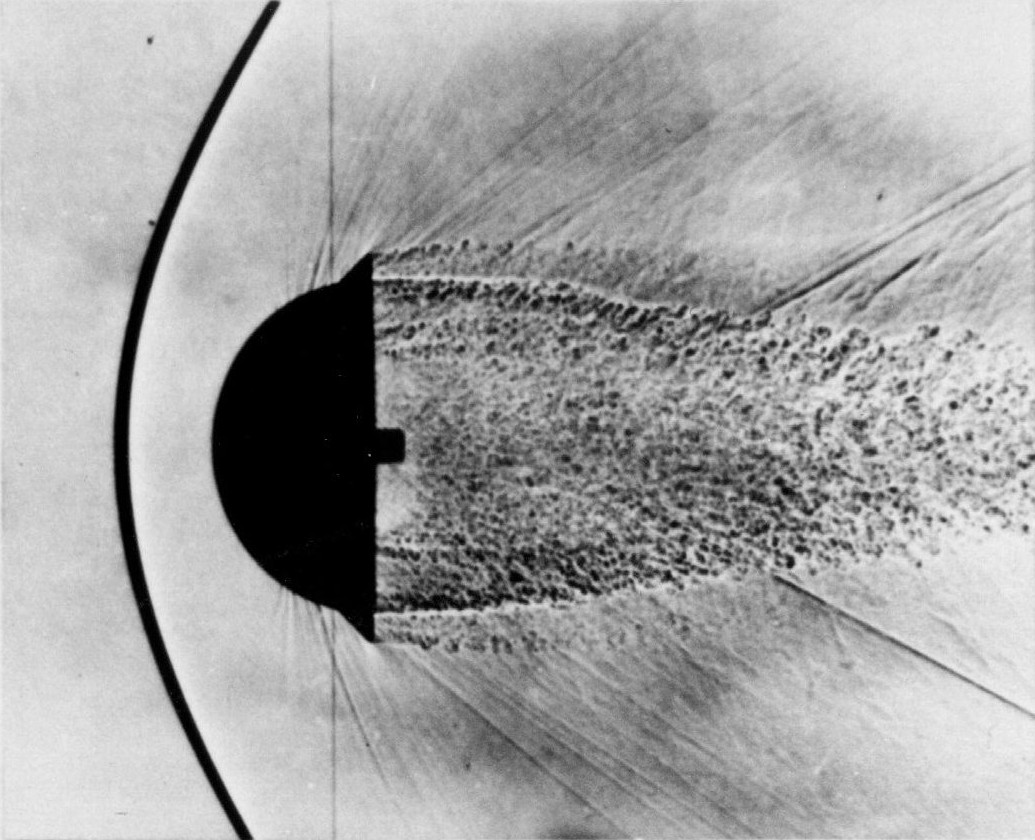}
	\end{minipage}
	\caption[Bow shock around a model of the blunt re-entry body]
	{Bow shock around a model of the blunt re-entry body (credit: NASA, 1960)}.
	\label{fig:bowshock_air}
\end{figure}

Typically, there are three types of shock waves around a moving solid object (\autoref{fig:shock_definition}). A shock wave is called \textit{normal} if its front is perpendicular to the direction of the entrance velocity. In this case, the flow direction does not change. However, during the motion of the object, it may not remain perpendicular to the flow direction. When the shock wave front is inclined with respect to the flow direction, it is called an \textit{oblique shock}. 
 Oblique shocks are more easily generated by pointy parts of an object such as the nose, the edge of the wing, and the trailing edges of the supersonic plane shown in \autoref{fig:obliqueshock_air}. 
 Oblique shocks are not always the preferred form around supersonic objects. 
If we consider a supersonic infinite-wedge object with size-angle $\delta$, the possible oblique shock is defined by the angle $\theta$ (\autoref{fig:shock_definition}, \textit{middle}), which differs from the supersonic angle $\mu$ above. At Mach number $M>1$, the existence of the oblique shock around this infinite-wedge object can be determined via \autoref{fig:oblique_angle} (more details can be found in the lecture of \textit{Daniel Guggenheim School of Aerospace Engineering}\footnote{http://seitzman.gatech.edu/classes/ae3450/outline.html}). For example, if the object is moving with $M=3$, and its size-angle is larger than $34^{o}$, there is no oblique shock around it. In this case, the solution is a \textit{bow shock} (or \textit{detached shock}), which sits ahead and does not attach to the object (\autoref{fig:shock_definition}, \textit{right}). Bow shocks cover all ranges of oblique shocks from the strongest normal shock at the centerline and to weaker shocks in the curving wings of the bow. In practice, bow shocks need to be considered in the design for return capsules from space missions (\autoref{fig:bowshock_air}) for two reasons: (1) the drag of the capsule in supersonic motion is significantly increased by the surrounding bow shock; and (2) the capsule is not directly in contact with the bow shock, so that its temperature is kept below the melting point.
    
\section{Astrophysical shocks probe stellar evolution} 
\label{sec:shocks_vs_stellar_evolution}
In astrophysics, shocks are ubiquitous. They form at three stages in stellar evolution: at early stages, when stars are born, shocks are formed by the interaction of young stellar objects outflows with the ISM; at near the end of the lives of low- and intermediate-mass stars, stellar winds produce shocks; and at the end of the life of high mass stars, the supernova phase generates extremely high velocity shocks.

This mass loss behavior leads to chemical enrichment of galaxies, reprocessing of matter, and generation of turbulence; it also influences star-formation processes, and thus impacts the further evolution of stellar systems and galaxies. 

\begin{figure}
	\centering
	\includegraphics[width=1.\textwidth, height=0.3\textheight]
	{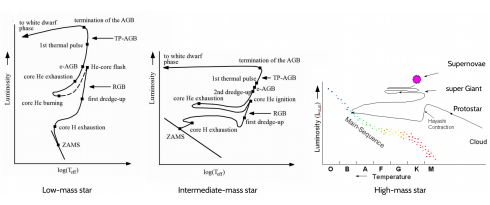}	
	\caption[Stellar evolution diagram]
	{Stellar evolution diagram. 
	(\textit{Left}) Evolution of a low-mass star. 
	(\textit{Middle}) Evolution of an intermediate-mass star. 
	(\textit{Right}) Evolution of a high-mass star.}
	\label{fig:Stellar_evolution}
\end{figure}

\subsection{Early stellar evolution and outflow shocks}
\begin{figure}
	\centering
	\includegraphics[width=0.8\textwidth]
	{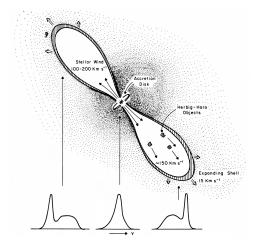}	
	\caption[First outflow model in the L1551 cloud]
	{Outflow model in the L1551 cloud (\citealt{Snell_1980}).}
	\label{fig:Snell_model}
\end{figure}

In general, stars form in dense molecular regions such as the cores inside the interstellar clouds, which contain gas and dust. 
At some point, these regions cannot resist their own gravity and they collapse. 
While collapsing, the density of the core increases, 
the inner region becomes optically thick, and the core is heated by the released gravitational energy. 
Once more material concentrates on the center, the increasing pressure stops 
the free fall to the central point and the core reaches a quasi-hydrostatic equilibrium, 
thus forming a protostar. 
Gradually, the envelope matter is depleted by accretion processes onto the new stellar surface.
The protostar is further heated by the released gravitational energy.
There, the thermal energy is converted into radiant energy 
that contributes to the luminosity of these objects. 
The angular momentum of the collapsing envelope is reduced by magnetic breaking the ejected outflow along the polar direction of the protostar, as confirmed by observation (e.g., \citealt{Konigl_2000}). 
The first model of outflow was derived by \citet{Snell_1980} 
who discovered 2 lines of CO from a large molecular outflow in the L1551. 
\autoref{fig:Snell_model} shows two outflows in opposite directions, 
the outflow sweeps out most of the ambient gas into a dense shell 
supported by the strong stellar wind and 
the shell itself also moves through the molecular cloud.            

\begin{figure}
	\centering
	\includegraphics[width=0.9\textwidth]
	{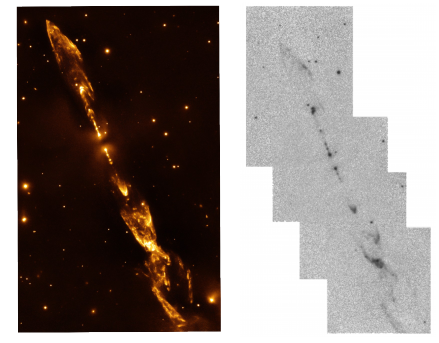}	
	\caption[HH212 outflow in the Orion constellation]
	{HH212 outflow in the Orion constellation. 
	(\textit{Left}) captured by a ground telescope \textit{VLT},
	and created by ESO/M, McCaughrean in 2015 
	(\textit{http://www.eso.org/public/images/potw1541a/}). 
	(\textit{Right}) 2.12 $\mu m$ H$_{2}$ 1-0S(1) emission
	captured by the space craft \textit{IRAS} (\citealt{Zinnecker_1996}).}
	\label{fig:HH212}
\end{figure}

Since then, many observational evidences of outflows have been observed in young stellar objects with higher resolution (\textit{http://casa.colorado.edu/hhcat/}). These outflows have supersonic motion and are driven by jets, which are narrow and difficult to detect. However, supersonic jets interact with the surrounding ambient medium and create one of the most beautiful astrophysical phenomenon: a \textit{shock} (see \autoref{sec:shock_define}). These shocks are easier to detect and to observe, and their properties allow us to deduce the properties of jets or even further of the protostar. \autoref{fig:HH212} displays the bipolar jet from the Herbig-Haro object HH212 in the Orion cloud.
The left side panel shows an infrared image observed by the ground telescope of the \textit{European Southern Observatory} (ESO). The right side panel shows the map of 2.12 $\mu m$ H$_{2}$ emission, captured by the \textit{Infrared Astronomical Satellite} (IRAS) (\citealt{Zinnecker_1996}). One can see the bipolar structure of the jet traced by the shock-excited rovibrational v=1-0, J=3-2 line of molecular hydrogen. This delineated bipolar structure is an important tool for revealing protostars. First, it allows us to determine their locations, where they are obscured, due to the drop of gas and dust density away from the source (\citealt{McCaughrean_1994}). Second, it allows us to determine the proper motion of the jet without using another reference star (e.g., \citealt{McCaughrean_2002},\citealt{Correia_2009}).      

\subsection{Low and intermediate mass late stellar evolution and shocked wind}
\label{sec:ABG_introduction}
When the temperature of a protostar  exceeds $8\;10^{6}$ K, hydrogen fusion reactions start. Hydrogen is burnt into helium and energy is released out. Then stars begin their life on the main-sequence. After central hydrogen is exhausted, the helium core shrinks, and is heated again by the released gravitationally energy. The hydrogen is then continuously burnt, surrounding an inactive helium core. During this stage, the star approaches the \textit{red giant branch} (RGB). From this point on, the lifetime and shock strength depend on its initial mass.

\begin{figure}
	\centering
	\includegraphics[width=0.9\textwidth]
	{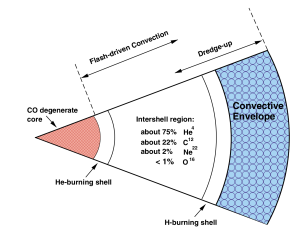}	
	\caption[Not to scale schematic structure of a low-mass AGB star]
	{Not to scale schematic structure of a low-mass AGB star 
	showing the He-burning shell above a degenerate C/O core, 
	and the H-burning shell below a deep convective envelop. 
	Between the two shells is an intershell region rich in helium and carbon 
	(\citealt{Karakas_2002}).}
	\label{fig:inner_structure}
\end{figure}

\textbf{Low-mass stars ($M_{i} \; \leq \; 2M_{\odot}$)}: fusion gradually exhausts hydrogen during stellar evolution in the main-sequence, but hydrogen burning still continues in a thick shell, moving outwards through the envelope. The still dormant helium core becomes electron degenerate and remains continuously fed by additional helium from that hydrogen burning shell. As the degeneracy sets in during the main-sequence phase, the temperature of the helium core is minimum, close to the surrounding H burning shell. Then the temperature decreases due to the degenerate electrons. The star globally starts to expand its own envelope. The He-core becomes denser, but the temperature did not reach yet the critical value required for helium fusion. During this phase, the luminosity increases drastically and the outer layers become convective. The convective region can reach down to the hydrogen burning shell, converted to helium and nitrogen via a CNO cycle. The newly formed elements are then mixed upwards to the upper layers through convection. This convection process mixing nuclear processed materials into the outer layer is called \textit{first dredge-up}. It leads to the enrichment of the surface layers. Finally, the thermal pressure from fusion is no longer sufficient to counter the gravity. The stars start to contract and to increase in temperature until the stars eventually becomes compressed enough so that the helium core becomes highly electron degenerate. This degeneracy pressure is finally sufficient to stop further collapse of the most central material. When the temperature reaches around $10^{8}$ K, the helium ignites and starts to fuse at the center through the triple-alpha process, by which three \isotope[4]{He} nuclei transform into \isotope[12]{C} and other heavier elements, \isotope[16]{O}, \isotope[20]{Ne}, \isotope[24]{Mg}. When the helium fusion begins with the triple-alpha process, the fusion rate raises rapidly, which again increases the temperature. This thermal run-away process is called \textit{He-core flash}. However, the total pressure only weakly depends on the temperature, since the degeneracy pressure (which is only a function of density) dominates thermal pressure that is proportional to the product of density and temperature. Therefore, the steep increase in temperature only causes a slight increase in pressure, so that the core cannot cool by expansion. However, the run-away process can make the temperature quickly rises to the point that thermal pressure is again dominant, eliminating the degeneracy. From then on, the core can expand and cool down, maintaining temperature to the critical value of $10^{8}$ K, where stable He-burning starts.
During the phase of core He burning, the central He supply gradually exhausts and an oxygen-carbon core develops. After the exhaustion of the central helium, the star evolves to the early asymptotic giant branch (AGB) phase. In this phase, the stellar luminosity of the star increases at almost constant temperature, and the stellar radius strongly increases. Surrounding the carbon-oxygen core is a helium-burning shell and a hydrogen-burning shell. These provide the energy output of the star. Above the He- and H-burning shells lies a deep convective stellar envelope (\autoref{fig:inner_structure}). At this stage, most of the luminosity of the star is provided by the H-burning shell. From now, the evolution of the low-mass stars is similar to the intermediate-mass star. 
This entire process is displayed in \autoref{fig:Stellar_evolution} (left). 

\begin{figure}
	\centering
	\includegraphics[width=0.9\textwidth]
	{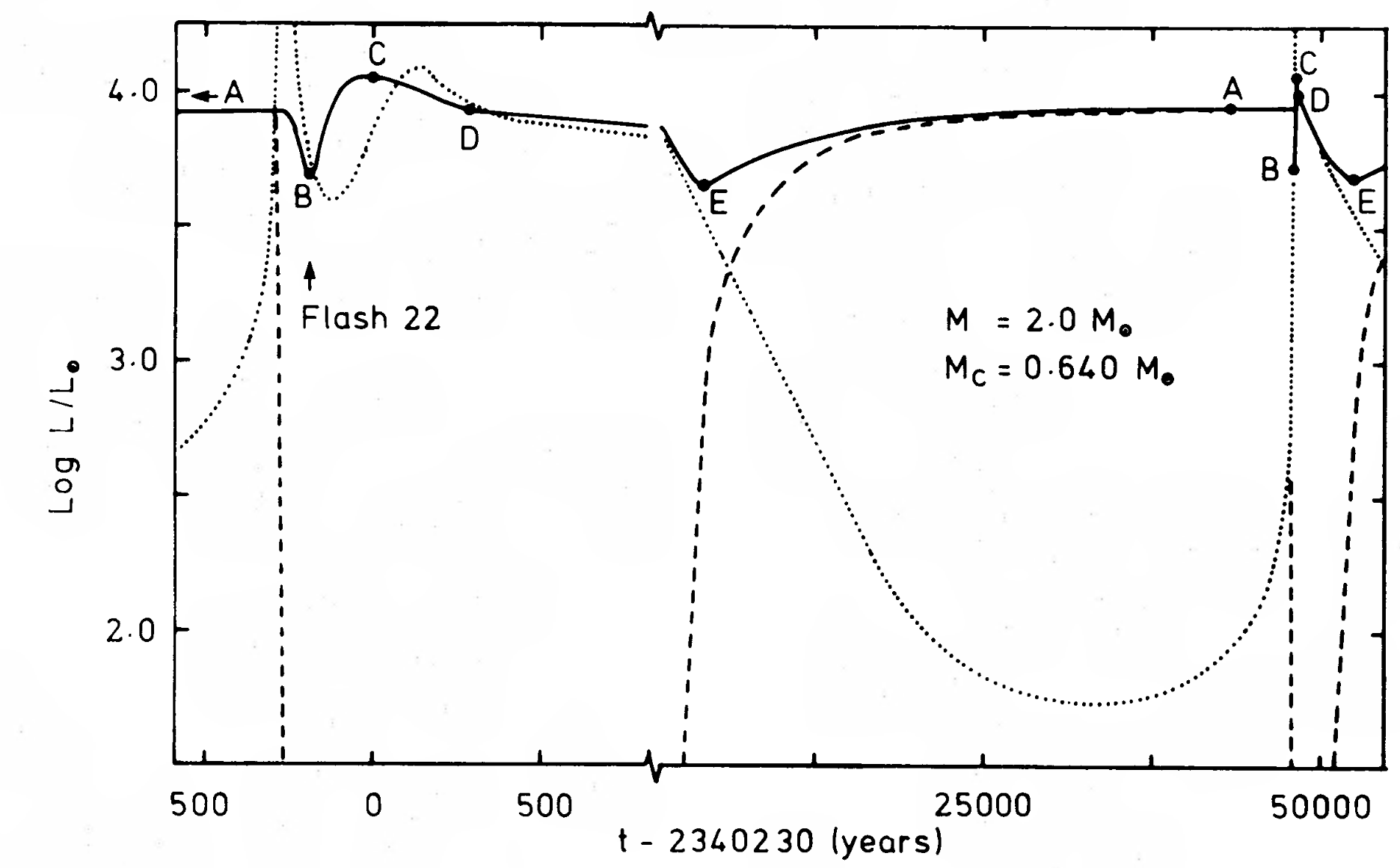}	
	\caption[Variation in surface luminosity, hydrogen-burning luminosity 
	and helium-burning luminosity during a flash cycle]
	{Variation in surface luminosity (solid line), 
	hydrogen-burning luminosity (dashed line) 
	and helium-burning luminosity (dotted line) 
	during a flash cycle for a 2M$_{\odot}$ star (\citealt{Wood_1981}).}
	\label{fig:thermal_pulse}
\end{figure} 

\begin{figure}
	\centering
	\begin{minipage}[c]{0.9\textwidth}
		\includegraphics[width=1.\textwidth]
		{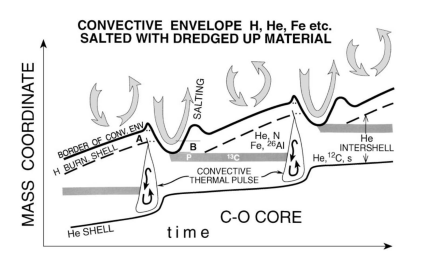}	
	\end{minipage}
	\caption[Inner structure of an AGB and dredge-up process due to thermal pulse]
	{Inner structure of an AGB and dredge-up process due to thermal pulse. 
	There are two convective zones, which mix the nuclear products 
	to the stellar surfaces during TP-AGB phase (\citealt{Busso_1999}).}
	\label{fig:convective_structure}
\end{figure} 

\textbf{Intermediate-mass stars ($2\;M_{\odot}\;<\;M\;<\;8\;M_{\odot}$)}: due to higher mass, and hence higher temperature, a convective core has developed because the nuclear burning in the core is sensitive to the temperature. This convective core contracts as hydrogen converts to helium. After H-core exhaustion, the convective He core remains and the stellar envelope expands, but H-burning continues in a shell. In this phase, the first dredge-up also appears. From now, the star evolves upward on the red giant branch (RGB) at a nearly constant surface temperature, and its radius also increases. During this phase, the central He core is contracting and heated by the gravitational energy. Again, when the central temperature exceeds  $10^{8}\; K$, the helium is ignited at the central region and forms carbon nuclei \isotope[12]{C} through the triple-alpha process and other heavier elements. Contrary to low-mass stars, the He-core has burnt under non-degenerate conditions, which avoids the He-core flash. After the ignition of helium, the star starts moving to the left on the Hertzprung-Russel Diagram (HRD), to higher surface temperature and higher luminosity. When temperature at the center is lower than the critical value, the He core-burning stops but the He still continues to burn in the thick shell. The He-exhausted core again contracts and heats, while the hydrogen envelope expands and cools down. In the HRD, the star evolves again toward to the giant branch. The convective envelope penetrates the dormant hydrogen shell and mixes \isotope[4]{He} and \isotope[14]{N} upwards to the outer layers. This mechanism is called \textit{second dredge-up}. The He shell burning heats up the base of the convective envelope and then makes the H burning to be reignited on top the He-shell. In the HRD, the star has reached the \textit{asymptotic griant branch} (\autoref{fig:Stellar_evolution}, middle).

Since the He-shell is thin compared to the radius of the shell 
(\autoref{fig:inner_structure}), its expansion is essentially isobaric. The temperature of the shell, therefore, must increase. This makes the He-shell thermally unstable (\citealt{Schwarzchild_1965}). A slight increase in temperature leads to a steep increase in the release of the nuclear energy through triple-alpha process, which further increases the temperature since the shell is extending. This thermal run-away process is able to increase the luminosity of the He-shell upto $10^{5}L_{\odot}$.  Upon reaching the luminosity peak, the He-shell is widely extended and thermally stable. Then the whole region contracts again, the H-shell is reactivated, and the flash cycle is repeated. This increase in luminosity is referred to as \textit{He-shell flash} or \textit{thermal pulse}. The star, therefore, is now located on the thermally pulsing AGB phase (TP-AGB). The thermal pulse process is shown in \autoref{fig:thermal_pulse}. During the TP-AGB phase, there are two convective zones: the inner convective zone is located in the intershell convection zone and mixes the processed matters from the He-shell (mainly \isotope[12]{C}) upwards to the H-burning shell (\autoref{fig:convective_structure}). After the He-shell flash and before the next shell flash, the outer convection zone reaches down to the intershell region and convects the material from this region upwards to the stellar surface. This mechanism enriches the newly processed matter from the inner region out to the outer envelope. This is called the \textit{third dredge-up}. During this dredge-up process, \isotope[12]{C} is enriched outward, and the C/O ratio increases from a value lower than $1$ to a value higher than $1$. Therefore, the third dredge-up is responsible for the formation of carbon-rich stars. The timescale of the star on AGB and TP-AGB phases depends on its initial mass and its metallicity. A star with $M_{\ast} = 1\,M_{\odot}$ and $Z = 0.006$, for example, spends 
$\sim 10^{7}\,$ yr on the early AGB phase and $\sim 10^{6}\,$yr on the TP-AGB phase (\citealt{Rosenfield_2014}).

\begin{figure}
	\centering
	\begin{minipage}[c]{0.9\textwidth}
		\includegraphics[width=1.\textwidth]
		{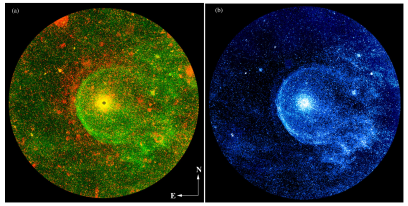}	
	\end{minipage}
	\caption[Shocked wind around the carbon star IRC +10216]
	{Shocked wind around the carbon star IRC +10216. 
	(a) Composite of FUV (green) and NUV (red). (b) FUV image itself. The position of the star is indicated by the $\star$ symbol. The asymmetry of the ring manifests the motion of the IRC +10216 star eastward to the ISM. The strong emission in FUV band is proposed to trace the collisional excitation of H$_{2}$ with electrons in the shocked gas (\citealt{Sahai_2010}).}
	\label{fig:shock_IRC10216}
\end{figure} 

The star has lost most of its own mass during the AGB phase, mainly due to stellar winds that are supersonic, and therefore generate shocks when they interact with the ambient gas (\citealt{Lamers_1999}). \autoref{fig:shock_IRC10216} illustrates the shock created by the wind from an AGB star named IRC +10216 (\citealt{Sahai_2010}). The observation is performed by the \textit{Galaxy Evolution Explorer} (GALEX) satellite in two wavelength ranges: 1344–1786 $\AA$ (near ultraviolet band - NUV) and 1771–2831 $\AA$ (far ultraviolet band - FUV). The asymmetry of the ring from east to west direction demonstrates that the IRC +10216 star moves eastward into the ISM. In fact, \autoref{fig:shock_IRC10216} shows the emission of the extended ring in the FUV band that is not visible in the NUV band. The strong FUV emission ring delineates the shock caused by the interaction between the wind from IRC +10216 with the surrounding ISM rather than by the dust scattering. Three are two reasons: first, in the case of dust scattering, the FUV/NUV ratio is expected to be $\sim$ 2.4 (\citealt{Whittet_1992}), where the observed value is $\sim$ 6. Second, the collisional excitation of the molecular hydrogen with the electrons in the shocked gas is the mechanism that best produces detectable FUV radiation, but no detectable NUV radiation. The region between the ring and the star position is a freely expanding stellar wind (unshocked wind). In this region, the emission is seen in both the NUV and FUV bands due to the scattering of ambient galactic starlight on dust particle in the stellar wind.

After the thin nuclear active shell burning around the central core stops because the fuel supply runs out, the core of the star moves to the left on the HRD and can be observed as a \textit{planetary nebula}. The star is then deceased. The remnant core becomes a \textit{white dwarf} and cools down.

\begin{figure}
	\centering
    \includegraphics[width=0.75\linewidth]
    {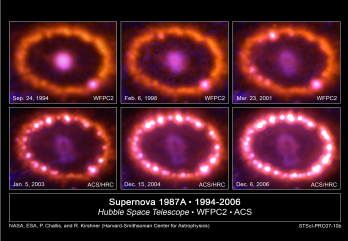}
	\caption[Shock around the supernovae remnant 1987A]
	{Image of a shock in the supernovae remnant 1987A from NASA's Hubble Telescope 
	(credit: NASA, ESA, P. Challis and R. Kirshner, Harvard-Smithsonian Center for Astrophysics).}
	\label{fig:shock_SNR}
\end{figure}

\subsection{High mass late stellar evolution and supernovae shocks}
\textbf{High-mass stars ($M\;>\;8\;M_{\odot}$)}: the helium-core of such stars is ignited before they reach the RGB, which leads to the production of Fe, the strongest bound nucleus. Then, the stars no longer produce energy through fusion reactions and cannot hold up the gravitational forces any more. Eventually, electron captures on iron nuclei suppress the pressure support, with subsequent implosion and rebound leaving either a neutron star or a black hole, depending on the mass of the star (\autoref{fig:Stellar_evolution}, \textit{right}).

The explosion of massive stars creates one of  the brightest phenomenon in the Universe, known as  \textit{supernovae}. The huge energy ($\approx 10^{51}\; erg$) produced by the explosion is able to create a tremendous shock in the surrounding medium (\citealt{Nadyozhin_2008}). 1987A is the brightest supernova blast observed from earth in more than 400 years (\autoref{fig:shock_SNR}). The shock velocity ranges from 300 to 1700 km$\,$s$^{-1}$ (\citealt{Zhekov_2006}).

As we don't have the tools to describe such powerful shocks, we will not hereafter study the shocks in supernovae remnants.  
\begin{figure}
	\centering
    \includegraphics[width=0.8\linewidth]
    {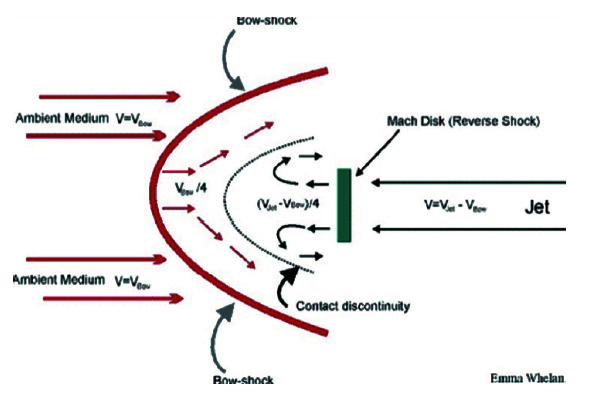}
	\caption[Jet-driven bow shock configuration]
	{Jet-driven bow shock configuration \citep{Gusdorf_2008}.}
	\label{fig:Jet_bowshock}
\end{figure}

\section{Jet-driven and stellar wind-driven bow shock models} 
\label{sec:bowshock_configuration}
\subsection{Jet-driven bow shock configuration}
The configuration of the jet-driven bow shock model is described in \autoref{fig:Jet_bowshock} that shows a strong supersonic jet propagating in the surrounding interstellar medium, and the interaction between the jet and the ambient medium creating a thin outflow around the jet. Ahead of the jet, two shocks are also created: a jet shock (or termination shock) and a bow shock (or ambient shock). The impacted gas in between the shocks has a high pressure and is ejected out, thus creating an outflow cavity around the jet. The properties of the jet-driven bow shock model are detailed in \cite{Arce_2007} and \cite{Gusdorf_2008}.
   
\begin{figure}
	\centering
    \includegraphics[width=0.7\linewidth]
    {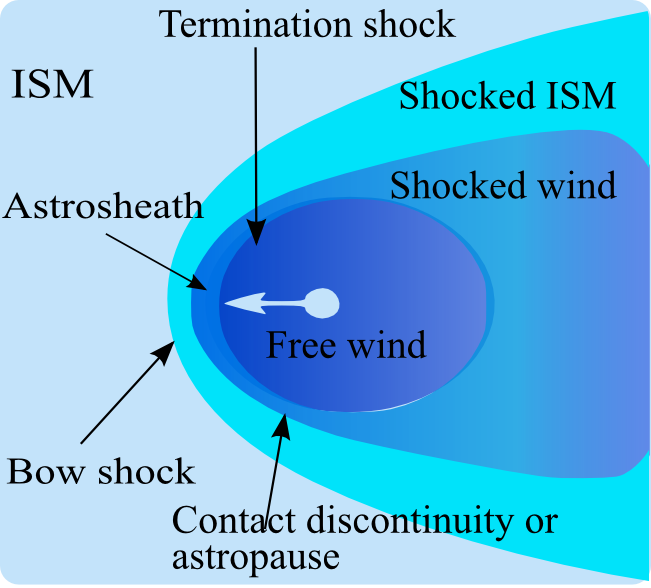}
	\caption[Stellar wind-driven bow shock configuration]
	{Stellar wind-driven bow shock configuration\\ (credit: N. Cox, KU Leuven).}
	\label{fig:wind_bowshock}
\end{figure}

\subsection{Stellar wind-driven bow shock configuration}
The stellar wind-driven bow shock model is described in \autoref{fig:wind_bowshock}. Basically, the physical process is similar to the case of the jet-driven bow shock model. The freely supersonic stellar wind sweeps up the surrounding interstellar materials, causing the development of an astrosphere. At the inner edge of the astrosphere, the free flowing stellar wind switches from supersonic to subsonic through the wind shock (or termination shock). The wind material in the astropause is separated from the interstellar matter by a contact discontinuity, where turbulent features form due to shear forces and density differences between the two fluids. If the speed of the star relative to the ambient medium is supersonic, a bow shock (or ambient shock) is formed at the outer edge of the astropause.

\section{Outline of the thesis} \label{sec:entire_structure}
The structure of this thesis follows the delineation of the bow shock model in \autoref{sec:bowshock_configuration}. It consists of three parts. First, we build a bow shock model with a three dimensional morphology characterizing the shocked ambient material in the ISM, and we compare it to observations. Second, we study the stellar wind-driven termination shock. From an observational point of view, both jets and stellar winds create a bright termination shock which can be studied, but the launch mechanism of the jet is unclear and debatable, while the mechanisms and initial conditions that generate the wind are well studied (e.g., \citealt{DeGreve_1997}, \citealt{LeBertre_1999}, \citealt{Gail_2013}) and the outcomes match well to observations. Furthermore, we are collaborating with observers who have studied stellar winds, such as \citet{Lebertre_2004}, \citet{Matthews_2013}, and \citet{Hoai_2017}, so we focus on winds rather than on jets. Third, we develop a spherical termination shock model, which physically and chemically couples the freely expanding stellar wind model (described in the second part) and the surrounding ISM.

\setstretch{1.1} 
\chapter{BOW SHOCKS} \label{chapter:bow_shock}
\lhead{\emph{INTRODUCTION}} 
\section{Molecular hydrogen: one of the best shock tracers}
\label{sec:shocked_H2}

Molecular hydrogen H$_{2}$, the most abundant molecule in the universe, naturally exists in shocked regions. Since molecular hydrogen is homonuclear, it has no dipole moment and the rovibrational transitions only occur by electric quadrupole radiation ($\Delta J = 0, \pm 2$). No dipole moment leads to the weakness of the quadrupole transitions. Consequently, the first observable rotational transition (J=2) state lies at 509 K (28.2 $\mu m$) above the ground state, while the first vibrational transition (v=1) approximately lies at 6330 K (2.2 $\mu m$) above the ground state. The full rovibrational levels of $H_{2}$ used in this thesis are given in \aref{app:tab_H2_excitation}. 

\begin{figure}
	\centering
    \includegraphics[width=0.8\linewidth]
    {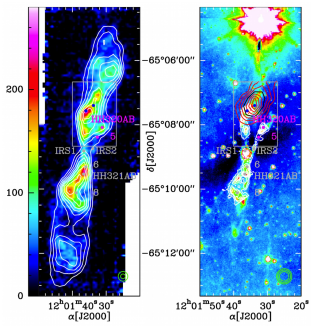}
   	\caption[BHR71 outflow]{The whole BHR71 outflow. 
   	(\textit{Left}) CO(6-5) in white contours associated with CO(3-2) in color, 
   	observed by APEX telescope. 
   	(\textit{Right}) 8 $\mu m$ emission in color with H$_{2}$ 0-0S(5) emission 
   	in white contour both detected by the Spitzer satellite \citep{G15}.}
	\label{fig:BHR71_outflow}
\end{figure}

Molecular hydrogen is a particularly important tracer, the mass fraction of which is important enough to determine the density of the gas. H$_{2}$ is one of the necessary element in order to help define the chemical state because it is at the origin of almost all chemical reaction chains that produce other molecules. In shocked regions, the temperature can rise up quickly and generate excitation both for rotational and vibrational levels as mentioned above. This makes H$_{2}$ a major coolant for the shocked gas. The strong emission of rovibrational lines, therefore, is a good tracer for the shock structure. In addition, because of rapid cooling, molecular hydrogen can be complemented by the excitation of other molecules such as CO, SiO, etc that have lower energy levels. As an example, \autoref{fig:BHR71_outflow} is the BHR71 outflow composed of several data sets \citep{Giannini04}. The structure of the outflow is mapped by CO on the left hand side and by H$_{2}$ on the right hand side. This figure visually indicates that CO and H$_{2}$ clearly probe the entire structure of the whole BHR71 outflow.  

\section{Excitation diagram} 
H$_{2}$ is one of the main tracers in shocked regions. In the following, we explain how to use it to deduce information on shocked regions. One effective way is to study the integrated intensity of rovibrational transitions to provide a good visualization of the physical conditions of the medium. This tool is known as the excitation diagram. The latter is a way to visualize the molecular hydrogen excitation state, by showing the logarithm of the column density of the excited rovibrational levels, divided by their statistical weight ($ln N_{vJ}/g_{vJ}$ with $N_{vJ}$ in cm$^{-2}$) against their excitation energy $E_{vJ} (K)$. Here, $vJ$ denotes the excited rovibrational levels and the statistical weight $g_{vJ} = I_{s}(2J+1)$, where the nuclear spin statistical weight $I_{s}$ equals 1 (even rotational level $J$), and 3 (odd rotational level $J$).

The column density N$_{vJ}$ of a rovibrational level of H$_{2}$ is deduced from its line intensity I$_{vJ}$ through the spontaneous probability of deexcitation given by the Einstein coefficient A$_{vJ}$. If one assumes that a given line of H$_{2}$ emission is optically thin (which is usually the case given the very smal values of $A_{vj}$) the column density is then calculated by
\begin{equation}
	N_{vJ} = \frac{4\pi}{h c} \frac{\lambda_{vJ}}{A_{vJ}}I_{vJ}
\end{equation}
where $\lambda_{vJ}$ is the central wavelength of the line transition, $h=6.626\, 10^{-27}\,$erg$\,$s is the Planck constant and $c=2.998\, 10^{10}\,$cm$\,$s$^{-1}$ is the speed of light in vacuum. If the gas is thermally excited at temperature $T_{ex}$, the column density $N_{vJ}$ is proportional to the product between the statistical weight $g_{vJ}$ and the Boltzmann factor $e^{-\frac{E_{vJ}}{k_{B}T_{ex}}}$. If $T_{ex}$ is constant, ln$N_{vJ}/g_{vJ}$ and $E_{vJ}$ should be proportional with a slope equal to $T^{-1}_{ex}$. Therefore, this diagram allows us to roughly estimate the excitation temperature. In the situation of local thermal equilibrium, the excitation temperature $T_{ex}$ is equal to the gas temperature. Hereafter, we introduce how to use the H$_{2}$ excitation diagram to interpret observations. 

%

\section{Single shock model to interpret observations} \label{sec:1D_model_limitation}
H$_{2}$ emissions from  pure shocked regions are the most interesting targets to study shock properties and to test shock models. Some shocks have been studied extensively, such as in the Orion Molecular Cloud - Peak1 (hereafter OMC-1 Peak1) \citep{Rosenthal00} (the brightest source of H$_{2}$ emission in the sky) and the BHR71 outflows \citep{G15}. 

\begin{figure}
	\centering
    \includegraphics[width=0.8\linewidth]
    {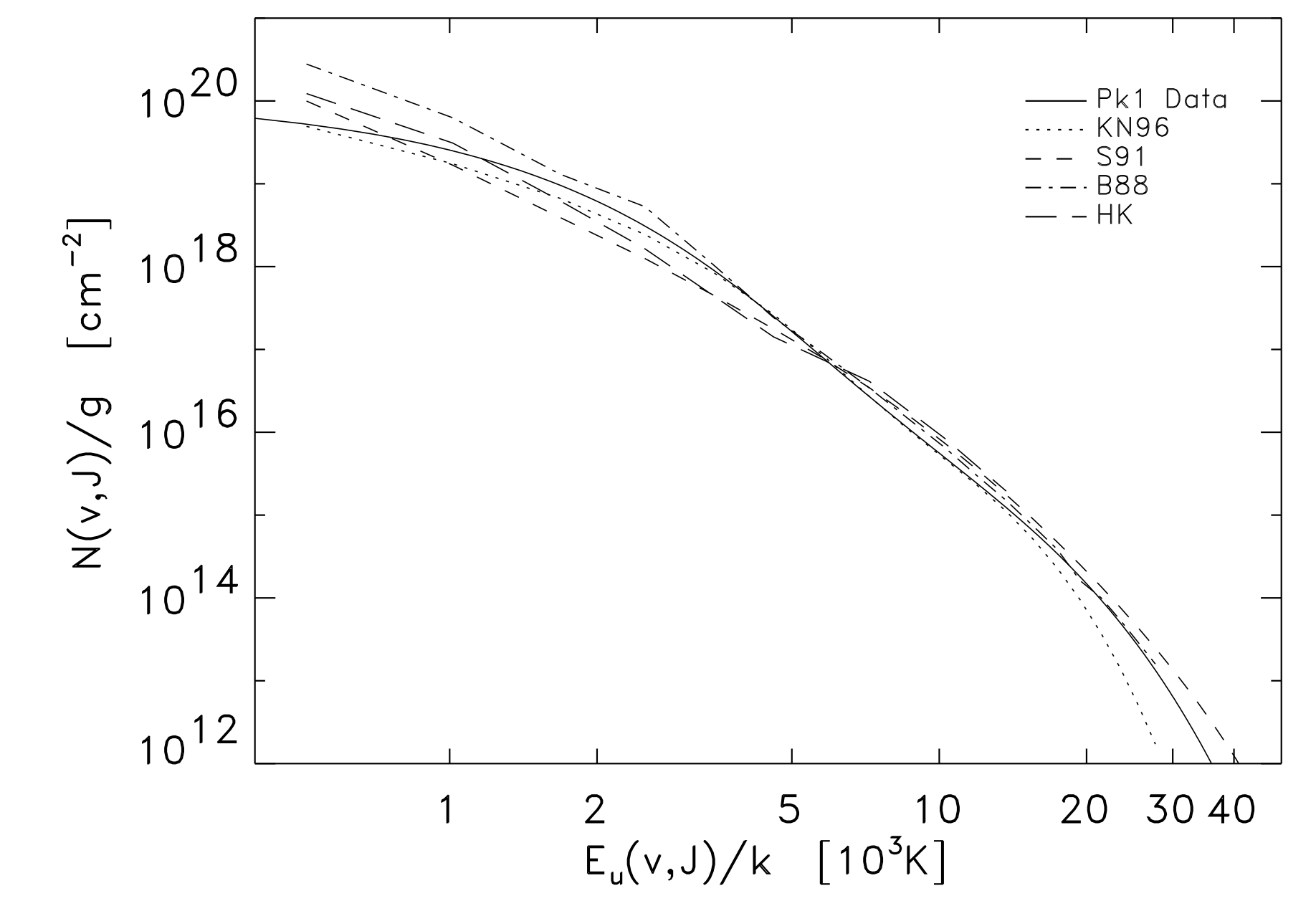}
   	\caption[Fit of single shock models to OMC-1 Peak1 shock]
   	{Fit of single shock models to OMC-1 Peak1 shock \citep{Rosenthal00}, 
   	except the S91 (bow shock model) \citep{S91}.}
	\label{fig:OMC_Rosenthal_fit}
\end{figure}

 H$_{2}$ emission from the OMC-1 Peak1 is suggested to arise from shocks (\citealt{Gautier_1976}, \citealt{B89}). Over decades, single shock models have been investigated to answer the question of the physical nature of shocks, and the calculated excitation diagram has been widely used to fit to the observable shocks. However, \citet{Rosenthal00} came to the conclusion that such models cannot fit the low and high excitation population levels simultaneously as shown in \autoref{fig:OMC_Rosenthal_fit}. A combination of two single planar C-shocks \citep{KN96} provides a good fit of the low excitation population levels corresponding to $v$ = 0, $J$ = 3 to 9, while it overestimates populations of higher levels. On the contrary, a single planar J-shock model \citep{B88} can match the medium and high excitation population levels, although it overestimates the population of lower levels. To conclude, no single \textit{stationary} planar shock model can reproduce the observed H$_{2}$ level populations for the OMC-1 Peak1. Hence, a combination of at least two different shock models, one for the low excitation level populations and one for the higher excitation levels, may be required. 

\begin{figure}
	\centering
    \includegraphics[width=0.8\linewidth]
    {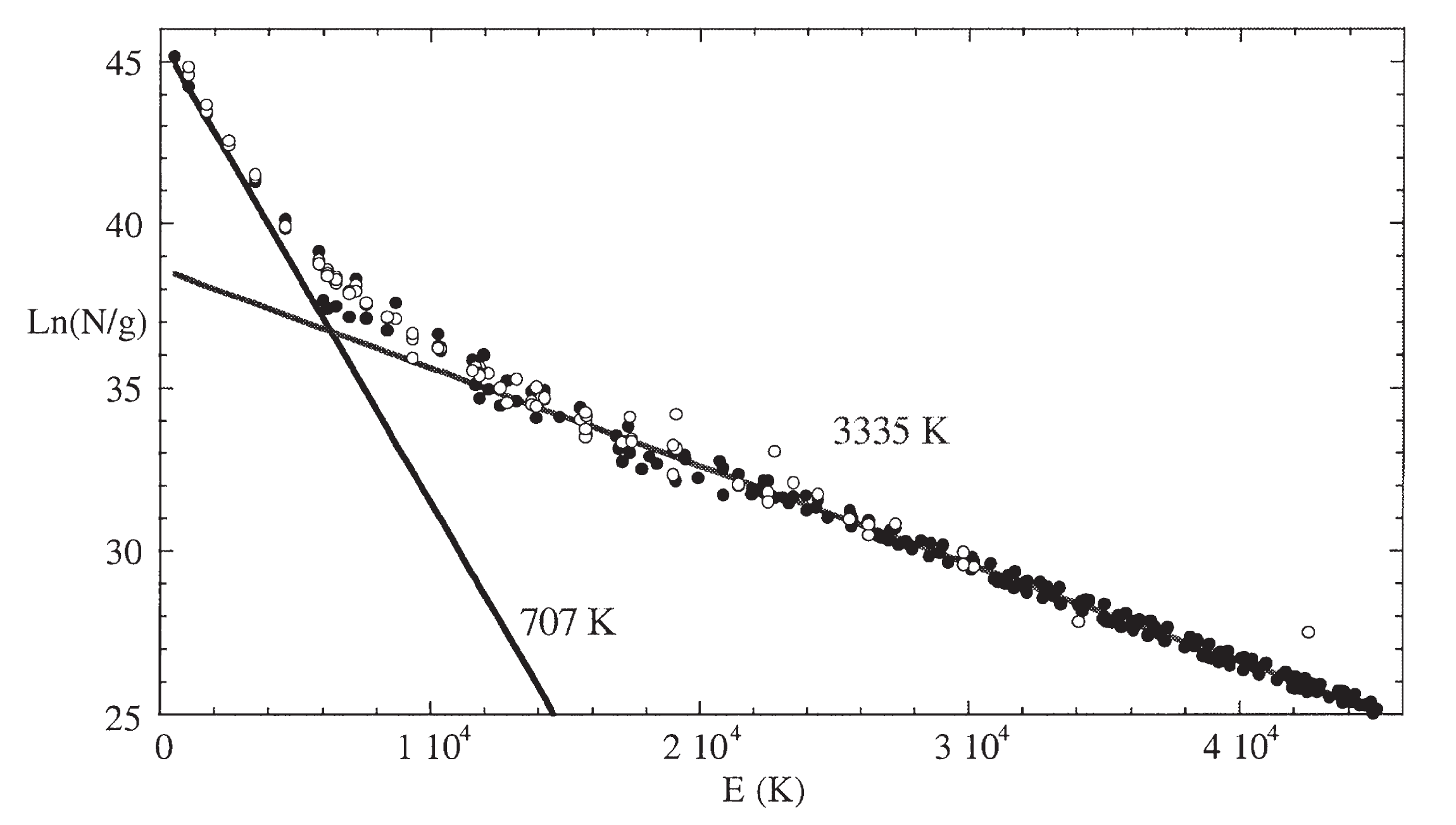}
   	\caption[Double shock models fits to OMC-1 Peak1 shock]
   	{Double shock models fits to OMC-1 Peak1 shock. 
   	The observational data is symboled by the empty circles, 
   	and the best diagrams is symboled by the black circles 
   	and plotted in solid lines \citep{Bourlot02}.}
	\label{fig:OMC_Bourlot_fit}
\end{figure}

\citet{Bourlot02} indicate that a two-component shock model including two planar shocks with different speeds, magnetized media and initial abundances of H can match well the observed H$_{2}$ from the OMC-1 Peak1 extending upto the rotational level $v=0,\ J=27$, which corresponds to an excitation energy of 42515 K (\autoref{fig:OMC_Bourlot_fit}). Specifically, the model with $v_{s}=40$ km$\,$s$^{-1}$, $B_{0}=400\,\mu$G, $n(H)/n(H_{2})=7.4\,10^{-4}$ fits well the lower excitation level populations $v=0,\ J < 7$ and the higher level populations are in good agreement with the model characterized by $v_{s}= 60\,$km$\,$s$^{-1}$, $B_{0}=100\,\mu$G, $n(H)/n(H_{2})= 0.5$. Despite the good fit, the origin of the difference between the two compounded shocks and why they should be linked remain unclear, as well as the properties of the ambient gas. Furthermore, the retrieved pre-shock density ($10^{4}$ cm$^{-3}$), corresponding to the best fit is lower by 2 orders of magnitude than the value ($\sim 10^{6}$ cm$^{-3}$) derived by (\citealt{Draine_1982}, \citealt{White86}, \citealt{B88}, \citealt{HM89}, \citealt{KN96}, \citealt{Kristensen08}). 

\begin{figure}
	\centering
    \includegraphics[width=0.9\linewidth]
    {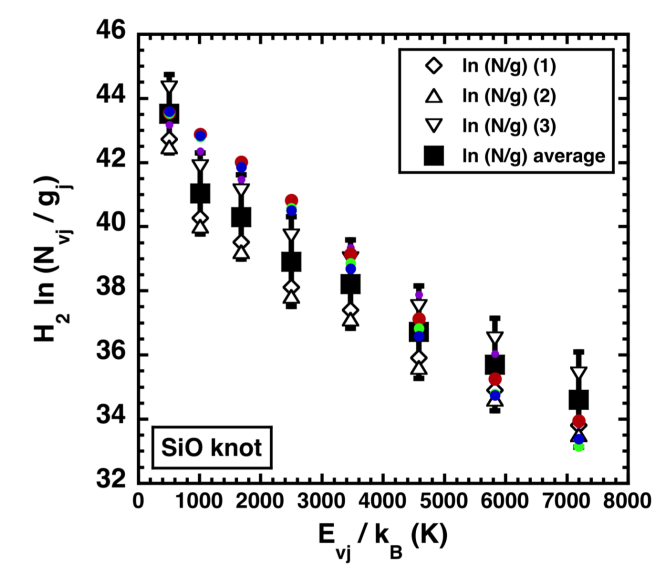}
   	\caption[Single shock models fits to BHR-71 shock]
   	{Single planar shock models fits to BHR-71 shock. 
   	The black symbols denote the observational data, 
   	the red circles indicate the best fit \citep{G15}.}
	\label{fig:BHR_single_fit}
\end{figure}

Can a \textit{non-stationary} planar shock model match better the observations? To constrain the physical conditions of the shocked gas from the BHR71 outflow, \citet{G15} calculate the pure low rotational H$_{2}$ excitation diagram for 1200 models \citep{F03}, comprising both stationary shocks and non-stationary shocks and then they compare them with diagram observed from the outflow. These authors figured out that the best fit is a non-stationary shock, and they estimated its age (\autoref{fig:BHR_single_fit}). However, the best diagram is different from that of the observed diagram: it falls down and crosses the observed one. That means that the non-stationary planar model overestimates the excited H$_{2}$ column densities of the BHR71 outflow for levels at excitation energy less than that of the intersection point, otherwise it underestimates for the rest.      

\section{Bow shock models to observations}
\label{sec:bowshock_promissing}
\begin{figure}
	\centering
    \includegraphics[width=0.8\linewidth]
    {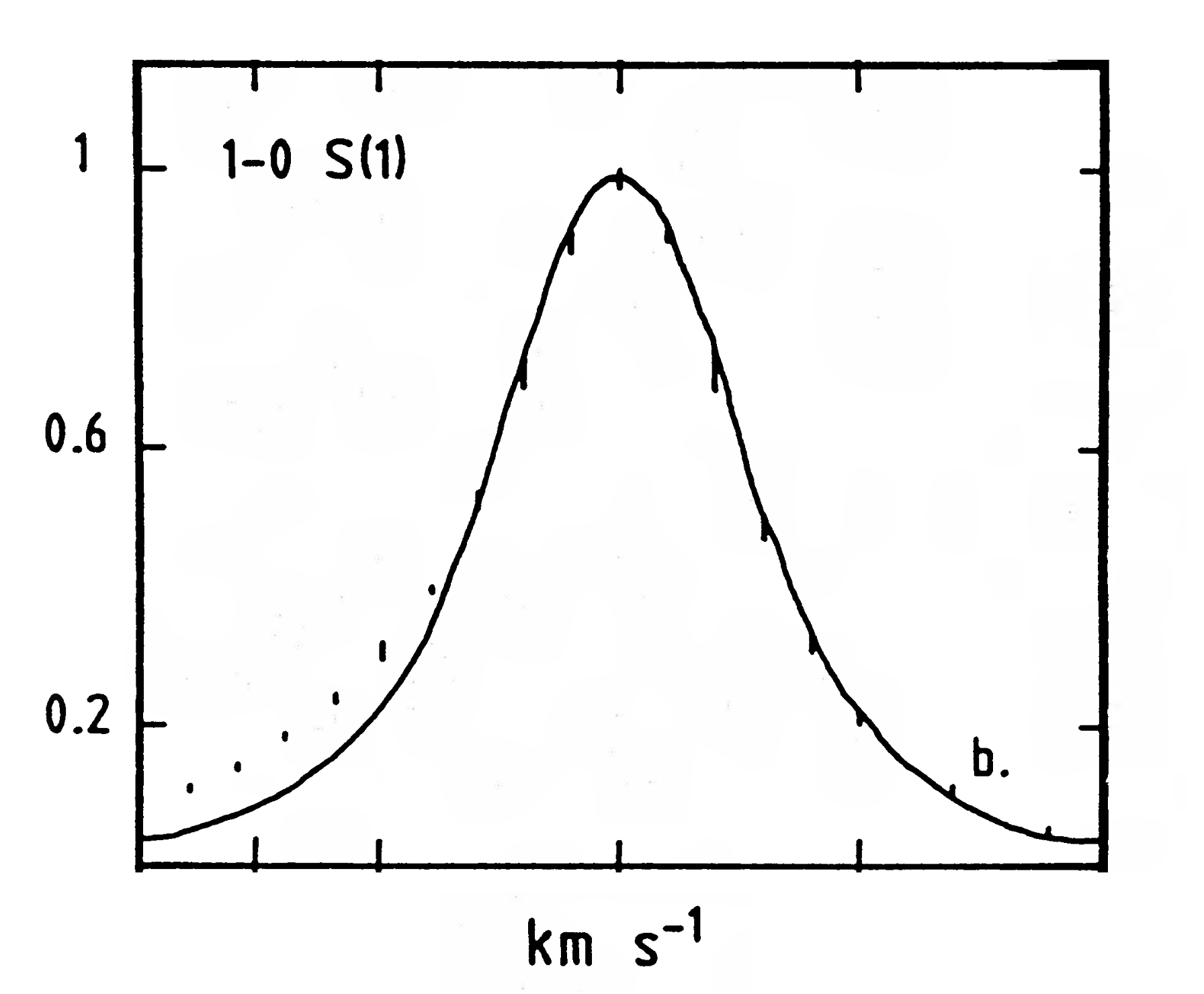}
   	\caption[Bow shock models fits to OMC-1 Peak1 shock]
   	{Bow shock models fits to OMC-1 Peak1 shock. 
   	The points denote the observational data, 
   	the solid line is the best fit \citep{S91_Bfield}.}
	\label{fig:OMC_bow_fit}
\end{figure}

To go beyond the discussion in \autoref{sec:1D_model_limitation}, it is fair to examine more complex shock models with a higher number of spatial dimensions. One solution is to run 2D or 3D numerical simulations, but they have been so far limited to single-fluid "jump" bow shocks, J-type (e.g., \citealt{Suttner97}, \citealt{Raga02}). 
Up to now  multidimensional bow shocks with "continuous" C-type shocks, where ion-neutral decoupling occurs in a magnetic precursor \citep{Draine93}, have not been modeled. However, orthogonal and oblique planar shocks have been treated in simulations by \citet{Mac_Low_1995}, \citet{Toth_1995}, and \citet{Stone_1997}. Such a situation is encountered in the bow shock whenever the entrance speed drops below the magnetosonic speed in the charged fluid. To address this case, one can predict H$_2$ emission from bow shocks by prescribing a bow shape and treat each surface element as an independent 1D plane-parallel J-type or C-type shock, assuming that the emission zone remains small with respect to the local curvature. This approach was first proposed by \citet{S90} and \citet{S91} who used simplified equations only for the 1D C-shock structure and cooling. In the same way, \cite{S91_Bfield} reproduced the line profile of H$_{2}$ emission 
from OMC-1 Peak1, observed by \citet{Moorhouse_1990} (\autoref{fig:OMC_bow_fit}). However, this model requires an extremely high magnetic field ($>$50 mG), when independent measurements show that it should range from 3 mG \citep{Norris_84} to 10 mG \citep{Chrysostomou_94} in the same region.  

The validity of this approach was actually recently investigated by \citet{Kristensen08} and \citet{Gustafsson10} who used refined 1D steady-state shock models from \citet{FP03} that solve the full set of magneto-hydrodynamical equations with non-equilibrium chemistry, ionization, and cooling.

\citet{Kristensen08} studied high angular resolution H$_2$ images of a bow shock in the Orion BN-KL outflow region, performing 
several 1D cuts orthogonal to the bow trace in the plane of the sky. They 
fitted each cut separately with 1D
steady shock models. They found that the resolved width, combined with the peak brightness 
required C-shocks, and that the variation of the fitted shock velocity and the transverse magnetic field 
along the bow surface was consistent with a steady bow shock propagating in a uniform medium. This result provided some validation for the
"local 1D-shock approximation" when modeling H$_2$ emission in bow shocks, at least for this parameter regime. Following this idea,
\citet{Gustafsson10}  built 3D stationary bow shock models
by stitching together 1D shock models. 
Then they projected them to produce maps of the H$_2$ emission in several lines that they compared to
observations. They obtained better results than
\citet{Kristensen08} thanks to the ability of the 3D model to account
both for the inclination of the shock surface, with respect to the line of sight, and the multiple shocks included in the depth of their 1D cuts. 
The width of the emission maps was better reproduced. The best fit density, bow shock inclination and ambient magnetic field all agreed with independent constraints.

\section{Power-law statistical equilibrium assumption}
\label{sec:power_law_assumption}
\begin{figure}
	\centering
    \includegraphics[width=0.8\linewidth]
    {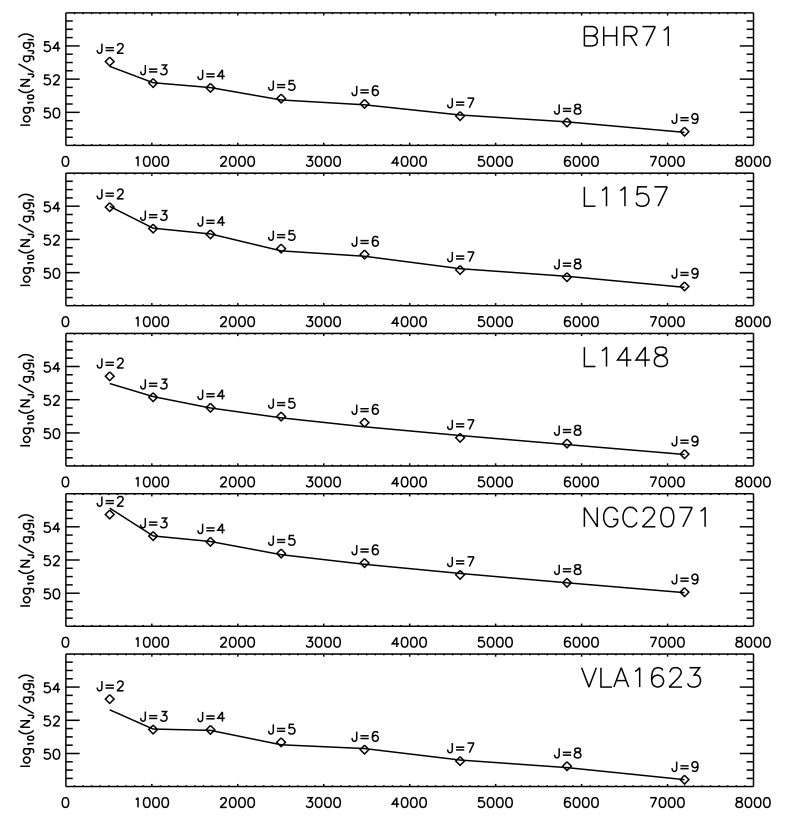}
   	\caption[Best fitting of the power-law statistical equilibrium assumption]
   	{Best fitting of the power-law statistical equilibrium assumption.   
   	   	Diamonds indicate the observed values, and the solid lines 
   	    are the best fitting procedure by the power-law statistical 
   	    equilibrium assumption from \cite{Neufeld08} (\autoref{eq:power_law}). 
   	    In this studies, the values of the power-index is in the range 2.3-3.3 \citep{Neufeld09}.}
	\label{fig:equilibrium_1D}
\end{figure}

\citet{Neufeld08} (hereafter NY08) and \citet{Neufeld09, Neufeld14} came up with a simple model assuming statistical equilibrium for a power-law temperature distribution $T^{-b}dT$. The corresponding column density of gas at temperature between T and dT is 
\begin{equation} \label{eq:power_law}
	dN = a T^{-b} dT
\end{equation}
with $a, b$ adjustable parameters. The temperature ranges between 100 K and 4000 K. This assumption turns out to be very effective at reproducing the pure rotational lines of H$_{2}$ (\autoref{fig:equilibrium_1D}). To interpret their results, these authors proposed the effect of the three-dimensional bow shock geometry. Owing to an accumulation of the bow shock surface, the mass of material crossing the working surface dA with velocity from v to v + dv perpendicular to the shock surface should be
\begin{equation} \label{eq:dM}
	dM \propto N(H_{2}) dA \mbox{.}
\end{equation}

For a parabolic shape of shock, \citet{SB90} showed that dA $\propto$ v$^{-4}$ 
($\forall$ v $<$ v$_{bow}$ terminal velocity at the head of bow shock). \citet{Neufeld06} found that the column density of H$_{2}$ was proportional to velocity as v$^{-0.75}$ and the velocity was related to temperature as T$^{1/1.35}$ for a single C-shock. Combining all of those relations, \autoref{eq:dM} yields
\begin{equation}
	dM \propto v^{-0.75} v^{-4} dv \propto T^{-3.77} dT \mbox{.}
\end{equation} 

Therefore, in the specific case of a parabolic shock shape, the power-index is expected to be b=3.77.

\section{Aims and outline}

Several models have been designed to reproduce the properties of bow shocks, most of them are one-dimensional (see \autoref{sec:1D_model_limitation}). In order to better match the observations, we will investigate shock models with more complex geometries. Based on the method of \cite{Kristensen08} and \cite{Gustafsson10}, we have built a 3D shock model made of 1D shock models stitched together. To extend the scope of the works of \citet{Gustafsson10}, we provide a general way to encode the 3D geometry of a bow-shock as a distribution of shock models. In addition, we consider the effect of young shock ages, where the shock is not stationary, and we investigate thoroughly the impact of various shock characteristics on the excitation diagram and line profiles integrated over the bow of the molecular hydrogen. Then we compare our 3D bow shock model with observations. The best fit provides us with constraints on some physical parameters of the bow shock.

We structure this part as below:  

\begin{itemize}
	\item Chapter 4: we recall the principles of the 1D Paris-Durham shock model (\citealt{F03,FP15}) and we introduce the physical and chemical input parameters.
	\item Chapter 5: we describe how to build the 3D bow shock by stitching several 1D Paris-Durham shock models.
	\item Chapter 6: we describe the procedure to fit the 3D bow shock model to the observations.
	\item Chapter 7: we summarize the achievements of our model and we sketch the prospects for future improvements and applications.
\end{itemize}

Chapters 5 and 6 follow very closely \citet{Tram_2018}, with only a few additions.

\setstretch{1.1} 
\chapter{WIND AND TERMINATION SHOCKS}

\lhead{\emph{Introduction}} 
Beside bow shocks occurring in the ambient material (\autoref{chapter:bow_shock}), a termination shock also forms at the head of jet outflows and in the bulk of the stellar wind surrounding the stars. As mentioned in \autoref{sec:entire_structure}, our study is focused on the termination shocks around Asymptotic Giant Branch (AGB) stars. In this chapter, we introduce the characters of  AGB star winds and their interaction with the ISM.

\section{Stellar winds from AGB stars}
\label{sec:stellar_wind}
As described earlier, low- and intermediate-mass stars $(1M_{\odot} \leq \;M_{\ast}\;\leq 8M_{\odot})$ reach the AGB phase, which is the last stage of their evolution before they become a white dwarf. During this phase, the star has lost most of its material throughout mass loss mechanisms. Material can be lost only when its flow exceeds the star's gravity. In the absence of a pressure gradient, for example when it has accelerated and its speed exceeded the escape speed, there is no turning back.

\begin{figure}
	\centering
    \includegraphics[width=0.9\linewidth]
    {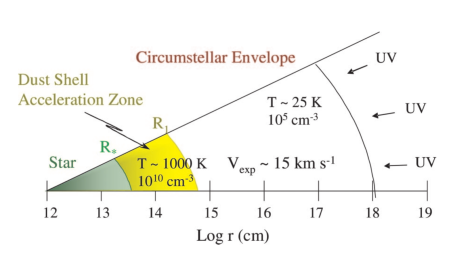}
   	\caption[Schematic physical structure of the CSE around a AGB star]
   	{Schematic physical structure of the CSE around an AGB star (\citealt{Ziurys_2006}).}
	\label{fig:CSE_cross_physics}
\end{figure}

While the flow remains subsonic, several mechanisms for initiating winds close to the star have been suggested: gradient of gas pressure (thermal wind), acceleration through waves (sound wave, Alfvén wave), or pulsations. Pulsations are currently the dominant paradigm 
(e.g., \citealt{Hoefner_1997}, \citealt{Willson_2000}) to lift up materials from the stellar surface into cooler regions (dust shell acceleration zone in \autoref{fig:CSE_cross_physics}), where molecules and dust grains can form. The latter scatter and absorb the stellar photons, which leads to a net force pushing them away from the star. Then they move through the gas and transfer momentum to gas molecules due to collisions. \citet{Tielens_1983} and \citet{Krueger_1994} found that the dust grains always move with their equilibrium drift velocity with respect to the gas, which is of the order of the isothermal sound speed or higher. Therefore, while the grains are not position-coupled to the gas, they are momentum-coupled to the gas.
Those collisions produce a drag force (\citealt{Gilman_1972}), which acts as an additional force sufficient for the gas to overcome the gravitational well of the star. In this case, the wind is called a \textit{dust-driven wind} or a \textit{radiation-driven wind}.  

\section{Circumstellar envelopes around AGB stars}
\label{sec:CSE}
Mass-loss from stars builds up an expanding circumstellar envelope (CSE) 
around the star, containing dust and gas. The mass-loss mechanism affects the geometry 
of the CSE. Most of the time, the CSE is not observed as a spherical symmetric or a homogeneous envelope, which hints that the mass-loss is not an isotropic process. 

Circumstellar envelopes of AGB stars can be considered as the most 
significant chemical laboratories in the universe (\autoref{tab:species_observed} and \autoref{fig:observed_species}). The effective temperature of those stars is usually low ($T_{\ast} \approx$ 2000 K - 3500 K) (comes from Infrared observations), and the timescale of 
the mass-loss is long, so that molecules and dust can form in the envelope through chemical and physical processes. Then they are blown into the interstellar medium. 
This material can dominate about 80$\%$ of the ISM by mass \citep{Jorgensen_1994}. 

\begin{table}
\centering
\begin{tabular}{cccccc}
\toprule 
    \multicolumn{2}{c}{Carbon-rich star} & \multicolumn{1}{c}{} &  
    \multicolumn{1}{c}{} &\multicolumn{2}{c}{Oxygen-rich star} \\
    \midrule
	CO       & SiC$_{2}$ & & & & CO          \\
	SiO      & CCH	     & & & & SiO         \\
	SiS		 & NaCN		 & & & & SiS         \\
	CS    	 & l-C$_{3}$H& &	& &	CS			\\
	CN    	 & c-C$_{3}$H& & & &	CN			\\
	HCN   	 & H$_{2}$C0 & & & &	HCN			\\
	HNC   	 & H$_{2}$CS & &	& & HNC			\\
	NaCl  	 & HC$_{3}$N & &	& & NaCl		\\
	PN    	 & C$_{4}$H	 & & & &	PN			\\
	HCO$_{+}$ &	CH$_{3}$CN&	& &	& HCO$_{+}$	\\
	PH$_{3}$  &	CH$_{3}$CCH&	& &	& NS		\\
	CH$_{2}$NH & Unidentified		   & & &	& PO		\\
	CP      &		& & &	& AlO			    \\
	SiC     &		& & &	& AlOH		        \\
	AlCl	&		& & &	& SO			    \\
	KCl		&		& & &	& H$_{2}$O	        \\
	AlF		&		& & &	& SiO			    \\
	SiN		&		& & &	& H$_{2}$S	        \\
	HCP		&		& & &	& Unidentified			        \\
    \bottomrule
\end{tabular}
\caption[Chemical species observed toward IRC +10216 and VY CMA by ARO SMT observation]
{Chemical species observed by the \textit{Submillimeter Telescope} (SMT) 
of the \textit{Arizona Radio Observatory} (ARO) toward IRC +10216 and VY CMa in \autoref{fig:observed_species} \citep{Tenenbaum_2010}.}
\label{tab:species_observed}
\end{table}

	\subsection{Circumstellar gas molecules}
		The origin of the circumstellar gas lies inside the stellar core 
	through its evolution stages (see \autoref{sec:shocks_vs_stellar_evolution}).
	Briefly, \isotope[12]{C}, \isotope[14]{N} and \isotope[16]{O} 
	are produced through the fusion of helium and alpha process. 
	The dredged-up processes then bring those nuclear products up to the stellar surface.
	   	   	   	   	   		
\begin{figure}
		\centering
    	\includegraphics[width=0.8\linewidth]
    	{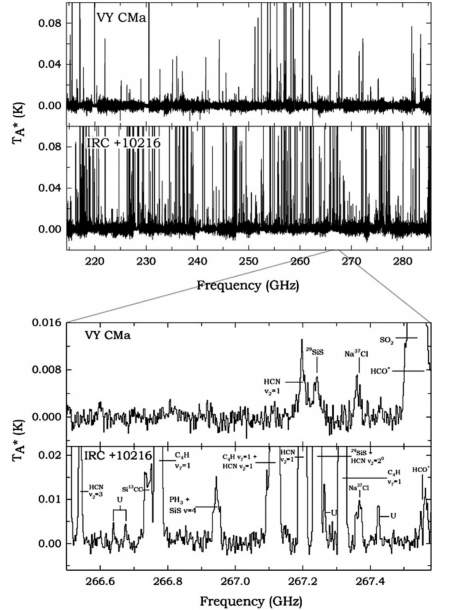}
   		\caption[Chemical species observed by the \textit{ARO SMT} toward IRC +10216 
   		and CY CMa]
   		{Chemical species observed by the \textit{Submillimeter Telescope (SMT)} 
   		of the \textit{Arizona Radio Observatory (ARO)} toward IRC +10216 and CY CMa. 
   		(\textit{Upper panel}) Complete spectra of the \textit{ARO SMT} survey 
   		in 214.5-285 GHz.
   		(\textit{Lower panel}) Detailed 1 GHz selection centered at 267 GHz of 
   		the survey \citep{Tenenbaum_2010}.}
		\label{fig:observed_species}
	\end{figure}

	 Most of the known circumstellar gas molecules are detected in 
	 carbon-rich stars 
	(mostly only in IRC +10216). 
	Observations toward IRC +10216, have detected about 71 chemical components 
	in its CSE (e.g., \citealt{Cernicharo_2000}, \citealt{He_2008}). 
	However, despite speculations that the oxygen-rich CSEs 
	are less chemically diverse \citep{Olofsson_2005}, 
	recent observations of VY CMa star 
	\citep{Tenenbaum_2010} demonstrate that oxygen-rich stars 
	are also chemically complex: about 32 different chemical species have been identified 
	in their CSEs. \autoref{fig:observed_species} 
	shows the spectral line survey 
	 of the \textit{Submillimeter Telescope} (SMT) 
	of the \textit{Arizona Radio Observatory} (ARO) toward the carbon-rich star 
	(IRC +10216) and oxygen-rich star (VY CMa). The names of  detected species
	are listed in \autoref{tab:species_observed}.

		Chemical models have been created to explain the formation mechanisms of those species in order to understand the 
	chemical processes in the ISM 
	(e.g., \citealt{WC_1998}, \citealt{Agundez_2006}, \citealt{Cherchneff_2006}, \citealt{Decin_2010}, \citealt{Li_2016}).  
	Based on these studies, the authors demonstrate that	
	the temperature and density in the inner envelope (r$\leq$5R$_{\ast}$), 
	although high, does not satisfy
	 the thermal equilibrium conditions as a result of shock propagation, and chemistry is also out of equilibrium. 
	"In any case, molecular abundances derived from TE calculations 
	should not be used in the interpretation of observational data which 
	are not of the photosphere" \citep{Cherchneff_2006}. 
	
		\begin{figure}
			\centering
    		\includegraphics[width=0.9\linewidth]
    		{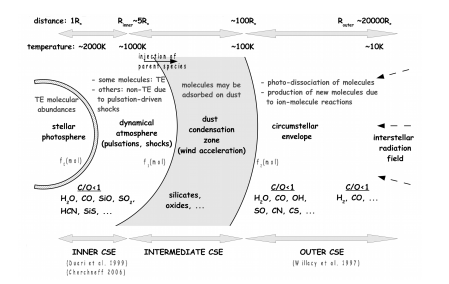}
   			\caption[Schematic chemical structure of the CSE of an Oxygen-rich AGB star]
   			{Schematic chemical structure of the CSE of an oxygen-rich AGB star 
   			\citep{Decin_2010}.}
			\label{fig:CSE_cross_chemistry}
		\end{figure}
		
	 	The chemistry strongly depends on the radius and remarkably varies through the
	 circumstellar envelope as indicated in \autoref{fig:CSE_cross_chemistry}.   
	In the inner region, the inner shocks trigger the formation of molecules and dust. 
	Those molecules are called \textit{"parent"} abundances.
	As molecules flow outward to the outer envelope, the 
	\textit{"parent"} abundances freeze out, and photons, cosmic rays and interstellar 
	radiation field initiate new types of chemical processes, 
	such as ion-molecule, photo-dissociation/ionization reactions 
	that create new molecules
	(e.g., \citealt{Millar_2000}, \citealt{Decin_2010}, \citealt{Li_2016}). 
	Those newly formed molecules are 
	called \textit{"daughter"}. 
	Beyond $\sim$ 1000 AU, the photo-dissociation by the interstellar radiation field 
	is so strong that the gas 
	molecules cannot subsist (\autoref{fig:CSE_cross_physics}). The dissociation radius is different for each molecule depending on the efficiency of its screening to photo-dissociation,
	and it also depends on the mass-loss rate and expansion velocity. 

	\subsection{Circumstellar dust}
	\label{sec:CSE_dust}
	Beside molecules, the circumstellar envelope is made of a various 
	circumstellar dust particles and is identified 
	by their properties. 
	The dust is thought to be formed by a mechanism of 
	gas-phase molecule condensation \citep{Kwok_2004} 
	during the expansion of the CSE (\autoref{fig:CSE_cross_physics}). 
	The conditions for dust 
	formation are low temperature (to allow for condensation) and 
	high density (to allow for sufficient interaction rate). Typical 
	condensation radii of dust range from 5 to 10 stellar radii, corresponding to a 
	temperature varying from 1000 K down to 600 K and a total number density varying 
	from $10^{10}$ to $10^{8}\,$cm$^{-3}$. 
	
	Since oxygen and silicon are amongst the most abundant molecules in the universe, 
	silicates are believed to be reasonably common 
	in the CSE of AGB stars. Most of the identified silicates are amorphous, 
	which satisfies the expectation of rapid formation of amorphous material 
	in gas-phase environment. Some materials, in particular, have high 
	condensation temperatures and can condense at $\sim$2 photospheric radii and act like 
	seed particles for further grain growth \citep{Lorentz-Martins_2000}.
	
	Dust grains are classified by their spectral features 
	and they correspond to a special kind of envelope properties. 
	Amorphous silicates, identified at 9.7 and 18 $\mu m$, 
	have been detected in more than 4000 oxygen-rich stars \citep{Kwok_1997}, 
	thus they are considered as a major feature of oxygen-rich stars. 
	In addition, \citet{Jaeger_1998_IV} found clear evidence for 
	the existence of crystalline silicates in the spectra measured by the 
	\textit{Short Wavelength Spectrometer} (SWS) of the 
	\textit{Infrared Space Observatory} (ISO). 
	The crystalline silicates are found in two forms: 
	olivine (Mg$_{2y}$Fe$_{2-2y}$SiO$_{4}$) and pyroxene (Mg$_{x}$Fe$_{1-x}$SiO$_{3}$) 
	\citep{Dorschner_1995}.
	\citet{Jaeger_1998} also point out that crystalline silicate in the CSE 
	of the oxygen-rich stars is magnesium-rich, which means that x, y are close to unity. 
	However, the abundance of the crystalline form is smaller than that of the amorphous form 
	\citep{Kwok_2004}.   
	
	Silicate carbide, which has a 11.3 $\mu m$-feature 
	is the most common dust grain condensed in the CSE of carbon-rich stars. 
	It has been detected in over 700 carbon-rich stars
	\citep{Kwok_1997}. In more evolved carbon-stars (the abundance of C is much larger than O), 
	the silicon carbide, however, becomes weaker and the amorphous carbon 
	increasingly dominates \citep{Kwok_2004}. 
	In addition to silicon carbide and amorphous carbon, 
	the \textit{Infrared Astronomical Satellite} (IRAS) observations with 
	\textit{Low Resolution Spectrometer} (LRS) toward carbon-rich stars 
	find an evidence for 21 $\mu m$ emission \citep{Kwok_1989}. 
	The solid-state structure of this strong emission is uncertain. 
	Some possible candidates have been proposed, 
	such as large polycyclic hydrocarbon (PAH) ($>$ 100 C atoms) cluster, 
	hydrogenated amorphous carbon (HAC) grain \citep{Buss_1990}, 
	nanodiamonds \citep{Hill_1998}, etc.       
	
\begin{figure}
		\centering
    	\includegraphics[width=1.\linewidth]
    	{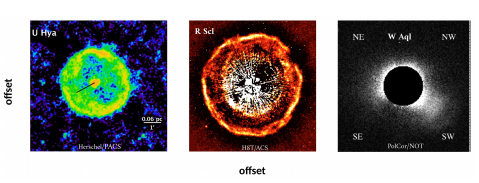}
   		\caption[Observation of  dust shells around  AGB stars]
   		{Observation of dust shells around AGB stars.  
   		(\textit{left}) IR emission of a dust shell around \textit{U Hya} star, 
   		observed by the \textit{Herschel} \citep{Cox_2012}.
   		(\textit{Middle}) Scattered light by a dust shell around \textit{R Scl} star, 
   		observed by the \textit{Hubble Space Telescope} \citep{Olofsson_2010}.
   		(\textit{Right}) Polarized light by dust shell around \textit{W Aql} star, 
   		observed by the \textit{Nordic Optical Telescope} \citep{Ramstedt_2011}.}
		\label{fig:dust_observation}
	\end{figure}

	Dust grains are opaque, and scatter the stellar light. 
	Therefore, the size of the condensation dust shell can be 
	determined by: (i) IR emission since stellar photons heat up grains, 
	which then produce  IR radiation by cooling
	(\autoref{fig:dust_observation}, \textit{left panel}), 
	(ii) the scattered light (\autoref{fig:dust_observation}, \textit{central panel}), 
	and (iii) polarized light since light becomes polarized 
	when it is scattered by grain particles 
	(\autoref{fig:dust_observation}, \textit{right panel}).


\section{Interaction with the ISM}
\label{sec:hydrogen}
As it reaches the ISM, the stellar wind interacts with it and sweeps up the surrounding materials. Thanks to infrared observations from \textit{Herschel}, \citet{Cox_2012} showed different kinds of morphology of the interaction modes. The hydrodynamic mechanisms are well studied 
(e.g., \citealt{Cox_2012}, \citealt{Villaver_2012}). 

As described in \autoref{sec:shocked_H2}, hydrogen is the best tracer for the interaction between the stellar wind and the ISM. Studies of HI 21 cm emission (e.g., \citealt{Gerard_2006, Gerard_2011}, \citealt{Libert_2007, Libert_2008, Libert_2010a, Libert_2010b}, \citealt{Matthews_2007, Matthews_2008, Matthews_2011, Matthews_2013}) conclude that neutral hydrogen is a good tracer of the extended CSEs. Since in the absence of strong UV, hydrogen is not easily ionized, 
its emission can therefore trace the very large scales of CSEs, larger than CO, which is easily dissociated by the \textit{interstellar radiation field} (ISRF) 
at a distance of $\sim$ 10$^{17}$ cm from the stars.    


Although part of the hydrogen is locked into non-linear molecules, such as H$_{2}$O, most of it is in either atomic or molecular form \citep{Gerard_2003}. The fractional ratio between atomic and molecular hydrogen in the CSEs has been discussed by \citet{Glassgold_1983}. For  \textit{"high"} stellar effective temperature (T$_{eff} >$ 2500 K), hydrogen should be mainly in atomic form. In contrast, for stars with \textit{"low"} effective temperature (T$_{eff} \leq$ 2500 K) it should be in molecular form in the upper atmosphere and in the inner CSE. 
This hypothesis seems to be confirmed by the detection of a 21 cm emission line in CSEs of \textit{"hot"} AGB stars, such as Mira \citep{Bowers_1988}, 
RS CnC \citep{Gerard_2003}, 
EP Aqr \citep{Lebertre_2004}, 
Xher (\citealt{Gardan_2006}, \citealt{Matthews_2011}), 
Y CVn \citep{Lebertre_2004}, and the detection of FUV emission from the \textit{"cold"} AGB star IRC +10216 \citep{Sahai_2010}, which is believed to trace the interaction between molecular hydrogen and electrons (see \autoref{sec:ABG_introduction}). 
However, \citet{Matthews_2015} recently discovered a thin shell of HI, the total mass of which is less than 1$\%$ compared with the total predicted mass of the CSE of the \textit{"low"} stellar effective temperature (IRC +10216). These authors suspect that this small amount of HI results from the photo-dissociation of H$_{2}$ by the ISRF as suggested by \citet{Glassgold_1983}'s model.

Despite  all the above, the physical-chemical mechanisms that transfer hydrogen from the stellar surface into the inner part of the CSEs, and then into its outer part, as well as its conversion processes, have not been well studied.

\begin{figure}
		\centering
    	\includegraphics[width=0.9\linewidth]
    	{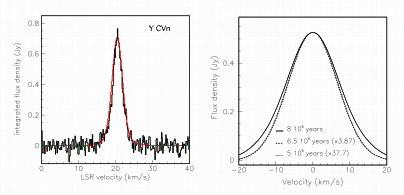}
   		\caption[Y CVn Integrated spectrum of HI and best fit]
   		{(\textit{left panel}) Y CVn integrated spectrum of HI and best fit results using 
   		\cite{Libert_2007}'s model. 
   		(\textit{right panel}) resulting HI profile when using the
   		time dependent model of \cite{Villaver_2002}. 
   		\citep{Hoai_2015}.}
		\label{fig:HI_best_fit}
	\end{figure}

\begin{figure}
		\centering
    	\includegraphics[width=0.9\linewidth]
    	{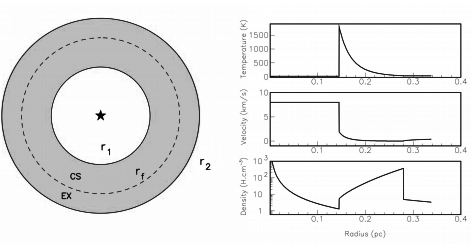}
   		\caption[Standard stationary model of the CSE around AGB star]
   		{Standard stationary model of the CSE around AGB star. 
		(\textit{Left panel}) Schematic view \citep{Libert_2007}. 
		(\textit{Right panel}) Wind properties \citep{Hoai_2015}.}
		\label{fig:Libert_model}
	\end{figure}

Some of the observed HI lines have been successfully interpreted by simple hydrodynamic models (\citealt{Libert_2007}, \citealt{Hoai_2015, Hoai_2017}). 
Their "standard" stationary model 
is described in \autoref{fig:Libert_model}.  
The free wind expansion takes place at $R_{\ast} < r < r_{1}$. 
The termination shock is located at $r_{1}$. 
The bow-shock is located at $r_{2}$. 
The wind and ambient materials are separated at $r_{f}$.
For the region of freely expanding wind, 
the temperature and hydrogen number density are assumed to depend 
on radius as a power-law, $T \sim r^{-0.5}$, and $n_{H} \sim r^{-2}$.
For the terminal shock region ($r_{1} \leq r \leq r_{f}$), 
the temperature, the velocity and the density 
are derived by solving the set of fluid dynamic equations for ideal gases, 
 adopting the upstream conditions: velocity is obtained from observations, 
density is calculated from the mass-loss rate, 
and temperature is equal to $T_{0} = 20$ K \citep{Libert_2007}.
For the external region ($r_{f} < r < r_{2}$), 
the density is again assumed to be $r^{-2}$, and temperature is constant.     
  
This simplified model that only accounts for the main hydrodynamic 
processes already nicely reproduces the spectrum of HI, 
such as for Y CVn (\autoref{fig:HI_best_fit}, \textit{left panel}). In this work we will attempt to improve the dynamical treatment by adding the coupling between the dust grains and gas and by including heating and cooling processes as done in steady-state wind models (\citealt{Justtanont_1994}, \citealt{Winters_1994}, \citealt{Decin_2006}). Finally, we will also include time-dependent chemistry, in a hope to predict the fractional abundance of HI in the CSE.

In addition, \citet{Villaver_2002} carefully studied the time dependent hydrodynamics of the circumstellar envelope. These authors took into account the \textit{thermal pulsation} effect and the influence of the external ISM. \citet{Hoai_2015} use this model to reproduce the Y CVn HI line shape. They compute the model at three different epochs corresponding to the first two thermal pulses and to the end of the last thermal pulse. However, the gas temperature in the CSE remains large ($> 5000$ K), which makes thermal broadening dominating the line profile. Consequently, the full-width at half-maximum (FWHM) is larger than the observed one (\autoref{fig:HI_best_fit}, \textit{right panel}).




     
\section{Aims and outline}
During the AGB phase, dredge-up processes mix the nuclear products deep inside the core up to the surface, and a mass-loss mechanism ejects them into the ambient medium. AGB stars lose most of their material through stellar winds, which eventually make up the circumstellar material around the star. The mechanisms that launch the material from the stellar surface into the CSE are well studied (see \autoref{sec:stellar_wind}). Since the temperature dramatically cools down further away from the stellar surface, the parent molecules and dust form (see \autoref{sec:CSE}). Then the dust absorbs stellar radiation, and it couples and transfers momentum to the gas. This impact acts like an acceleration processes which pushes the gas away from star. The gas flow thus crosses the \textit{"critical point"} where its speed exceeds the thermal sound speed, and the wind becomes supersonic. This supersonic wind eventually interacts with the ISM.

The whole collection of processes which take place in the CSE makes it look like a chemical factory. Among chemical species, hydrogen turns up as an important tool for tracing the larges scale of the CSE (see \autoref{sec:hydrogen}). The hydrodynamical models, whose outcomes match well the existing HI observations, could be improved to interpret better the hydrogen atomic and molecular fractions. Hence, in this part, we aim at studying the hydrogen chemistry in the CSE. 
\begin{itemize}
	\item Chapter 8: from the Paris-Durham shock code, we re-create a stationary hydrodynamic wind model in 1D spherical geometry. Although the preferred driving mechanism in the sub-sonic region is thermal pulsations, we assume that pressure gradient that lifts material up from the stellar surface. We also introduce a chemical network, which is coupled with the hydrodynamic model above.
	\item Chapter 9: we calculate the line profiles of the atomic hydrogen, including the termination shock, and compare them to observations.
	\item Chapter 10: we discuss the results and future prospects of our model.
\end{itemize}

%
\makeatletter
\part{BOW SHOCK MODEL}
\setstretch{1.1} 
\chapter{1D-SHOCK MODEL: PARIS-DURHAM}
\label{Chapter1}
\lhead{Chapter 1. \emph{1D-Shock model: Paris-Durham shock code}} 

The Paris-Durham\footnote{Also known as the Durham-Paris shock code on the other side of the Channel} shock code is born from a long term collaboration between David Flower in Durham and G. Pineau des For\^ets in Paris. 
The first version of the Paris-Durham code was introduced by \citet{Flower_1985} with the main objective of simulating \textit{1D steady-state} shocks propagating through the interstellar medium. That version included gas-phase chemical processes, studied by Flower and Pineau des For\^ets in a series of article published from 1985 to 1989. The solid-phase chemical processes were included in the next series of papers (e.g., \citealt{FP94, FP95}, \citealt{FP96}). As shown by \cite{PL04a, PL04b}, the Paris Durham shock code can also compute approximations to 1D \textit{non steady-state} magnetohydrodynamical shocks by glueing together pieces of steady-state models. Over the time, motivated by spectroscopy data acquired from satellites (ISO, Herschel, and Spitzer), the code has been improved to study the intensities of the molecular lines in sub-mm and in the infrared. In its recent state-of-the-art version, the Paris-Durham code is mainly written in FORTRAN 90, except for a few routines that are coded in FORTRAN 77. It uses the DVODE algorithm\footnote{https://computation.llnl.gov/casc/odepack/} to solve the ODE equations. 
\citet{FP15} presents the official up to date version.


\section{Magnetohydrodynamic shock wave}
Magnetohydrodynamic (MHD) shock waves have been well studied in astrophysics because the astrophysical gas is usually magnetized, with a magnetic pressure comparable to the turbulent pressure of the gas. That kind of wave is very common in the interplanetary medium, the interstellar medium and in the star formation regions. The ionization fraction of the gas is very important to the study of shock waves, because the magnetic field directly interacts with the ionized gas and indirectly with the neutral gas via the collisions between the charged particles and the neutral particles. If the gas is ionized enough in a shock wave, the coupling between the charged particles and the neutral particles is strong so that the gas behaves like a single-fluid. Conversely, if the gas is weakly ionized, the collisions occur and the gas behaves like a multi-fluid. The principles of MHD shock waves and the main differences between those fluids are discussed below. 
\subsection{Set of conservation equations}
\label{sec:hydro_eqs}
In general, the dynamical state of the gas is identified by the number density $n$, the mass density $\rho$, the velocity $v$ and the temperature $T$, which are calculated from a set of conservation equations of number density, mass density, momentum and energy of neutral and charged fluids. The subscript $"n"$ is used for the neutral particles and $"i"$ for the ionized particles. In the shock plane, let us denote: (1) \textit{z} an independent variable, which defines the positive coordinate of the gas flow with respect to an arbitrary reference point in the pre-shock gas, (2) \textit{t} the corresponding traveling time of the flow, and (3) $B$ the transverse magnetic field perpendicular to the flow. With this simplified hypothesis, we can ignore the inherent complication of the oblique model, in which the magnetic field and the shock propagation creates an angle different from $90^{o}$ with respect to the $z$-direction.

The conservation equation for the number density of neutral particles is
\begin{equation} \label{eq:Nn}
	\frac{\partial n_{n}}{\partial t} + \frac{\partial}{\partial z}(n_{n}v_{n}) = N_{n} \mbox{.}
\end{equation}
where $N_{n}$ is the number of neutral particles created per unit volume and time. A corresponding equation holds for the charged particles
\begin{equation}\label{eq:N+}
	\frac{\partial n_{i}}{\partial t} + \frac{\partial}{\partial z}(n_{i}v_{i}) = N_{i} \mbox{.}
\end{equation}
 
The mass conservation of neutral fluid is written by
\begin{equation} \label{eq:mass_cons_n}
   	\frac{\partial \rho_{n}}{\partial t} + \frac{\partial}{\partial z}(\rho_{n}v_{n}) = S_{n}
\end{equation}
where $S_{n}$ is the neutral mass change due to chemical reactions. The corresponding equation for the positive charged fluid is
\begin{equation} \label{eq:mass_cons_i}
   \frac{\partial\rho_{i}}{\partial t} + \frac{\partial}{\partial z}(\rho_{i}v_{i}) 
   = -S_{n} \mbox{.}
\end{equation}

The momentum of the fluid is also conserved. For the neutral fluid, the equation of momentum conservation is
\begin{equation} \label{eq:moment_cons_n}
   \frac{\partial}{\partial t}(\rho_{n} v_{n}) + \frac{\partial}{\partial z}\left(\rho_{n} v^{2}_{n} + n_{n}k_{B}T_{n}+\pi_{v}\right) = A_{n}
\end{equation}
where $A_{n}$ denotes the change of momentum of the neutral fluid per unit volume and time. $k_{B}$ is the Boltzmann constant and $n_{n}k_{B}T_{n}$ is the thermal pressure of the neutral fluid. $\pi_{v}=\rho_{n}\lambda_{0}c_{0}\frac{\partial v_{n}}{\partial z}$
is a viscous pressure built with a constant viscous length and velocity
$\lambda_{0}=3.10^{14}$ cm and $c_{0}=1$ km$\,$s$^{-1}$. The viscous
term hence diffuses momentum over a typical length scale $\lambda_{0}$
at a dispersion speed $c_{0}$. It is switched on when we want to
trigger a viscous discontinuity (J-type, see \autoref{sec:types-def}) in the flow, and it is switched off whenever
$\pi_{v}$ gets back below one part per million of the thermal pressure. In effect, viscosity dissipates ordered kinetic energy into heat.
   
  If the magnetic field is accounted for, it acts directly onto the charged fluids and indirectly onto the neutral fluid through collisions and it adds a  magnetic pressure term $B^{2}/8\pi$  to the equation of momentum conservation. Thereby, the equation of momentum conservation for ion-election fluid is
\begin{equation} \label{eq:moment_cons_i}
   \frac{\partial}{\partial t}(\rho_{i} v_{i}) + \frac{\partial}{\partial z}\left[\rho_{i} v^{2}_{i} + n_{i}k_{B}(T_{i}+T_{e}) +\frac{B^{2}}{8\pi}\right] = -A_{n} \mbox{.}
\end{equation}


    
The equation of conservation of energy for the neutral fluid yields
   	\begin{equation} \label{eq:energy_cons}
   		\frac{\partial}{\partial t}\left(\frac{1}{2}\rho_{n}v^{2}_{n} + \epsilon\right)+ \frac{\partial}{\partial z}\left(\frac{1}{2}\rho_{n} v^{3}_{n} 
   		+ \frac{\gamma}{\gamma -1}v_{n}n_{n}k_{B}T_{n} + v_{n}\pi_{v}\right) = B_{n}
   	\end{equation}
where $B_{n}$ is the change of energy of the neutral fluid per unit volume and time, $\epsilon$ is the internal specific energy, and $\gamma$ is the adiabatic index. For the ion-electron fluid, similarly, the magnetic field adds one more term $B^{2}/4\pi$ due the magnetic energy flux:   
   	\begin{equation} \label{eq:energy_i}
   		\frac{\partial}{\partial t}\left(\frac{1}{2}\rho_{i}v^{2}_{i} + \epsilon\right) + \frac{\partial}{\partial z}[\frac{1}{2}\rho_{i} v^{3}_{i} 
   		+ \frac{\gamma}{\gamma -1}v_{i}n_{i}k_{B}(T_{i}+T_{e}) + \frac{B^{2}v_{i}}{4\pi}] = B_{i} + B_{e} \mbox{.}
   	\end{equation}
  
\subsection{Source terms}
The source terms $N$, $S$, $A$ and $B$ which appear on the right hand side of the equations of conservation respectively represent the rate of change in number density, mass, momentum and energy per unit volume of the neutral to charged fluids through irreversible micro-physics processes. These mechanisms in fact depend on the context being considered. In this section, we summarize some of the main source terms that may appear in interstellar molecular clouds. Further details can be found in \citet{Flower_1985} and \citet{Flower_2015}.

\subsubsection{Number and mass of particles source terms}
If $\alpha$ is a particular atomic or molecular species, and $C_{\alpha}$ is the net production of species $\alpha$ per unit volume, the rates of change of the total number of neutral species and positive ion per unit volume are
\begin{equation}
	N_{n} = \sum _{\substack{\alpha \\ (neutral\ species)}} C_{\alpha}
\end{equation}
\begin{equation}
	N_{i} = \sum_{\substack{\alpha \\ (ionized\ species)}} C_{\alpha} \mbox{.}
\end{equation}

The changing rate of neutral and positive ion mass are then 
\begin{equation}
	S_{n} = \sum_{\substack{\alpha \\ (neutral\ species)}} C_{\alpha} m_{\alpha}
\end{equation}
\begin{equation}
	S_{i} = \sum_{\substack{\alpha \\ (ionized\ species)}} C_{\alpha} m_{\alpha} \mbox{.}
\end{equation}
 
\subsubsection{Momentum source terms}
Let us denote $C_{\alpha \beta}$ the creation ($C_{\alpha\beta} > 0$) or the destruction ($C_{\alpha\beta} < 0$) rates of species $\alpha$ through the reaction $\beta$. Therefore,
\begin{equation}
	C_{\alpha} = \sum_{\beta} C_{\alpha \beta} \mbox{.}
\end{equation}

 Through the ion-neutral reactions, the charged fluid transfers momentum to the neutral fluid at rate $A^{(1)}_{n}$
\begin{equation} \label{eq:A1}
	A^{(1)}_{n} = \sum_{\substack{\alpha \\ (neutral\ speci)}} \sum_{\beta} C_{\alpha \beta}  m_{\alpha} v_{\beta}(CM)  
\end{equation} 
where $v_{\beta}(CM)$ is the  collision center-of-mass velocity defined as
\begin{equation}
	v_{\beta}(CM) = \frac{m_{i}v_{i}+m_{n}v_{n}}{m_{i}+m_{n}}
\end{equation} 
where $m_{i}$, $m_{n}$ are the mass of ions and neutral reactants, and $\beta$ is the dummy index for ion-neutral reactions with  net rate $C_{\alpha \beta}$. \autoref{eq:A1} indicates that species are created and destroyed at the center-of-mass collision velocity $v_{\beta}(CM)$.

Owing to elastic scattering on the ions, the neutral fluid gains momentum at a rate 
$A^{(2)}_{n}$
\begin{equation} \label{eq:A2}
	A^{(2)}_{n} = \frac{\rho_{n}\rho_{i}}{\mu_{n}+\mu_{i}}<\sigma v>_{in}(v_{i}-v_{n})
\end{equation}
where the rate coefficient is defined by
\begin{equation}
	<\sigma v>_{in} =2.41 e \left(\frac{\alpha_{n}}{\mu_{in}}\right)^{\frac{1}{2}}
\end{equation} 
in unit of cm$^{3}\,$s$^{-1}$, with $\alpha_{n}$ the polarizability of the neutral fluid, $\mu$ the mean molecular weight and $\mu_{in}=\mu_{i}\mu_{n}/(\mu_{i}+\mu_{n})$ the reduced mass.
 
In the case of a dense cloud medium where the ionization degree of the gas is small, momentum  transfer between the neutral fluid and the charged grains is important. The collision cross-section can be approximated by the grain cross-section $\pi a^{2}_{g}$, where $a_{g}$ is the grain radius, and the collision speed is close to the ion-neutral drift $|v_{i}-v_{n}|$. Hence, the rate of  momentum transfer between the neutral fluid and the charged grains  derives from  \autoref{eq:A2} with  $\mu_{g}\gg \mu_{n}$ as
\begin{equation} \label{eq:A3}
	A^{(3)}_{n} = \rho_{n}n_{g}\pi a^{2}_{g}|v_{i}-v_{n}|(v_{i}-v_{n}) \mbox{.}
\end{equation}  

The total rate of change for the  neutral fluid momentum is then the sum of momentum transfer from those processes $A_{n} = A^{(1)}_{n} + A^{(2)}_{n} + A^{(3)}_{n}$.

\subsubsection{Energy source terms}
The micro-physical processes along the shock also lead to energy exchanges between the charged and the neutral fluids, as well as between the charged grains and the neutral fluid. Chemical reactions are responsible for part of the kinetic energy transfer from the charged to the neutral fluids. The exchange rate per unit volume through the chemical reactions $\beta$ is derived from \autoref{eq:A1}
\begin{equation}
	B^{(1)}_{n} = \sum_{\substack{\alpha \\ (neutral\ species)}} \sum_{\beta} C_{\alpha \beta} \frac{1}{2}m_{\alpha} v^{2}_{\beta}(CM) \mbox{.}
\end{equation}

When an ion at temperature $T_{i}$ dissociatively recombines with an electron at temperature $T_{e}$ to form two neutral species, an amount of energy $3/2 k_{B}(T_{i}+T_{e})$ is transferred to the neutral fluid. On the contrary, when a neutral is photo-ionized, it loses an amount of heat $3/2 k_{B}T_{n}$. The heat rate transfer to the neutral fluid per unit volume is then
\begin{equation} \label{eq:energy_reactants}
	B^{(2)}_{n} = \sum_{\substack{\alpha \\ (neutral\ species)}} 
	\left[\sum_{\substack{\beta \\ (C_{\alpha \beta}>0)}} 
	C_{\alpha \beta} \frac{3}{2}k_{B}(T_{i}+T_{e}) 
	+ \sum_{\substack{\beta \\ (C_{\alpha \beta}<0)}} C_{\alpha \beta}\frac{3}{2}k_{B}T_{n}\right] 
	\mbox{.}
\end{equation}
 
 The chemical reactions can also affect the thermal balance of the medium via the chemical energy released $\Delta E$. This heats the neutral fluid with a corresponding rate
\begin{equation}
	B^{(3)}_{n} = \sum_{\substack{\alpha \\ (neutral\ species)}} \sum_{\beta} C_{\alpha \beta}\frac{M_{\beta}-m_{\alpha}}{M_{\beta}}\Delta_{\beta}
\end{equation}
where $M_{\beta}$ is the total mass of the products from reaction $\beta$ and $\Delta_{\beta}$ is the net chemical energy released by this reaction.

The elastic scattering of the neutral fluid on the ions results to a rate of heating for the neutral fluid as
\begin{equation}
	B^{(4)}_{n}=\frac{\rho_{n}\rho_{i}}{\mu_{n}\mu_{i}}<\sigma v>_{in}
	\frac{2\mu_{n}\mu_{i}}{(\mu_{n}+\mu_{i})^{2}}\left[\frac{3}{2}k_{B}(T_{n}+T_{i})+
	\frac{1}{2}(v_{i}-v_{n})(\mu_{i}v_{i}+\mu_{n}v_{n})\right] \mbox{.}
\end{equation}

The elastic scattering of the neutral fluid on the electrons results to the same rate of heating for the neutral fluid, except for the fact that $m_{e} \ll m_{n}$
\begin{equation}
	B^{(5)}_{n}=\frac{\rho_{n}\rho_{e}}{\mu_{n}\mu_{e}}<\sigma v>_{en}
	\frac{2\mu_{e}}{\mu_{n}}\left[\frac{3}{2}k_{B}(T_{n}+T_{e})+
	\frac{1}{2}(v_{i}-v_{n})\mu_{n}v_{n}\right]
\end{equation}
where the scattering cross section is
\begin{equation}
	<\sigma v>_{en} = 10^{-15}\left(\frac{8k_{B}T_{e}}{\pi m_{e}}\right)^{\frac{1}{2}} \mbox{.}	
\end{equation}

The rate of energy transfer from the charged grains to the neutral fluid is derived from  \autoref{eq:A3}
\begin{equation}
	B^{(6)}_{n} = \rho_{n}n_{g}\pi a^{2}_{g}|v_{i}-v_{n}|(v_{i}-v_{n})v_{i}
\end{equation} 

The total rate of energetic change for the  neutral fluid ($B_{n}$) is also the sum all of those processes $B_{n} = B^{(1)}_{n} + B^{(2)}_{n} + B^{(3)}_{n}+ B^{(4)}_{n}+ B^{(5)}_{n} + B^{(6)}_{n}$. The total rate of energetic change for the ionized fluid ($B_{i}$) proceeds similarly to the neutral fluid.

The electron particles can transfer energy via three main processes: (1) dissociatively recombining with an ion, (2) scattering on ions and (3) through photo-ionization. 

As described in \autoref{eq:energy_reactants}, when an electron at temperature T$_{e}$ dissociatively recombines with an ion to create neutral species, it loses an amount of heat
\begin{equation}
	B^{(1)}_{e} = \sum_{\substack{\alpha \\ (ionized\ species)}} \sum_{\substack{\beta \\ (C_{\alpha \beta} < 0)}} C_{\alpha,\beta}\frac{3}{2}k_{B}T_{e}.
\end{equation}    
     
The heat can also be transferred between the fluid of electrons and the fluid of ions through collisions. The heating rate can be determined as
\begin{equation}
	B^{(2)}_{e} = \frac{4e^{4}}{\mu_{i}k_{B}T_{e}} \left(\frac{2\pi m_{e}}{k_{B}T_{e}}\right)^{\frac{1}{2}} ln\Lambda \left(\frac{\rho_{i}}{\mu_{i}}\right)^{2}k_{B}(T_{i}+T_{e}) 
\end{equation}
where
\begin{equation}
	\Lambda = \frac{3}{2e^{3}}\left(\frac{k^{3}_{B}T^{3}_{e}\mu_{i}}{\pi \rho_{i}}\right)^{\frac{1}{2}} \mbox{.}
\end{equation}

The rate of heating through photo-ionization should be
\begin{equation}
	B^{(3)}_{e} = \sum_{\alpha} \delta E_{\alpha} \gamma_{\alpha}n_{\alpha}
\end{equation}
where $\delta E_{\alpha}$ is the mean energy of the photo-electron created by the photo-ionization of the species $\alpha$, with density $n_{\alpha}$ and photo-ionization rate $\gamma_{\alpha}$.

The total rate of energetic change for the electron fluid is $B_{e} = B^{(1)}_{e} + B^{(2)}_{e} + B^{(3)}_{e}$.

In addition, the Paris-Durham code incorporates a wide range of cooling and heating processes relevant to the ISM. Lyman $\alpha$ cooling is included as well as line excitation cooling from neutral atoms and ions: C, N, O, S, Si, C$^{+}$, N$^{+}$, O$^{+}$, S$^{+}$, Si$^{+}$, Fe$^{+}$. We use tables for the line cooling from molecules: H$_{2}$O, OH and CO from \citet{Neufeld_Kaufman_1993}. H$_{2}$ line cooling is treated thanks to the level by level time-dependent treatment of all populations. Photo-electric heating from dust grains and cosmic ray ionization heating are also included.

\subsection{Transverse stationary shock wave}
The stationary hypothesis is a simplified way to analyze the shock structure. An MHD shock wave is called stationary if its structure does not change in time, so that the time derivative in all conservation equations above vanishes in the frame of motion of the structure. In addition, the MHD shock wave is called transverse if the direction of the ambient magnetic field is perpendicular to the direction of the shock propagation. 
%
%

\subsubsection{Rankine-Hugoniot relation}
For a single fluid, mass and momentum are conserved. The source terms $S$, $B$ and $A$, therefore, on the right hand side of \autoref{eq:mass_cons_n} and \autoref{eq:moment_cons_n} are equal to zero. In general, number density and energy can vary because of the neutral-neutral reactions, such as the collisional dissociation of $H_{2}$. However, those chemical collisional processes are \textit{all inelastic}, for which the time (and distance) scales are larger compared to the corresponding elastic collision process. Therefore, the first few mean free-paths of the shock, where the viscous transition takes place, qualify as adiabatic, which means that the shock does not exchange energy with the shock's ambient medium.

Owing to all of those approaches, we enable relations to be obtained between the pre-shock (upstream) and the postshock (downstream) gas. Those relations are referred to as the Rankine-Hugoniot relations

\begin{equation} \label{eq:mass_cons_adia}
	\rho_{1}v_{1} = \rho_{2}v_{2}
\end{equation}
\begin{equation}
	\rho_{1}v^{2}_{1} + n_{1}k_{B}T_{1} + \frac{B^{2}_{1}}{8\pi}=
	\rho_{2}v^{2}_{2} + n_{2}k_{B}T_{2} + \frac{B^{2}_{2}}{8\pi}	
\end{equation}
\begin{equation} \label{eq:energy_cons_adia}
	\left(\frac{1}{2}\rho_{1}v^{2}_{1}+\frac{\gamma}{\gamma-1}n_{1}k_{B}T_{1}+\frac{B^{2}_{1}}{4\pi}
\right) v_{1} =
	\left(\frac{1}{2}\rho_{2}v^{2}_{2}+\frac{\gamma}{\gamma-1}n_{2}k_{B}T_{2}+\frac{B^{2}_{2}}{4\pi}\right) v_{2}
\end{equation}
\begin{equation} \label{eq:frozen_mag}
	B_{1}v_{1}=B_{2}v_{2}
\end{equation}

where the subscripts (1) and (2) represent  the pre-shock and the post-shock gas, respectively; and $\gamma$ adopts the value 5/3. The combination of  equations  \ref{eq:mass_cons_adia}-\ref{eq:frozen_mag} yields an equation for the compression ratio $\rho_{2}/\rho_{1}$ across the adiabatic shock front
\begin{equation} \label{eq:compress_eq}
	2(2-\gamma)b\left(\frac{\rho_{2}}{\rho_{1}}\right)^{2} +
	\left[(\gamma -1)M^{2}+2\gamma (1+b)\right]\frac{\rho_{2}}{\rho_{1}} -
	(\gamma +1)M^{2} = 0 \mbox{.}
\end{equation} 

In \autoref{eq:compress_eq}, $M = v_{s}/c_{1}$ is the Mach number, which is the ratio of the shock speed to the isothermal sound speed in the pre-shock medium corresponding to the pressure $p_{1} = n_{1}k_{B}T_{1}$; and $b=B^{2}_{1}/(8\pi p_{1})$ is the ratio of the magnetic pressure to the pre-shock pressure. The positive solution of the quadratic \autoref{eq:compress_eq} yields an analytical expression for the compression ratio of the gas caused by a discontinuity adiabatic shock:
\begin{equation} \label{eq:compress_ratio}
	\frac{\rho_{2}}{\rho_{1}} = \frac{2M^{2}(\gamma +1)}{D+\sqrt{D^{2}+8bM^{2}(2-\gamma)(\gamma +1)}}
\end{equation}   
where  $D$ is
\begin{equation}
	D = (\gamma -1)M^{2}+2\gamma(1+b) \mbox{.}
\end{equation}                                             

When there is no magnetic field ($b_{1}=0$), the Mach number is
\begin{equation}
	M^{2} = \frac{\gamma +1}{2}\frac{p_{2}}{p_{1}} + \frac{\gamma -1}{2}
\end{equation}
and \autoref{eq:compress_eq} gives the simplified expression of the compression ratio:
\begin{equation} \label{eq:compress_ratio_reduce}
	\frac{\rho_{2}}{\rho_{1}} = \frac{M^{2}(\gamma +1)}{(\gamma -1)M^{2} + 2\gamma}
	= \frac{p_{1} + hp_{2}}{p_{2}+hp_{1}}
\end{equation}
 where $h=(\gamma +1)/(\gamma -1)$. When the value of $\gamma$ is $5/3$, the value of $h$ is 4. In the shock region, the shock transition process leads to an increase of entropy, this increase consequently forces $p_{2}>p_{1}$. We can demonstrate that the gas density in the post-shock region is always greater than in the pre-shock region from \autoref{eq:compress_ratio_reduce}.
 
 \autoref{eq:compress_ratio_reduce} also shows that in the extreme case where $M \gg 1$ (strong shock), 
 $\rho_{2}/\rho_{1} = h = 4$. Then, from \autoref{eq:mass_cons_adia}, we come up with
 \begin{equation}
 	\frac{v_{2}}{v_{1}} = \frac{\rho_{1}}{\rho_{2}} = \frac{1}{4}
 \end{equation}
   
 and the temperature change across the adiabatic front is given by 
 \begin{equation}
 	\frac{T_{2}}{T_{1}} = \frac{p_{2}}{p_{1}}\frac{\rho_{1}}{\rho_{2}}
	= \left[\frac{p_{2} + hp_{1}}{p_{1}+hp_{2}}\right]\frac{p_{2}}{p_{1}} \mbox{.}
\end{equation}
In the extreme limit case where $p_{2} \gg p_{1}$: 
\begin{equation}
	\begin{split}
	&T_{2}/T_{1} = p_{2}/hp_{1} \rightarrow \infty \\
	&T_{2} = \frac{5}{4} \frac{\rho_{1}}{n_{2}k_{B}}v^{2}_{1}
	\end{split}
\end{equation}

To summarize, across the viscous discontinuity, the gas is compressed, the gas pressure and temperature increase, while the velocity of the gas decreases in the shock frame.

\subsubsection{C-type and J-type shocks}
\label{sec:types-def}
As seen in \autoref{eq:mass_cons_adia}-\ref{eq:frozen_mag}, the existence of the magnetic field affects the structure of the fluid. 
Two main approximations accordingly apply: single or multi-fluids. 

\subsubsection*{Single fluid flow: J-type shock wave}
If the magnetic field is weak or absent, all components (neutral, ion and electron) are assumed to have the same velocity and the fluid behaves like a single flow. The shock caused by the supersonic propagation is sometimes called \textit{"hydrodynamic"} with an extra contribution from the magnetic pressure. If the speed of the shock is greater than the signal speed in the pre-shock medium. The latter cannot "feel" the shock wave
before it arrives. Across the shock front, the variables (pressure,
density, velocity, etc.) of the fluid vary as a viscous discontinuity
jump (the so-called $J$-type shock). After being heated, accelerated and compressed by the shock wave, the gas cools down through radiative emission.

\subsubsection*{Multi-fluid flow: C-type shock wave}
If the magnetic field is significant, its interaction with the charged component (including the grains) leads to the multifluid situation, where the neutral and charged components have different velocities. The magnitude difference strongly depends on the collisional coupling efficiency between the neutral and charged fluids. 

When the ionization fraction is small, the magnetosonic speed $v_{m}$ in the charges in the direction of shock propagation is defined as
\begin{equation} \label{eq:magnetosonic_speed}
	v_{m} = \sqrt{c^{2}_{s} + \frac{B^{2}}{4\pi \rho_{c}}} \simeq \frac{B}{\sqrt{4\pi \rho_{c}}}
\end{equation}
where $c_{s}$ and $B^{2}/4\pi \rho_{c}$, the speed of sound and the Alfv\'en speed of the {\it charged} fluid, 
can be greater than the shock entrance velocity. Then a magnetic precursor forms
upstream of the discontinuity, where the charged and neutral fluids dynamically decouple. The resulting friction between the two fluids
heats up and accelerates the neutral fluid. If the intensity of the magnetic field keeps increasing, the precursor size also increases, and
the neutrals are compressed sooner before the arrival of the shock front. This leads eventually to the disappearance of the discontinuity, and the shock variables change continuously (the so-called C-type shock). Because of friction between the neutral and charged components, the kinetic energy dissipation is a much more gradual process and is spread over a much larger volume. 

\begin{figure}
	\centering
	\includegraphics[width=0.8\linewidth]
	{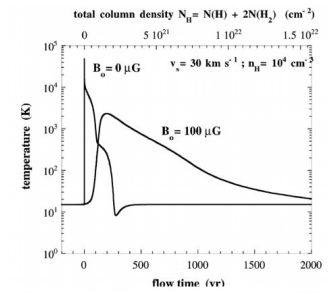}
	\caption[Thermal profile of the stationary (C-type and J-type) shocks]
	{Thermal profile of the stationary (J-type and C-type) shocks (\citealt{Flower_2003}). 
	The calculations were implemented with $30\,$km$\,$s$^{-1}$ 
	for the shock speed, 
	$10^{4}\,$cm$^{-3}$ for the pre-shock density,
	 and $0\,\mu$G/$100\,\mu$G for the initial strength of the magnetic field 
	 in the J-type and C-type shock respectively.}
	\label{fig:C_J_shock}
\end{figure}

\autoref{fig:C_J_shock} illustrates the difference of the thermal profile between J-type and C-type shocks. 
 
\subsection{Transverse non-stationary shock wave: CJ-type}
\label{sec:CJ-type}
In the previous section, the shock properties were described in the case of steady state ($\partial/\partial t = 0$). This assumption is also satisfied if the time to reach the steady state is short compared to the age of the shock wave. \citet{Ch1998} provided the time scale for a MHD shock reach steady state related to the shock speed, the initial gas density and the initial magnetic induction as
\begin{equation}
	\left |\frac{dv_{n}}{dz}\right| ^{-1} \approx \left( n_{i}<\sigma v>_{in}\right)^{-1}
\end{equation}
However, the authors demonstrated that the shock speed and the magnetic induction have tiny influence on this time scale, of which is mostly influenced by the initial gas density. For instance, the time required to attend at steady state of a shock with the initial conditions of n$_{H}$=$10^{3}\,$ cm$^{-3}$, v$_{n}$=$10\,$km$\,$s$^{-1}$ and B=$10\,\mu$G is about $5\,10^{5}\,$yr. That definitely raises a need to develop a $non-stationary$ shock model. In addition, some previous studies pointed out the need of non stationary shock
models to explain observational H$_{2}$ emissions originating from outflows (i.e., \citealt{Giannini04}; \citealt{Gusdorf_2008}; \citealt{G15}).

\begin{figure}
   \begin{minipage}[c]{1\textwidth}
      \includegraphics[width=1\linewidth]
      {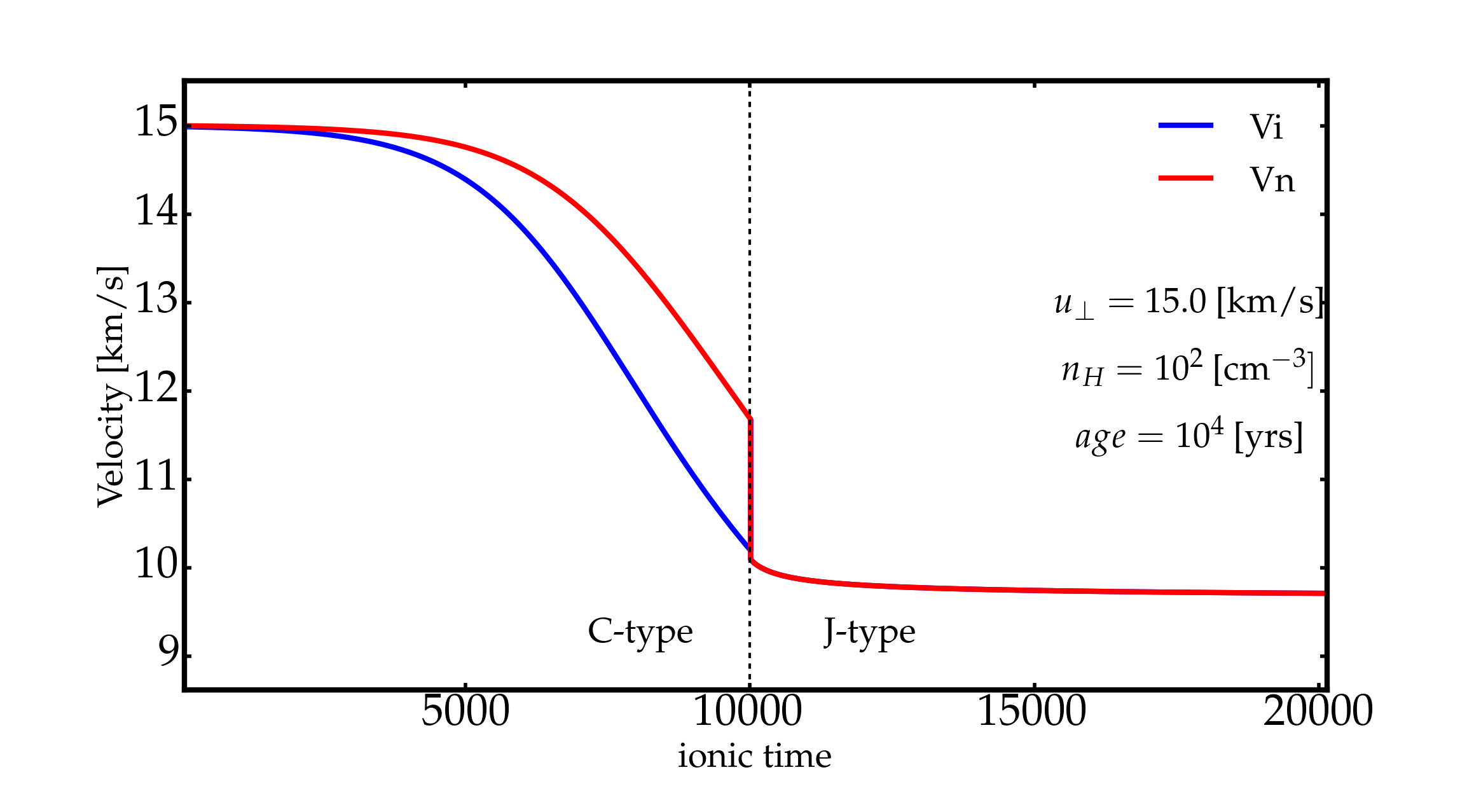}
   \end{minipage} \\
   \begin{minipage}[c]{1\textwidth}
      \includegraphics[width=1\linewidth]
      {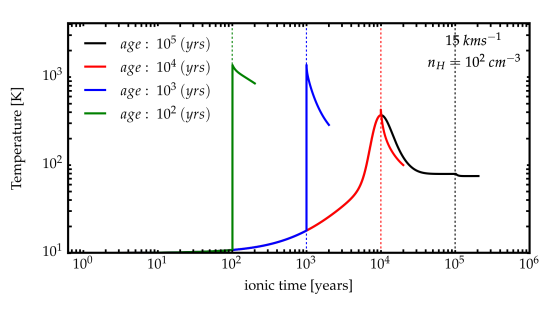}
   \end{minipage}\\

	\caption[Velocity and thermal profiles of the non-stationary (CJ-type) shock]
	{Velocity and thermal profiles of the non-stationary (CJ-type) shock. 
	(\textit{Top}) Velocity profile of a non-stationary shock 
	with a shock speed of 15 km$\,$s$^{-1}$, 
	a pre-shock density of $10^{2}\,$cm$^{-3}$, 
	and a dynamical shock age of $10^{4}\,$yr. 
	(\textit{Bottom}) Thermal profiles of the non-stationary shock 
	(same shock speed and pre-shock density) for the various dynamical shock ages.} 
	\label{fig:CJ_profile}
\end{figure}

\citet{Smith_1997} studied the formation and time-evolution of a 1D C-type shock. In this simple case, the chemistry was ignored and the ionization fraction was a power law of gas density. They found that the evolution of C-shock approached the analytic steady-state solution in all cases. \cite{Ch1998} also studied the time-dependent evolution of C-type and J-type shock, including the state of ionization of the gas by taking into account a chemical network. They showed that young C-type shocks looked like truncated steady-state: this yielded techniques to produce time-dependent snapshots from pieces of steady-state models \citep{FP99,PL04b}. Following the approach of \citet{PL04b} in the large compression case, the J-type front in a young C-type shock is thus inserted when the flow time in the charged fluid is equal to the age of the shock (so-called CJ-type shock). This approach is illustrated by \autoref{fig:CJ_profile}. The J-type shock is truncated when the total neutral flow time across the J-type part reaches the age of the shock (the same holds for young J-type shocks). As the shock gets older, the magnetic precursor grows larger and the velocity entrance into the J-type front decreases due to the ion-neutral drag. As a result, the maximum temperature at the beginning of the J-type front decreases with age, as illustrated in the bottom panel of \autoref{fig:CJ_profile}. If the magnetic field is strong enough, the J-type tail eventually disappears and the shock becomes
stationary. The resulting structure forms a continuous transition
between the pre-shock and the post-shock gas (a stationary C-type shock).

\subsection{Influence of chemistry}
\begin{figure}
	\includegraphics[width=1\linewidth]{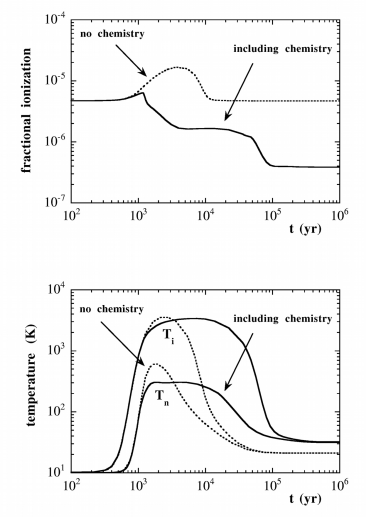}
	\caption[Effect of chemistry]
	{Effect of chemistry. (\textit{Top}) Ionization fraction predicted by a model 
	in which $v_{s}=10\,$km$\,$s$^{-1}$, $n_{H}=10^{3}\,$cm$^{-3}$ and $B_{0}=25\,\mu$G. 
	(\textit{Bottom}) Corresponding temperature profile of 
	the ionized and the neutral fluids \citep{PDF_1997}.}
	\label{fig:chem_effect}
\end{figure}

Shocks play an important role in the interstellar gas evolution from both a dynamical and a chemical point of view. Through  chemical processes, species are either formed or destroyed, ions and neutrals in fluid react, which affects the gas thermal balance. \citet{PDF_1997} investigated the effect of the chemistry on the time-dependent shock calculations: when the chemistry is switched off, the ionization fraction changes only through the differential compression of the ionized and neutral fluids. So this fraction is the same in the pre-shock and post-shock regions. When the chemistry is accounted for, the ionization fraction is much lower which leads to a weaker ion-neutral coupling. Consequently, the shock region is broader than in the case of no chemistry as shown in the top panel of \autoref{fig:chem_effect}. Therefore, the maximum of the temperature is also lower because the energy is dissipated over a larger region, as shown in the bottom panel of \autoref{fig:chem_effect}. It is also clear that the dynamics and the chemistry are closely linked and that they need to be treated in parallel.        

\section{Input parameters}
In this section, we present the input parameters of the Paris-Durham shock code, with typical ranges, as well as the initial species abundances and chemical networks.
\subsection{Physical inputs}
\begin{itemize}
	\item[1-] Type of shocks
		\begin{itemize}
			\item[•] C represents the stationary C-type,
			\item[•] J represents the stationary J-type,
			\item[•] S represents the isochore evolution towards thermal and chemical steady state (for pre-shock computation).
			\item[•] T represents the evolution at constant temperature and density.  
		\end{itemize}
	\item[2-] Number of fluids
		\begin{itemize}
			\item[•] 1 used for J and S single-fluid computations,
			\item[•] 2 used for C computation, with neutral 
					and charged fluid separated,
			\item[•] 3 same as 2 but the temperature of the positive and negative fluids are decoupled (although they keep the same velocity).
		\end{itemize}
	\item[3-] Magnetic field
		\begin{itemize}
			\item[•] b = B$_{0}(\mu G)$/$\sqrt{n_{H}\,(\textrm{cm}^{-3})}$, 
				  with $B_{0}$ the ambient traverse magnetic field.
		\end{itemize}
	\item[4-] Initial density of hydrogen nuclei (cm$^{-3}$)
		\begin{itemize}
			\item[•] $n_{H}=n(H)+2n(H_{2})+n(H^{+})+...$
		\end{itemize}
	\item[5-] Shock speed (km$\,$s$^{-1}$)
	\item[6-] Initial difference in velocity (cm$\,$s$^{-1}$)
			  between the ionized and neutral fluid: 
			  $\Delta v = 10^{-3}\,$cm$\,$s$^{-1}$
	\item[7-] Kinetic temperature of the gas: $T_{0}$ taken from the steady state computation.
	\item[8-] Grain surface temperature: $T_{g}=15\,$K
	\item[9-] Method of molecular cooling calculation (except H$_2$): \citet{Neufeld_Kaufman_1993} LVG tables or simple analytic formula (low density regime).
	\item[10-] Environment
		\begin{itemize}
			\item[•] cosmic ray ionization rate: $3\,10^{-17}\,$s$^{-1}$
			\item[•] local radiation field: multiplicative factor with respect to the \citet{D78} field $G_0=$1
			\item[•] visual extinction of the incident radiation field: 
					 A$_{v}=0.1$
		\end{itemize}
	\item[11-] $H_{2}$ parameters
		\begin{itemize}
			\item[•] initial value of the ortho/para ratio of $H_{2}$: usually O/P=3
			\item[•] number of rovibrational levels of $H_{2}$: usually 150 levels
			\item[•] number of $H_{2}$ transitions to be output: 50 lines
			\item[•] method of determination for the internal 
					 energy distribution
					 of $H_{2}$ when it forms on grains, 
					 usually assumed proportional to the Boltzmann
					 distribution 
					 at 17249 K (4.48eV/3).
			\item[•] method of determination for the kinetic energy 
					 of $H_{2}$ newly formed on grains, usually a third 
					 of the formation energy.
		\end{itemize}
	\item[12-] Parameters of the numerical methods
		\begin{itemize}
			\item[•] maximum number of integration steps: $10^{5}$
			\item[•] inverse of the collision cross-section  to characterize the viscosity: 
					 $3\,10^{14}\,$cm$^{-2}$
			\item[•] tolerance on DVODE's numerical integration: 
					 $\epsilon = 10^{-8}$
			\item[•] dynamical age (years) of shock in CJ-type shock.
			\item[•] maximum evolutionary age (years) of the shock: 
					 $10^{6}$ years. Note that if this parameter is lower than 
					 the dynamical age of shock, no CJ-type shock is calculated 
		\end{itemize}
\end{itemize}

\subsection{Chemical inputs}
\label{sec:chemical_PD}
We use 134 chemical species consisting of neutrals, positive ions, negative ions and grains (mantles and cores), the list of which is given in \aref{app:chemical_species}. The initial elemental compositions of the most abundant species are listed in \autoref{tab:elemental_compostion}. The net formation and destruction rate of the chemical species per unit volume is carried out by a chemical network with 1180 chemical reactions, including: cosmic-ray ionization or dissociation, cosmic-ray-induced desorption from grains, H$_{2}$ formation on grains, three-body reactions on grain surfaces, sputtering of grain mantles, erosion of grain cores, adsorption onto grain surfaces, collisional dissociation of H$_{2}$, all other reactions, and reverse (endoergic). The form of the rate coefficient varies from on reaction to another, but all the rate coefficients are described by three parameters $\alpha, \beta, \gamma$, which are parameterized in the chemical network through an Arrhenius form: $k=\gamma (T/300)^{\alpha} e^{-T/\beta}$. \autoref{tab:five_reactions} shows the first five reactions of the full chemical network.

\newpage
\begin{table}
\centering
\begin{tabular}{l l l l l l l}
\hline \hline 
element & \hspace{1mm} Xtot & \hspace{1mm} gas & \hspace{1mm} PAH &  Mantle\hspace{1mm} &  Core\hspace{1mm} \tabularnewline
\hline 
                            
H      &       1.000E+00  &  1.000E+00  &  1.800E-05  &  2.300E-14  &   \tabularnewline

He     &       9.999E-02  &  9.999E-02  &    &    &   \tabularnewline
C      &       3.549E-04  &  1.380E-04  &  5.400E-05  &  6.999E-15  &  1.629E-04 \tabularnewline
N      &       7.939E-05  &  7.939E-05  &    &  3.000E-15  &   \tabularnewline
O      &       4.419E-04  &  3.020E-04  &    &  1.400E-14  &  1.399E-04 \tabularnewline
Mg     &       3.700E-05  &  9.999E-16  &    &   &  3.700E-05 \tabularnewline
Si     &       3.707E-05  &  3.370E-06  &   &  3.000E-15  &  3.370E-05 \tabularnewline
S      &       1.860E-05  &  1.860E-05  &    &  2.000E-15  &   \tabularnewline
Fe     &       3.231E-05  &  1.500E-08  &    &  9.999E-16  &  3.230E-05 \tabularnewline
G      &       4.627E-11  &  4.627E-11  &    &    &   \tabularnewline

\hline \hline 
\end{tabular}
\caption[Initial elemental compositions of the most abundant species]
{Initial elemental compositions of the most abundant species in the ISM (\citealt{FP03}).}
\label{tab:elemental_compostion}
\end{table}

\begin{table}
\centering
\begin{tabular}{c c c c c c c c}
\hline \hline 
R1 & \hspace{1mm} R2 & \hspace{1mm} P1 & \hspace{1mm} P2 & \hspace{1mm} P3 & \hspace{1mm} 
$\gamma$ & \hspace{1mm} $\alpha$ & \hspace{1mm} $\beta$ \tabularnewline
\hline 
                            
H      &    H       &  H$_{2}$      &         &         & 8.14E-17 & 0.5 &          \tabularnewline
H      &    e$^{-}$ &  H$^{+}$      & e$^{-}$ & e$^{-}$ & 9.20E-10 & 0.5 & 157890.0 \tabularnewline
H2     &    e$^{-}$ &  H$^{+}_{2}$  & e$^{-}$ & e$^{-}$ & 1.40E-09 & 0.5 & 179160.0 \tabularnewline
H      &    H$^{+}$ &  H$^{+}$      & H$^{+}$ & e$^{-}$ & 1.30E-13 & 0.5 & 157890.0 \tabularnewline
H      &    H$^{+}_{3}$ &  H$^{+}_{3}$ &  H$^{+}$  & e$^{-}$  &   1.30E-13 & 0.5 & 157890.0 \tabularnewline

\hline \hline 
\end{tabular}
\caption[First five reactions of the full chemical network]
{First five reactions of the full chemical network.}
\label{tab:five_reactions}
\end{table}

\setstretch{1.1} 
\chapter{BOW SHOCK MODEL}
\label{Chapter5}

\lhead{Chapter 5. \emph{Bow shock model}} 
As in \cite{Gustafsson10}, we assume that the 3D bow shock is made of independent planar shocks. 
We actually neglect the curvature effects and the friction between different 1D shock layers, the gradients of entrance conditions in the planar shock models, and the possible geometrical dilution in the post-shock: 
our approximation is valid as long as the curvature radius of the bow shock is large with respect to the emitting thickness of the working surface.

\section{Geometry and coordinate system} \label{sec:geometry_shock}
\begin{figure}
	\includegraphics[width=1\linewidth]
	{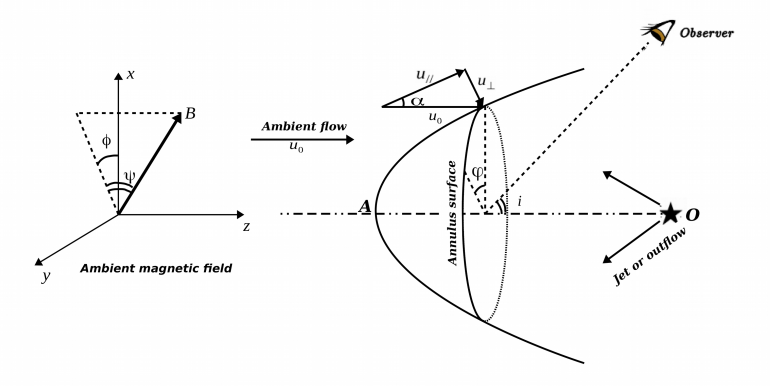}
	\caption[Morphology of a magnetized bow shock]
	{Morphology of a magnetized bow shock in the frame
	of a star or a jet. The direction of the magnetic field is expressed
	by the angles $\psi$ and $\phi$. 
	The observer lies at an angle $i$ to the z-axis
	in the Oxz plane.}
	\label{fig:bowshock_sketch}
\end{figure}

Let's consider an axisymmetric 3D bow shock around a supersonic star (or
a jet) traveling at the speed of $-\textbf{u}_{0}$ relative to an
ambient molecular cloud assumed to be at rest. In the frame of the
star, the impinging velocity of ambient gas is uniform and equal to
$\textbf{u}_{0}$. The apex of the bow shock is at position A and the
star at position O
(\autoref{fig:bowshock_sketch}). The axis of
symmetry (z-axis) along the direction (AO). The observer is assumed to stand in the (Oxz) plane
The axisymmetric shape of the
bow shock is completely determined by the function $z=f(x)$. The local position
along the planar shock can be specified by the angle between the
incoming flow and the tangent to the surface $\alpha =
\arcsin(u_{\bot}/u_{0})$ \cite[see][figure 1]{S90}, and by the angle $\varphi$ between the radius and
the x-axis in the (xy) plane of projection.

The impinging velocity can be expressed  as $\textbf{u}_{0} = \hat{\textbf{t}} u_{\parallel} + \hat{\textbf{n}} u_{\bot} = u_{0}(\hat{\textbf{t}} \cos\alpha + \hat{\textbf{n}} \sin\alpha)$, where $\hat{\textbf{n}}(-\cos\alpha \cos\varphi,-\cos\alpha \sin\varphi, \sin\alpha)$ is the unit normal vector pointing inside the bow and $\hat{\textbf{t}}(\sin\alpha \cos\varphi, \sin\alpha \sin\varphi, \cos\alpha)$ is the unit tangent vector along the working surface and directed away from apex. The effective shock speed at the local point is $v_s=u_{\bot}=u_0 \sin \alpha$. Away from the axis of symmetry, the effective entrance velocity into the shock decreases down to the sound speed $c_s$ in the ambient medium. Beyond this point, the shock working surface is a cone of opening angle $\alpha_0=\arcsin(c_s/u_0)$, wider as the terminal velocity is closer to the sound speed. Here, we mainly focus on the ``nose'' of the bow shock where $u_{\bot}>c_s$, and we neglect the very weak emission from these sonic conical ``wings'', assuming that they fall outside the observing beam.

The orientation of the line-of-sight of the observer in the $(x,z)$ plane is defined by the inclination angle $i$: $\hat{\textbf{l}}(\sin i, 0, \cos i)$. The ambient uniform magnetic field is identified by the obliqueness $\psi$ and the rotation $\phi$:  $\textbf{B}/ B_0 = (\cos \psi \cos\phi,\cos\psi \sin\phi, \sin\psi)$. $\psi$ and $\phi$ are fixed for each bow shock. 

\section{Distribution of effective 1D-shock velocity}
\label{sec:1D_distribution}
This section aims at computing the fraction $P(u_{\bot})\mathrm{d}u_{\bot}$ of planar shocks
with an entrance shock speed $u_{\bot}$ within
$\mathrm{d}u_{\bot}$ in a given bow shock shape. This will help us
building a model for the full bow shock from a grid of planar shocks.

Considering the shock geometry as described in \autoref{sec:geometry_shock}, we aim at obtaining the formula for the unit area $ds$
corresponding to these shocks as a function of $du_{\bot}$. 

The norm of a segment $dl$ on the $(x,z)$ section of the bow shock surface is: 
\begin{equation} \label{eq:dl}
	dl = \sqrt{dx^{2} + dz^{2}} = \sqrt{1 + f'^{2}(x)}dx.
\end{equation} 

Now, take that segment and rotate it  around the $z$-axis, over a circle of radius $x$. The
area ($ds$) of the bow shock's surface swept by this segment can be expressed as:
\begin{equation} \label{eq:ds}
	ds = 2\pi x dl = 2\pi x \sqrt{1 + f'^{2}(x)}dx \mbox{.}
\end{equation}

Note that the angle $\alpha$ defined in \autoref{fig:bowshock_sketch}
is also the angle between the segment $dl$ and the differential
length $dz$ along the $z$-axis. The tangent of the angle $\alpha$ can then 
be set as
\begin{equation} \label{eq:tan_alpha}
	\tan \alpha = \frac{dx}{dz} = \frac{1}{f'(x)} \mbox{.}
\end{equation}

The relationship between $\alpha$ and $u_{\bot}$ will be
realized according to whether we consider the shock in the ambient medium or in the
stellar wind or jet. Then, $ds$ can be obtained as a function of $du_{\bot}$ by
replacing that relation into \autoref{eq:ds}. 
However, we will only focus here on the bow shock in the ambient material. In that case, the norm of the
effective velocity (i.e., the effective normal velocity $u_{\bot}$) is 
related to the norm of the incident velocity $\textbf{u}_{0}$ through
the angle $\alpha$
\begin{equation} \label{eq:alpha}
	u_{\bot} = u_{0} \sin \alpha \rightarrow \alpha = \arcsin(\frac{u_{\bot}}{u_{0}}) \mbox{.}
\end{equation}

$x$ can now be expressed as a function of $u_{\bot}$
by substituting \autoref{eq:alpha} into \autoref{eq:tan_alpha}:
\begin{equation} \label{eq:x_vs_u}
	\tan[\arcsin(\frac{u_{\bot}}{u_{0}})] = \frac{1}{f'(x)} \rightarrow x = f'^{-1}\lbrace  \cot[\arcsin(\frac{u_{\bot}}{u_{0}})] \rbrace = g(u_{\bot}) \mbox{.}
\end{equation} 

In \autoref{eq:ds}, the unit area $ds$ of the shock is a function of the
coordinate $x$, while in \autoref{eq:x_vs_u}, the coordinate $x$ is a
function of the effective shock velocity $u_{\bot}$. To sum up, we
can obtain $ds$ as a function of $u_{\bot}$:
\begin{equation}\label{eq:ds_ambient}
	\begin{split}
	ds(u_{\bot}) & = 2\pi g(u_{\bot}) \sqrt{1 + \cot^{2}[\arcsin(\frac{u_{\bot}}{u_{0}})]}g'(u_{\bot})du_{\bot} \\ 
	            & = \pi \sqrt{1 + \cot^{2}[\arcsin(\frac{u_{\bot}}{u_{0}})]}\ d[g^{2}(u_{\bot})]
	\end{split}
\end{equation}

Finally, the distribution function of shock velocities is simply defined as
\begin{equation} \label{eq:pdf} P(u_{\bot}) =
  \frac{ds(u_{\bot})}{\int_{c_{s}}^{u_{0}}ds},
\end{equation}
where the integral of $P(u_{\bot})$ is normalized to unity. Note that the lower limit of the integral is the sound speed in the ambient medium. This implicitly assumes that we only focus on the ``nose'' of the bow shock, where $u_\bot<c_{s}$. One could include the conical ``wings'' by adding a Dirac distribution $\delta(u_\bot=c_s)$. Conversely, one could also narrow down the integration domain if the beam intersects a smaller fraction of the bow. We implemented this mathematical formulation numerically to compute the distribution $P$ from an arbitrary input function $f$. The results we obtained agree with those obtained using the analytical expressions when the shape assumes a power-law dependence $z \sim x^\beta$ (see \aref{app:pdf_accuracy} for detail).

In an elegant and concise article, \cite{W96} derived an analytical
description of the shape of a bow shock around a stellar wind when it is dominated by the
ram pressure of the gas. When dust grains control
the dynamics of the gas, the main forces are the gravitation pull and
the radiation pressure from the star, therefore the shape of the shock should be very close to the grains avoidance parabola derived in \cite{A97}. 
In fact, the ISM mixes gases and dust grains, so the actual bow shock shape should lie 
in-between.  

If dust dominates, the bow shock shape is
the Artymowicz parabola expressed as $z = \frac{1}{4R_{0}}x^{2} -R_{0}$ with $R_{c} = 2R_{0}$ the curvature radius at apex, $R_0$ being the star-apex distance. 
If gas dominates, the bow shock shape follows the Wilkin formula 
$R = \frac{R_{0}}{\sin \theta}\sqrt{3}\sqrt{1-\theta \cot \theta}$ with $R_{c} = 5/4R_{0}$ the curvature radius at the apex and $\theta$ the polar angle from the axis of symmetry as seem by star.

\begin{figure}
	\begin{minipage}[c]{1\textwidth}
		\includegraphics[width=1\linewidth]{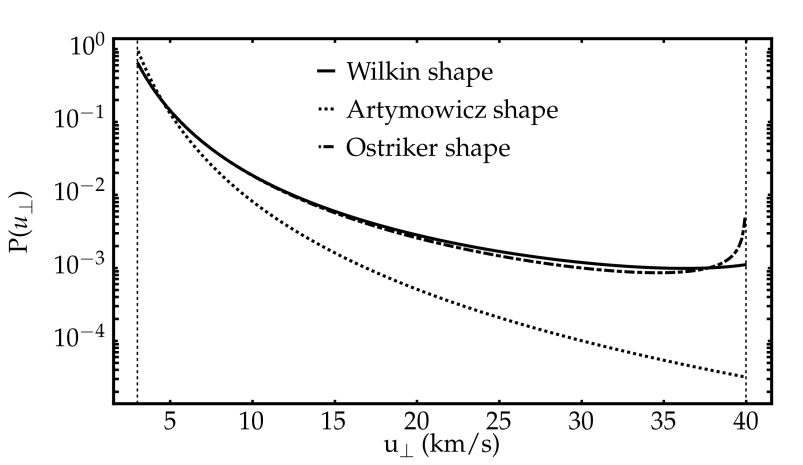}
	\end{minipage} \\
	\caption[Statistical distributions of 1D planar shock along the bow shock 
	obtained for various bow shock shapes]
	{Statistical distributions of 1D planar shock along the bow shock 
	obtained for various bow shock shapes. 
	These distributions are dominated by low-velocity shocks.}
	\label{fig:pdf}	
\end{figure}

\begin{figure}
	\begin{minipage}[c]{1\textwidth}
		\includegraphics[width=1\linewidth]{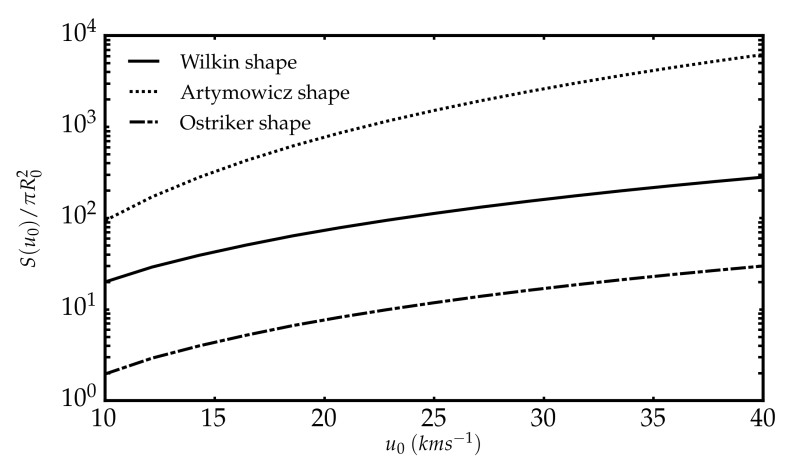}
	\end{minipage}\\
	\caption[Total surface of the bow shock for various bow shock shapes 
	and terminal velocities]
	{Total surface of the bow shock for various bow shock shapes and terminal velocities, 
	in units of $\pi R^{2}_{0}$, where $R_{0}$ is the length-scale parameter 
	of the bow (on the order of the nose’s curvature radius).}
	\label{fig:S}
\end{figure}

Finally, in the case of the tip of a jet, \cite{Ostriker01} showed that the shape of the bow shock was cubic $z=x^{3}/R_0^2 - z_{j}$ with an infinite curvature radius at apex ($R_0$ and $z_j$ are length-scale parameters). \autoref{fig:pdf} displays the distributions obtained for
various bow shock shapes.  Note that low-velocity shocks (u$_{\bot}$ $\leq$ 15 km$\,$s$^{-1}$) always dominate
the distribution: this stems from the fact that the corresponding surface increases further away from the
axis of symmetry, where entrance velocities decrease. The distribution for the cubic shape has a spike
due to its flatness (infinite curvature radius) near the apex. The Wilkin shape has a cubic tail but a parabolic nose. 
 In \autoref{fig:S} we display the dimensionless surface $S/ \pi R_0^2$ where $S$ is the total surface of the bow shock out to $u_{\perp}$=$c_{s}$, and $R_0$ an estimate of the radius of the nose of the bow. For elongated shapes such as the parabolic shape, the total surface can be much bigger than the nose cross-section  $\pi R_0^2$. We will subsequently essentially consider an ambient shock with a parabolic shape (Artymowicz shape).

\section{Distribution of effective magnetic field}
\begin{figure}
   \begin{minipage}[c]{1\textwidth}
      \includegraphics[width=0.9\linewidth]
      {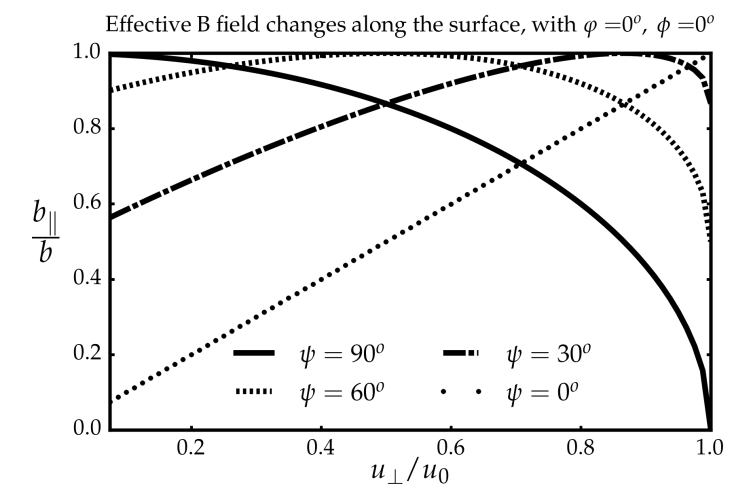}
   \end{minipage} \\
   \begin{minipage}[c]{1\textwidth}
      \includegraphics[width=0.9\linewidth]
      {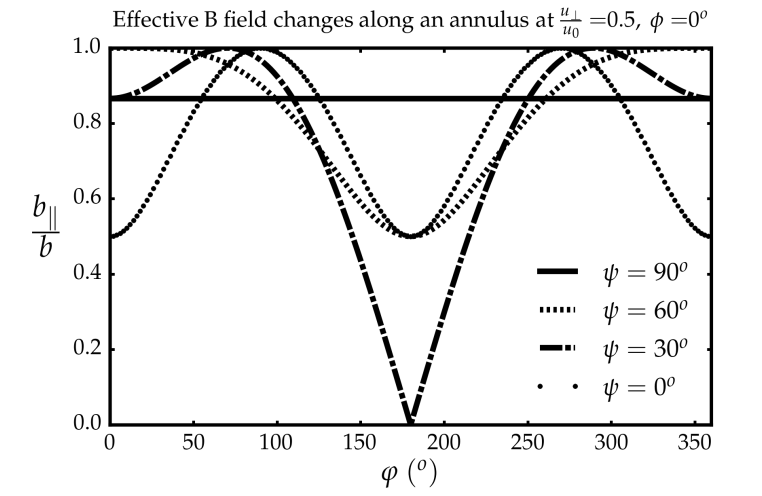}
   \end{minipage}\\

	\caption[Variation of the effective transverse magnetic field 
	$b_{\parallel}$ along the bow shock surface]
	{Variation of the effective transverse magnetic field 
	$b_{\parallel}$ along the bow shock surface for various directions 
	of \textbf{b} (\textit{top}) and a fixed direction 
	depending on the position on an annulus circle (\textit{bottom}).} 
	\label{fig:orientation_b}
\end{figure}

The magnetic field  decouples the ions from the neutral fluid in the shock. However, as discussed by \cite{S92}, the effective magnetic field is the component of the field parallel to the shock surface. The vector of the magnetic field can be expressed as 
$\textbf{B} = \hat{\textbf{t}}B_{\parallel} + \hat{\textbf{n}}B_{\bot}$, where the perpendicular component is defined by $B_{\bot} = \hat{\textbf{n}}.\textbf{B}$. 
If the homogeneous pre-shock density is $n_H$, the strength scale factor of the ambient uniform magnetic field is defined as $b_0 = B_0(\mu G)/\sqrt{n_{H} [cm^{-3}]}$.
The component of the field parallel to the working surface 
(scale factor $b_{\parallel} = B_{\parallel}(\mu G)/\sqrt{n_{H} [cm^{-3}]}$ ) is therefore given by 
\begin{equation} \label{eq:b-parallel}
	\left(\frac{b_{\parallel}}{b_0}\right)^2 = \cos^2 \alpha \sin^2 (\varphi-\phi)+ [\cos \psi \sin \alpha+ \sin \psi \cos \alpha \cos (\varphi - \phi) ]^2
\end{equation}

where the angles $\alpha$ and $\varphi$ monitor the position in the bow shock (this expression is actually valid regardless of the bow shock shape). 
\autoref{fig:orientation_b} displays how this component ($b_{\parallel}$) changes along the shock surface in a few cases. For each given direction of the ambient magnetic field ($\phi, \psi$ fixed), 
$b_{\parallel}$ actually varies along two directions: along the bow shock shape ($\varphi$ fixed, $\alpha$ varied), visualized in the \textit{top panel}, and along the annulus surface at one position on the bow shock surface ($\alpha$ fixed, $\varphi$ varied), visualized in the \textit{bottom panel}. 

\section{Grid of 1D-shock models}
We set all the parameters to values corresponding to typical conditions encountered in the molecular interstellar gas in our Galaxy (\autoref{tab:input_parameters}).
We assume that the ambient gas is initially at chemical and thermal equilibrium and we compute the initial state as in \cite{PL13} by evolving the gas at constant density  during $10^{12}/n_{H}\,$yr. Our initial elemental abundances in the gas, grain cores and ice mantles are the same as in \cite{FP03} (\autoref{tab:elemental_compostion}). We also include PAHs with ratio $n(PAH)/n_{H} = 10^{-6}$. The irradiation conditions are for a standard external irradiation field ($G_0=1$) but an additional  buffer of $A_{v 0}=0.1$, $N_0({\rm H}_2)=10^{20}\,$cm$^{-2}$ and $N_0(CO)=0\,$cm$^{-2}$ is set between the source and the shock so that the gas is mainly molecular \citep{PL13}. In our calculations, the atomic hydrogen fractions $n(H)/n_{H}$ are $7.85\ 10^{-2}$, $5.94\ 10^{-4}$ and  $5.89\ 10^{-6}$ for pre-shock gas densities of $10^{2}$, $10^{4}$ and $10^{6}\,$cm$^{-3}$, respectively. These initial conditions at steady state are then used as pre-shock conditions to compute the grid of planar shock models. 

 The shock velocities range from 3 to 40 km$\,$s$^{-1}$ as in \cite{PL13}, with a step of $\Delta u = 1\,$km$\,$s$^{-1}$. However, we take into account the effect of the finite shock age by considering snapshots at 4 different values of age: $10^{2}$, $10^{3}$, $10^{4}$, and $10^{5}$ years for a density of $n_H=10^2~$cm$^{-3}$, and a hundred times shorter for a density of $n_H=10^4~$cm$^{-3}$. Note that the typical time to reach the steady-state in a C-type shock with $G_0=1$ is about $t_s=10^6 $yr$ / (n_H/10^2\textrm{cm}^{-3})$ \cite[with little or no magnetic field dependence][]{PL04a}. 

\begin{figure}
	\includegraphics[width=1.\linewidth]
     	  {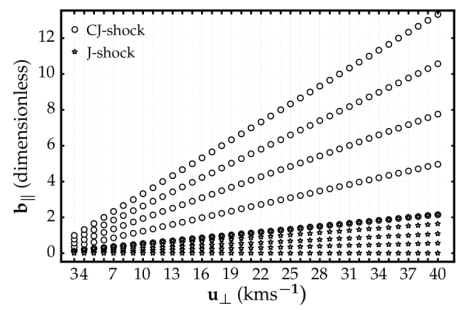}   
          \caption[Grid input paramters]
          {Grid of 1D models in the parameter space$(u_{\bot},b_{\parallel})$.}
          \label{fig:grid_model}
\end{figure}

  The projected value of the magnetic field parallel to the shock $B_{\parallel}$ varies along the shock surface, so we need to sample the range of attainable values in our grid. The first constraint for a shock to exist is that its entrance velocity $u_\bot$ be greater than the Alfv{\'e}n velocity $v_{A} =\frac{B_{\parallel}}{\sqrt{4\pi\rho}} \simeq b_{\parallel}\,1.85\,$km$\,$s$^{-1}$ where we defined the dimensionless value of the transverse magnetic field using the standard scaling $b_{\parallel}=B_{\parallel}/\mu $G$ / (n_H/{\rm cm}^{-3})^{1/2}$. The condition $u_\bot>v_{A}$ translates as $b_{\parallel}<u_\bot/1.85\,$km$\,$s$^{-1}$, and we use as upper limit of our grid
$b_{\parallel}<b_{\parallel{\rm max}}=u_{\bot}$/3 km$\,$s$^{-1}$ (\autoref{fig:grid_model}).

  Another important parameter is the magnetosonic speed in the charged fluid (\autoref{eq:magnetosonic_speed}) defined as the fastest signal speed in a partially ionized medium. Due to the low ionization degree in the molecular ISM, it is almost proportional to the local magnetization parameter: $v_{m} \simeq \sqrt{B_{\parallel}^2/4\pi\rho_c}=b_{\parallel}v_{m1}$ where $v_{m1}$ is the magnetosonic speed obtained when the magnetization parameter is equal to unity. In our calculations, we find $v_{m1}= 18.5$  km$\,$s$^{-1}$ or $v_{m1}= 19.2$  km$\,$s$^{-1}$ for respective densities of $n_{H}= 10^{2}\,$cm$^{-3}$ or $n_{H}= 10^{4}\,$cm$^{-3}$. The charged fluid mass is dominated by the dust grains: the gas-to-dust ratio turns out to be $\rho/\rho_{d} = 180$ for the cores and mantle composition used in our simulations.

\begin{table}
\centering
\begin{tabular}{l c l}
\hline \hline 
Parameter & \hspace{10mm} Value & \hspace{1mm} Note \tabularnewline
\hline 
$n_{H}$ & $10^{2}$cm$^{-3}$, $10^{4}$cm$^{-3}$  & Pre-shock density of H nuclei \tabularnewline

$A_{\nu}$ & 0.1 & Extinction shield \tabularnewline

$N_{0}(H_2)$ & $10^{20}$cm$^{-2}$ & Buffer H$_{2}$ column density \tabularnewline

$N_{0}(CO)$ & 0 cm$^{-2}$ & Buffer CO column density \tabularnewline

$G_{0}$ & 1 & External radiation field \tabularnewline

$\zeta$ & $3.10^{-17}$ s$^{-1}$ & Cosmic ray flux \tabularnewline

OPR & 3 & Pre-shock H$_{2}$ ortho/para ratio \tabularnewline

$u_{\bot}$ & $3, 4, 5, \dots 40$ km$\,$s$^{-1}$ & Effective shock velocity \tabularnewline

$b_{\parallel}v_{m1}/u_{\bot} $ & $0, \dots, 1$   & Range of $b_{\parallel}$ parameter for J-type shocks \tabularnewline

$b_{\parallel}v_{m1}/u_{\bot} $ & $1, \dots, \frac{v_{m1}}{3\,\textrm{km\,s}^{-1}}$ & Range of $b_{\parallel}$ parameter for CJ-type shocks \tabularnewline

age $\times n_H/$100 cm$^{-3}$ & $10^{2}$, $10^{3}$, $10^{4}$, $10^{5}\,$yr& Shock age \tabularnewline
\hline \hline 
\end{tabular}
\caption[Main input parameters of bow shock model]
{Main input parameters of model.}
\label{tab:input_parameters}
\end{table}

\section{H$_{2}$ excitation diagram}
The average column-density of a given excited level of H$_2$ along the bow shock can be expressed as:
\begin{equation}
N^{\rm tot}_{vJ}(age,u_0,b_0,\psi)= \int_0^{2\pi}\frac{d\varphi}{2 \pi} \int_{c_{s}}^{u_{0}}  P_{u_0}(u_\bot)  N_{vJ}(age,u_{\bot},b_{\parallel}) du_{\bot}
\end{equation}
where $P_{u_0(u_\bot)}$ is the distribution computed in section 2 and $N^{\rm tot}_{vJ}$ and $N_{vJ}$ are the column-densities
of H$_{2}$ in the excited level $(v,J)$ in the whole bow shock and in each planar shock, respectively.

\subsection{H$_{2}$ excitation in C-type and J-type shocks}
\begin{figure}
 	\begin{minipage}[c]{.5\textwidth}
 	 	\includegraphics[height=0.25\textheight]
 	 	{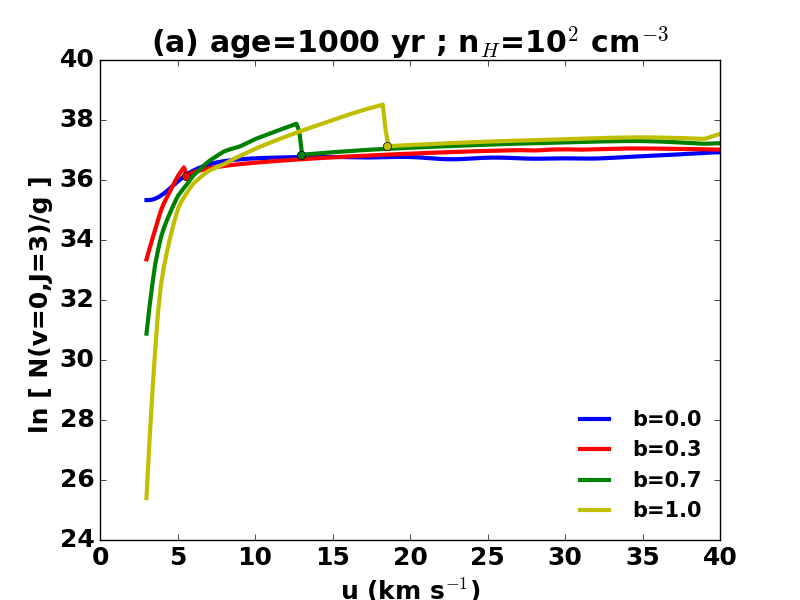}
 	\end{minipage} \hfill
 	\begin{minipage}[c]{.5\textwidth}
 	 	\includegraphics[height=0.25\textheight]
 	 	{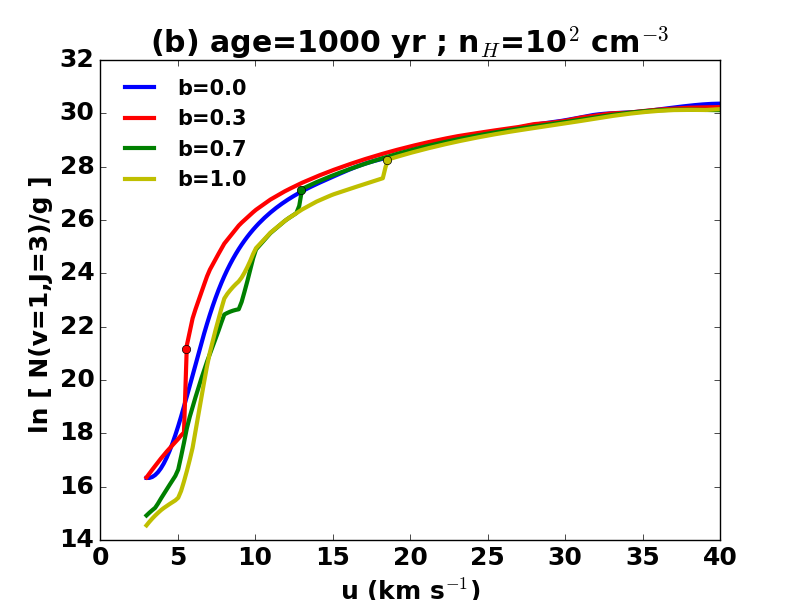}
 	\end{minipage} \hfill
 	\begin{minipage}[c]{.5\textwidth}
 	 	\includegraphics[height=0.25\textheight]
 	 	{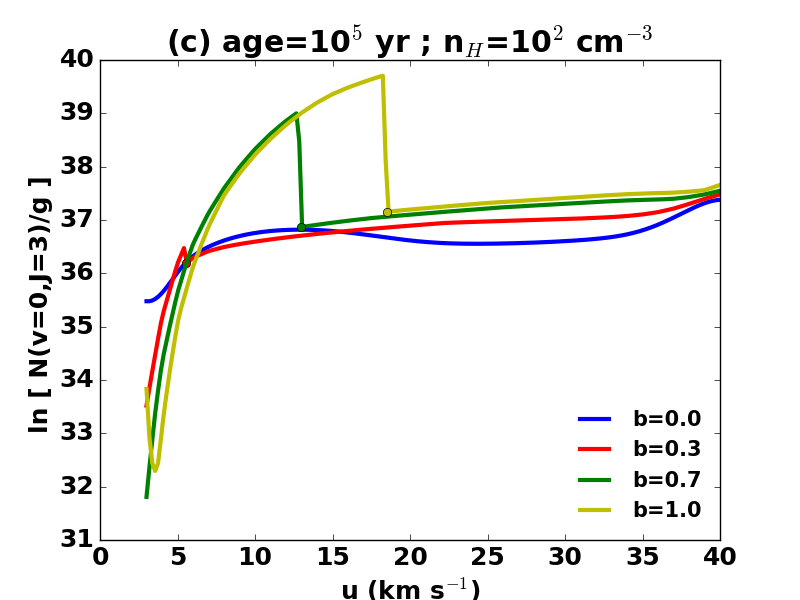}
 	\end{minipage} \hfill
	\begin{minipage}[c]{.5\textwidth}
 	 	\includegraphics[height=0.25\textheight]
 	 	{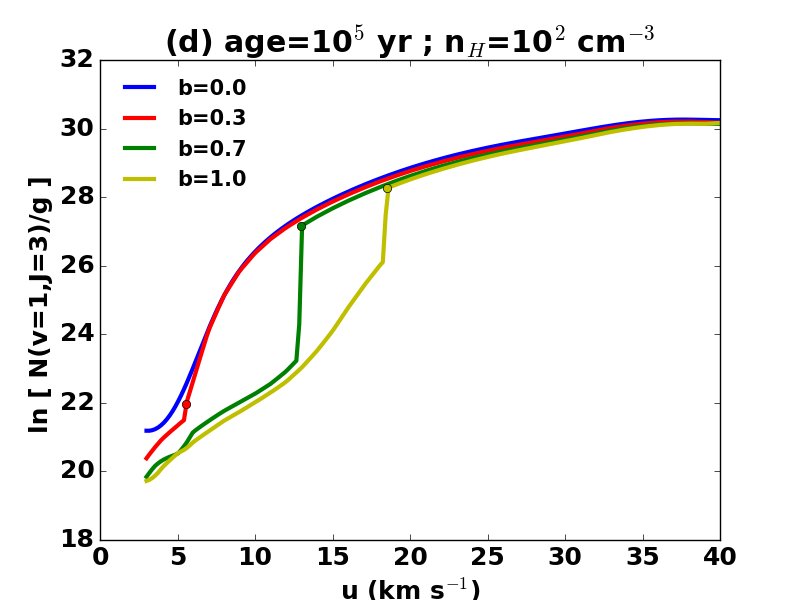}
 	\end{minipage} \hfill	 
   \caption[Natural logarithm of the integrated 
   column-densities of H$_2$ populations normalized by their 
   statistical weight as a function of the velocity 
   $u$ for various values of the  magnetic field parameter $b_{\parallel}$ 
   for a pre-shock density 
   $n_H=10^2\,$cm$^{-3}$]
   {Overview of our model results for a pre-shock density 
   $n_H=10^2$/cm$^3$. The natural logarithm of the integrated 
   column-densities of H$_2$ populations is normalized by their 
   statistical weight. They are given as a function of the velocity 
   $u$ for various values of the  magnetic field parameter $b_{\parallel}$. 
   {\it Left} panels are for the level $(v,J)=(0,3)$, the upper level of the 
   0-0S(1) line and the {\it right} panels are for the level $(v,J)=(1,3)$, 
   the upper level of the 1-0S(1) line. {\it Upper} panels are for a 
   young age of 10$^3$ yr while {\it bottom} panels are nearly 
   steady-state at an age of $10^5$ yr. In each panel, 
   the symbol 'o' marks the transition between CJ-type shocks 
   (on the left-hand side) and J-type shocks (on the right hand side), 
   when the velocity $u$ is equal to the magnetosonic speed $b v_{m1}$.}
  \label{fig:H2excited-nH1e2.png}
\end{figure}

A H$_2$ rovibrational level $(v,J)$ can be populated after a collision with another species, mainly H$_{2}$, H, He and e$^{-}$ provided that the temperature yields more energy per particle than the energy level $E_{vJ}$. In a J-type shock, the sudden surge of viscous heat in the adiabatic shock front easily leads to high temperatures ($T_J=53 K (u/$km$\,$s$^{-1})^2$, \cite{PL13}) which are able to excite high energy levels. The plots a and b of \autoref{fig:H2excited-nH1e2.png} show the level populations for young ages, where even CJ-type shocks are dominated by their J-type tail contribution. These figures illustrate the threshold effect for two different energy levels: their population quickly rises and reaches a plateau when $u>u_{vJ}$, with $u_{vJ}$ a critical velocity depending on the energy level. Note the weak dependence of the plateau on the shock magnetization for J-type shocks, as magnetic pressure only marginally affects their thermal properties. The critical velocity $u_{vJ}$ mainly depends on the energy level ($E_{vJ}\simeq k_BT_J$) and only weakly depends on the magnetization. 

On the other hand, C-type shocks dissipate their energy through ion-neutral friction, a process much slower than viscous dissipation: at identical velocity, C-type shocks are much cooler than J-type shocks, but their thickness is much larger. C-type shocks dominate the emission of old CJ-type shocks, when the J-type front contribution almost disappears (\autoref{fig:H2excited-nH1e2.png}c in C-type shocks). Due to their low temperature, high energy levels can never be populated (\autoref{fig:H2excited-nH1e2.png}d). This enhances the threshold effect, with a discontinuous jump at $u=b v_{m1}$. On the contrary, energy levels lower than $k_B T_C$, with $T_C$ the typical temperature of a C-type shock, will be much more populated in a C-type shock than in a J-type shock due to the overall larger column-density. This is illustrated in the \autoref{fig:H2excited-nH1e2.png}c for a low energy level. The discontinuous jump at  $u=b v_{m1}$ becomes a drop instead of a surge and a peak appears in the level population. Magnetization in C-type shocks controls the compressive heating which, in turn, impacts the temperature: excitation of  H$_2$ low-energy levels in C-type shocks decreases systematically with larger magnetization, but the effect remains weak within C-type shocks. However, the magnetization is important insofar as it controls the transition between C-type and J-type shocks, which have very different emission properties.

 To summarize, at a density of $n_H=10^2\,$cm$^{-3}$, the column density excitation of a given H$_2$ level follows a threshold in velocity after which a plateau is reached, with little or no magnetic field dependence. However, low energy levels at old ages, for velocities below the magnetosonic speed, can be dominated by C-type shock emission. 
In that case, the H$_2$ level population peaks at the magnetosonic speed before reaching a plateau. Therefore, H$_2$ emission in bow shocks is likely to be mostly dominated by J-type shocks. 

\begin{figure}
	\begin{minipage}[c]{.5\textwidth}
 	 	\includegraphics[height=0.24\textheight]
 	 	{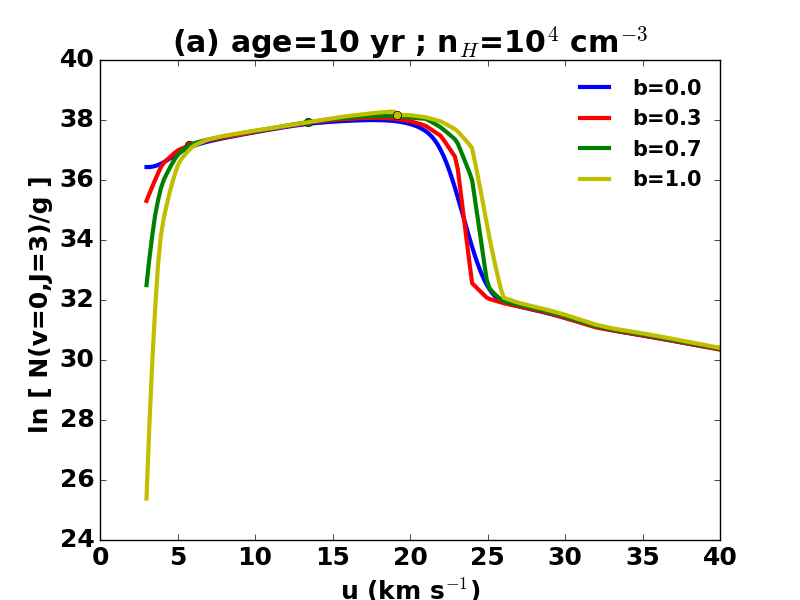}
 	\end{minipage} \hfill
 	\begin{minipage}[c]{.5\textwidth}
 	 	\includegraphics[height=0.24\textheight]
 	 	{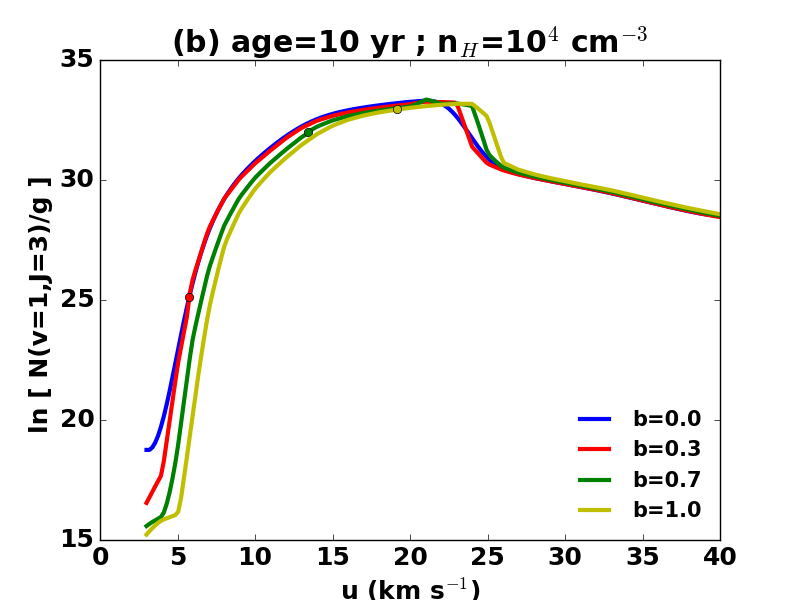}
 	\end{minipage} \hfill
  	\begin{minipage}[c]{.5\textwidth}
 	 	 \includegraphics[height=0.24\textheight]
 	 	 {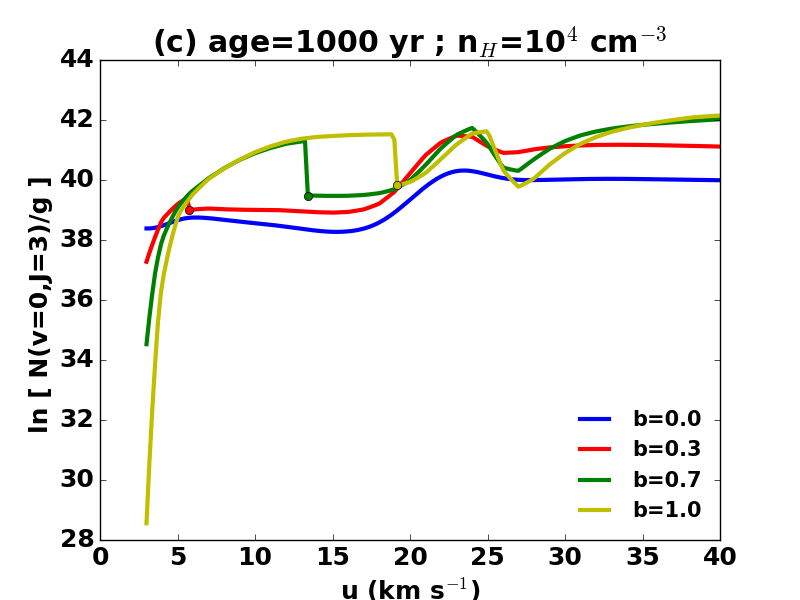}
 	\end{minipage} \hfill
 	\begin{minipage}[c]{.5\textwidth}
 	 	\includegraphics[height=0.24\textheight]
 	 	{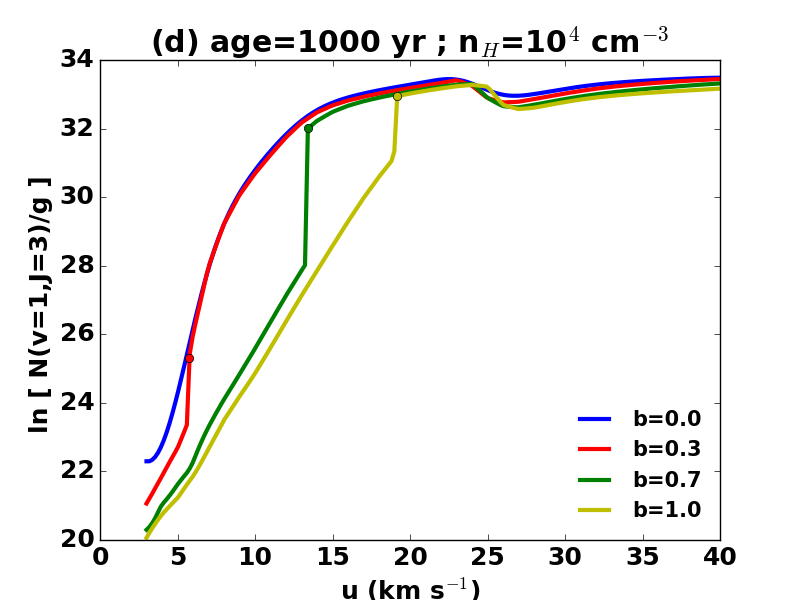}
 	\end{minipage} \hfill
  	\caption[Natural logarithm of the integrated 
   			column-densities of H$_2$ populations normalized by their 
   			statistical weight as a function of the velocity 
   			$u$ for various values of the 
   			magnetic field parameter $b_{\parallel}$ 
   			for a pre-shock density 
   			$n_H=10^4\,$cm$^{-3}$]
  			{Same as \autoref{fig:H2excited-nH1e2.png} but for 
  			the denser case $n_H=10^4\,$cm$^{-3}$. 
  			The corresponding ages are: {\it upper} panels at 
  			a young age of 10 yr while the {\it bottom} panels 
  			are nearly steady-state at an age of $10^{3}\,$yrs.}
  \label{fig:H2excited-nH1e4.png}
\end{figure}

  At higher density of $10^{4}\,$cm$^{-3}$, the picture is essentially unchanged, except for the effect of H$_2$ dissociation which is felt when the velocity is larger than the H$_2$ dissociation velocity ($v_{s}\sim 25\,$km$\,$s$^{-1}$): the value of the plateau decreases beyond this velocity (see the right half of each panel in \autoref{fig:H2excited-nH1e4.png}, which is in other respects similar to \autoref{fig:H2excited-nH1e2.png}).  At even higher densities, H$_2$ dissociation completely shuts off H$_2$ emission in J-type shocks, and we reach a situation where the bow shock emission is dominated by C-type shocks, as in \cite{Gustafsson10}.

\begin{figure}
	\includegraphics[width=1\linewidth]
	{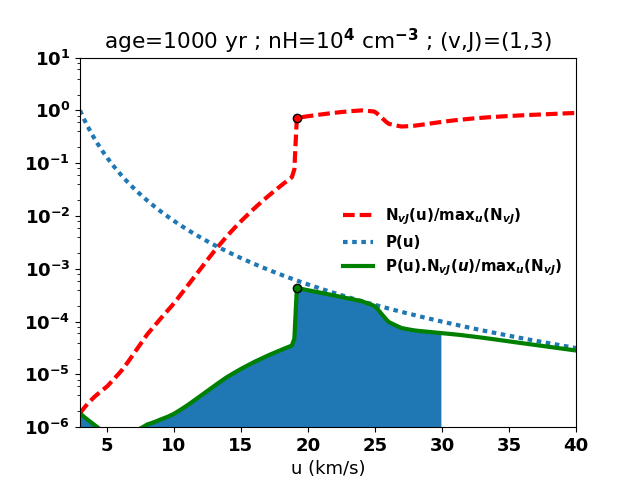}
	\caption[Illustration of the ``Gamow-peak'' effect on 
        the integration of the total column densities of the 
        H$_2$]{Illustration of the ``Gamow-peak'' effect on 
        the integration of the total column densities of the 
        H$_2$ level $(v,J)=(1,3)$ in a bow shock with 
        terminal velocity $u_0=30\,$km$\,$s$^{-1}$, $n_H=10^2$cm$^{-3}$, 
        and the age is 10$^{5}\,$yr.} 
        \label{fig:illustration-convolution}
\end{figure}

  As demonstrated, $N_{vJ}$ sharply increases as a function of $u_{\bot}$ at a given threshold velocity $u_{vJ}$ before reaching a plateau. We also showed that the statistical distribution of shock speeds P(u$_{\perp}$) in a bow shock was steeply decreasing as a function of $u_{\bot}$. As a result, the product of the two peaks at around $u_{vJ}$ and its integral over $u_{\bot}$ is a step function around $u_{vJ}$ (\autoref{fig:illustration-convolution}). This situation is reminiscent of the Gamow peak for nuclear reactions. Then, $N^{\rm tot}_{vJ}(u_0)$ tends to a finite value when $u_0$ is much greater than the threshold velocity $u_{vJ}$. The final value depends both on magnetization and age.

\begin{figure}
	\begin{minipage}[c]{.5\textwidth}
 	 	\includegraphics[height=0.24\textheight]
 	 	{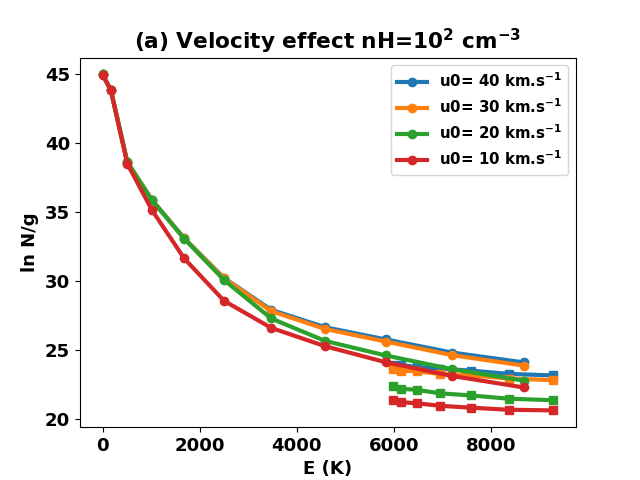}
 	\end{minipage}	\hfill
	\begin{minipage}[c]{.5\textwidth}
 	 	 \includegraphics[height=0.24\textheight]
 	 	 {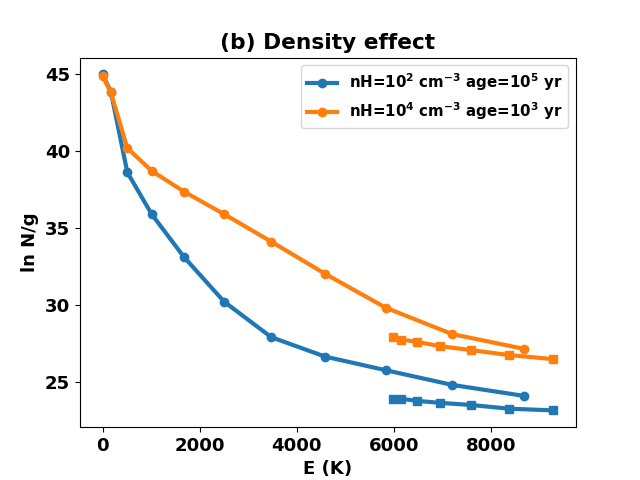}
 	\end{minipage} \hfill 
 	\begin{minipage}[c]{.5\textwidth}
 	 	 \includegraphics[height=0.24\textheight]
 	 	 {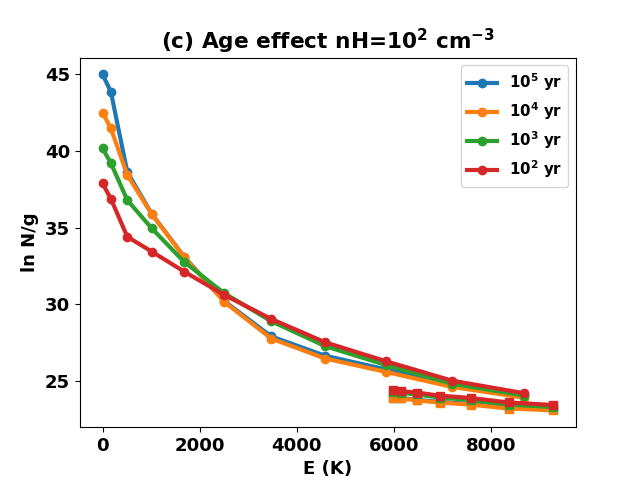}
 	\end{minipage} \hfill	
 	\begin{minipage}[c]{.5\textwidth}
 	 	 \includegraphics[height=0.24\textheight]
 	 	 {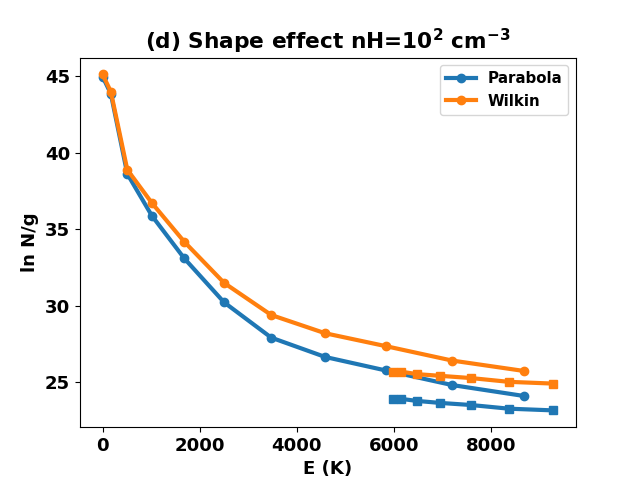}
 	\end{minipage} \hfill 
 	\begin{minipage}[c]{.5\textwidth}
 	 	 \includegraphics[height=0.24\textheight]
 	 	 {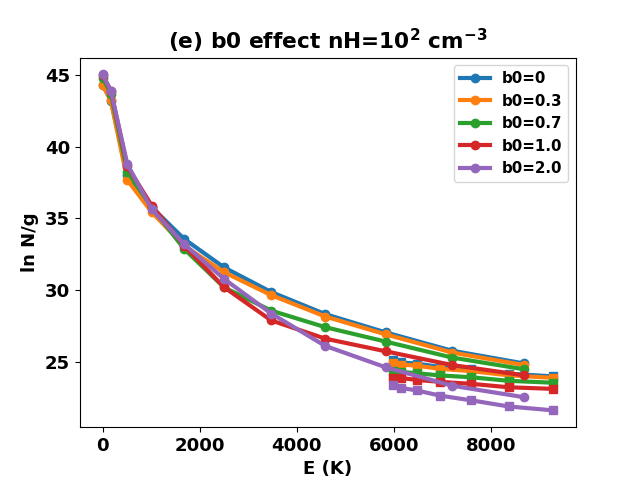}
 	\end{minipage} \hfill
 	\begin{minipage}[c]{.5\textwidth}
 	 	 \includegraphics[height=0.24\textheight]
 	 	 {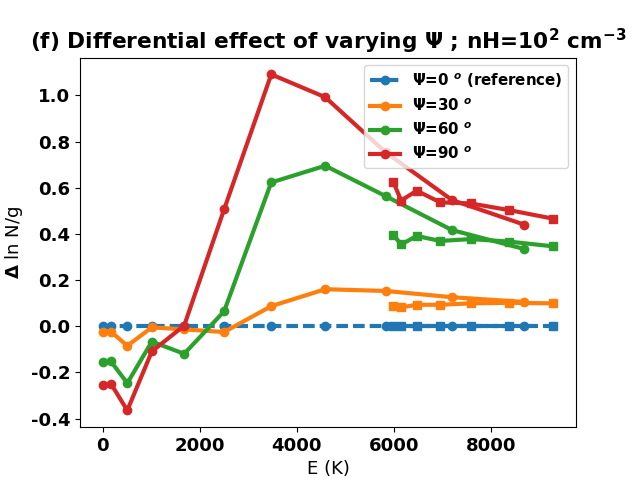}
 	\end{minipage} \hfill
    \caption[Excitation diagrams of H$_{2}$ 
    showing the effect of varying some of the parameters of the model]
    {Excitation diagrams of H$_{2}$ 
    showing the effect of varying some of the parameters of the model. 
    The reference model (n$_{H}$=10$^{2}\,$cm$^{-3}$, age=10$^{5}\,$yr, 
    b$_{0}$=1, $\Psi$=0, 
    parabola shape, u$_{0}$=40 km$\,$s$^{-1}$) is displayed in blue. 
    The circle symbols correspond to $v=0$ (pure rotational levels) 
    and the square symbols to $v=1$.}
      \label{fig:h2_diagram}
\end{figure}

\subsection{Effect of the terminal velocity}
\autoref{fig:h2_diagram}a shows the influence of the terminal velocity on the excitation diagrams of H$_{2}$ at an age of $10^{4}\,$yr. As expected, the excitation diagram saturates at large velocity, when $u_0$ is larger than all the individual $u_{vJ}$ of the levels considered. That saturation occurs quicker at low energy levels, as the corresponding critical velocity is lower.

\subsection{Effect of the ambient density}
\autoref{fig:h2_diagram}b illustrates the effect of density on the excitation diagram.  Roughly speaking, the column-densities are proportional to the density, but in this example (40 km$\,$s$^{-1}$ bow shock), higher energy levels are subject to H$_2$ collisional dissociation, and they are slightly less populated relative to their low energy counter part.

\subsection{Effect of shock age}
At young ages, shocks are dominated by the emission properties of J-shock: as time passes, C-type shocks increase the emission of low energy levels and the excitation diagram of the bow shock is slightly steeper at the origin 
(\autoref{fig:h2_diagram}c). Interestingly, the energy level just above 2000K does not seem to be affected by age (it is also weakly affected by all of the other parameters) and all the curves converge on this point.

\subsection{Effect of shock shape}
As mentioned in \autoref{sec:1D_distribution}, the shape of bow shocks affects the velocity distribution and the relative weight of the large velocities increases when one moves from a parabola to a Wilkin shape. As a result, a bow shock with a Wilkin shape has more excited high energy levels than a parabolic bow shock (\autoref{fig:h2_diagram}d).

\subsection{Effect of ambient magnetic field}
The magnetic field tends to shift the transition between C-type and J-type shocks in the bow shock to larger velocities. At early age, it does not matter much, since both C-type and J-type shocks are dominated by J-type shock emission. At later ages, though, the low energy levels get an increasing contribution from C-type shocks and see their excitation increase.  
Conversely, high energy levels are less excited because the overall temperature of the shock decreases (\autoref{fig:h2_diagram}e). The orientation of the magnetic field azimuthally affects the range of values of $b$ (as $\varphi$ varies) but its main systematic effect is to shift the maximum magnetization from low velocities to large velocities as it gets more and more perpendicular to the axis of symmetry (\autoref{fig:orientation_b}). \autoref{fig:h2_diagram}f shows the differential effect caused by varying the angle $\Psi$: tending $\Psi$ to 0$^{o}$ amounts to increasing $b$ (high energy levels are less excited, whereas low energy levels are more excited). The resulting change is subtle but we show below that it might still be probed by observations.

\subsection{Bias between 1D- and 3D-shock models}
\label{sec:bias}
Observations often consider low energy transitions (pure rotational or low vib-rotational levels): although we included the first 150 levels in our calculations, here we mainly consider the levels with an energy up to $10^{4}$K. The two lowest rotational states (J=0 and 1) are, of course, unobservable in emission. The \textit{James Webb Space Telescope} (JWST) will observe pure rotational transitions up to energies of about 5900K (seven levels involved). This is similar to the performances of the \textit{Infrared Space Observatory} (ISO) and the \textit{Spitzer} telescope. These two telescopes that have been used to observe shocked regions generate excitation diagrams and maps around Young Stellar Objects (YSOs) (i.e., \citealt{Giannini04}; \citealt{Neufeld09}) or supernova remnants (SNRs) (i.e., \citealt{Cesarsky99}; \citealt{Neufeld14}) shocks. The AKARI mission has also been used for similar purposes in SNRs environments (i.e., \citealt{Shinn11}). The JWST will also target rovibrational transitions. Finally, the \textit{Echelon-Cross-Echelle Spectrograph} (EXES) on board the \textit{Stratospheric Observatory For Infrared Astronomy} (SOFIA) operating between 4.5 and 28.3 $\mu$m \citep{Dewitt14} should allow observations of pure rotational transitions of H$_2$, but no program has been explicitly dedicated to the observation of shocked H$_2$ with this instrument so far.

Most observations are unable to resolve all details of a bow shock, and the beam of the telescope often encompasses large portions of it, therefore mixing together planar shocks with a large range of parameters. 
However, it is customary to use 1D models to interpret observed excitation diagrams. In addition, previous work (NY08; \citealt{Neufeld09}; \citealt{Neufeld14}) have also shown that statistical equilibrium for a power-law temperature distribution T$^{-b_{SE}}$ dT could be quite efficient at reproducing the observed H$_{2}$ pure rotational lines 
(see \autoref{sec:power_law_assumption}). We thus seek to explore how accurately these two simple models perform compared to 3D bow shocks. We consider the worst case scenario where the whole nose of a parabolic bow shock is seen by the telescope: the effective entrance velocity $u_{\bot}$ varies from the speed of sound $c_{s}$ (in the wings of the bow shock) to the terminal velocity $u_{0}$ (at the apex of the bow shock).

The following $\chi$ function is used to estimate the distance between 1D and 3D models:

\begin{equation}
	\chi^{2} = \frac{1}{L}\sum_{vj} [\ln(\frac{N^{\rm tot}_{vj}}{g_{vj}}) - \ln(\frac{N^{u_{\bot}}_{vj}}{g_{vj}}) - C]^{2}
\end{equation}
with $L$ the number of observed rovibrational levels $(v,j)$, and $g_{vj}$ the statistical weight of each level $(v,j)$. The constant $C$ reflects the fact that the beam surface at the distance of the object may not match the actual emitting surface of the bow-shock, either because of a beam filling factor effect or because the bow-shock surface is curved. We assume here that the observer has a perfect knowledge of the geometry and we take $C=0$, which means that the 1D shock model has the same surface as the 3D bow-shock to which it is compared with. The best 1D model and power-law assumption selected is the one yielding the smallest $\chi^{2}$ value on our grid of 1D models. 

 \begin{figure}
	\begin{minipage}[c]{.5\textwidth}
		\includegraphics[height=0.22\textheight]
		{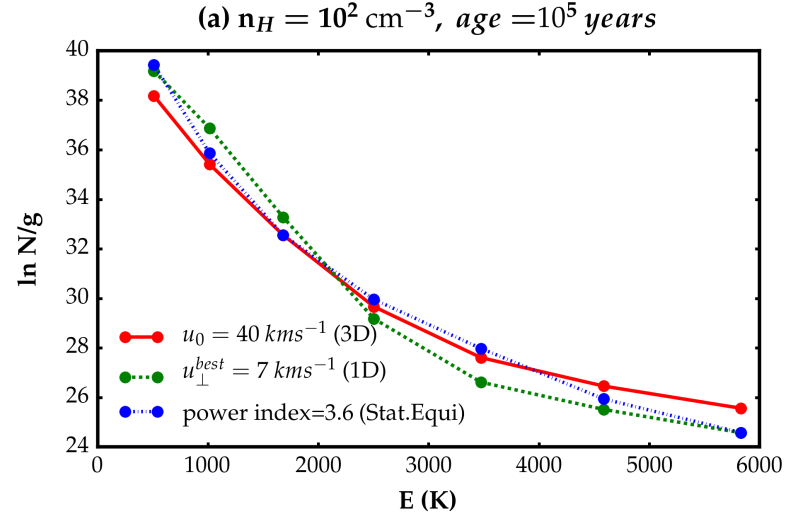}
 	\end{minipage} \hfill
 	\begin{minipage}[c]{.5\textwidth}
 		\includegraphics[height=0.22\textheight]
 		{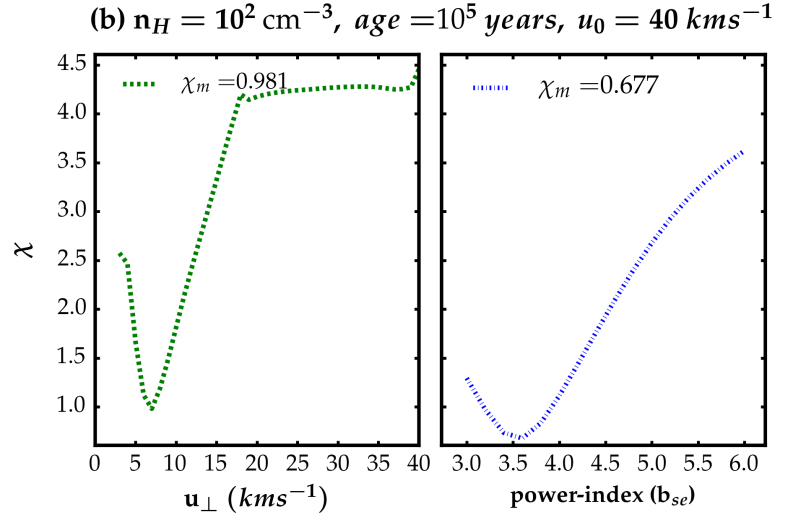}
 	\end{minipage} \hfill
	\begin{minipage}[c]{.5\textwidth}
		\includegraphics[height=0.22\textheight]
		{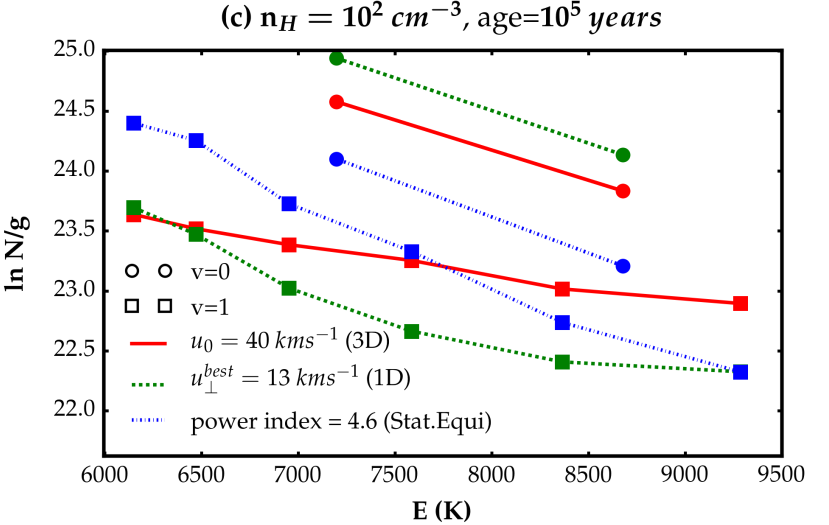}
 	\end{minipage} \hfill
	\begin{minipage}[c]{.5\textwidth}
		\includegraphics[height=0.22\textheight]
		{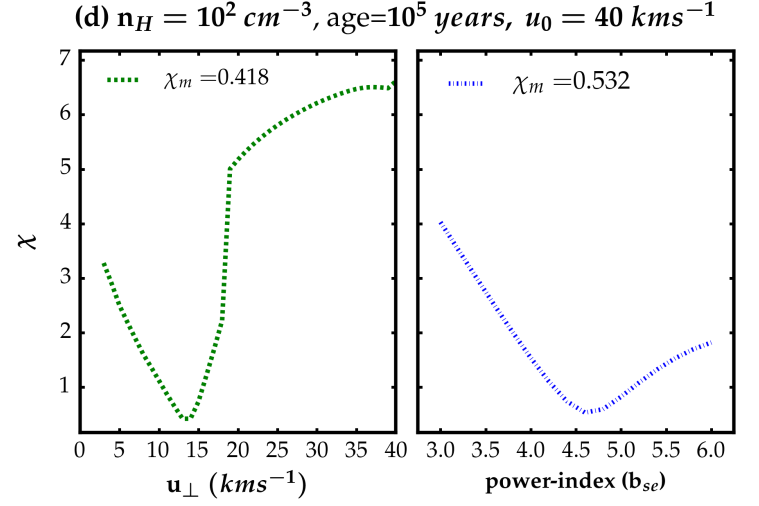}
 	\end{minipage} \hfill
 	\caption[Fit of 1D models 
 	and statistical equilibrium approximations to a 3D bow shock]
 	{Results of the fit of 1D models 
 	and statistical equilibrium approximations to a 3D bow shock. 
 	(a-b) Transitions with upper level $E_{vj} < 5900~$K (JWST-like) 
 	are used. (c-d) Fitted transitions have 5900 K$<E_{vj} <10\,000~$K. 
 	(a-c) Comparison of the 
 	excitation diagrams of the bow shock to the best 1D fit 
 	and the best NY08 fit , 
 	(b-d) standard deviation of the natural logarithm difference 
 	between the two diagrams ($\chi=\sqrt{\chi^2}$) as the entrance 
 	velocity in the 1D model 
 	and the power-index in NY08 assumption vary. 
 	The bow shock parameters are: 
 	pre-shock density $10^{2}\,$cm$^{-3}$, $b_0=1$, $\psi=90^o$, 
 	and the age is 10$^5$ yr. 
 	Connected circle symbols all have $v=0$ (pure rotational levels) 
    while square symbols have $v=1$.}	
      \label{fig:fit}
\end{figure}

  \autoref{fig:fit} shows the result of the fit on a 30 km$\,$s$^{-1}$ bow shock at age $10^{5}$ years, density $n_H=10^2$ cm$^{-3}$, and magnetization parameter $b_0=1$ ($\Psi=90^o$). 1D models have the same parameters (same age, pre-shock density and $b_{\parallel}=1$) except the entrance velocity $u_\bot$. We find that the best velocity is either 7 or 13 km$\,$s$^{-1}$ depending on the range of lines considered. This is way below the terminal velocity and this illustrates again the fact that the resulting 3D excitation diagram is dominated by low velocity shocks. As a consequence, the use of higher energy lines reduces the bias, and a cubic shape for the bow shock yields less bias towards low velocity than a parabolic shape (not shown here). In the left hand sides of the panels (b)-(d), the resulting $\chi^2$ is around one in all cases: it corresponds to an average mismatch of about a factor of 3 between the 3D and 1D column-densities, a common result when comparing 1D models and observations. 

\begin{figure}
	\centering
	\includegraphics[width=0.9\textwidth]
	{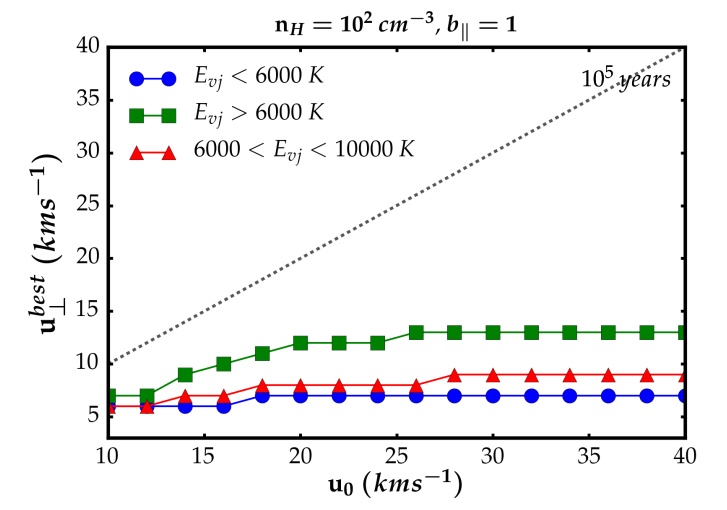}    	
     \caption[Velocity bias between 1D and 3D model]
     {Velocity bias between 1D and 3D model. 
     Blue circle symbols fit only $E_{vj} < 5900\,$K (JWST-like), 
     the green square symbols fit only $E_{vj} > 5900 K$ (ground based) 
     and the red triangles fit both ranges. The parameters of the bow shock 
     are the same as for \autoref{fig:fit}. 
     The dotted black line is $u^{best}_{\perp} = u_{0}$.}	
      \label{fig:bias_v}
\end{figure}

 \autoref{fig:bias_v} systematically explores this bias as a function of the bow shock terminal velocity: the best 1D model usually has an entrance velocity smaller than the terminal velocity of the 3D bow-shock. Moreover, when the 3D excitation diagram saturates at large $u_{0}$, the best 1D model does not change. 
  
 Following the approach of NY08 (described in \autoref{sec:power_law_assumption}), we calculate the H$_{2}$ levels population in statistical equilibrium for temperatures ranging from 100K to 4000K and we convolve this with a power-law distribution of the gas temperature. We explore power-indices ($b_{SE}$) varying from 3 to 6 (as in NY08) with steps of 0.2. We confirm that that the NY08 approximation performs very well in the low energy regime of pure rotation. In the case displayed in \autoref{fig:fit}(a), the optimal power-index is 3.6, close to the estimation of 3.77 for parabolic bow-shocks calculated by equation (4) in NY08. However, \autoref{fig:fit}(c) shows that this simple approach fails for higher energy vibrational or rotational levels.
  
\begin{figure}
   	\begin{minipage}[c]{0.9\textwidth}
     	 \includegraphics[width=1\textwidth]
     	 {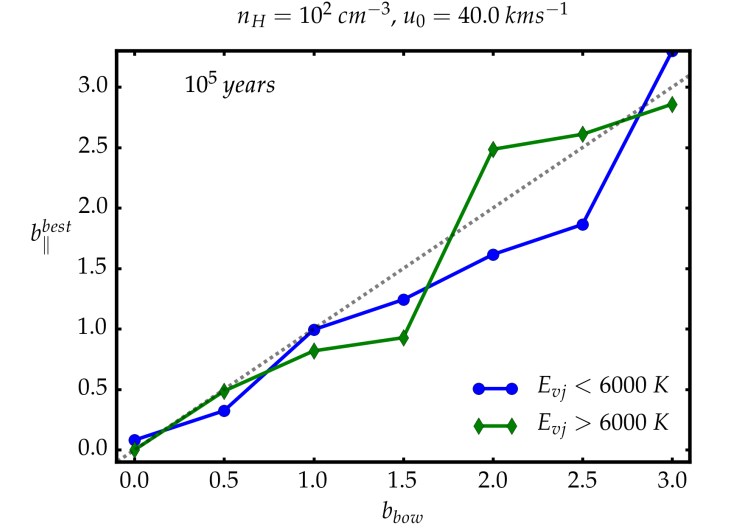}
	\end{minipage} \\
	\begin{minipage}[c]{1\textwidth}
     	 \includegraphics[width=1\textwidth]
     	 {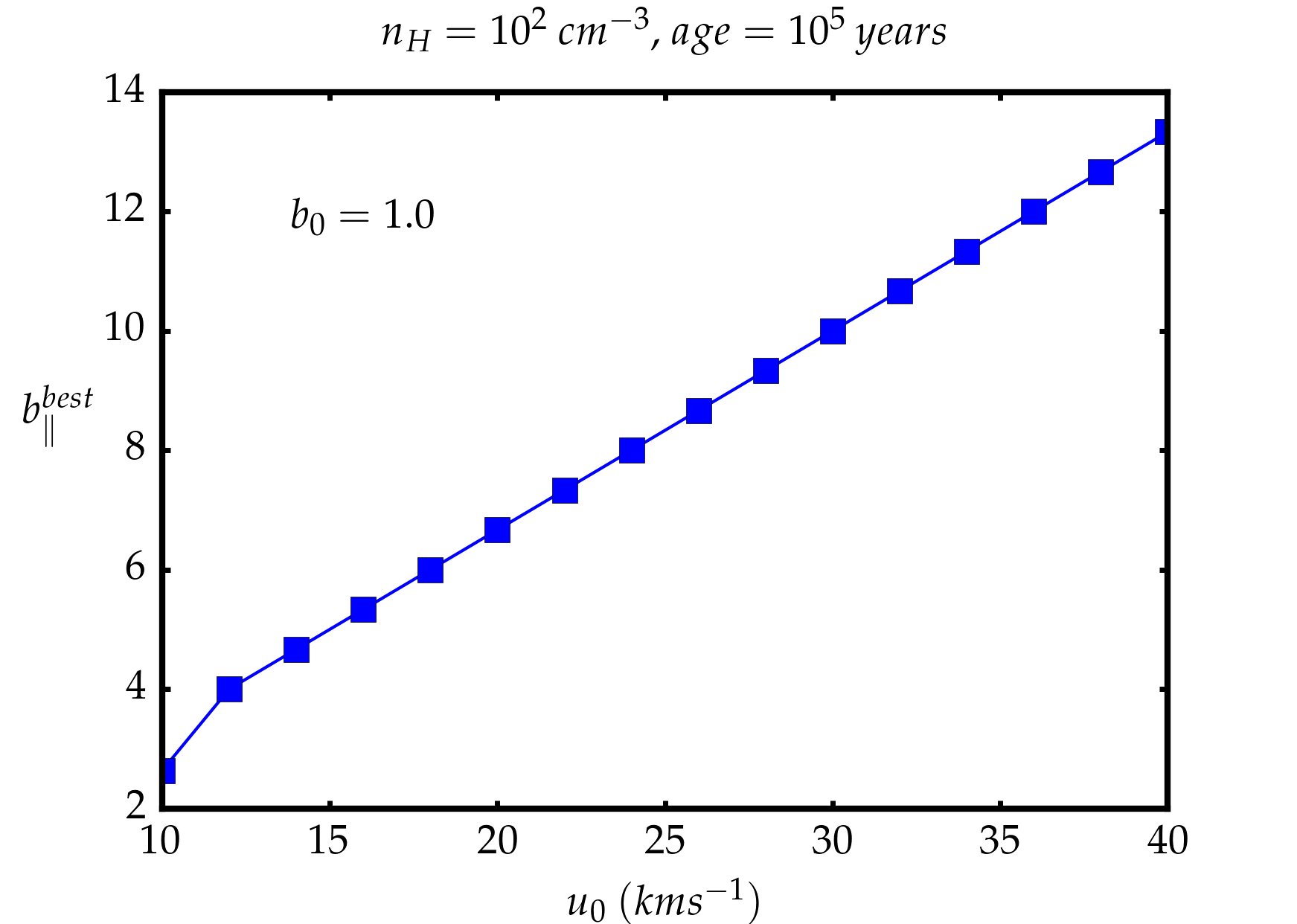}
	\end{minipage} \hfill
     \caption[Magnetization bias between 1D and 3D models]
     {Magnetization bias between 1D and 3D models. 
      The {\it top panel} is at $u_0=40$ km$\,$s$^{-1}$ and 
      for each value of $b_0$, it gives the best $b_\parallel$ 
      after the best $u_\bot$ has been determined. 
      The {\it bottom panel} is at $b_0=1$ and for 
      each value of $u_0$, it gives the best matching 
      $b_{\parallel}$ when $u_{\bot}=u_0$ is assumed. 
      Symbols are the same as in \autoref{fig:bias_v}. 
      The remaining parameters of the bow shock are 
      the same as in \autoref{fig:fit}.}
      \label{fig:bias_b}
\end{figure}

  We then turn on recovering magnetization from 1D models. We first fix the terminal velocity of the bow shock to $u_0=40$ km$\,$s$^{-1}$ and explore several values of the magnetization $b_0$, while keeping 
$\Psi=90^o$. Once the best matching 1D velocity is found, we further let the magnetization parameter $b_{\\}$ of the 1D model vary freely and explore which value best fits the 3D model (while keeping $u_\bot$ fixed). The result of this second adjustment is shown in \autoref{fig:bias_b}: the magnetization parameter of the best 1D model is only slightly below and represents a good match to the original magnetization parameter of the bow shock.
 Next, we assume that a priori information about the bow shock velocity (usually by looking at some molecular line width, for example) is available. We now fix $b_0=1$ for the underlying 3D model and assume that $u_\bot=u_0$ in the 1D models while searching for the best $b_\parallel$ value. The retrieved magnetization parameter is usually too high, which may lead to an overestimation of the magnetization parameter when the dynamics have been constrained independently.

\section{H$_{2}$ line shape}
\label{sec:H2_profile}
\cite{Smith90_H2} pioneered the study of the emission-line profile of molecular hydrogen from a simple C-type bow shock. We revisit their work with our models that better takes into account shock age, charge/neutrals momentum exchange, cooling/heating functions, the coupling of chemistry to dynamics, and the time-dependent treatment of the excitation of H$_2$ molecules. We also introduce line broadening due to the thermal Doppler effect.

 In the shock's frame, the velocity of the gas equals $\textbf{v}(r,u_{\bot}, \varphi) = \hat{\textbf{t}}\ u_{\parallel} + \ \hat{\textbf{n}}\ u(r,u_{\bot},b_{\parallel})$, where $r$ is the distance within the shock thickness (orthogonal to the bow shock surface) and $u(r,u_{\bot},b_{\parallel})$ is the shock orthogonal velocity profile as computed in the 1D model. Because of the large compression, the shock frame is moving very slowly with respect to the star, which use adopt as the observer's frame. In the observer's frame, the emission velocity becomes 
 \begin{equation}
 \begin{split}
 	\bf{v}_{\rm obs} \sim \bf{v} = \bf{v} - \bf{u}_{0} + \bf{u}_{0}\\
 	&= \hat{\bf{n}}u(r,u_{\bot},b_{\parallel})-\hat{\bf{n}}u_{\bot}+\bf{u}_{0}\\
 	&= \hat{\textbf{n}}u_{\bot}(\xi-1) + \bf{u}_{0}
 \end{split}
 \end{equation}
 where $0\leq\;\xi(r,u_{\perp},b_{\parallel})\;\leq 1$ is the ratio between the local velocity $u(r,u_{\bot},b_{\parallel})$ to the orthogonal entrance velocity $u_{\bot}$.
   
 However, the observer only senses the component along the line of sight: $\textbf{v}_{\rm obs}.\hat{\textbf{l}}$ with $\hat{\textbf{l}}$ a unit vector on the line of sight, pointing {\it towards} the observer. Adopting the geometric symmetry, that unit vector relates to the viewing angle ($i$) as 
 \begin{equation}
 	\hat{\textbf{l}} = \sin(i)\;\hat{\textbf{i}} + \cos(i)\;\hat{\textbf{k}}
 \end{equation}
where $\hat{\textbf{i}},\;\hat{\textbf{k}}$ are the unit vectors on the x and z axes.
 
When this is expressed in the observer's frame, the emission velocity becomes 
\begin{equation}
	\begin{split}
	v_{\rm rad}&=-\bf{v}_{\rm obs}.\hat{\bf{l}} \\
		       &= \hat{\bf{n}}u_{\bot}(1-\xi).\hat{\bf{l}} 
		       -\bf{u}_{0}\hat{\bf{l}}\\
		       &= 0.5\;u_{0}(\xi-1)\cos(\varphi)\;\sin(2\alpha)\sin(i) +
		         (1-\xi)\frac{u_{\bot}^{2}}{u_{0}}\cos(i) - u_{0}\cos(i)
	\end{split}
\end{equation}
with $\alpha$ defined at \autoref{eq:alpha}, $u_{\perp}/u_{0}=\sin \alpha$ and we have used $\hat{\textbf{n}}$ 
at above \autoref{sec:geometry_shock}.

 We assume the H$_2$ emission to be optically thin. Then the line profile is defined by integration over the whole volume of the bow shock, including the  emission coming from each unit volume inside each planar shock composing the bow shock. The line emission at velocity 
V$_{r}$ can be computed as follows:
	
\begin{multline}  \label{eq:line_profile}	
	f(V_r,i) = \int_{u_{\bot}}P(u_{\bot}){\rm d}u_{\bot} 
\int_{\varphi}\frac{{\rm d}\varphi}{2\pi} \\
\int_{r}{\rm d}r 
\frac{R^{2}_{0}}{\sqrt{2\pi} \sigma_{T}(r,u_{\bot},b_{\parallel})} \epsilon (r,u_{\bot},b_{\parallel}) e^{-\frac{[v_{\rm rad}(r,u_{\bot},b_{\parallel}) - V_r]^{2}}{2\sigma^{2}_{T}(r,u_{\bot},b_{\parallel})}} 
\end{multline}

which includes Doppler broadening with $\sigma^2_{T}(r,\alpha) = (k_{B}/m_{H_{2}}) T_{H_{2}}(r,u_{\bot},b_{\parallel})$, the thermal velocity of the H$_2$ molecule. 
Note that the azimuthal angle $\varphi$ occurs both in the expression of $b_{\parallel}$  (see \autoref{eq:b-parallel}) and in the projection of $v_{\rm obs}$ onto the line-of-sight direction $\hat{\textbf{l}}$.

\subsection{Effect of viewing angle}
\begin{figure}
	\centering
	\begin{minipage}[c]{1.\textwidth}	
		\includegraphics[width=1\textwidth]
		{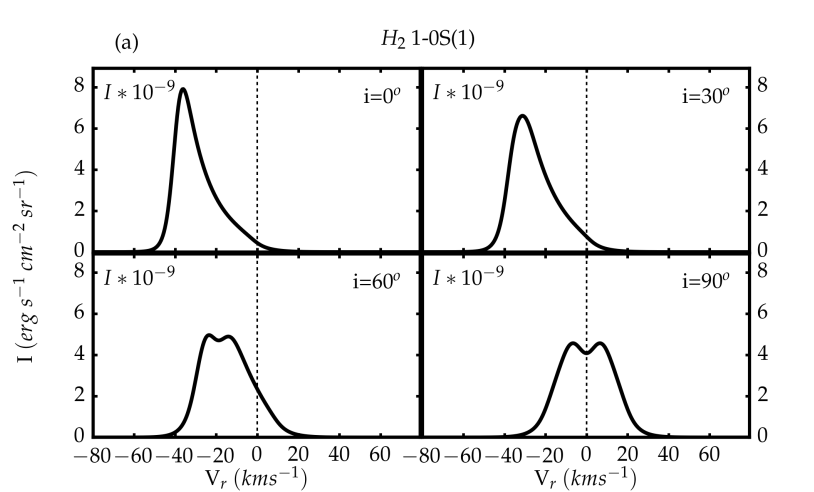}	
	\end{minipage} \hfill
	\begin{minipage}[c]{1.\textwidth}
		\includegraphics[width=1\textwidth]
		{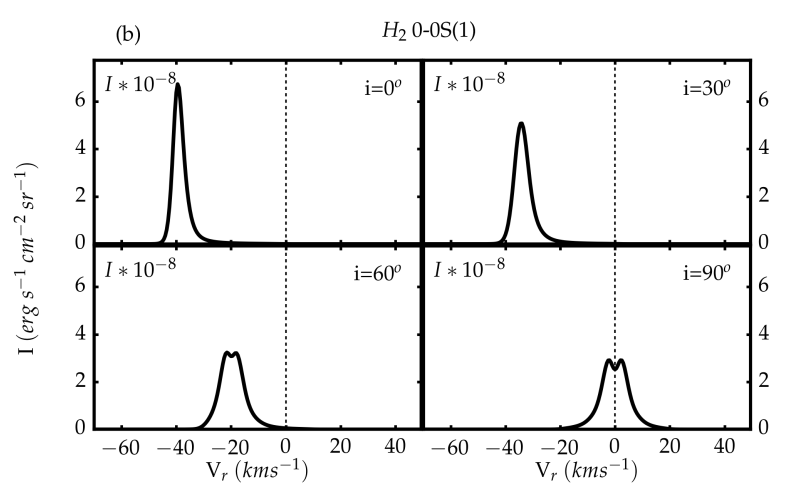}
	\end{minipage} \hfill
	\caption[Effect of viewing angle on H$_{2}$ line profile]
	{Line profiles of a whole bow shock parameterized by 
	u$_{0}$ = 40 km$\,$s$^{-1}$, age = 10$^{2}$ years, b$_{0}$ = 1 
	and $\phi$ = 0$^{o}$. \textit{(a)} for the H$_{2}$ 1-0S(1) line and 
	\textit{(b)} for the H$_{2}$ 0-0S(1) line.}
	\label{fig:profile_vs_i}
\end{figure}

 \autoref{fig:profile_vs_i}a shows the effect of the viewing angle $i$ on the 1-0S(1) line shape. When the observer looks at the bow-shock from the point of view of the star ($i=0^o$), all the emission is blue-shifted, with a stronger emission at a slightly positive velocity, coming from the part of the shock structure closest to the star, close to the J-type front where this line is excited. As $i$ increases, the line of sight intercepts two sides of the working surface, one going away and the other going towards the observer. The line profile then becomes doubly peaked. We checked that the integrated line emission did not vary with the viewing angle.
 
\subsection{Effect of shock age}
\begin{figure}
	\begin{minipage}[c]{1.\textwidth}
	  	\includegraphics[width=1\textwidth]
	  	{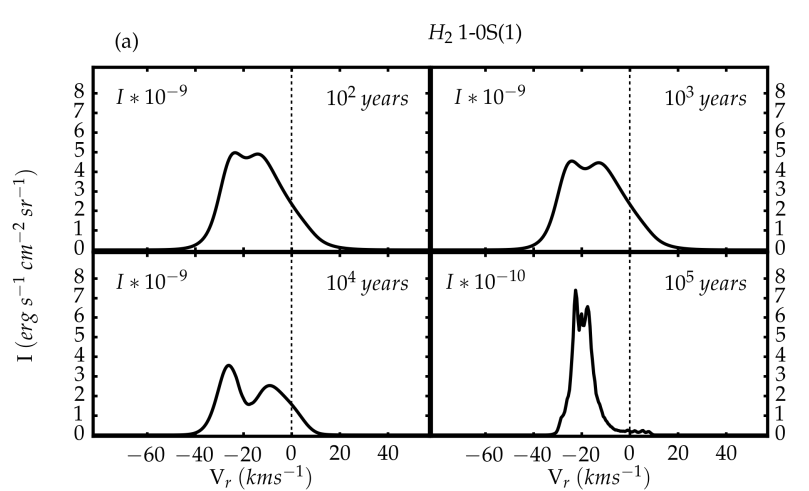}
	\end{minipage} \hfill
	\begin{minipage}[c]{1.\textwidth}
 	 	\includegraphics[width=1\textwidth]
 	 	{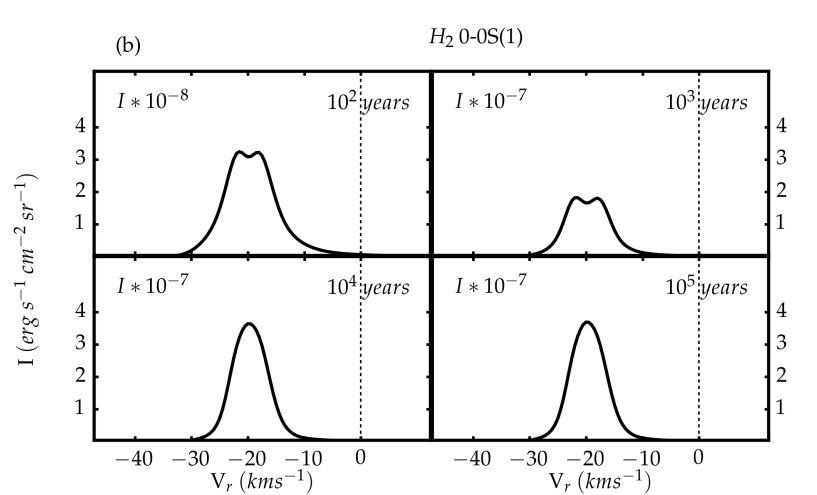}
 	\end{minipage} \hfill
	\caption[Effect of age on H$_{2}$ line profiles]
	{Line profiles of a whole bow shock parameterized by 
	u$_{0}$ = 40 km$\,$s$^{-1}$, i = 60$^{o}$, b$_{0}$ = 1 and $\phi$ = 0$^{o}$. 
	\textit{(a)} for the H$_{2}$ 1-0S(1) line and \textit{(b)} for the 
	H$_{2}$ 0-0S(1) line.}
	\label{fig:profile_vs_age}
\end{figure}

\autoref{fig:profile_vs_age}a shows how the age affects the 1-0S(1) line profile at the viewing angle  $i$=60$^o$. As the shock becomes older, the J-tail entrance velocity decreases: this explains why the two peaks of the line profile get closer to each other as age proceeds. The velocity interval between the two peaks is proportional to the entrance velocity in the J-type tail of the shocks. Furthermore, as the entrance velocity decreases, the temperature inside the J-shock decreases accordingly and the Doppler broadening follows: the line gets narrower as time progresses. The width of the 1-0S(1) could thus serve as an age indicator, provided that the shock velocity is well known.
  
The 0-0S(1) line corresponds to a much lower energy level than the 1-0S(1) line: while the 1-0S(1) is sensitive to temperature and shines mostly around the J-type front, the 0-0S(1) line emits in the bulk of the shock, where gas is cooler. Since the 0-0S(1) line probes a colder medium, the resulting profiles are much narrower (\autoref{fig:profile_vs_i}b). For early ages (100 and 1000 yr), one can however still notice the double peak signature of the J-front (\autoref{fig:profile_vs_age}b). Because at these early ages the temperature in the magnetic precursor is much colder than the transition's upper level temperature of 1015 K for level (0,3), the 0-0S(1) line is shut off in the magnetic precursor (see the bottom panel of \autoref{fig:CJ_profile}, for example) and it therefore probes the J-shock part. 

\setstretch{1.1} 
\chapter{BOW SHOCK MODELS TO INTERPRET OBSERVATIONS}
\label{Chapter6}

\lhead{Chapter 6. \emph{Bow shock model with observations}} 
\section{H$_{2}$ excitation diagram}
In this section, we briefly show how 3D bow shock models can be used to interpret and constrain the parameters of observations.
\subsection{BHR71}
\begin{figure}
	\includegraphics[width=1.\textwidth]
	{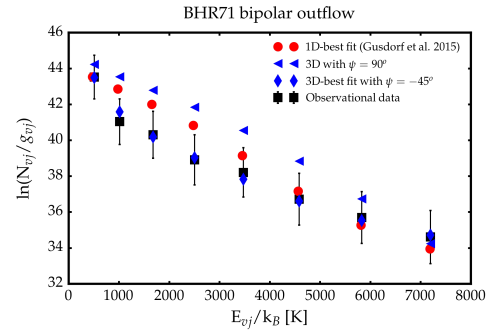}   
   	\caption[Comparison between BHR71 observations and several 
    		bow shock models]
    		{Comparison between BHR71 observations and several 
    		bow shock models. {\it Red circles}: best fit with the 1D model 
    		of \protect\cite{G15}, {\it blue triangles}: our own corresponding
     		1D model (a 3D model with velocity close to 22 km$\,$s$^{-1}$, 
     		$\psi=90^o$, $age=1000\,$yr, $G_{0}=1$ and $b_{0}=1.5$ 
     		so that the transverse magnetic field is uniform), 
     		{\it blue diamonds}: best fit with our 3D model 
     		(same as the previous model, but with
     	 	magnetic field orientation $\psi=-45^o$).}
    \label{fig:bhr71}
\end{figure}

Located at a distance of about 175 pc \citep{Bourke95}, BHR71 is a double bipolar outflow (\citealt{Bourke97}, \citealt{Bourke01}) emerging from a Bok Globule visible in the southern sky. The two outflows are spectrally distinguishable \citep{Parise06}. Their driving protostars, IRS 1 and IRS 2, have luminosities of 13.5 and 0.5 L$_{\odot}$ \citep{Chen08} and are separated by about 3400 AU. For this double star system, the time since collapse has been evaluated to about 36000 yr \citep{Yang17}. Many observations have been performed from infrared to sub-millimeter wavelength ranges. Bright HH objects HH320 and HH321 have been detected \citep{Corporon97}, as well as chemical enhancement spots \citep{Garay98} and several other knots of shocked gas \citep{Giannini04}. By combining H$_2$ observations performed by \textit{Spitzer} (\citealt{Neufeld09}, \citealt{Giannini11}) and SiO observations obtained from the APEX telescope, \citet{Gusdorf11} were able to characterize the non-stationary CJ-type shock waves propagating in the northern lobe of the biggest outflow. They more tightly constrained the input parameters of Paris-Durham shock models by means of successive observations of low- to higher-$J_{\rm up}$ CO (\citealt{G15}) using APEX and SOFIA. The most recent studies based on \textit{Herschel} observation report the presence of an atomic jet arising from the driving IRS1 protostar (\citealt{Nisini15}, \citealt{Benedettini17}). This does not challenge the existence of a molecular bow-shock around the so-called SiO knot position in the northern lobe of the main outflow, where most attempts have been made to compare shock models with observations (\citealt{Gusdorf11, G15}, \citealt{Benedettini17}). These studies have placed constraints on shock models of the H$_2$ emission over a beam of 24" centered on this position: pre-shock density $n_{\rm{H}} = 10^4$~cm$^{3}$, magnetic field parameter $b = 1.5$, shock velocity $v_{\rm s} = 22\ \rm km\ s^{-1}$, and age of 3800 years. The influence of the external ISRF or from the driving protostar was neglected, with an equivalent $G_{0}$ factor set to 0. The excitation diagram that was used can be seen in \autoref{fig:bhr71}, where the large error-bars reflect the uncertainty on the filling factor and the proximity of the targeted region to the edge of the \textit{Spitzer}-IRS H$_2$ map. 
    
Here we attempt to reproduce the same H$_{2}$ emission data around the SiO knot position as in \cite{G15}. To fit a 3D model to this data, we should in principle adjust all the parameters in \autoref{tab:bhr71_parameters}, which would be a bit tedious and very likely underconstrained by the observations. Instead, we started up from already published parameters and expanded around these values. We hence use a narrow range of velocities around $u_0=22$ km$\,$s$^{-1}$, $b_{0} = 1.5$ and $n_{H}=10^{4} $cm$^{-3}$ as indicated by \cite{G15}.
 These authors found an age of 3800 yr, so we took our grid models at an age of 1000 yr, as 10$^4$yr would not be compatible with the extent of the shock. A speed of 22 km$\,$s$^{-1}$ during 1000 yr already results in a shock width of 0.02 pc, about the same size of the beam (24" at 200pc according to \cite{G15}), although the H$_2$ lines emission region is a factor of a few smaller.  

\begin{table}
\centering
\begin{tabular}{l l l}
\hline \hline 
Parameter & \hspace{1mm} Value & \hspace{1mm} Description \tabularnewline
\hline 
$n_{H}$ & $10^{4}$ cm$^{-3}$  & Pre-shock density of H nuclei \tabularnewline

$age $ & $10^{3}$ yr & Shock age \tabularnewline

$\Delta u_{\bot}$ & 21-23 km$\,$s$^{-1}$ & Range of $u_{\bot}$\tabularnewline	

$b_{0}$ & 1.5 & Strengh of the magnetic field\tabularnewline

$\psi$ & $-50^{o} \pm 20^{o}$ & Orientation of the magnetic field \tabularnewline

$u_{0}$ and $\beta$ & N.A.  & Bow shock terminal velocity and \tabularnewline
$ $ & $ $ & shape are irrelevant because of \tabularnewline
$ $ & $ $ & the narrow range of velocities \tabularnewline
\hline \hline 
\end{tabular}

\caption[Best fit parameters in BHR71]
{Parameters that best reproduce the excitation diagram in BHR71. We also give a $3\sigma$ uncertainty range for the parameter $\Psi$ (see text).}
\label{tab:bhr71_parameters}
\end{table}

\begin{figure}
	\includegraphics[width=1.\textwidth]
	{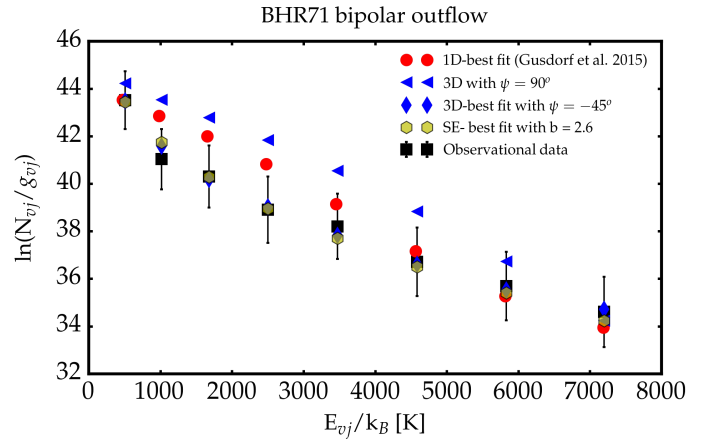}   
   	\caption[Comparison between BHR71 observations 
		   	and power-law statistical equilibrium assumption models]
    		{Same as \autoref{fig:bhr71}, 
    		with in addition the best-fitting 
    		power-law statistical equilibrium assumption model (see text). 
    		{\it yellow hexagons}: best fit with power-index $b_{SE} = 2.6$. 
    		The other symbols are the same as in \autoref{fig:bhr71}.}
    \label{fig:bhr71_2}
\end{figure}

  \autoref{fig:bhr71} illustrates the comparison between our models and the observational values. We first restrict the velocity range in the bow shock velocity distribution to the narrow interval 
[21,23] km$\,$s$^{-1}$ that is close to the solution of \cite{G15}. This also accounts for the fact that the beam selects a local portion of the bow-shock and one might expect to find a privileged velocity.

 First, we examine the case $\psi=90^o$ when the magnetization is close to $b_0$ and is uniform throughout a transverse annulus of the bow-shock. Technically this is still a 3D model, but it is very close to the model in our grid of planar shocks with similar parameters because we use a very narrow range of velocities combined with uniform magnetization. The excitation diagram for this model is noted as the blue triangles in \autoref{fig:bhr71}. Although it slightly differs from the best model of \citet{G15}, it is not much further away from the observational constraints ($\chi=1.0$ in the model in \citet{G15} and $\chi=1.5$ in our model at $\psi=90^o$). 

  Second, we leave the orientation of the magnetic field $\Psi$ free and we find the best model at $\psi=-45^o$: this greatly improves the comparison with observations ($\chi=0.2$). In particular, the curvature of the excitation diagram that was difficult to model, is now almost perfectly reproduced. At this orientation, the model is a mixture of planar shocks with transverse magnetization between $b_0$ and a small minimum value. Because we limited the velocity to such a narrow range, this model is effectively a 2D model.
  
  Third, we checked that increasing the velocity range, changing the shock shape, or limiting the integration range for the angle $\varphi$ (to account for the fact that the observational beam probably intersects only one flank of the bow shock) did not improve the fit: the interpretation capabilities of our 3D model seems to be reached. \autoref{tab:bhr71_parameters} sums up our constraints on the parameters of our model. We estimate 3-$\sigma$ error bars for $\Psi$ by investigating the shape of the $\chi^2$ well around the best value: we vary $\Psi$ with all other parameters kept fixed and we quote the range of values where $\chi^2$ is below four times its minimum value. 
  
  Finally, we checked the NY08 approximation. As mentioned in \autoref{sec:bias}, that simple assumption surprisingly works well in the case of low pure rotational excitation. \autoref{fig:bhr71_2} shows the best fit from the NY08 assumption with the value of the power-index at $b_{SE}=2.6$, consistent with the value 2.5 in \cite{Neufeld09} for the same object. The accuracy obtained as close to the data as our 3D model, with $\chi = 0.2$.
      	
\subsection{Orion BN-KL outflow}
The BN-KL region in the Orion molecular cloud (OMC-1) is one of the well studied massive star forming regions. A central young stellar object generates a strong outflow that shocks the surrounding gas and yields a wealth of 
H$_2$ infrared emission lines previously observed by \citet{Rosenthal00}. These authors however indicated that the full range of H$_{2}$ level population could not be reproduced by a single shock model. In fact, \citet{Bourlot02} showed that only a mixture between two C-type shock models could account for the population of both the low and the high energy levels (see \autoref{sec:1D_model_limitation}). 
In this work, we try to reproduce the observed excitation diagram of H$_{2}$ and strongest H$_{2}$ 1-0S(1) line profile from the OMC-1 Peak1 with one of our bow shock models. 

  We ran a new grid of models at the pre-shock conditions in Orion, 
  $n_{H}= 10^{6}$ cm$^{-3}$ (\citealt{White86}, \citealt{B88}, \citealt{HM89}, \citealt{KN96},  \citealt{Kristensen08}). We limited the age to 1000 yr, which roughly corresponds to the dynamical age of the outflow \citep{Kristensen08}. At these densities, the shocks should have reached steady-state long ago.

  Then we explore the parameter space of all possible bow-shocks and seek
the best fitting model.  We considered $u_0$ between 20 and 100 km$\,$s$^{-1}$ and we varied $b_{0}$ from 1 to 6 with step 0.5. For each value of $b_{0}$,
we let the angle $\psi$ vary from $0^{o}$ to $90^{o}$ with step
$5^{o}$. Finally we explore the shape of the shock for $\beta$ in the interval from 1.0 to 3.0 with step of 0.2.

\begin{figure}
	\includegraphics[width=1.01\linewidth, height=0.7\textwidth]
   	{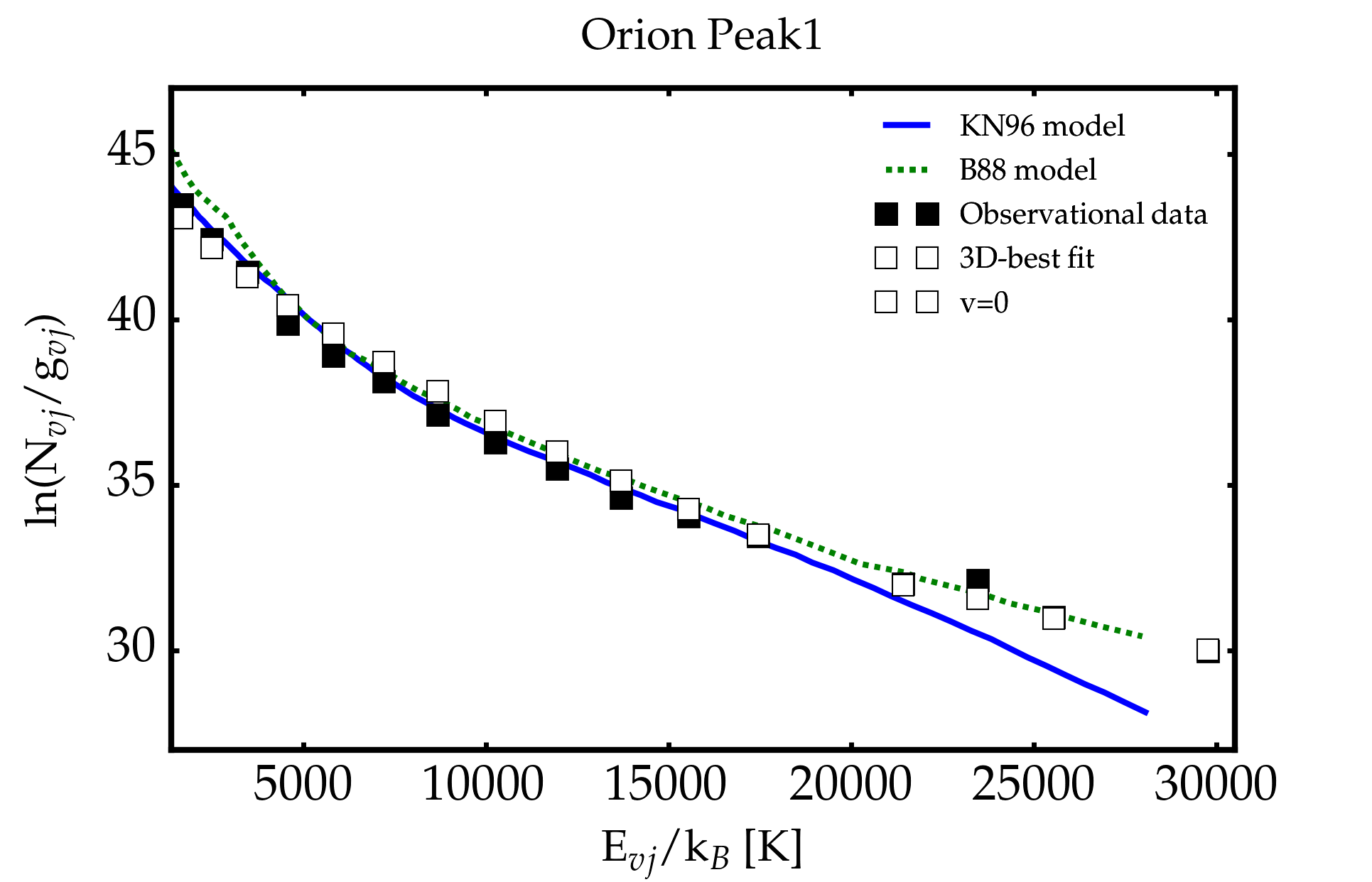}   
   	\caption[Pure rotational H$_{2}$ excitation diagram observed in OMC-1 Peak1
   	\citep{Rosenthal00} compared with various models]
   	{Pure rotational H$_{2}$ excitation diagram observed in OMC-1 Peak1
   	\citep{Rosenthal00} compared with various models: 
   	our best-fit 3D-model of bow shock (open symbols), 
   	and the best fit models from \citet{Rosenthal00}: 
   	a combination of two planar C-shocks models from 
   	\citet{KN96} (KN96) and one J-type shock model 
   	from \citet{B88} (B88).}
	\label{fig:orion_peak1}
\end{figure}

\begin{figure}
	\includegraphics[width=1.01\linewidth, height=0.7\textwidth]
   	{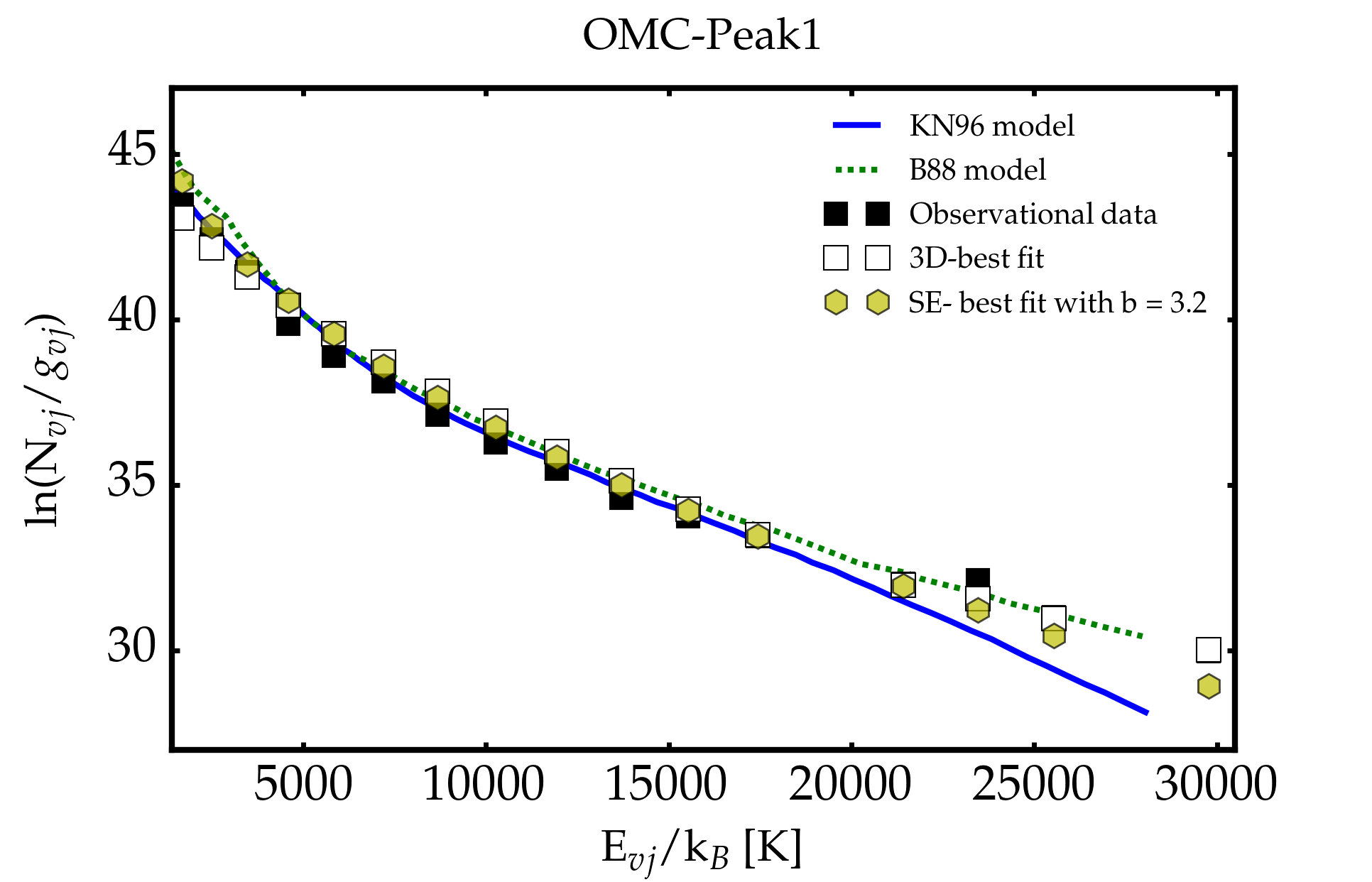}   
   	\caption[Comparison between pure rotational H$_{2}$ excitation diagram observed 
   	in OMC-1 Peak1 and power-law statistical equilibrium assumption models]
	{Same as \autoref{fig:orion_peak1}, 
	with in addition the best-fitting 
	power-law statistical equilibrium assumption models (see text). 
    {\it yellow hexagons}: best fit with power-index $b_{SE} =3.2$, 
    while the other symbols are the same.}
	\label{fig:orion_peak1_2}
\end{figure}

  In the simplified case, we compute the $\chi^2$ for the 17 pure rotational transitions, with the vibrational levels $v=0$ among the 55 transitions which have been measured, discarding the upper limits (table 3 of \citealt{Rosenthal00}). The parameters that best fit the excitation diagram are listed in \autoref{tab:orion_parameters}. We also provide an estimation of the 3$-\sigma$ uncertainty range for some parameters by investigating the shape of the $\chi^2$ well around the best value, as we did above for the parameter $\psi$ in the case of BHR71. The best model convincingly reproduces most of the lines ($\chi=0.4$), as long as the terminal velocity is greater than 30 km$\,$s$^{-1}$. The comparison to the observations is displayed in 
\autoref{fig:orion_peak1}: both the low and high energy regimes of the excitation diagram are simulated by the same model. Two best matching models found by \citet{Rosenthal00}, which are a mixture of two C-type shock models from \citet{KN96} and a single J-type shock model from \citet{B88}, are also displayed for comparison. 
We also checked the NY08 approximation as shown in \autoref{fig:orion_peak1_2}. Our best fit value is obtained at $b_{SE}=3.2$ for $\chi=0.6$. Again, this approach yields satisfying results for levels with a low excitation energy but tends to deviate at high excitation energy. 

\begin{table}
\centering
\begin{tabular}{l l l}
\hline \hline 
Parameter & \hspace{10mm} Value & \hspace{1mm} Description \tabularnewline
\hline 
$n_{H}$ & $10^{6}$ cm$^{-3}$  & Pre-shock density of H nuclei \tabularnewline

$b_{0}$ & $4.5 \pm 0.9$ & Strength of the magnetic field\tabularnewline

$u_{0}$ & $\geq$ 30 km$\,$s$^{-1}$ & 3D terminal velocity\tabularnewline	

$age $ & $10^{3}$ yr & shock's age \tabularnewline

$\psi$ & $90^{o} \pm 30^{o}$ & Orientation of the magnetic field \tabularnewline

$\beta$ & $2.1 \pm 0.2$ & Shock shape \tabularnewline
\hline \hline 
\end{tabular}
\caption[Optimal parameters of the OMC-1 Peak1 (only pure 17 rotational levels) found with bow shock model]
{Optimal parameters of the OMC-1 Peak1 
(pure rotational levels) found with our model (see \autoref{fig:orion_peak1}).}
\label{tab:orion_parameters}
\end{table}

Finally, we extend our computation of $\chi^{2}$ until 44 rovibrational transitions, for which the vibrational levels $v$ varies from 0 to 4 (the upper limits are also discarded). The best comparison to observation ($\chi=0.45$) is showed in \autoref{fig:orion_peak1_2}. 
The parameters that reproduce this best fit are listed in \autoref{tab:orion_parameters_allvj}. 
The 3-$\sigma$ uncertainty range for some parameters is also provided.  

\begin{table}
\centering
\begin{tabular}{l l l}
\hline \hline 
Parameter & \hspace{10mm} Value & \hspace{1mm} Description \tabularnewline
\hline 
$n_{H}$ & $10^{6}$ cm$^{-3}$  & Pre-shock density of H nuclei \tabularnewline

$b_{0}$ & $3^{+2.5}_{-1.5}$ & Strength of the magnetic field\tabularnewline

$u_{0}$ & $\geq$ 30 km$\,$s$^{-1}$ & 3D terminal velocity\tabularnewline	

$age $ & $10^{3}$ yr & shock's age \tabularnewline

$\psi$ & $70^{o} \pm 25^{o}$ & Orientation of the magnetic field \tabularnewline

$\beta$ & $2.4 \pm 0.27$ & Shock shape \tabularnewline
\hline \hline 
\end{tabular}
\caption[Optimal Best-fit parameters of the OMC-1 Peak1 (44 rovibrational levels) found with bow shock model]
{Optimal parameters of the OMC-1 Peak1 
(44 rovibrational levels) found with our model (see \autoref{fig:orion_peak1_3}).}
\label{tab:orion_parameters_allvj}
\end{table}

\begin{figure}
	\includegraphics[width=1.01\linewidth, height=0.7\textwidth]
   	{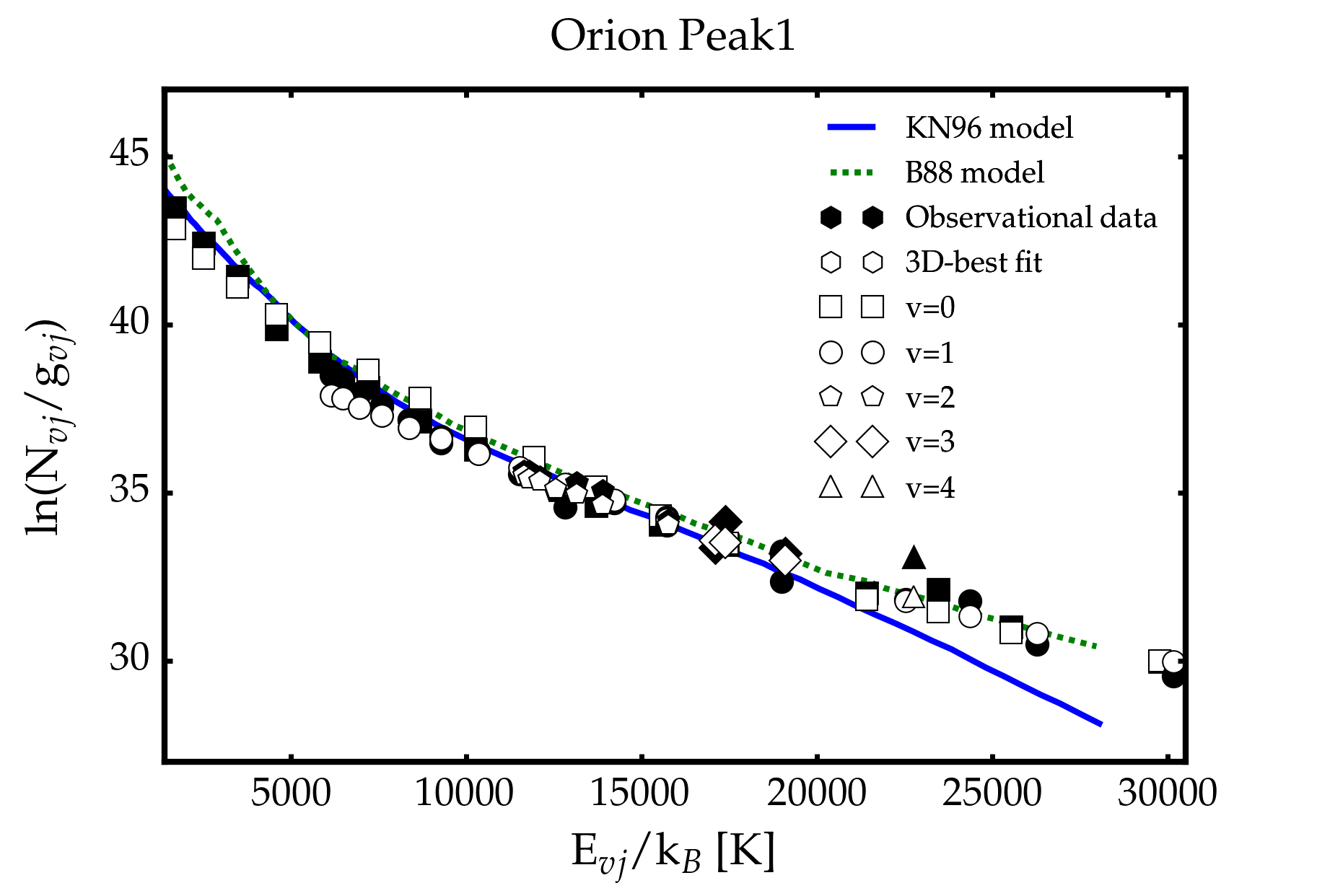}   
   	\caption[Comparison between rotvibrational H$_{2}$ excitation diagram observed 
   	in OMC-1 Peak1 with various models]
	{Rovibrational H$_{2}$ Excitation diagram observed in OMC-1 Peak1
   	\citep{Rosenthal00} compared with various models: 
   	our best-fit 3D-model of bow shock is indicated by the open symbols.}
	\label{fig:orion_peak1_3}
\end{figure}    
  
\section{H$_{2}$ line shape}
The previous \autoref{sec:H2_profile} shows that a wealth of dynamical information is contained in the line shapes. However, this information is difficult to retrieve, as the line shaping process is quite convoluted. In particular, each line probes different regions of the shock depending on the upper level sensitivity to temperature. 
\subsection{HH54}
\begin{figure}	
  	\includegraphics[width=1\linewidth]
  	{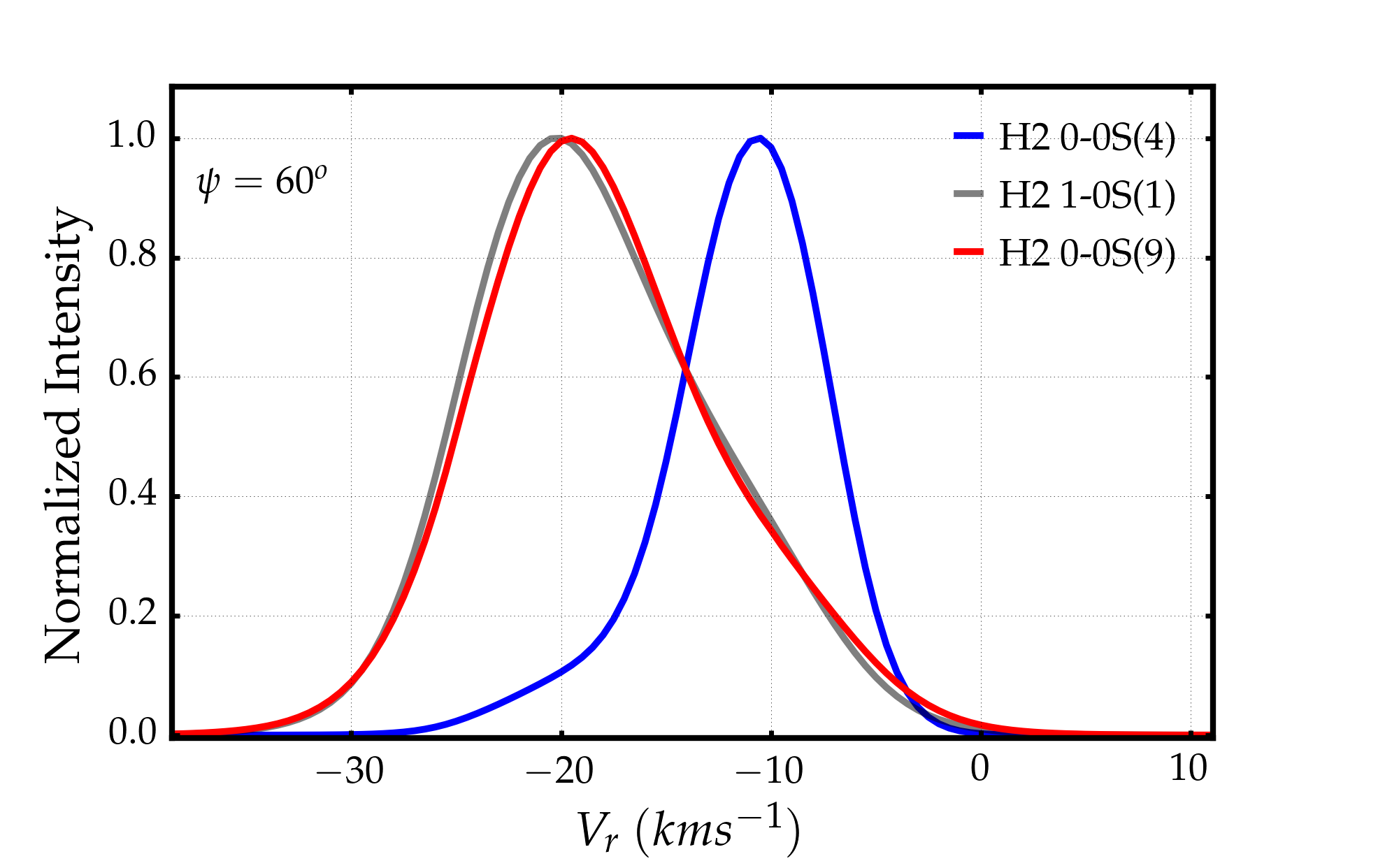}		
    \caption[Line profiles of HH54-like bow shock]
    {Line profiles of three different transitions in a bow shock 
        at age 100 yr with parameters u$_0$=20 km$\,$s$^{-1}$, 
        n$_H$=10$^4$ cm$^{-3}$, b$_0$=1, 
        and viewing angle i = -60$^{o}$.}
        \label{fig:santangelo}
\end{figure}

We plot the normalized line shapes for three different transitions in a 20 km$\,$s$^{-1}$ bow shock with pre-shock density 10$^4$ cm$^{-3}$, age 1000 yr and $b_0=1$ (\autoref{fig:santangelo}). This figure compares well with the figure 2 in \cite{San14}, which plots resolved observations of H$_2$ lines in HH54. These observations come from two different slit positions: a CRIRES slit for 1-0S(1) and 0-0S(9) near the tip of the bow, orthogonal to the outflow axis, and a VISIR slit for the 0-0S(4) line along this axis. On the other hand, our models cover the whole extent of our bow shock, which questions the validity of the comparison. Despite this, some similarities are striking: the two lines 1-0S(1) and 0-0S(9) perfectly match and are blue-shifted. The insight from our computations allows us to link the good match between the line profiles of 1-0S(1) and 0-0S(9) to the very similar energy of the upper level of the two transitions. Furthermore, we checked that the emission from the low energy 0-0S(4) in our model is completely dominated by the C-type parts of our shocks, where the velocity is still close to the ambient medium velocity: this explains why this line peaks around $V_r=0$. This C-type component should shine all over the working surface of the bow shock, and the VISIR slit along the axis probably samples it adequately. Conversely, we checked that the emission coming from lines 1-0S(1) and 0-0S(9) is completely dominated by the J-type parts of our shocks. Hence they should shine near the tip of the bow shock (traversed by the CRIRES slit) at a velocity close to that of the star and its observed radial speed should lie around $-u_0 \cos(i)$, blue-shifted for an acute angle $i$.
\begin{figure}
  	\includegraphics[width=1\linewidth]
  	{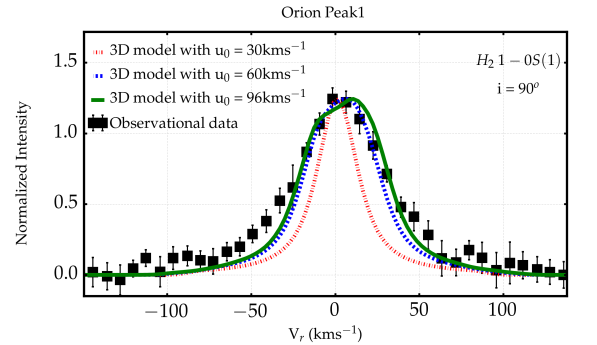}			
        \caption[Comparison of the H$_2$ line profile between OMC-1 Peak1
         observation and bow shock model]
         {Comparison of the H$_2$ line profile between OMC-1 Peak1
         observation and a bow shock model. \textit{Black square:} 
         the observational data \citep{B89}. 
         \textit{Solid lines:} our 3D model 
         using parameters in \autoref{tab:orion_parameters} with 
         different values of u$_{0}$. The best 3D model constrains 
         the terminal shock velocity to about 100 km$\,$s$^{-1}$.}
        \label{fig:Orion_profile}
\end{figure}
\subsection{Orion BN-KL outflow}
\cite{B89} managed to observe a few wide H$_{2}$ line profiles from OMC-1 Peak1 by using the UKIRT telescope, configured at a  5$''$ sky aperture and with a resolution of 12 km$\,$s$^{-1}$ full width at half maximum (FWHM). A single shock model was not able to reproduce these wide observed lines (as indicated by \citealt{B89} and \citealt{Rosenthal00}). A C-type bow shock model of \cite{S91_Bfield} could reproduce these lines and widths, but this assumed a extremely high magnetic field strength of $\geq$ 50 mG (which amounts to b$_{\parallel}$ $\geq$ 50 for $n_{H}\sim 10^{6}\,$cm$^{-3}$) while independent measurements in the same region gave much lower values: 3 mG by Zeeman splitting \citep{Norris_84} or 10mG by polarization \citep{Chrysostomou_94}. Here we use the best parameters listed in \autoref{tab:orion_parameters} to try and reproduce the profile of the H$_{2}$ 1-0S(1) line with a more reasonable magnetization. As mentioned in the previous subsection, the excitation diagram alone did not allow to constrain the terminal shock velocity. Now, the width of the profile allows us to constrain the velocity to about $u_0=100$ km$\,$s$^{-1}$ as illustrated by 
\autoref{fig:Orion_profile}.  The viewing angle $i\simeq$90$^{o}$ can be adjusted to the position of the peak of the line profile. Note that shock models with $u_{\bot}>$40km$\,$s$^{-1}$ are not included in these line shape models. They should contribute little to the emission since H$_2$ molecules are dissociated at high shock velocities (both due to the high temperatures experienced in these shocks and to their radiative precursors).

\setstretch{1.1} 
\chapter{CONCLUSIONS AND PERSPECTIVES}
\label{Chapter7}

\lhead{Chapter 7. \emph{Conclusions and perspectives}} 
\section{Conclusions and remarks}
In this study, we provide a mathematical formulation which links an arbitrarily shaped bow shock to a distribution of planar shocks. Then, a simple convolution of this distribution with a grid of planar shocks allows to produce intensities and line shapes for any transition of the H$_2$ molecule.

  We used that property to explain the dependence of the excitation diagram of a bow shock to its parameters: terminal velocity, density, shape, age, and magnetization properties (magnitude and orientation). The combination of a steeply decreasing distribution with a threshold effect linked to the energy of the upper level of each transition yields a ``Gamow-peak'' effect. A given H$_2$ level then reaches a saturation value of column density when the terminal velocity is above a threshold which depends directly on the energy of the level. The magnetic field  and the age dependence enter through the transition between the J-type and the C-type part of a time-dependent magnetized shock.

  The wings of a bow shock usually have a larger surface than its nose. From this, it follows that the distribution and hence the global emission properties of a bow shock are generally dominated by low-velocity shocks. A direct consequence is that the excitation diagram of a whole bow shock resembles a 1D planar shock with a lower velocity: data interpretation with 1D models is likely to be biased towards low velocity. However, if the terminal velocity of the bow shock was estimated independently (from line Doppler broadening measurements, for example), we suggest that a magnetization adjustment from 1D models to the excitation diagram will over estimate the magnetization parameter. Previous authors \citep[NY08,][]{Neufeld09} have suggested that the statistical equilibrium approximation could accurately reproduce observed intensities of low-energy pure rotational levels. We confirm this result, and its probable link to the distribution of entrance velocities as pointed out by NY08. However, we remark that this simple model does not satisfyingly reproduce the observations of the higher-lying transitions. A possible interpretation is that these levels are more sensitive to J-type shocks, where the sudden temperature jump is more likely to put the gas away from statistical equilibrium.

  We provide some illustrations of how our results could improve the match between model and observations in BHR71 and Orion OMC-1. We show that 3D models largely improve the interpretation. In particular, we are able to obtain much better match than in previous works with relatively little effort (and with the addition of only one or two parameters compared to the 1D models: the magnetic field orientation and the shape of the bow shock). 

  We compute line shapes with an unprecedented care and examine their dependence on age and viewing angle. Although line shapes result from a convoluted process, they contain a wealth of dynamical information. In particular, we link the double peaked structure of 1-0S(1) in young bow shocks to the dynamics of their J-type part components. The line width results from the combined effects of geometry, terminal velocity, and thermal Doppler effect. We show how different lines probe different parts of the shocks depending on the temperature sensitivity of the excitation of their upper level. We show how our 3D model can reproduce the broad velocity profile of the H$_{2}$ 1-0S(1) line in Orion Peak1 with a magnetization compatible with other measurements. The excitation diagram fails to recover dynamical information on the velocity (it only gives a minimum value), but the line shape width provides the missing constraint, which agrees with proper motion $\sim 100\,$km$\,$s$^{-1}$ of the tips of the H$_{2}$ "fingers" in the region.
  
\section{Perspectives}
\begin{figure}
	\centering
    \includegraphics[width=0.9\linewidth]
    {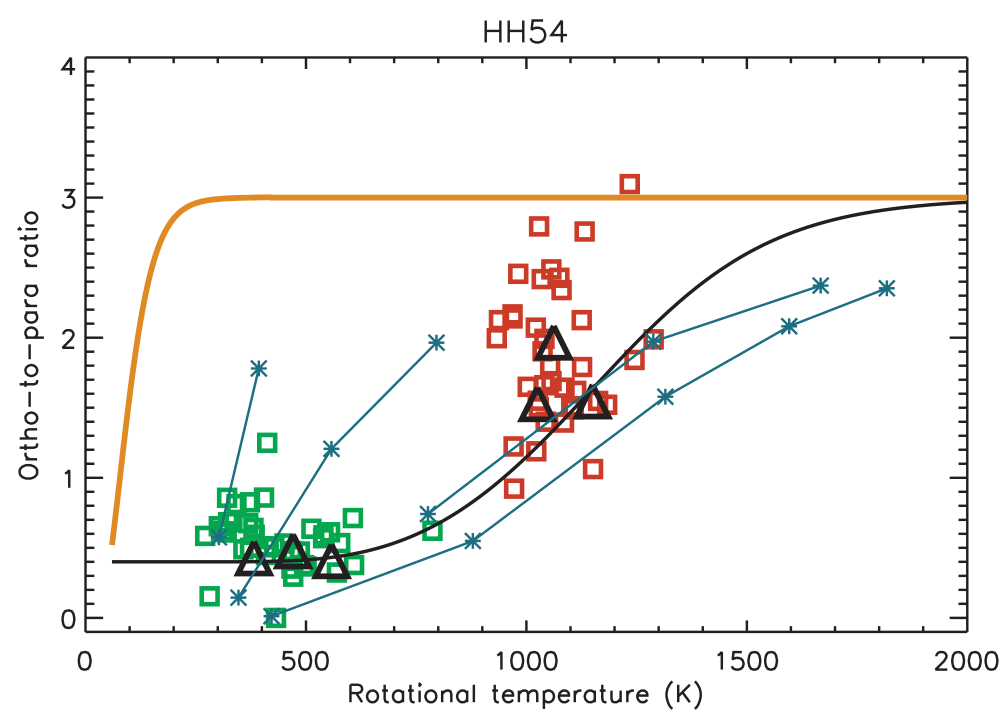}
	\caption[Correlation between H2 ortho/para ratio and 
	rotational temperature for the HH54 object]
	{Correlation between H2 ortho/para ratio and 
	rotational temperature for the HH54 object. 
	The black curve shows the behavior for gas with an initial 
	ortho/para ratio of 0.4. 
	The orange curve shows the ortho/para ratio in LTE. 
	The cyan curves and asterisks show the model predictions 
	of \citet{Wilgenbus_2000}.} 
	\label{fig:O/P_ratio}
\end{figure}

 All models presented here were run for a pre-shock ortho-para ratio of 3 (\aref{app:othor_para}). However, the dilute ISM is known to experience much lower ratios and that should vary as a function of the excitation temperature \citep{Neufeld06}. This variation is illustrated in \autoref{fig:O/P_ratio}. 
David Neufeld suggested we should explore the effect of this parameter on the excitation diagrams of bow-shocks in future work and compare again to the
observable data from shocks in BHR71 and OMC-1 Peak1 (\autoref{Chapter6}). 

\begin{figure}
	\centering
    \includegraphics[width=0.9\linewidth]
    {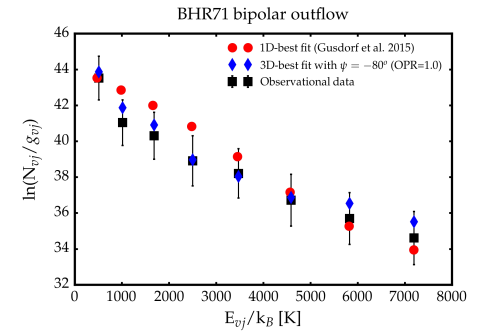}
	\caption[Comparison between BHR71 observations 
	and bow shock model, whose initial ortho/para ratio is 1]
	{Comparison between BHR71 observations and 
	a bow shock model, whose initial ortho/para ratio is 1. 
    The symbols are the same as \autoref{fig:bhr71}.} 
	\label{fig:BHR_OP1}
\end{figure}

We have started to investigate a grid of models with a lower ortho/para ratio. 
For example, \autoref{fig:BHR_OP1} shows the comparison between 
BHR71 observations and the best bow shock model, 
for which the initial ortho/para ratio is 1. Due to the jagged variation of 
the statistical weight of H$_2$ levels (which differ for even and odd rotational number $J$), the excitation diagram at low ortho-para ratios show a characteristic oscillation between ortho and para levels.
Contrary to the one in \autoref{fig:bhr71}, the best fit diagram is not a smooth curve anymore and displays the similar weak oscillations as the observations. 
At low excitation energy, the fit is better than in the previous case with ortho/para=3, 
while it looks worse for higher excitation levels. We will investigate it further with David Neufeld and examine systematically the effect of the ortho-para ratio on 3D models of bow-shocks.
 
We are co-Is of two observational program with SOFIA-EXES, that will target pure rotational H$_{2}$ lines: S(1) and S(5)) in a sample of Galactic supernova remnants (\textit{IC443, W28, W44, 3C391}), while S(4) and S(7) in a sample of molecular cloud (HH7). These observations will provide the first velocity resolved H$_{2}$ spectra in SNRs and in molecular cloud 
(velocity resolution of a few km$\,$s$^{-1}$). Our objective for those studies is to confront our models with H$_{2}$ spectra in simple geometries (spherical, in the case of \textit{IC443}), in order to pinpoint the kind of shocks that are propagating in these environments (whether they are magnetized or not, stationary or not, dissociative or not), and to formulate diagnostics for their physical conditions.


Previous H$_{2}$ maps of 3D bow-shocks \citep{Gustafsson10} assumed 
that the direction of the flow was unperturbed (z-axis approximation). 
This is valid for C-type shocks, but not for J-type shocks. 
We plan to model both C-type and J-type shock
trajectories by using a local planar approximation and to
compare the resulting H$_{2}$ maps to previous work.

Our methods could be used to model other molecules of interest, provided that we know their excitation properties throughout the shock and that their emission remains optically thin. We expect that such developments will improve considerably the predictive and interpretative power of shock models in a number of astrophysical cases. In particular, if some excited CO lines can be assumed optically thin, they will allow direct probes of the dynamics of the gas. For the optically thick lines, one will need to be much more careful with the radiative transfer, but we could think using our method to build a 3D map with the correct CO abundance, and post-process it with a 3D radiative tool such as RADMC \citep[][or seelook at \textit{http://www.ita.uni-heidelberg.de/~dullemond/software/radmc-3d/}]{Dullemond_2012}.
  
 Further work will address some of the shortcomings of our method. First, it will be straightforward to apply similar techniques to the shocked stellar wind side of the bow shock working surface, helping us to access knowledge on the reverse shock (see \autoref{Chapter9}). Second, the different tangential velocities experienced on the outside and on the inner side of the working surface will very likely lead to Kelvin-Helmoltz instabilities, generate turbulence and hence mixing, as multidimensional simulations of J-type bow shocks show. A challenge of the simplified models such as the ones presented here will be to include the mixing inside the working surface. One way of proceeding will be to bracket the true behaviour between the two extreme situations. In the first situation, the contact discontinuity between the two forward and reverse shocks remain stable, and the observations can be simply modeled through the same methods presented here. In the second situation, one can assume the fluid in between the two forward and reverse shocks is completely mixed, just as \citet{W96} did for mass and momentum, but extending this property to energy and chemical composition. This will allow to compute self-consistently the shape of the working surface, its chemical composition, temperature, emission properties, etc... 
 Note the same method can be utilized in the context of shocks in binary winds, where we can expect to build powerful tools to synthesize observations (see \autoref{sec:partIII_perspectives}).
 
%
\makeatletter
\part{STELLAR WIND MODEL}
\setstretch{1.1} 
\chapter{GOVERNING EQUATIONS}
\label{Chapter8}
\lhead{Chapter 8. \emph{Basic equations}} 

 The hydrodynamic model of the stellar wind from a AGB star is the solution 
 of a set of hydrodynamic equations (\autoref{sec:hydro_eqs}), 
 associated with a chemical network. 
 In order to achieve a set
 of simplified numerical equations of that system, 
 we assume \textit{spherical symmetry} of the central 
 star and its circumstellar envelope. This assumption 
 satisfies observations at least in the outer region of 
 the circumstellar envelope (e.g., \autoref{fig:shock_IRC10216}), 
 but is uncertain very 
 close to the star. 
 Observing the CO lines from AGB stars 
 (e.g., RS Cancri and EP Aquarii, Mira Ceti, Red Rectangle), 
 our collaborators \footnote{https://vnsc.org.vn/dap/} 
 found significant evidences for an asymmetric morphology close to the star. 
 (e.g., \citealt{Anh_2015}, \citealt{Nhung_2015a}, \citealt{Diep_2016}, \citealt{Hoai_2016}) 

\section{Hydrodynamics} \label{sec:hydrodynamic}
 The velocity and density profiles of the circumstellar outflow 
 are determined by the physical laws of mass conservation and momentum
 conservation. Throughout this work, we mostly consider an ideal fluid,
 so that the viscous and conducting phenomena are neglected, except in the case of a terminal wind shock, where we trigger a viscous jump.
 For the hydrodynamics of this gas, we consider a multicomponent fluid 
 made of several gaseous chemical species and  micron scale solid particles. Under these assumptions,
 the continuity equation of the gas can be derived from mass conservation as
 \begin{equation} \label{eq:continuity}
 	\frac{\partial \rho}{\partial t} + \frac{1}{r^{2}}\frac{\partial}{\partial r} (r^{2} \rho v) = 0	
 \end{equation}
 with $\rho$ the total mass density, $t$ the time, 
 $r$ the radial coordinate, and $v$ the gas velocity. Note that the mass conservation equation for the neutrals is equivalent to \autoref{eq:mass_cons_n} with $z=r-r_{0}$ ($r_{0}$ is the starting radius of the computation) and $S_{n} \rightarrow S_{n}-2\rho_{n} v_{n}/r$: the only additional contribution compared to the original planar mass conservation equation in the Paris-Durham shock code is geometrical dilution. Similarly, chemical equations have their source term transformed as $N_{n} \rightarrow N_{n}-2 n_{n}v_{n} /r$ for the neutrals  and $N_{i} \rightarrow N_{i}-2 n_{i}v_{i} /r$ for the ions. The neutral momentum source has $A_{n} \rightarrow A_{n} - 2 \rho_{n} v_{n}^2 / r$ and the corresponding energy source has $B_n \rightarrow B_{n} - 2 (\frac{1}{2} \rho v_{n}^2+\frac{\gamma}{\gamma-1}n_{n}k_{B}T_{n}) v_{n}/r$, etc... This is the convenient way we use to introduce geometrical dilution due to spherical geometry in the Paris-Durham planar shock code. 
 
 In the case of stationarity, \autoref{eq:continuity} can be integrated to yield
 \begin{equation} \label{eq:mass_conservation}
 	\dot{M} = 4\pi r^{2} \rho v 
 \end{equation}      
 where the mass loss rate $\dot{M}$ is one of 
 the fundamental model parameters and does not depend on the radius.
 
 Similarly, the stationary equation of motion in spherical symmetry 
 is derived from the law of momentum conservation
 \begin{equation} \label{eq:motion}
 	v\frac{\partial v}{\partial r} = \frac{1}{\rho} f - \frac{1}{\rho} \frac{\partial P}{\partial r}
 \end{equation}
 with $f$ the algebraic sum of the \textit{external} forces acting on 
 a unit volume and $P$  the \textit{internal} gas pressure. 
 Once dust-grains form, the radiation force from the star acting on 
 the dust particles accelerates them outward. The surrounding gas particles 
 then will be dragged along by collisions with the dust. 
 Therefore, the external force 
 $f = f_{grav} + f_{drag}$. 
 Gravitation force ($f_{grav}$) attracts 
 the gas inward to the stellar center, while 
 drag force ($f_{drag}$) drives it outward. 

 The gravitation force per a unit volume is simply proportional to $r^{-2}$
 \begin{equation}
 	f_{grav} = -G\frac{M_{\ast}\rho}{r^{2}}
 \end{equation}
 
 The drag force, on the other hand, is somewhat more complicated. 
 This force depends on the relative  
 \textit{drift speed, v$_{drift}$} between the gas and dust components
 \begin{equation}
 	v_{drift} = v_{d} - v	
 \end{equation}  
 where $v_{d}$ is the flow speed of the dust and $v_{d} > v$. 
 There are two expressions for the drag force depending on 
 the thermal sound speed of gas
 \begin{equation} \label{eq:Cs}
 	c_{s} = \sqrt{\gamma \frac{P}{\rho}} = 
 	\sqrt{\frac{\gamma k_{B}T}{\mu m_{H}}}
 \end{equation}
 where $\gamma$ is the ratio the specific heats,
 $\mu$ is molecular weight of gas and $m_{H}$ is the mass of a proton. 
 Depending on the composition of the gas, 
 its molecular weight can be different. 
 For example, $\mu \approx 2.33$ for a 
 mixture between helium and molecular hydrogen, 
 while a mixture with atomic hydrogen has $\mu \approx 1.4$.

 If the drift speed is much faster than the thermal sound speed of 
 the gas particles, the ram pressure acting on the gas is 
 $\rho v^{2}_{drift}$; and the drag force per unit volume of gas, therefore, 
 is the product of that ram pressure times the 
 cross-section of the dust grain: 
 $f_{drag}(a) = \pi a^{2} \rho v^{2}_{drift} n_{g}$, 
 where $a$ is the radius of a single grain. 
 On the other hand, if the drift speed is lower than the thermal speed of the gas particles, 
 the drag force $f_{drag}(a) = \pi a^{2} \rho c_{s} v_{drift} n_{g}$. 
 To combine two those limits, we can express the drag force as below
 \begin{equation}
 	f_{drag}(a) = \pi a^{2} n_{g} \rho v_{drift} \sqrt{v^{2}_{drift} + c^{2}_{s}} \mbox{.}
 \end{equation}  
 
  Since the mean free path of the gas is higher than the typical dust radii 
  and the velocities of gas and dust are different from each other, 
  the grains are not position coupled to the gas. 
  Despite the fact that the grains collide with only a small fraction 
  of the gas particles,
  \citet{Gilman_1972} indicates that the subsequent collisions among the gas
  molecules allow the momentum that they receive from the radiation field 
  to be transfered to the gas. 
  \citet{Gilman_1972} also demonstrates that the small grains rapidly 
  reach the \textit{terminal drift velocity}. 
  The grains move at the terminal drift velocity 
  when the radiation force balances with the drag force: 
\begin{equation}
\label{eq:force-balance}
        f_{drag}=f_{rad} 
\end{equation}
  where $n_{d}$ is the dust number density and $f_{rad}$ is the radiation force acting on one grain, defined by
 \begin{equation} \label{eq:f_rad}
 	f_{rad} = \frac{\sigma_d \bar{Q_{rp}}L_{\ast} }{4\pi r^{2} c_{l}}
 \end{equation}   
 where $\sigma_{d}=\pi a^2$ is the grain cross-section (assumed circular here) and $c_{l}$ is the speed of light. $Q_{rp}$ is the radiation pressure efficiency and $\bar{Q}_{rp}$ is the wavelength averaged radiation pressure efficiency weighted by the stellar spectrum. $L_{\ast}$ is the total stellar luminosity. It is determined through the absorption $Q_{ext}(a,\lambda)$ and scattering $Q_{sca}(a,\lambda)$ coefficients which are calculated by using the Mie theory with complex radiative indices (\autoref{sec:Mie_theory}).

  Let us assume that the grains are moving at terminal drift velocity, 
  the momentum \autoref{eq:motion} now expands as
  \begin{equation} \label{eq:wind_eq1}
  	v\frac{\partial v}{\partial r} +\frac{1}{\rho}\frac{\partial P}{\partial r} =  - \frac{GM_{\ast}}{r^{2}} + n_{d}\frac{\sigma_{d}}{\rho}\frac{\bar{Q}_{rp}L_{\ast}}{4\pi r^{2}c_{l}} 
  \end{equation}
  
  The two terms on the right hand side in \autoref{eq:wind_eq1} vary as $r^{-2}$ , it is hence convenient to group them in a simple form of the momentum equation
  \begin{equation} \label{eq:wind_eq2}
  	v\frac{\partial v}{\partial r} + \frac{1}{\rho}\frac{\partial P}{\partial r} = (\Gamma - 1)\frac{GM_{\ast}}{r^{2}}
  \end{equation}   	        
  with the radiative acceleration on one spherical species of dust  
  \begin{equation}
  	\Gamma(a) = n_{d} \frac{\sigma_d\bar{Q}_{rp}(a) L_{\ast}}{4\pi c_{l} G M_{\ast} \rho} \mbox{.}
  \end{equation}     
 
  \autoref{eq:wind_eq2} 
  can be rewritten in the form of the standard \textit{wind equation}
  \begin{equation} \label{eq:wind_iso_equ}
  	(\frac{v^{2} - c^{2}_{i}}{v})\frac{\partial v}{\partial r} = 
  	\frac{2 c^{2}_{i}}{r} - \frac{\partial c^{2}_{i}}{\partial r} + 
  	(\Gamma - 1)\frac{GM_{\ast}}{r^{2}}	
  \end{equation}
  with $c^{2}_{i}=P/\rho$. When the temperature profile (hence the isothermal sound speed profile) is prescribed, this equation shows the existence of a critical point at $r = r_{ci}$, 
  where the speed of the gas reaches the isothermal sound speed. 
  That point is called the \textit{isothermal sonic point}.

  In the more realistic case where the temperature evolution is solved along the wind, we can also get an expression for the velocity gradient by combining mass, momentum and energy conservation so as to eliminate the pressure gradients. We arrive at
\begin{equation} \label{gen_adiab_equ}
\frac{\partial v}{\partial r} = 
\frac 
{ \frac12 \frac{\gamma+1}{\gamma-1} S_n v^2 - \frac{\gamma}{\gamma-1} A_n v + B_n}
{ \frac{\gamma}{\gamma-1}P - \frac1{\gamma-1} \rho v^2 } 
\end{equation}
  which is expressed after some simplifications as
\begin{equation} \label{eq:wind_adiab_equ}
(\frac{v^{2} - c^{2}_{s}}{v})\frac{\partial v}{\partial r} = \frac {2v^2}{r} + (\Gamma - 1)\frac{GM_{\ast}}{r^{2}} + (\gamma-1) \frac{\Lambda}{\rho v}
\end{equation}
  where $\Lambda$ is the net radiative cooling. This last form closely resembles \autoref{eq:wind_iso_equ} and applies in the general case.
It shows that the sonic point when cooling and heating are introduced occurs when the velocity crosses the {\it adiabatic} sound speed.

  However, we experienced numerical issues when integrating \autoref{eq:wind_adiab_equ}, and we were never able to cross the sonic point with this form. We found a compromise by using \autoref{eq:wind_iso_equ} with $c_s$ in place of $c_i$ for the space derivative of the velocity from the stellar surface to the adiabatic sonic point $r=r_c$. We revert to the more proper \autoref{eq:wind_adiab_equ} after the sonic point has been crossed. The standard temperature gradient is used throughout to control the temperature profile. 

  Once the grains form, the thermal sound speed 
  and its radial derivative are small compared 
  to the last term on the right hand side of \autoref{eq:wind_iso_equ}. 
  Therefore, the sonic point occurs at a radius just before 
  the point where $\Gamma$ starts to be greater than unity.   

  We now consider a collection of spherical grain particles, with a constant size distribution $dn_{d}=f(a)da$. The resulting equivalent $\bar{\Gamma}$ of the radiative acceleration on dust is obtained through the relation
\begin{equation}
        \bar{\Gamma}=\int da \Gamma(a) f(a)/ n_{d}
\end{equation}
  where $n_{d}=\int da f(a)$ is the {\it total} density of grain particles, 
  we adopt a standard spectrum of grain-size distribution $f(a) = A a^{b} n_H$. For the interstellar dust, the spectral index $b$ is usually assumed to be equal to -3.5 from the famous MRN law (\citealt{MRN_1977}). For the circumstellar dust shell, \citep{Dominik_1989} deduced a steeper slope for the spectral distribution and estimated the spectral index $b \sim -5$. However, \citet{Decin_2006} suggest that the choice of the slope has little influence on the resulting dynamics. Thus, we consider the term $A$, which is the factor giving the number of dust particles per $H$ atom, to be a constant for the case of the interstellar dust. Its value is estimated to be $10^{-25.10}\  $cm$^{2.5}/H$ for Silicate grains and $10^{-25.13}\ $cm$^{2.5}/H$ for carbon grains with size minimum and maximum boundaries $a_{\rm min}=0.005 \mu$m and $a_{\rm max}=0.25 \mu$m (\citealt{Draine_1984}).

\section{Thermodynamics}
  The temperature profile of the multicomponent outflow 
  circumstellar envelope, consisting of gas molecules and 
  solid grain particles, is determined by the laws 
  of thermodynamics. The gas molecules are assumed to be 
characterized by their local kinetic 
  temperature $T(r)$. Each grain particle characterized by its
  radius $a$ is also assumed  to be a thermal emitter, 
  characterized by the size-dependent temperature $T_{d}(a,r)$. 

\subsection{Gas temperature}

In this section, we describe only the cooling and heating terms specific to the stellar wind situation. The standard heating and cooling terms of the Paris-Durham shock code as described in \autoref{Chapter1} still apply and are considered in our wind models.
%

 \subsubsection{Grain-gas collisional heating}
 As mentioned in \autoref{sec:hydrodynamic}, the grains rapidly reach their terminal drift velocity. They move at the terminal velocity if the drag force balances with the radiation force. The balance of these terms for grain of size $a$ leads to
 \begin{equation} \label{eq:vdrift_1}
 	\frac{\bar{Q}_{rp}(a)L_{\ast}}{4\pi r^{2} c_{l}} = \rho v_{drift}\sqrt{v^{2}_{drift}(a)+c^{2}_{s}} 
 \end{equation}
 The expression for the drift velocity is derived by taking the square at both sides of \autoref{eq:vdrift_1}
 \begin{equation} \label{eq:vdrift_2}
 	v^{4}_{drift}(a) + c^{2}_{s}v^{2}_{drift}(a) 
 	- \left(\frac{\bar{Q}_{rp}(a)L_{\ast}}{4\pi r^{2} \rho c_{l}}\right)^{2}=0 \mbox{.}
 \end{equation}
 
 Solving the second order \autoref{eq:vdrift_2} in $v_{drift}$ and combining the solution with  \autoref{eq:mass_conservation} give
 \begin{equation}
 	v_{drift}(a) = \frac{1}{2}\left[\left\{\left(\frac{2v}
 	{\dot{M}c_{l}}\bar{Q}_{rp}(a)L_{\ast}\right)^{2} + c^{4}_{s}\right\}^{0.5} 
 	- c^{2}_{s}\right]^{0.5} \mbox{.}
 \end{equation} 
 In the limiting case where the sound speed $c_{s}$ is small compared to the outflow speed $v$, this expression reduces to
 \begin{equation}
 	v_{drift}(a) = \sqrt{\frac{\bar{Q}_{rp}(a)L_{\ast}v}{\dot{M}c_{l}}} \mbox{.}
 \end{equation}
 This expression for the drift velocity shows the dependence on the mass-loss rate $\dot{M}$ and the outflow speed $v$. The gas-grain collisions, therefore, become increasingly important as the  distance from the star increases and $v$ reaches its terminal value.
  The resulting heating rate corresponds to the work done by the drag force. For the whole distribution of grains, this yields:
\begin{equation}
\label{eq:qdrift}
 q_{drift}=\rho \int da f(a) \frac{1}{2} \sigma_{d} v^{3}_{drift}(a) \mbox{.}
\end{equation}

 \subsubsection{Molecular pumping from stellar radiation}
 \label{sec:pumping}

 Close to the star, the molecular gas receives energy from the radiation field that is able to excite the molecules into higher excitation levels. Then, molecules cool down by spontaneous de-excitation processes, or by collisional de-excitation, in which case the radiation energy is transferred to the gas as thermal energy. In this section, we study the pumping effect from the stellar radiation to molecules. 
 
 Let's consider a system of two levels, with $n_{i}$ and $n_{j}$ column densities of the upper and lower levels, respectively. The corresponding energies are $E_{i}$ and $E_{j}$, where $\Delta E = E_{i} - E_{j} = h\nu_{ij}$. The number of density of atoms in the level $i$ changes basically due to: (1) spontaneous emission, (2) stimulated emission, (3) photon absorption, and (4) collisions. This variation per unit time is defined as:   
 \begin{equation} \label{eq:origin}
 	\frac{dn_{i}}{dt} = -n_{i}\left[A_{ij}+B_{ij}\rho(\nu_{ij})+C_{ij}\right] +
 	                    n_{j}\left[B_{ji}\rho(\nu_{ij})+C_{ji}\right] 
 \end{equation}

where $\rho(\nu)$ is the spectral energy density at frequency $\nu$ of the radiation field and $T_{ij} = \Delta E/ k_{B}$. If we consider the star as a black-body, $\rho(\nu_{ij})$ is defined  by Planck's law as:
\begin{equation}
	\rho(\nu_{ij}) = \frac{1}{4}\left(\frac{R_{\ast}}{r}\right)^{2} \frac{F(\nu_{ij})}{e^{T_{ij}/T_{\ast}}-1}
\end{equation}
with $R_{\ast}$ the stellar radius, and $F(\nu_{ij})$ a function defined as
\begin{equation}
	F(\nu_{ij}) = \frac{8\pi h \nu^{3}_{ij}}{c^{3}_{l}}
\end{equation} 
 
where $A_{ij}$, $B_{ij}$, $B_{ji}$ are the Einstein coefficients. 
 $A_{ij}$ is the probability in unit time that a particle in state $i$ spontaneously decays to the stage $j$. 
 $B_{ij}$ is the probability per unit time per unit spectral energy density of the radiation field that a particle in the stage $i$ decays by stimulation to the stage $j$. 
 $B_{ji}$ is the probability per unit time per unit spectral energy density of the radiation field that a particle in state $j$ absorbs a photon to jump to state $i$. 
 $C_{ij}$ and $C_{ji}$ are the collional de-excitation and excitation coefficients. These coefficients are linked with each others by

\begin{equation} \label{eq:E1}
	A_{ij} = F(\nu_{ij}) B_{ij}
\end{equation}
\begin{equation} \label{eq:E2}
	B_{ij} = \frac{g_{j}}{g_{i}}B_{ji}
\end{equation}
\begin{equation}
	C_{ij} = \frac{n}{n_{C_{ij}}}A_{ij}
\end{equation}
\begin{equation}
 	C_{ji} = C_{ij}\frac{g_{i}}{g_{j}}e^{-T_{ij}/T_{kin}}
\end{equation}

with $n$ the total density of colliders, $n_{C_{ij}}$ the critical density of the transition, and $T$ the gas temperature. We can rewrite \autoref{eq:origin} in a simple form 
 \begin{equation} \label{eq:reform}
	 \frac{dn_{i}}{dt} = -n_{i}(A_{ij}+C_{ij}) + (n_{j}B_{ji} - n_{i}B_{ij})\rho(\nu_{ij}) + n_{j}C_{ji}      
 \end{equation}

Substituting those expressions into \autoref{eq:reform}, while keeping $A_{ij}$ as a reference coefficient, we have:  
\begin{equation} \label{eq:replace}
	\begin{split}
	\frac{dn_{i}}{dt} &= -n_{i}\left(A_{ij}+\frac{n}{n_{C_{ij}}}\right) + n_{j}\frac{n}{n_{C_{ij}}}A_{ij}\frac{g_{i}}{g_{j}}e^{-T_{ij}/T_{kin}} + \frac{A_{ij}}{F(\nu)}\left(n_{j}\frac{g_{i}}{g_{j}}-n_{i}\right)\frac{1}{4}\left(\frac{R_{\ast}}{r}\right)^{2}\frac{F(\nu)}{e^{T_{ij}/T_{\ast}}-1} \\	
							\\
	  &= A_{ij}\left[-n_{i}+\frac{n}{n_{C_{ij}}}\left(-n_{i}+n_{j}\frac{g_{i}}{g_{j}}e^{-T_{ij}/T_{kin}}\right) +\frac{n_{j}\frac{g_{i}}{g_{j}}-n_{i}}{e^{T_{ij}/T_{\ast}}-1}\frac{1}{4}\left(\frac{R_{\ast}}{r}\right)^{2}\right]
	\end{split}
\end{equation}

In thermodynamic equilibrium, the density $n_{i}$ ($n_{j}$) is proportional to the product between the statistical weight $g_{i}$ ($g_{j}$) and the Boltzmann factor at the  temperature $T$: 
\begin{equation} \label{eq:LTE}
	\begin{split}
	&n_{i}= n_{M}\frac{g_{i}}{Z}e^{-T_{i}/T_{ex}} \\
	\\
	&n_{j}= n_{M}\frac{g_{j}}{Z}e^{-T_{j}/T_{ex}} \\
	\\
	&\frac{n_{j}}{n_{i}} = \frac{g_{j}}{g_{i}}e^{(T_{i}-T_{j})/T_{ex}} 
	                     = \frac{g_{j}}{g_{i}}e^{T_{ij}/T_{ex}}
	\end{split}
\end{equation}
with $n_{M}$ the density of the molecule considered and 
$Z(T)=\sum_{i} g_{i} e^{-T_{i}/T}$ the partition function.
At high density we are close to steady-state and $dn_{i}/dt = 0$. Thus, the net rate of energy lose for the gas through collisions is then:
\begin{equation}
	\begin{split}
	\Lambda_{i} &= h\nu_{ij}(n_{j}C_{ji}-n_{i}C_{ij}) \\
	            &= A_{ij}h\nu_{ij}\frac{n}{n_{C_{ij}}}\left(-n_{i}+n_{j}\frac{g_{i}}{g_{j}}e^{-T_{ij}/T_{kin}}\right)\\
	            &= A_{ij}h\nu_{ij}\left[-n_{i}+\frac{n_{j}\frac{g_{i}}{g_{j}}-n_{i}}{e^{T_{ij}/T_{\ast}}-1}\left(\frac{R_{\ast}}{r}\right)^{2}\frac{1}{4}\right] \\
	            &= -A_{ij}n_{i}h\nu_{ij} \left[1-\frac{e^{T_{ij}/T_{ex}}-1}{e^{T_{ij}/T_{\ast}}-1}\left(\frac{R_{\ast}}{r}\right)^{2}\frac{1}{4}\right]\mbox{.}
	\end{split}              
\end{equation}

As long as the local thermal equilibrium (LTE) is satisfied, $T_{kin} = T_{ex}=T$. Then the rate of energy lose is:
\begin{equation} \label{eq:cooling}
	\Lambda_{i} = -A_{ij}n_{i}h\nu_{ij} \left[1-\frac{e^{T_{ij}/T}-1}{e^{T_{ij}/T_{\ast}}-1}\left(\frac{R_{\ast}}{r}\right)^{2}\frac{1}{4}\right] \mbox{.}
\end{equation}
 
If $T,\,T_{\ast} \ll T_{ij}$, one might replace $e^{T_{ij}/T}\,-1\,1$ and $e^{T_{ij}/T_{\ast}}\,-\,1$ by $e^{T_{ij}/T}$ and $e^{T_{ij}/T_{\ast}}$, $\Lambda_{i}$ approximates to
\begin{equation}
	\Lambda_{i} = -A_{ij}n_{i}h\nu_{ij} \left[1-\frac{1}{4}\left(\frac{R_{\ast}}{r}\right)^{2}e^{T_{ij}\left(\frac{1}{T}-\frac{1}{T_{\ast}}\right)}\right]
\end{equation}
If $T,\,T_{\ast} \gg T_{ij}$, one might replace $e^{T_{ij}/T}$ 
and $\exp(T_{ij}/T_{\ast})$ by $1\,+\,T_{ij}/T$ and $1\,+\,T_{ij}/T_{\ast}$, $\Lambda_{i}$ 
approximates to
\begin{equation}
	\Lambda_{i} = -A_{ij}n_{i}h\nu_{ij} \left[1-\frac{1}{4}\frac{T_{\ast}}{T}\left(\frac{R_{\ast}}{r}\right)^{2}\right]
\end{equation} 
This last case is valid next to the star where the temperature is high compared to the typical transition energies of the molecules.
Note that $-A_{ij}n_{i}h\nu_{ij}$ in unit of $erg\,$cm$^{-3}\,$s$^{-1}$ is the cooling term, which is already embedded into the Paris-Durham shock code (\autoref{Chapter1}). Therefore, close to the star, where LTE applies, the pumping of molecules by the stellar radiation field can be considered simply by multiplying by a factor of $\left[1 - \left(1/4\right)\left(T_{\ast}/T\right)\left(R_{\ast}/r\right)^{2}\right]$ the molecular cooling term. Far from the star, this factor is close to 1 and the standard ISM non-LTE cooling functions apply.
     
\subsection{Grain temperature and condensation radius}
	The grain temperature evolution is determined by a balance 
	between the heating and the cooling rate.
	In principle, the grains can be heated either by collisions 
	with the gas particles or by direct absorption  of 
	stellar or ambient radiation.
	The grains also can be cooled either by collisional energy transfer 
	or by thermal radiation.
	We assume here that the stellar radiation dominates. 
	The balance that determines the grain temperature $T_{d}$, henceforth, 
	is based on the \textit{radiative equilibrium} condition
	\begin{equation} \label{eq:radiative_equilibrium}
		\int_{0}^{\infty} k_{\lambda} B_{\lambda}(T_{d}) d\lambda =
		\int_{0}^{\infty} k_{\lambda}J_{\lambda} d\lambda
	\end{equation}
	with $k_{\lambda}$ the opacity related to the cross-section of a grain 
	$\sigma Q_{abs}(a,\lambda)$. 
	The left hand side of \autoref{eq:radiative_equilibrium} is the 
	radiative cooling of a grain assumed to be a black-body at wavelength 
	$\lambda$ and temperature $T_{d}$. The right hand side of 
	\autoref{eq:radiative_equilibrium} is the radiative heating from the 
	monochromatic mean intensity $J_{\lambda}$ of the stellar radiation
	field, for which the average of the radiation intensity $I_{\lambda}$ is
	\begin{equation}
		J_{\lambda} = \frac{1}{4\pi}\int^{4\pi}_{0} I_{\lambda} d\Omega	 \mbox{.}
	\end{equation}	 
	Far from to the star, the radiation is well approximated by a diluted Black-body radiation:
	\begin{equation}
		J_{\lambda} = W(r) B_{\lambda} (T_{\ast})
	\end{equation}
	where $W(r) \sim \frac{1}{4}(R_{\ast}/r)^{2}$ is the 
	\textit{geometrical dilution factor}.	  
	
	The radiation condition from \autoref{eq:radiative_equilibrium} 
	for a grain is expanded to 
	\begin{equation} \label{eq:radiative_equilibrium_expansion}
		\int_{0}^{\infty} \pi a^{2}Q_{abs}(a,T_{d})B_{\lambda}(T_{d}) 
		d\lambda =
		\int_{0}^{\infty} \pi a^{2}Q_{abs}(a,T_{\ast})B_{\lambda}(T_{\ast})W(r) d\lambda \mbox{.}
	\end{equation}
	
	Dividing both sides of \autoref{eq:radiative_equilibrium_expansion} by $\int^{\infty}_{0}B_{\lambda}(T)$, the left and right integrations are simply reformulated as the Planck mean efficiency $Q^{P}_{abs}(a,T)$. Therefore, the grain temperature is a function of the radius $r$, which satisfies the relation
	\begin{equation} \label{eq:Td}
	\begin{split}
		& T^{4}_{d} Q^{P}_{abs}(a,T_{d}) = T^{4}_{\ast}W(r)Q^{P}_{abs}(a,T_{\ast}),\ \mbox{or}\\
		\\
		& T_{d}=T_{\ast} W^{1/4} \left\{\frac{Q_P(a,T_{\ast})}{Q_P(a,T_{d})}\right\}^{1/4} 
	\end{split}
	\end{equation}	 
	Close to the star, the geometrical dilution factor $W$ equals $1/2$ and	
	\autoref{eq:Td} indicates that the grain
	temperature is almost as high as the stellar temperature. 
	In fact, at such high
	temperature, the grains can not exist because of sublimation. However, we 
	can find the innermost distance $r_{c}$ where the grains may form by 
	replacing the grain temperature $T_{d}$ by the grain condensation 
	temperature $T_{c}$ in the
	\autoref{eq:radiative_equilibrium}. In the simple case when the grain 
	absorption efficiency can be approximated as a
	power-law of wavelength ($Q_{abs} \approx \lambda^{-p}$), the
	innermost distance $r_{c}$ is determined as
	\begin{equation}
		r_{c} = 
		\frac{R_{\ast}}{2}\left(\frac{T_{\ast}}{T_{c}}\right)^{\frac{4+p}{2}} \mbox{.}
	\end{equation}	   
 
%
\section{Interaction between grain particles and\\ stellar radiation}
\subsection{Mie theory for spherical grains}
\label{sec:Mie_theory}
As shown in \autoref{eq:f_rad}, in order to calculate the radiative acceleration on dust $\bar{\Gamma}$, the quantity of the radiation pressure coefficient $Q_{rp}$ is required.

The radiation pressure coefficient is given by 
\begin{equation}
	Q_{rp}(a,\lambda) = Q_{abs}(a,\lambda) + (1-g)Q_{sca}(a,\lambda)
\end{equation}
where $Q_{abs}$, $Q_{sca}$ and $g$ are the absorption coefficient, scattering coefficient and anisotropy parameter. In the case of perfect forward-scattering $Q_{rp}=Q_{abs}$, isotropy scattering $Q_{rp}=Q_{abs}+Q_{sca}=Q_{ext}$ and perfect back-scattering $Q_{rp} = Q_{abs} + 2Q_{sca}$. In the more general case of scattering, the anisotropy parameter has to be in between $\pm 1$.  
 
The absorption and scattering coefficients are calculated by solving the appropriate boundary-value problem for Maxwell$'$s equation, which is known as the Mie theory 
(more detail can be found in \citealt{Gail_2013}). The result for those coefficients from Mie theory are given by the expression
  \begin{equation}
  	Q_{ext} = \frac{2}{x^{2}}\sum_{j=1}^{\infty} (2j+1)\ Re(a_j+b_j)	
  \end{equation}   
  \begin{equation}
  	Q_{sca} = \frac{2}{x^{2}}\sum_{j=1}^{\infty} (2j+1)\ (|a|^{2}_j+|b|^{2}_j)	
  \end{equation}
  \begin{equation}
  	Q_{ext} = Q_{abs} + Q_{sca} \mbox{.}
  \end{equation}
  
  The quantity $x$ is defined as
  \begin{equation}
  	x = \frac{2\pi a n_m}{\lambda}
  \end{equation}
  where $\lambda/n_m$ represents the wavelength in a surrounding medium of refractive index $n_m$ where the dust is embedded. In the vacuum  $n_m$ equals 1. 
  
  The coefficients $a_j$ and $b_j$ are defined by
  \begin{equation}
  \begin{split}
  	&a_j = \frac{m \psi_{j}(mx)\psi^{'}_{j}(x) - \psi_{j}(x)\psi^{'}_{j}(mx)}{m \psi_{j}(mx)\xi^{'}_{j}(x) - \xi_{j}(x)\psi^{'}_{j}(mx)}, \\
  	\\
  	&b_j= \frac{\psi_{j}(mx)\psi^{'}_{j}(x) - m\psi_{j}(x)\psi^{'}_{j}(mx)}{ \psi_{j}(mx)\xi^{'}_{j}(x) - m\xi_{j}(x)\psi^{'}_{j}(mx)} \mbox{.}
  \end{split}
  \end{equation}
  
  The quantity $m=n^{d}/n_m$ is the ratio between the complex indices of refraction of the dust material $n^{d}$ and the surrounding. $n^{d}$ depends on the optical properties of each grain (see \autoref{sec:dust_property}).     

 The wave functions $\psi$ and $\xi$ are determined by the recurrence relation
 \begin{equation}
 	\begin{split}
 	&\psi_{j+1}(x) = \frac{2j+1}{x}\psi_j(x)-\psi_{j-1}(x) \\
 	&\xi_{j+1}(x) = \frac{2j+1}{x}\xi_j(x)-\xi_{j-1}(x) \mbox{.}
	\end{split}
 \end{equation}
 
 The calculations start with
 \begin{equation}
 	\begin{split}
 		&\psi_{-1}(x) = \cos x,\ \ \ \ \psi_0(x)=\sin x \\
 		&\xi_{-1}(x)  = \cos x + i\sin x, \ \ \ \ \xi_0(x)=\sin x -i\cos x  \mbox{.}
 	\end{split}
 \end{equation}
 
 \begin{figure}
   \begin{minipage}[c]{.5\textwidth}
      \includegraphics[width=1\linewidth]
      {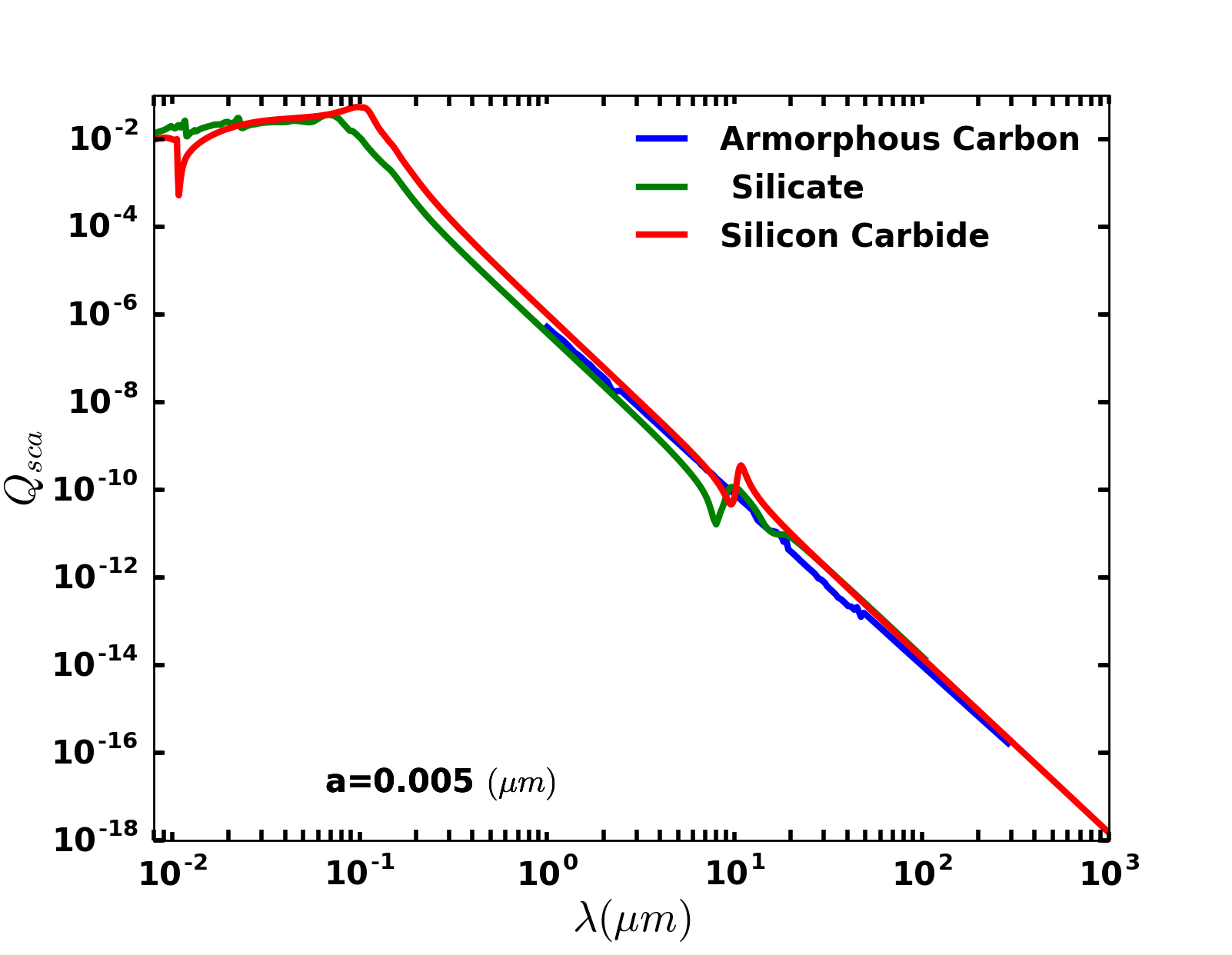}
   \end{minipage} 
   \begin{minipage}[c]{.5\textwidth}
      \includegraphics[width=1\linewidth]
      {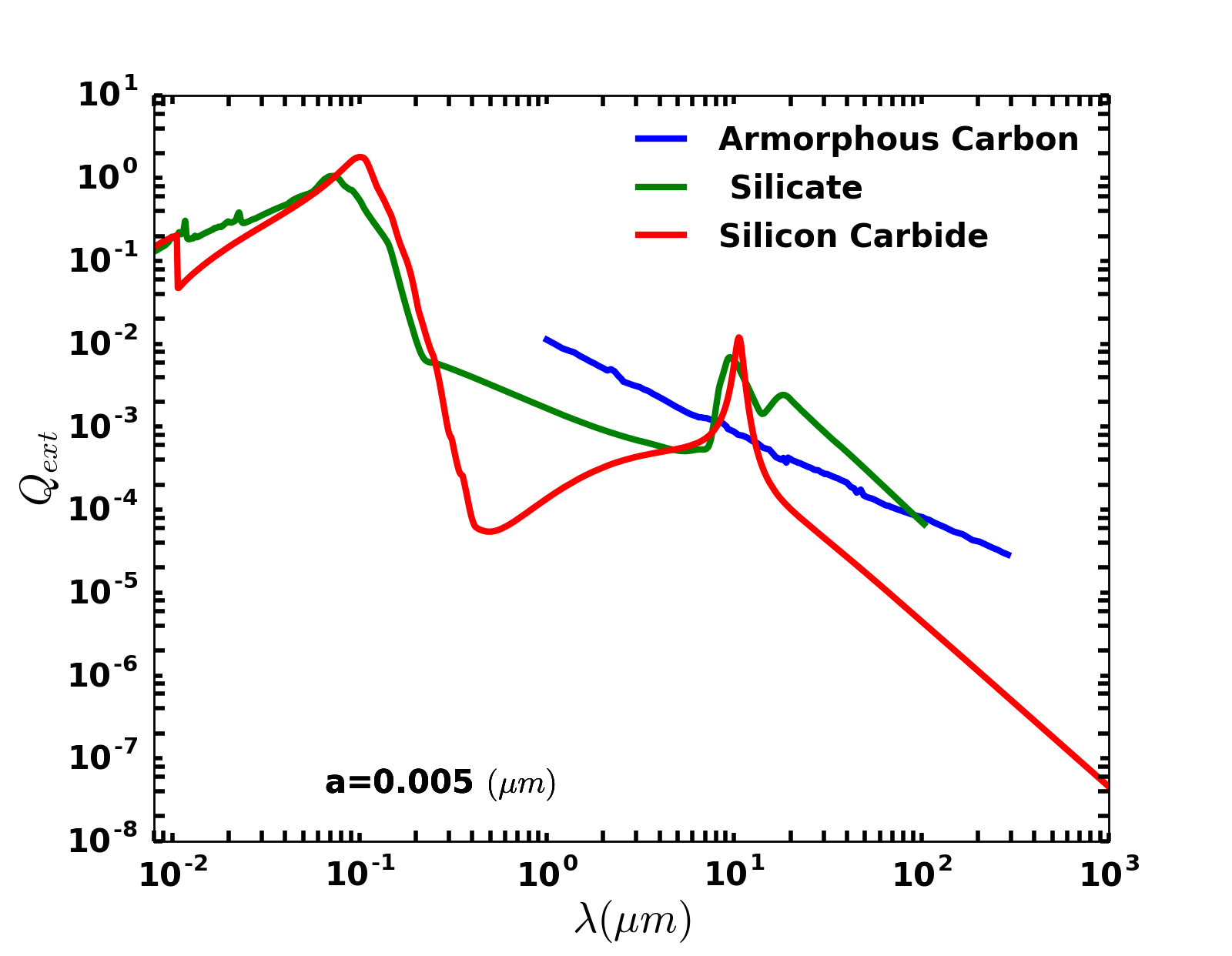}
   \end{minipage}\\

	\caption[Optical properties of grains]
	{(\textit{left panel}) Absorption coefficient and (\textit{right panel}) extinction coefficient of a spherical grain of $0.005 \mu$m radius. The indices of refraction $n^{d}$ are described in \autoref{sec:dust_property}}. 
	\label{fig:optical_properties_dust}
\end{figure}  

\subsection{Optical constants of dust materials} \label{sec:dust_property}
\begin{figure}
	\centering
      \includegraphics[width=0.8\linewidth]
      {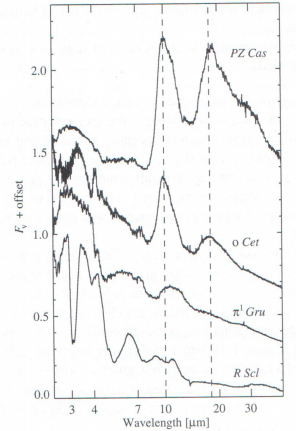}
	\caption[IR spectra from circumstellar dust shells]
	{Examples of IR spectra from circumstellar dust shells, 
	which represent the main different types of AGB stars forming dust shells 
	at different stages of their evolution: $o$ Ceti (M-type), 
	$\pi^1$ Gru (S-type) and RScl (C-type); 
	PZ Cas is a super-giant with an M type spectrum \citep{Gail_2013}.} 
	\label{fig:IR_spectra}
\end{figure}

In order to construct the optical properties of grains and calculate the radiation pressure  on the dust grains, the complex indices of refraction of the grain materials should be known. Although they are called optical 'constants', they are wavelength dependent and they vary depending on the grain type. They cannot be straightforwardly determined from observations. But they can be measured in the laboratory. The comparison between laboratory spectra and IR spectra from circumstellar shells (e.g., \autoref{fig:IR_spectra}) hence allows us to derive the possible features of grains. According to the elemental composition of the stars and the thermal stability of condensed states, the grain material can be identified. The selection of grain material consistent with the stellar material composition (C/O ratio) is essential (see \autoref{sec:CSE_dust}). The first classification of grain material was studied by \citet{Gilman_1969} and \citet{Ney_1969} by using the earliest IR spectra in circumstellar envelopes. \autoref{tab:list_of_dust_species} presents some dust species representative of oxygen-rich and carbon-rich stars. Based on their weak emission in the IR spectra (\autoref{fig:IR_spectra}), stars from the transition region between those two types (hereafter called $S$ stars) can be considered as oxygen-rich stars. 

\begin{table}
\centering
\begin{tabular}{l l l l}
\hline \hline 
Species & \hspace{2mm} Star& \hspace{2mm} References \tabularnewline
\hline 
                            
Armorphous olivine       &\hspace{2mm}    Oxygen-rich  &\hspace{2mm} \citet{Dorschner_1995}  &  \tabularnewline
\\
Armorphous pyroxene      &\hspace{2mm}    Oxygen-rich  &\hspace{2mm} \citet{Jaeger_1994}  &  \tabularnewline
\\
"Astronomical silicate"  &\hspace{2mm}    Oxygen-rich  &\hspace{2mm} \citet{Draine_1984}, \citet{Laor_1993}  &  \tabularnewline
\\
Graphite                 &\hspace{2mm}    Carbon-rich  &\hspace{2mm} \citet{Draine_1984}, \citet{Laor_1993}  &  \tabularnewline
\\
Armorphous carbon        &\hspace{2mm}    Carbon-rich  &\hspace{2mm} \citet{Maron_1990}, \citet{Jaeger_1998}  &  \tabularnewline
\\
Silicon carbide          &\hspace{2mm}    Carbon-rich  & \hspace{2mm} \citet{Laor_1993}  &  \tabularnewline

\hline \hline 
\end{tabular}
\caption[Dust species identified in circumstellar envelopes]
{Some dust species identified in circumstellar envelopes 
for oxygen-rich and carbon-rich stars.}
\label{tab:list_of_dust_species}
\end{table}

\section{Chemistry}
\subsection{Initial photospheric elemental compositions}
	Along the stellar wind, the evolution of chemical species essentially depends on the initial elemental composition at the stellar photosphere. We assume the elemental photospheric abundance to be the same as the solar abundance except for the C/O ratio. Each value of this ratio corresponds to an AGB type (\citealt{Cherchneff_2006}). Carbon-rich stars have C/O larger than 1 and on the contrary oxygen-rich stars have C/O lower than 1. 
	
	Most of the elemental abundances at the solar photosphere are deduced by spectroscopy: the emitted spectra are first calculated by atmospheric models; then calculated spectra are compared to the observed spectrum; and finally the abundances are returned based on those comparisons. However, the conversion from the spectrum to the value of abundances requires the theoretical knowledge of line position, transition probability and lifetime of excited levels. The local thermal equilibrium (LTE) for which excited levels are populated regarding the relations of Boltzmann and Saha is therefore usually used to calculate solar abundances. \autoref{tab:elemental_solar} shows most of the initial element compositions at the solar photosphere (\citealt{Lodders_2009}).
	
	\begin{table}
\centering
\begin{tabular}{l l c c}
\hline \hline 
Element & \hspace{1mm} Composition & \hspace{1mm} Initial density ($n/n_{H})$ & \hspace{1mm} Formation enthalpy \tabularnewline
\hline 
                            
H    &   01000000000000 & 1  &  51.634   \tabularnewline
He   &   00000010000000 & 8.414(-02)  &  0.0000   \tabularnewline
C    &   00050000000000 & 2.455(-04)  &  169.98   \tabularnewline
N    &   00001000000000 & 7.244(-05)  &  112.53   \tabularnewline
O    &   00000100000000 & 2.455(-04)  &  58.980   \tabularnewline
Mg   &   00000000100000 & 3.467(-05)  &  35.000   \tabularnewline
Si   &   00000000001000 & 3.311(-05)  &  106.70   \tabularnewline
S    &   00000000010000 & 1.380(-05)  &  65.600   \tabularnewline
Fe   &   00000000000010 & 2.818(-05)  &  98.700   \tabularnewline

\hline \hline 
\end{tabular}
\caption[Initial elemental composition of the most abundant species in the solar photosphere]
{Initial elemental composition of the most abundant species in the solar photosphere when C/O equals 1.5 (\citealt{Lodders_2009}). Numbers in parentheses are powers of 10.} 
\label{tab:elemental_solar}
\end{table}

\subsection{Chemical network}
\label{sec:3body}
	In this study, we use the network from the Paris-Durham code (\autoref{sec:chemical_PD}), but we will focus on hydrogen chemistry. The Paris-Durham chemical network is optimized and includes most of the recent bimolecular reactions relevant for the interstellar medium (\aref{app:chemical_species}). However, the gas density in the inner envelope is sufficiently high 
($n \gtrsim 10^{10}\,$cm$^{-3}$)  that the trimolecular processes can be initiated.  Through those  reactions, which matter for hydrogen, two atomic hydrogens react together, when another atomic hydrogen or a molecular hydrogen acts as a third body to evacuate the binding energy of the H$_{2}$ molecule
 
	\begin{equation} \label{eq:CD_H}
		\ce{H + H_{2} <=>[k_{1}][k_{2}] H + H + H}
	\end{equation}
	\begin{equation} \label{eq:CD_H2}
		\ce{H_{2} + H_{2} <=>[k_{3}][k_{4}] H + H + H_{2} \mbox{.} } 
	\end{equation}		

    In a such dense medium, molecular hydrogen is formed by three-body recombination of hydrogen with rate coefficients $k_{2}$ and $k_{4}$, while it is destroyed by the collisional dissociation with rate coefficients $k_{1}$ and $k_{3}$. The values of those coefficients are discussed in the next section.
	
    Finally, the formation of dust grains is not treated self-consistently throughout the model. Dust grains are turned on whenever the gas temperature is below the condensation temperature and then they remain constant. Thus, we turn off gas-grain reactions in the chemical network, except for the formation of H$_2$ at the grains' surface.

\subsection{Formulation of hydrogen chemistry on the stellar surface}
\label{sec:Formulation of hydrogen chemistry on the stellar surface}	
	\subsubsection{Discussion of selected reaction rate coefficients}
\begin{figure}
		\begin{center}
		\begin{minipage}{0.9\linewidth}
    		\includegraphics[width=1.\linewidth]
      		{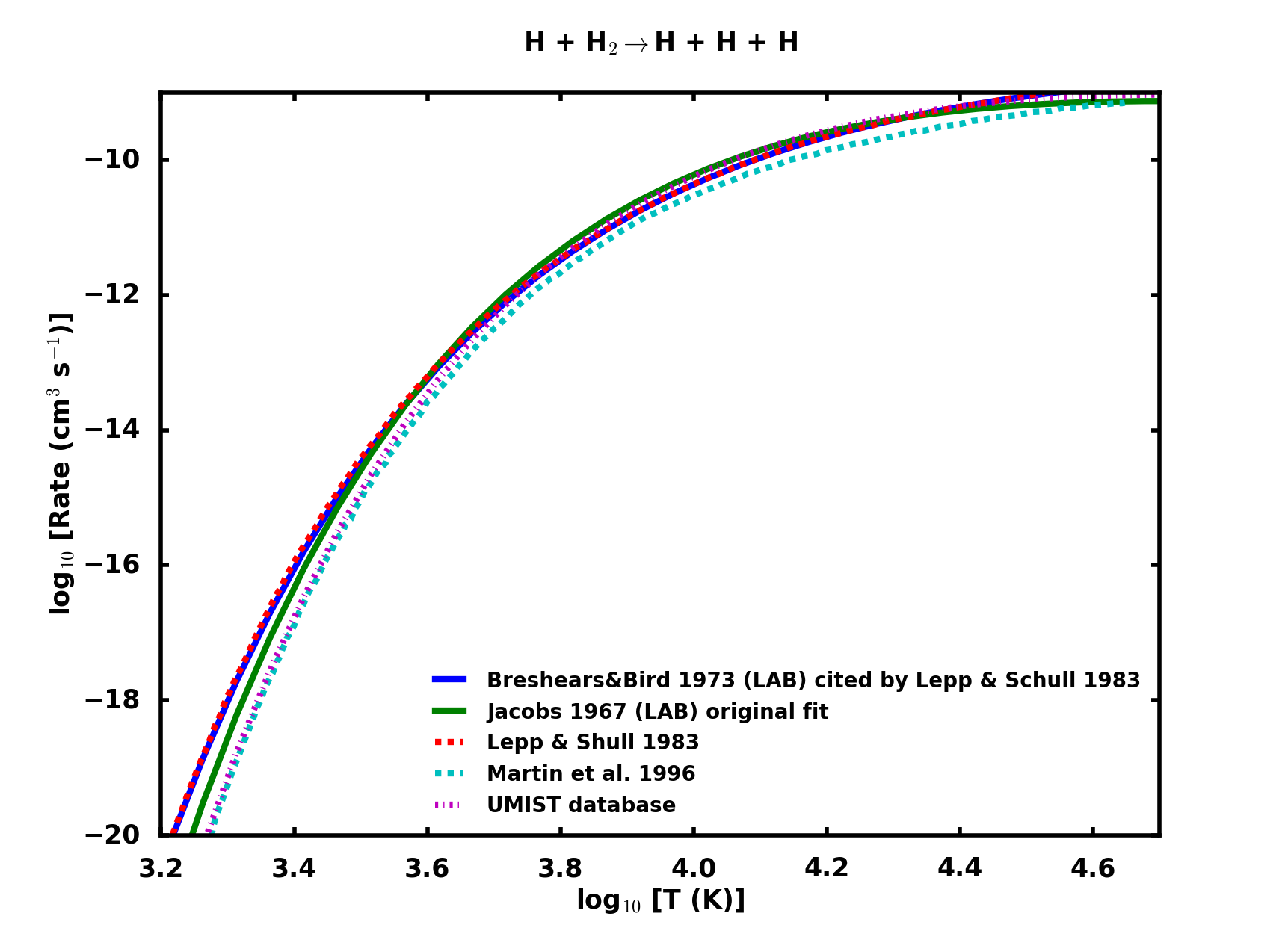}
      	\end{minipage} \hfill
      	\begin{minipage}{0.9\linewidth}
    		\includegraphics[width=1.\linewidth]
      		{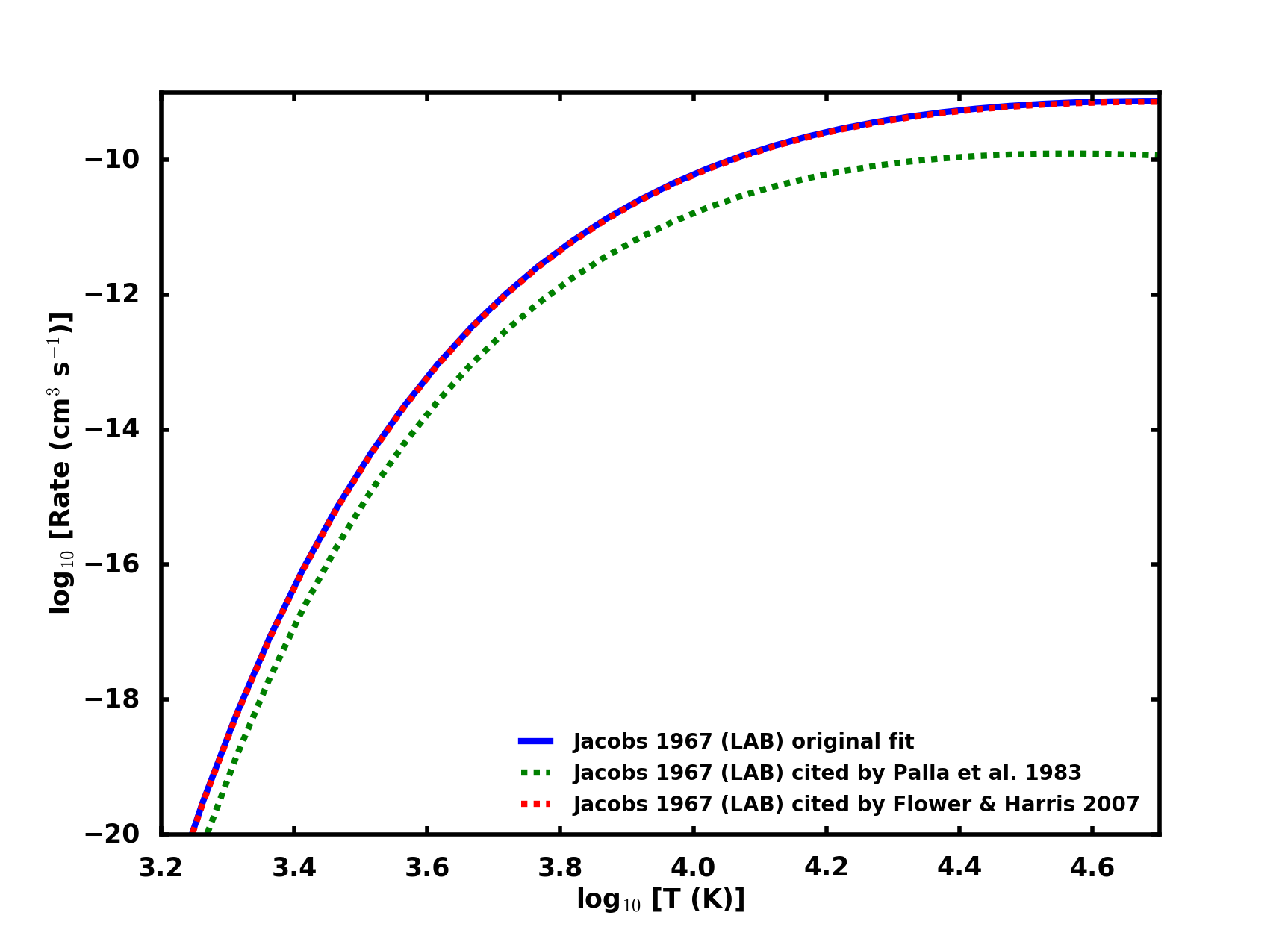}
      	\end{minipage}
		\caption[Studies of H-H$_{2}$ collisional dissociation]
		{Studies of H-H$_{2}$ collisional dissociation rate coefficients. 
		(\textit{Top}) Discrepancy of collisional dissociation rate coefficients 
		between experiments and theoretical studies. 
		(\textit{Bottom}) Discrepancy between citations of 
		\cite{Jacobs_1967}'s collisional dissociation rate coefficient.} 
		\label{fig:H_H2_CD}
		\end{center}
		\end{figure}

\begin{figure}
		\begin{center}
		\begin{minipage}{0.9\linewidth}
    		\includegraphics[width=1.\linewidth]
      		{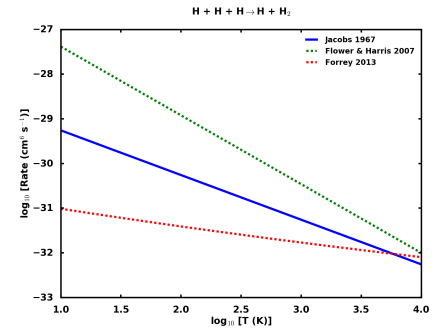}
      	\end{minipage} \hfill
		\caption[Three-body recombination of atomic hydrogen]
		{Three-body recombination of atomic hydrogen 
		(\citealt{Jacobs_1967}, \citealt{Flower_2007}, and \citealt{Forrey_2013}.} 
		\label{fig:3H}
		\end{center}
		\end{figure}

		The evolution of the abundance of molecular hydrogen 
		($x_{2} = n(H_{2})/n_{H}$)
		is generally expressed in the Lagrangian form (\citealt{Glassgold_1983}) as:
		\begin{equation} \label{eq:xH2_formulation}
			\frac{{\rm d} x_{2}}{{\rm d} t} = P - D\,x_{2}
		\end{equation}
		where $P$ and $D$ are the production and destruction rates 
		of H$_{2}$. When it reaches equilibrium, 
		these two rates must be equal. 
		In chemical equilibrium, thus, 
		the forward and backward rates in \autoref{eq:CD_H} balance:  	
		\begin{equation}
			n(H)n(H_{2})k_{1} = n^{3}(H)k_{2}
		\end{equation}
		and the Saha relation gives
		\begin{equation} \label{eq:Saha}
			\frac{n(H_{2})}{n^{2}(H)} = \frac{Z(H_{2})}{Z(H)^2}\left(\frac{h^{2}}{\pi m_{H}k_{B}T}\right)^{3/2} e^{\frac{E_{diss}}{k_{B}T}} = \frac{k_{2}}{k_{1}}
		\end{equation}
		where $E_{diss} = 4.48 eV$ is the dissociation energy of H$_{2}$. 
		The partition function of H is $Z(H) = 2$ \citep{Flower_2007} 
		or $Z(H)=4$ if discernibility and electron degeneracy 
		is taken into account \citep{Forrey_2013}. 
		The partition function of H$_{2}$ is defined by
		\begin{equation} 
			Z(H_{2}) = \sum_{i}g_{i} e^{-\frac{E_{i}}{k_{B}T}}
		\end{equation}
		where g$_{i}$ and $E_{i}$ are the statistical weight and 
		energy for the excitation level $i$. 
		According to the table of the excitation energy (\aref{app:tab_H2_excitation}), 
		\citet{Flower_2007} fit the partition $Z(H_{2})$ 
		as a function of temperature up to $\sim$ 2000 K. 
		The best fit gives the formula
		\begin{equation} \label{eq:Z_sim}
			Z(H_{2}) \simeq 0.028 T^{0.985} \mbox{.}		
		\end{equation}
		Knowing the rate coefficient k$_{1}$ allows us to derive the rate coefficient $k_{2}$. 
		\autoref{fig:H_H2_CD} shows a few rate coefficients of H$_{2}$ collisional dissociation. 
		The evolution of this rate coefficient with respect to the temperature differs between 
		laboratory experiments (e.g., \citealt{Jacobs_1967}, 
		\textit{Breshears \& Bird 1973} cited by \citealt{Lepp_1983}) 
		and theoretical studies (e.g., \citealt{Lepp_1983}, \citealt{Martin_1996}), 
		and even with the \textit{UMIST} database (\autoref{fig:H_H2_CD}, \textit{top}, 
		this coefficient in UMIST has no precise reference cited in but "literature search"). 
		As many previous authors, we used the experimental rate coefficient 
		from \citet{Jacobs_1967}. 
		However, the latter is not the same in different citations 
		as shown in the \textit{bottom} of \autoref{fig:H_H2_CD}. 
		We chose
		\begin{equation}
			k_{1}=1.38\,10^{-4}\,T^{-1.025}\;e^{(-52000/T)}
		\end{equation}		 
		in unit of cm$^{3}\,$s$^{-1}$ (\citealt{Jacobs_1967}; \citealt{Flower_2007}). 
		Then \autoref{eq:Saha} allows us to derive
		\begin{equation} \label{eq:k2_FH}
			k_{2} = 1.44\,10^{-26}\,T^{-1.54} 		
		\end{equation}		 
		in unit of cm$^{6}\,$s$^{-1}$. This derived rate coefficient, however, 
		differs from \citet{Jacobs_1967}'s deduced coefficient as
		\begin{equation}
			k_{2,J} = 5.52\,10^{-29}\,T^{-1}		
		\end{equation}
		and from the theoretical study of \citet{Forrey_2013}
		\begin{equation}
			k_{2,F} = 6\,10^{-32}\,T^{-1/4} + 2\,10^{-31}\,T^{-1/2} 				
		\end{equation}		 
		in the same unit.  \autoref{fig:3H} illustrates the discrepancy 
		between  those three rate coefficients. 
		In this work, we have chosen the rate coefficient 
		of the recombination of three atomic hydrogen ($k_{2}$) 
		as in \autoref{eq:k2_FH}.   
    		
		The connection of the rate coefficients between \autoref{eq:CD_H} 
		and \autoref{eq:CD_H2} is
		\begin{equation} \label{eq:rate_balance}
			\frac{k_{1}}{k_{2}} = \frac{k_{3}}{k_{4}}													
		\end{equation}
		
		In practice, the rate coefficient k$_{4}$ (or k$_{3}$) in \autoref{eq:CD_H2} 
		is easier to measure in the laboratory than 
		the rate coefficient k$_{2}$ (or k$_{1}$) in \autoref{eq:CD_H}. 
		\citet{Ham_1970} and \citet{Cohen_1983} made experiments 
		at low temperature (up to room temperature $\sim 300\,$K) 
		to measure k$_{4}$, 
		for which hydrogen is initially mainly molecular.  
		\citet{Cohen_1983} extrapolated it for high temperature and recommended the relation
		\begin{equation}
			k_{4,C} = 2.8\,10^{-31}\,T^{-0.6} 		
		\end{equation}
		in unit of cm$^{6}\,$s$^{-1}$. In addition, \citet{Jacobs_1967} showed that 
		this rate coefficient related to the rate coefficient k$_{2}$ as $k_{4} = k_{2}/8$, 
		and adopting \autoref{eq:k2_FH} for $k_{2}$, we get
		\begin{equation} \label{eq:k4_FH}  
			k_{4} = k_{2}/8 = 1.8\,10^{-27}\,T^{-1.54}
		\end{equation}
		in the same units. 
		Nevertheless, these two rate coefficients differ between each other 
		by a factor of 10 at 1000 K.
		That is $k_{4,C} \simeq 4.4\,10^{-33}\,$cm$^{6}\,$s$^{-1}$, 
		while $k_{4} \simeq 4\,10^{-32}\,$cm$^{6}\,$s$^{-1}$ at $T=1000\,$K. 
		Even so, in this work, we have chosen the rate coefficient k$_{4}$ 
		(\autoref{eq:k4_FH}). 
		This choice bases on two reasons, 
		which have been discussed by \citet{Flower_2007}. 
		First, the extrapolation of \citet{Cohen_1983} to higher temperature remains uncertain; 
		and second, the inverse reaction in \autoref{eq:CD_H} is less important 
		than in \autoref{eq:CD_H2} 
		since atomic hydrogen is initially dominant at stellar surface, 
		which is the opposite to the laboratory situation. 
		Finally, the last rate coefficient ($k_{3}$) in unit of cm$^{3}\,$s$^{-1}$ 
		is derived from \autoref{eq:rate_balance}, which yields
		\begin{equation}
			k_{3} = k_{1}/8 = 1.73\,10^{-5}\,T^{-1.025}\,e^{(-52000/T)} \mbox{.}	
		\end{equation}		 
		 
		The corresponding chemical timescales of these four reactions are
  		\begin{equation} \label{eq:chemical_timescale}
    		\begin{split}
    			t^{chem}_{1} &= \frac{1}{n_{H} k_{1}} 
    			       = \frac{0.72\,10^{4}}{n_{H}}\,T^{1.025}\,e^{52000/T}  \\
    			t^{chem}_{2} &= \frac{1}{n^{2}_{H} k_{2}}
    			       = \frac{0.69\,10^{26}}{n^{2}_{H}}\,T^{1.54}\\
    			t^{chem}_{3} &= \frac{1}{n_{H} k_{3}}
    			       = \frac{0.58\,10^{5}}{n_{H}}\,T^{1.025}\,e^{52000/T}   \\
        		t^{chem}_{4} &= \frac{1}{n^{2}_{H} k_{4}}
        		       = \frac{0.56\,10^{27}}{n^{2}_{H}}\,T^{1.54} \mbox{.}
    		\end{split}
    	\end{equation}
    	in unit of $s$. In general, the timescales are dependent on both density and temperature. 
		The timescale of the collisional dissociation $t_{1}$ and $t_{3}$ 
		are inversely proportional to the density, 
		while the ones of the three-body recombination 
		$t_{2}$ and $t_{4}$ are inversely proportional to density squared.  
	       		
	\subsubsection{Glassgold and Huggins critical effective temperature}
\begin{figure}
		\centering
    		\includegraphics[width=0.9\linewidth]
      		{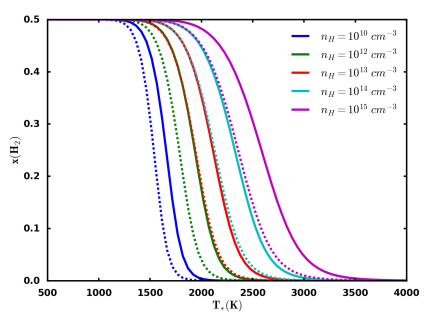}
			\caption[Thermal equilibrium of abundance of molecular hydrogen at stellar photosphere]
			{Thermal equilibrium of abundance of molecular hydrogen at stellar photosphere. 
			 Dotted lines correspond to the use of the rates 
			 by \citet{Forrey_2013} instead of \citet{Flower_2007}.} 
		\label{fig:glassgold_critical}
	\end{figure}

		As presented in equations \ref{eq:CD_H} and \ref{eq:CD_H2}, 
		at high densities close to the surface and in the absence of dust, 
		the hydrogen molecule is formed with rate coefficients 
		$k_{2}$ and $k_{4}$, 
		while it is destructed with rate coefficients $k_{1}$ and $k_{3}$. 
		Taking them into account, chemical equilibrium yields:
		\begin{equation} \label{eq:xH2_1}
			\begin{split}
			1 &= \frac{k_{2}x^{3} + k_{4}x^{2}x_{2}}{k_{1}x_{2}x+k_{3}x^{2}_{2}} n_{H}\\
			\\
			&=\frac{k_{2}(1-2x_{2})^{3} + k_{4}x_{2}(1-2x_{2})^{2}}{k_{1}x_{2}(1-2x_{2})+k_{3}x^{2}_{2}} n_{H} \mbox{.}
			\end{split}
		\end{equation}		  
		We rewrite \autoref{eq:xH2_1} as:
		\begin{equation}\label{eq:xH2}
			(1-2x_{2})^{2}x_{2} + \frac{k_{2}}{k_{4}} (1-2x_{2})^{3} - 
			\frac{1}{n_{H}}\left[\frac{k_{1}}{k_{4}}(1-2x_{2})x_{2} + \frac{k_{3}}{k_{4}x^{2}_{2}}\right] = 0	\mbox{.}		
		\end{equation}		 		
		Solving \autoref{eq:xH2} allows us to determine $x_{2}$, 
		the equilibrium fractional abundance 
		of molecular hydrogen $x_{2}$. 
		Since the reaction rate coefficients are a function of temperature, 
		the hydrogen fractional abundance at the stellar photosphere 
		 is a function of both stellar temperature and density. 
		\autoref{fig:glassgold_critical} shows the relation between 
		H$_{2}$ fractional abundance and stellar temperature 
		for several density values. 
		Depending on the stellar temperature, hydrogen can be either 
		in molecular or in atomic form
		at the stellar surface. The hydrogen is in molecular form for \textit{"cold"} stars, 
		and in atomic form for \textit{"hotter"} stars. This transition point varies over
		the density of stellar surface. For stars with $n_{H}\geq 10^{14}\,$cm$^{-3}$, 
		the transition is around at $2500\,$K, which means that stars 
		with $T_{\ast} \leq 2500\,$K contains mostly molecular hydrogen,
		and mostly atomic hydrogen with $T_{\ast} > 2500\,$K. 
		That was concluded also by \citep{Glassgold_1983}.
		The use of the rates by \citet{Forrey_2013} leads 
		to a shift of this critical temperature of about 200K towards lower temperatures. 
		This corresponds to a 4-fold increase of atomic hydrogen in the molecular side of 
		the diagram, which has a similar impact on the predicted HI emissivities for given 
		hot (i.e., hotter than the critical temperature) AGB stars observations of HI.
	    
    \subsection{Collisional dissociation of H$_2$ level by level}
  Collisional dissociation of H$_2$ is included in the Paris-Durham code, as it is important for dissociative shocks. In the dilute conditions of the ISM, LTE is not realized for the H$_2$ levels, and it is important to consider the dissociation level by level \citep{Bourlot02}. Our level by level implementation assumes that the dissociation energy barrier for a given level is lowered precisely by its excitation energy:
\begin{equation}
D_{i}=D_{00}e^{T_{i}/T}
\end{equation}
where $D_{00}$ is the rate for collisional dissociation {\it of the ground level} ($v=0,J=0$). In the former version of the Paris-Durham code, the input list of chemical reactions assumes that the rate for $D_{00}$ is given. For instance, the rate used for H + H$_2 \rightarrow 3H$ was $D_{00}=10^{-10}\exp(-52000/T)\,$cm$^{3}\,$s$^{-1}$ \citep{Dove_1986}. However, the rates we provide now are for the {\it total} rate at LTE:
\begin{equation}
D_{\rm LTE}=\sum_{i} D_{i}\frac{g_{i}}{Z(H_{2})}e^{-T_{i}/T}=D_{00}\frac{\sum_{i} g_{i}}{Z(H_{2})} \gg D_{00} \mbox{.}
\end{equation}

\begin{figure}
\centering
	\includegraphics[width=0.9\linewidth]
	{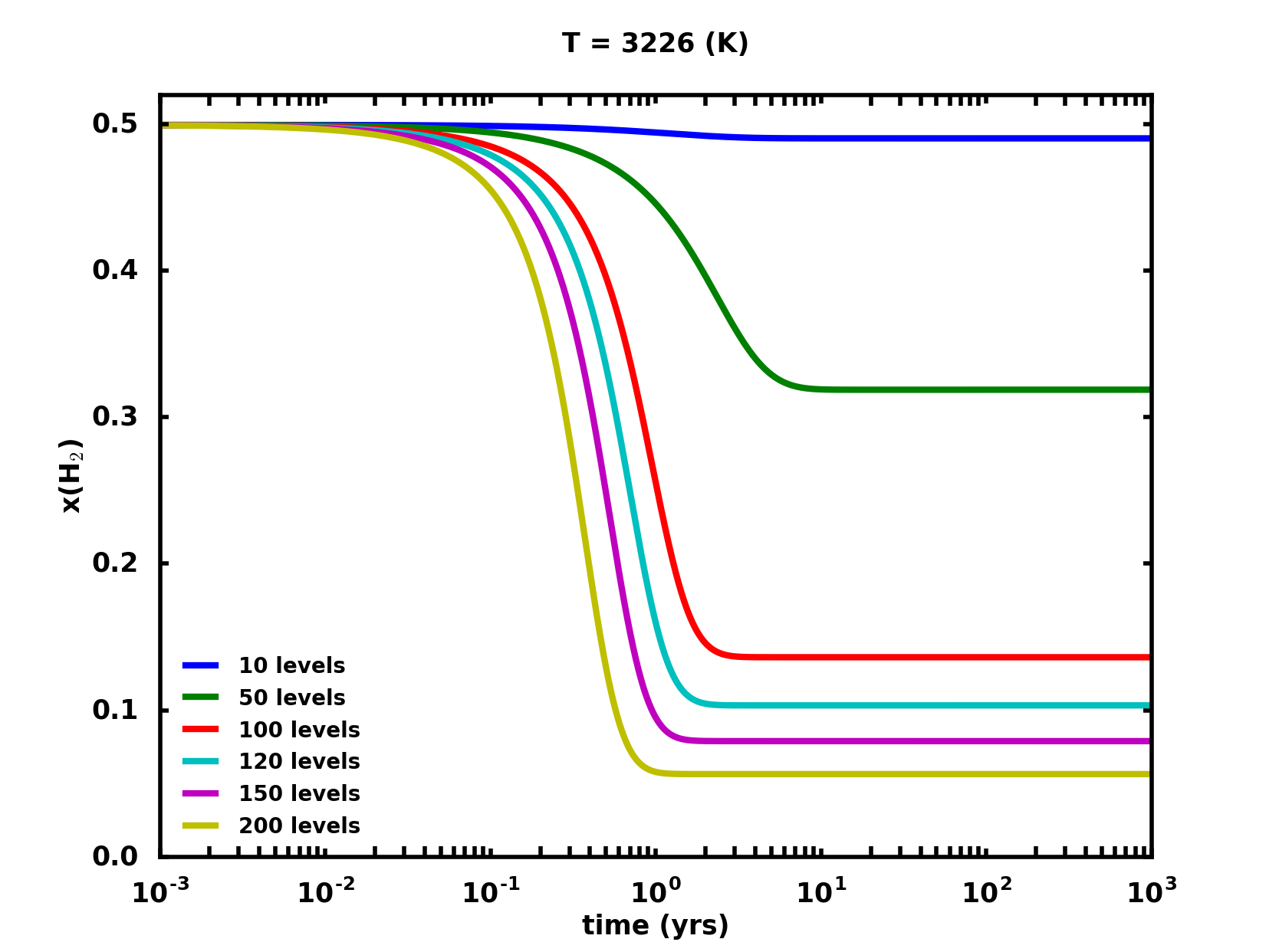}
	\caption[Convergence of H$_{2}$ abundance with respect 
	to the number of H$_{2}$ levels treated]
	{Convergence of H$_{2}$ abundance with respect 
	to the number of H$_{2}$ levels treated at $3226\,$K and $n_{H}=10^{9}\,$cm$^{-3}$.} 
	\label{fig:H2_convergence}
\end{figure}
	
  In order to recover the correct rate at LTE near the stellar surface, and to keep our assumed level by level behavior in the dilute ISM phase, we therefore need to use the new prescription:
\begin{equation}
D_{i}=D_{\rm LTE}\frac{Z(H_{2})}{\sum_{i} g_{i}} e^{T_{i}/T}
\end{equation}
which we implemented in the Paris-Durham code. Note that in our assumption, the barrier compensation due to excitation exactly compensates the Boltzmann factor: at LTE, each level contributes roughly in proportion of its statistical weight. This means in particular that we need to include a large number of H$_2$ levels in order get convergence for the hydrogen chemistry (see \autoref{fig:H2_convergence}). It would be desirable to implement a more efficient treatment for the discarded levels in order to achieve better computational efficiency for the winds. For example, one could assume that all discarded levels are populated as a Boltzmann population with respect to the last included level.

 \section{Extinction}

  Radiation from nearby stars is able to photo-ionize species and photo-dissociate molecules. Dust grains, however, will absorb the radiation and protect the chemical elements against photo-destruction. In the Paris-Durham shock code, the impact of the dust is modeled through the extinction variable $A_{v}$. This is computed with a plane-parallel geometry. We also assume the interstellar radiation comes from upstream \citep{PL13}, allowing us to integrate $A_{v}$ alongside the model calculation:
\begin{equation}
\label{eq:extinction-shock}
A_{v}^{\rm shock}=A_{v0}+\int_{z_{0}}^{z} dz \alpha_{A_{v}}n_{H}
\end{equation}
where $A_{v0}$ is the starting extinction upstream of the shock and $\alpha_{A_{v}}=5.34 \times 10^{-22}$cm$^{2}$ is a cross-section parameter which characterizes the optical properties of interstellar dust and their abundance with respect to $n_{H}$.

\begin{figure}
\centering
	\includegraphics[width=0.9\linewidth]
	{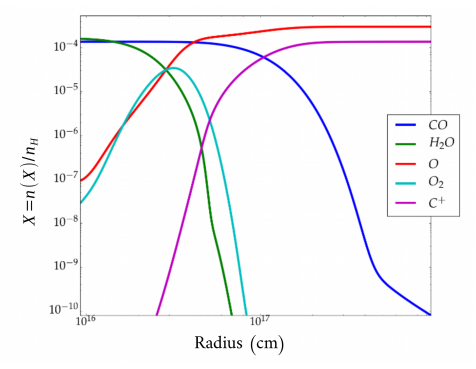}
	\caption[Some chemical abundances]
	{Some chemical abundances for comparison with figure(2) of \citet{Mamon_1987}.} 
	\label{fig:chem_abundance}
\end{figure}

  In the wind situation, however, the chemically active radiation comes from the outside, at the tip of the wind, and we don't know the density profile. However, we can nevertheless resort to approximations, as many authors before (e.g., \citealt{Mamon_1987}). We assume that the terminal velocity is already reached and that the density profile decays exactly as $1/r^2$. We also assume that the dust properties behave like standard ISM dust 
\begin{equation}
\label{eq:extinction-wind}
A_{v}^{\rm wind}(r)=A_{v0}+ r\alpha_{A_{v}}n_{H}(r) \mbox{.}
\end{equation}
As $n_{H}$ varies as $1/r^2$, this expression diverges close to the star, where the extinction is so large that the photo-reactions don't matter anymore. Far from the star, dust properties are more likely to behave as standard ISM dust, so the approximation is also appropriate. We checked that this yields photo-dissociation profiles comparable to the results found by \citet{Mamon_1987}, as illustrated in \autoref{fig:chem_abundance}.
  
  Self-shielding functions for the photo-dissociation of H$_2$ and CO was also included in the Paris-Durham code \citep{Lee_1996, PL13}. These functions require column-densities as entry parameters, and we made similar assumptions as for \autoref{eq:extinction-wind}:
\begin{equation}
N^{\rm shield}_{H_{2}}=N_{H_{2}0}+r.n(H_{2})
\end{equation}
and
\begin{equation}
N^{\rm shield}_{CO}=N_{CO}+r.n(CO) \mbox{.}
\end{equation}

  Extinction also occurs in the line cooling: when photons need to go through a large quantity of matter to find the exit to the circumstellar envelope, radiation becomes optically thick, and line cooling effectively shuts off. This is accounted for by a velocity-gradient parameter in the molecular cooling tables of \citet{Neufeld_Kaufman_1993}. We use their recommended value for spherical symmetry:
\begin{equation}
\frac{dv_{n}}{dz}=\frac{27}{2}\frac{c_{s}}{r}+2\frac{v_{n}}{r} \mbox{.}
\end{equation} 

\section{Validation}
\begin{figure}
   \begin{center}
   \begin{minipage}[c]{.9\textwidth}
      \includegraphics[width=0.91\linewidth]
      {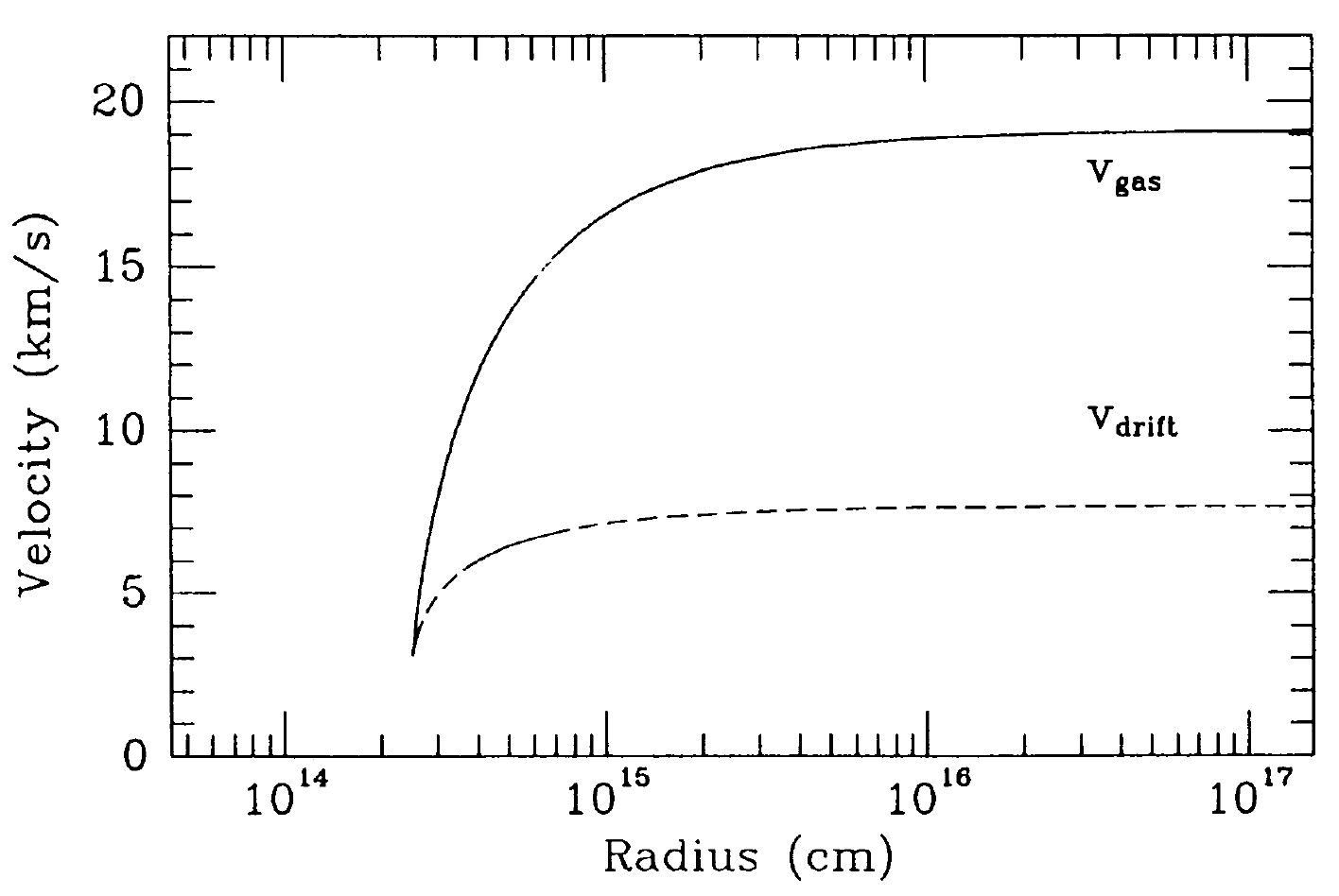}
   \end{minipage} 
   \begin{minipage}[c]{.9\textwidth}
      \includegraphics[width=1\linewidth]
      {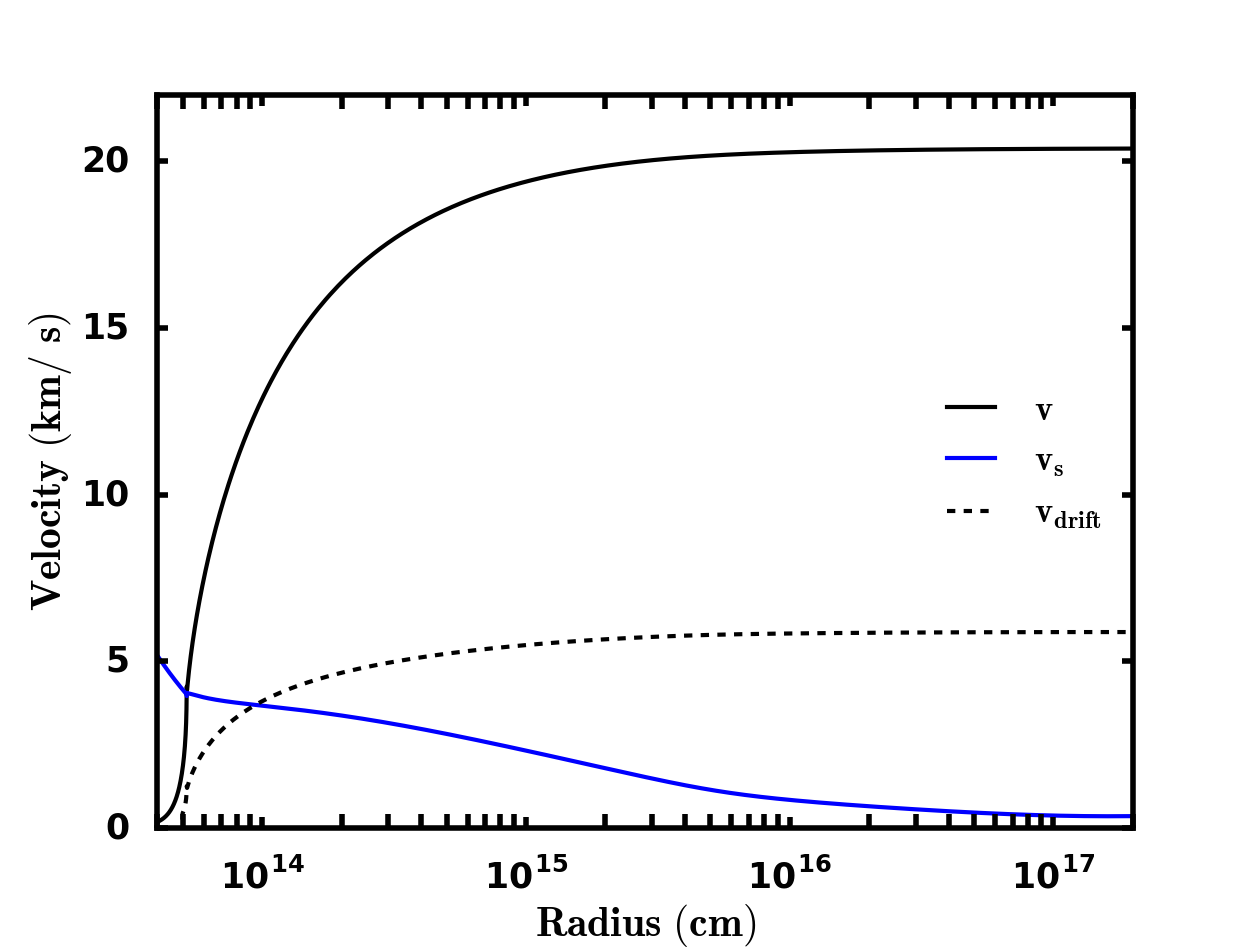}
   \end{minipage}\\
	\caption[Dynamical properties of gas in the CSE of GX Mon star]
	{Dynamical properties of gas in the CSE of GX Mon star (\textit{bottom}) 
	for comparison with \citet{Justtanont_1994}'s model (\textit{top}).
        The blue line in the \textit{bottom panel} is the speed of sound.
        GX Mon parameters: 
	$M_{\ast} = M_{\odot}$, $\dot{M}_{\ast} = 7.2\,10^{-6}\,$M$_{\odot}\,$yr$^{-1}$, 
	$T_{\ast} = 2500\,$K.} 
	\label{fig:GXMon}
   \end{center}
\end{figure}

Finally, we ran a stellar wind model for GX Mon star and we checked that the dynamical properties of the gas that we calculated are comparable to the results of \citet{Justtanont_1994}, as showed in \autoref{fig:GXMon}

\setstretch{1.0} 
\chapter{MODELS}
\label{Chapter9}
\lhead{Chapter 9. \emph{Models}} 
\section{IRC +10216 star}
	\subsection{Hydrodynamics}
	IRC +10216 or CW Leonis is the nearest carbon-rich AGB star, at a distance of $\sim$130 pc, 
	and it has been extensively observed.
	IRC +10216 is believed to be close to the stage where it becomes a protoplanetary nebula 
	(e.g., \citealt{Skinner_1998}). 
	Because of this, there is no doubt that IRC-10216 has high mass-loss rate 
	($\approx 2\,10^{-5}\,$M$_{\odot}\,$yr$^{-1}$; 
	i.e., \citealt{Crosas_1997}; \citealt{Groenewegen_1998}). 
	The main parameters of IRC +10216 are summarized in \autoref{tab:IRC_parameters} 
 
	The effective temperature is uncertain 
	due to the thickness of the dust envelope. 
	Quoted values are $2300\,$K \citep{Cohen_1979}, 
	$2330\,$K \citep{Ridgway_1988},
	$2200\,$K \citep{Ivezic_1996}, or
	$1915-2105\,$K \citep{Bergeat_2001}. 
	We, therefore, choose a mean value of $2200\,$K \citep{Matthews_2015}.
	
	The stellar mass is adopted to be $0.8\,$M$_{\odot}$ as a mean value 
	ranges from 
	$0.7$ to $0.9\,$M$_{\odot}$ \citep{Ladjal_2010}. 
	Thanks to VLA observations, \citet{Menten_2012} determines 
	the luminosity of IRC +10216 at about $8640\,$L$_{\odot}$.
	Adopting these stellar effective temperature and luminosity allows us 
	to derive its radius $R_{\ast} \approx 4.5\,10^{13}\,$cm.   
	The carbon-to-oxygen ratio is assumed to be 1.5 
	(e.g., \citealt{WC_1998}, \citealt{Cherchneff_2006}). 
	Since we are concerned with dust condensation, and since IRC +10216 
	is an extreme carbon-rich star, 
	we use the spherical amorphous carbon dust grains, with the refraction index 
	from \citet{Maron_1990} (see \autoref{sec:dust_property}). 
							
	\subsubsection{Freely expanding wind region}	  
\begin{figure}
		\centering
    		\includegraphics[width=0.9\linewidth]
      		{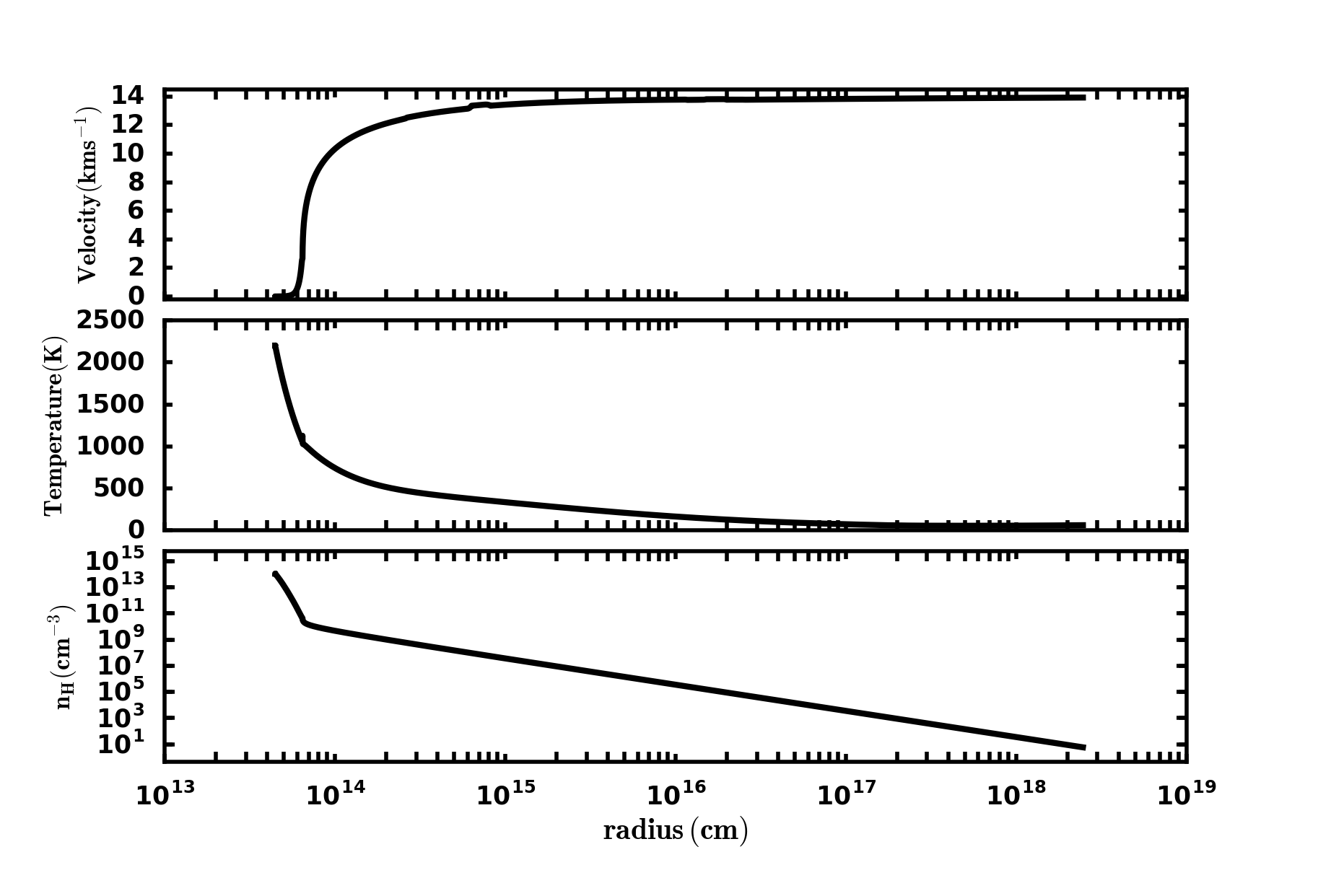}
			\caption[Physical profiles of gas in the freely expanding wind region 
			of the IRC+10216's CSE]
			{Physical profiles of gas in the freely expanding wind region 
			of the IRC+10216's CSE.} 
			\label{fig:free_IRC}
		\end{figure}

		Using the parameters of \autoref{tab:IRC_parameters}, the physical profiles of the gas 
		in the freely expanding wind region are shown in 
		\autoref{fig:free_IRC}. 
		The \textit{top panel} indicates that the terminal velocity 
		is about $14$ km$\,$s$^{-1}$, which is in good agreement with the observed value 
		for IRC +10216 
		(e.g., \citealt{Olofsson_1993}, \citealt{Knapp_1998}). 
		In addition, the gas flow starts reaching this stationary value 
		at $\sim$ 10$^{15}$ cm, 
		which is also in good agreement with \citet{Agundez_2012}.
		
		Inside the dust-free region, we fit the gas kinetic temperature and the gas number density 
		to a power-law of radius as $r^{\alpha}$. 
		The number density varies with $\alpha = -2$. 
		While the gas temperature varies 
		with $\alpha = -0.33$, where 
		$r \leq 6.5\,10^{13}\,$cm.
		
		\begin{table}
	\centering
	\begin{tabular}{l l l}
	\hline \hline 
	Parameter & \hspace{1mm} value \tabularnewline
	\hline 
                            
	Stellar radius ($R_{\ast}$)      &    $4.5\,10^{15}\,$cm     &   \tabularnewline
	Stellar effective temperature ($T_{\ast}$) &  $2200\,$K  &   \tabularnewline
	Stellar luminosity ($L_{\ast}$) &  $8640\,$L$_{\odot}$  &   \tabularnewline
	Stellar mass ($M_{\ast}$) &  $0.8\,$M$_{\odot}$  &   \tabularnewline
	Mass-loss rate ($\dot{M}_{\ast}$) &  $2\,10^{-5}\,$M$_{\odot}$ yr$^{-1}$ & 
	\tabularnewline
	Carbon/oxygen ratio ($C/O$) &  $1.5$ & \tabularnewline
	$\bar{\Gamma}$ & $\simeq 1.25$ & \tabularnewline
	\hline \hline 
	\end{tabular}
	\caption[Input model parameters for IRC +10216]{Input model parameters for IRC +10216.}
	\label{tab:IRC_parameters}
	\end{table}	
		
		\subsubsection{Detached shell region}
		\label{sec:IRC_detached_shell}
		\citet{Matthews_2015} reported the discovery of a faint HI shell at a radius 
		$\sim 1.2\,10^{18}\,$cm around the IRC +10216 star. 
		The kinematics of this shell are consistent with matter 
		that has been slowed down by interaction with the ISM. 
		In addition, the HI emission from the freely expanding wind is 
		broad \citep{Hoai_2015}, 
		therefore it is not the main reason for this detection. 
		Based on the FUV radial intensities obtained by \textit{GALEX} observations 
		(see \autoref{sec:ABG_introduction}), \citet{Sahai_2010} estimated  
		the detached shell defined by a termination shock with the inner radius 
		of $\sim 8.58\,10^{17}\,$cm  
		and the outer radius of $\sim 1.008\,10^{18}\,$cm.

	 	\begin{table}
		\centering
		\begin{tabular}{l c l}
		\hline \hline 
		Parameter & \hspace{10mm} Value & \hspace{1mm} Note \tabularnewline
		\hline 
		$n_{H}$ & $52\,$cm$^{-3}$  & Pre-shock density of H nuclei \tabularnewline
		$A_{\nu}$ & 0.022 & Extinction shield \tabularnewline
		$N_{0}(H_2)$ & $10^{20}$cm$^{-2}$ & Buffer H$_{2}$ column density \tabularnewline
		$N_{0}(CO)$ & 0 cm$^{-2}$ & Buffer CO column density \tabularnewline
		$G_{0}$ & 1 & External radiation field \tabularnewline
		$\zeta$ & $3.10^{-17}$ s$^{-1}$ & Cosmic ray flux \tabularnewline
		$OPR$ & 3 & Pre-shock H$_{2}$ ortho/para ratio \tabularnewline
		$v_{s}$ & $14$ km$\,$s$^{-1}$ & Effective shock velocity \tabularnewline
		$T$ & 56 K & Initial gas temperature \tabularnewline
		$T_{d}$ & 11 K & Initial grain temperature \tabularnewline
		$b_{\parallel}$ & $0$   & No magnetic field \tabularnewline
		\hline \hline 
		\end{tabular}
		\caption[Main input parameters of termination shock in the CSE of IRC +10216]
		{Main input parameters of termination shock in the CSE of IRC +10216. 
		Note that we neglect the motion of the termination shock.}
		\label{tab:IRC_input_termination_shock}
		\end{table}
		
		\begin{figure}
		\centering
    		\includegraphics[width=0.9\linewidth]
      		{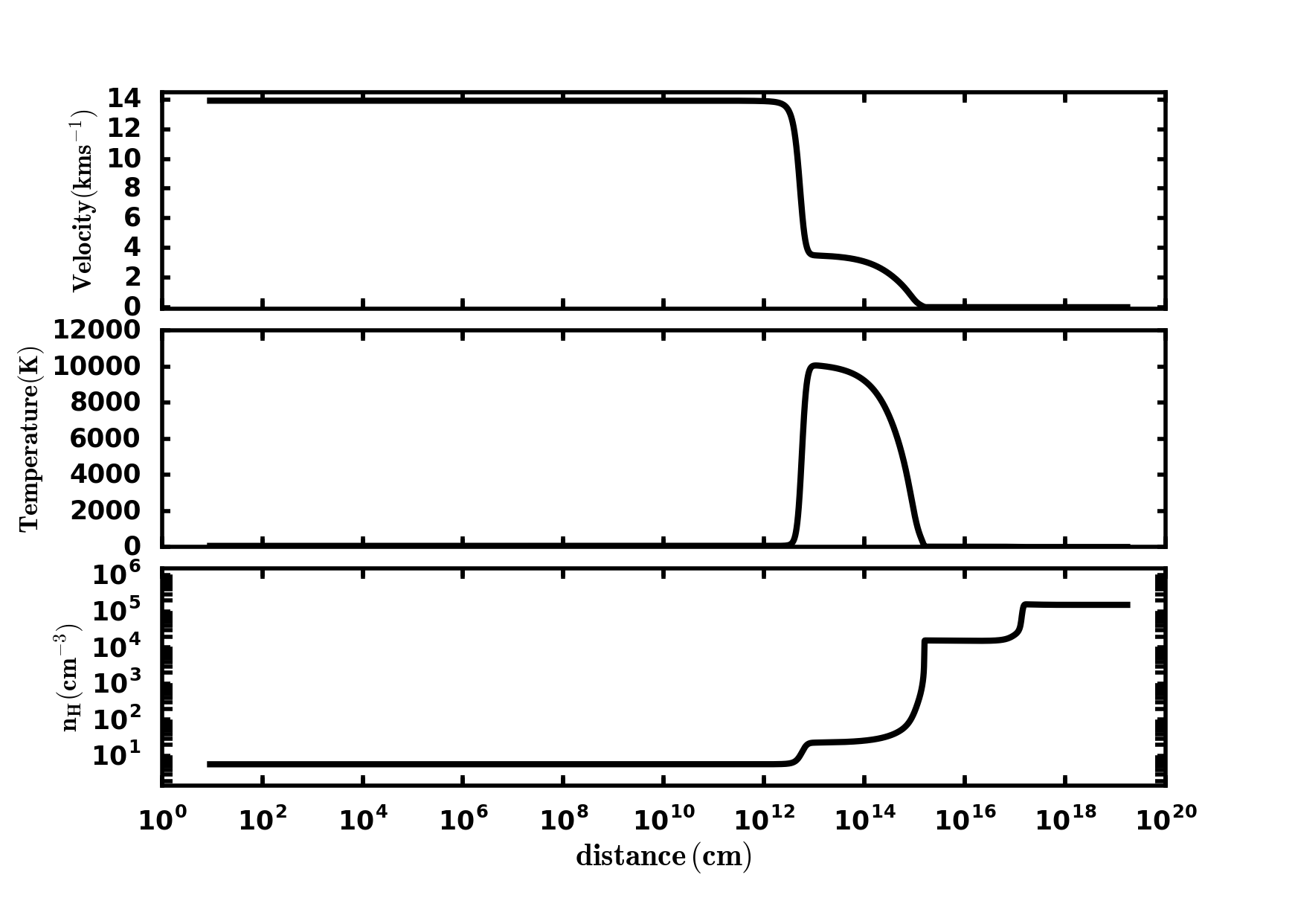}
			\caption[Physical profiles of gas in the termination shock region 
			of the IRC +10216 CSE]
			{Physical profiles of gas in the termination shock region 
			of the IRC +10216 CSE. These properties are plotted in shock frame.} 
		\label{fig:shock_IRC}
		\end{figure}

\begin{figure}
			\centering
    		\includegraphics[width=0.9\linewidth]
      		{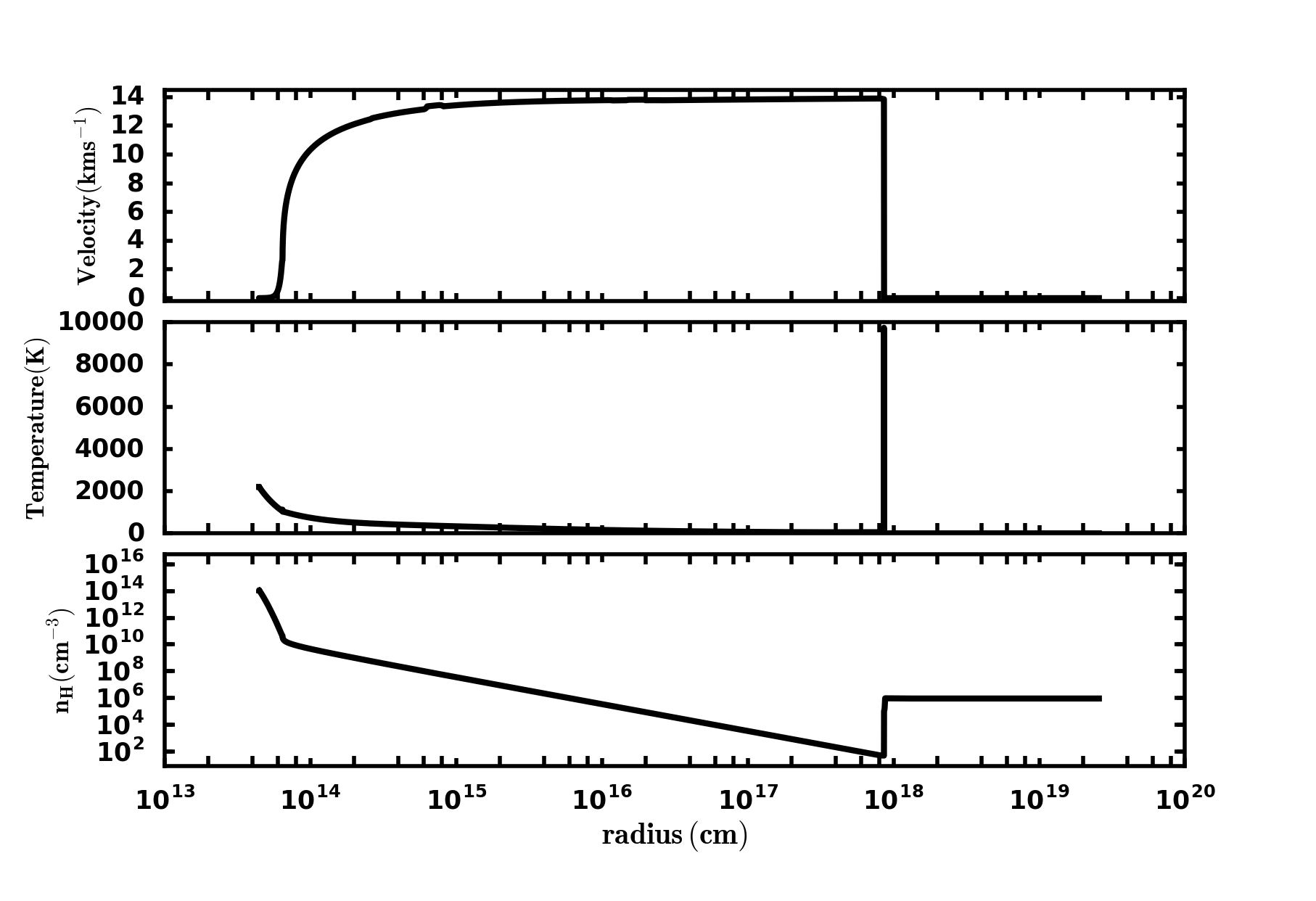}
			\caption[Physical profiles of gas in the IRC +10216 CSE]
			{Physical profiles of gas in the IRC +10216 CSE.} 
			\label{fig:full_IRC}
		\end{figure}

		Since we neglect the effect of the magnetic field, 
		we run a 1D J-type shock (see \autoref{sec:types-def}) 
		with  entrance conditions 
		provided in \autoref{tab:IRC_input_termination_shock}. 
		The gas profiles in the termination shock region, 
		plotted in the shock frame,
		are shown in \autoref{fig:shock_IRC}: 
		the gas is decelerated by a factor of $\sim 4$ 
		when it crosses the termination shock and is heated 
		as mentioned in \citet{Hoai_2015}. It cools down thereafter, 
		and its velocity continues to decrease via expansion, 
		while the density is increasing. 
		Due to the J-type shock character, the maximum of the gas temperature is large 
		$\sim 10^{4}\,$K, corresponding to the $14\,$km$\,$s$^{-1}$ shock 
		(see \citealt{PL13}, equation 10).     
		To sum up, \autoref{fig:full_IRC} displays the properties of gas 
		in the CSE of IRC+10216, 
		starting from the stellar surface through the CSE medium 
		until its interaction with the ISM.  
	
	\subsection{Hydrogen profile} \label{sec:IRC_Hydrogen}
\begin{figure}
		\centering
    		\includegraphics[width=0.9\linewidth]
      		{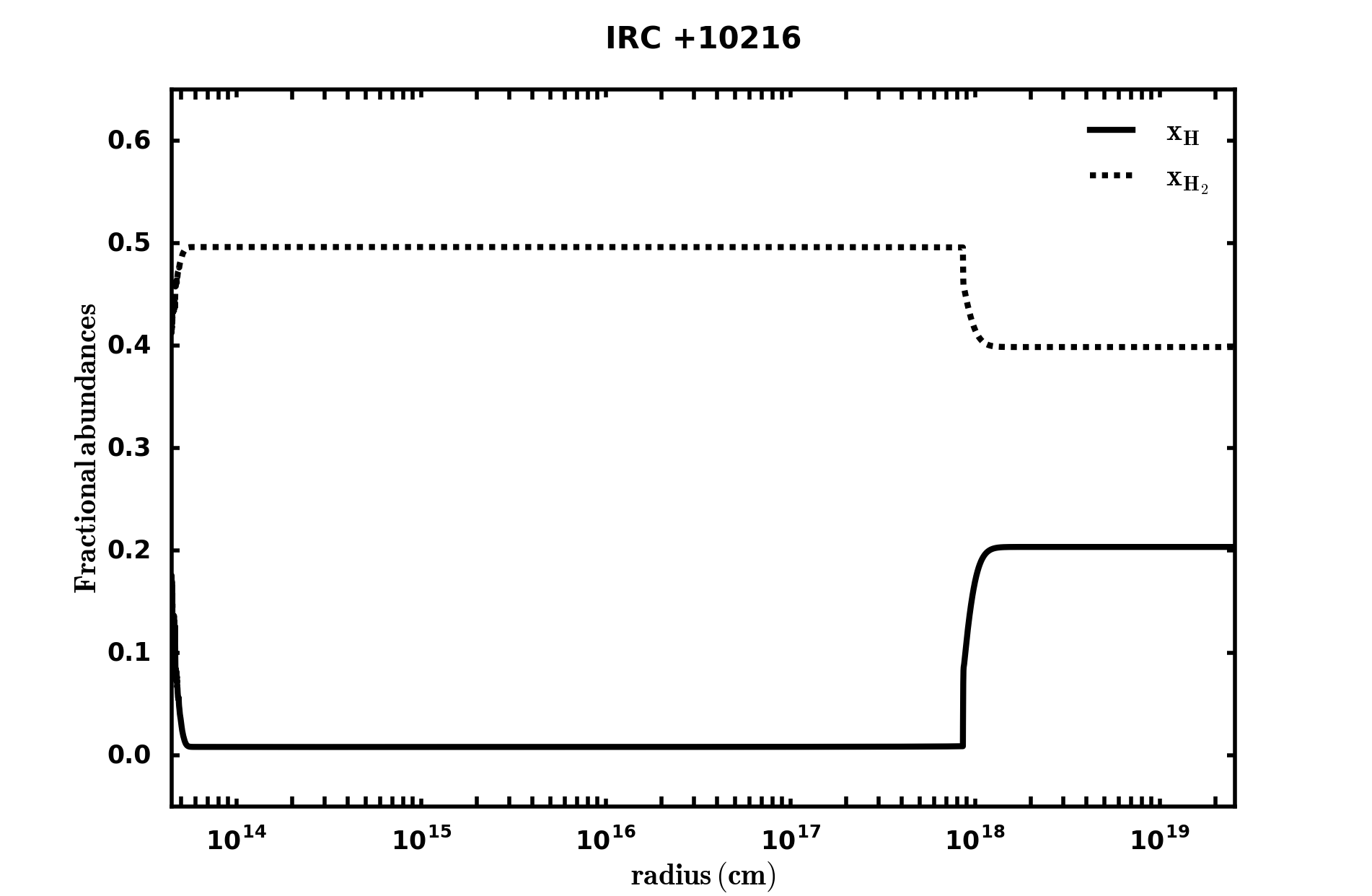}
			\caption[Abundance of hydrogen in CSE of IRC +10216]
			{Abundance of hydrogen in CSE of IRC +10216.} 
		\label{fig:IRC_hydrogen}
	\end{figure}

	The profile of hydrogen in the CSE of IRC +10216 is displayed in 
	\autoref{fig:IRC_hydrogen}.	At such a low effective temperature ($2200\,$K), the IRC+10216 star 
	should have a surrounding CSE mostly composed of molecular hydrogen. 
	\autoref{fig:full_IRC} indicates that the initial gas density ($n_{H}$) 
	in the stellar photosphere
	is	$\sim 10^{14}\,$cm$^{-3}$. 
	Reporting this value in \autoref{fig:glassgold_critical}, we see that
	the initial fractional abundance of atomic hydrogen in the stellar photosphere is $\sim 0.1$, 
	while that of molecular hydrogen is $\sim 0.45$. 
		
	Close to the star, where the wind starts to launch, 
	the velocity of the gas is increasing but its absolute value is still small, 
	so that the dynamical timescale remains longer than the chemical timescale. 
	Thus the chemistry including the three-body reactions 
	impacts on the variation of the hydrogen abundances. 
	The chemical timescale depends on both the density and the temperature of the gas. 
	However, these timescales weakly depend on the density 
	since its decreasing slope is less steep than the temperature (\autoref{fig:free_IRC}).
	Therefore, the timescales of the three-body recombination 
	$t^{chem}_{2}$ and $t^{chem}_{4}$ are shorter 
	than the ones of the collisional dissociation $t^{chem}_{1}$ and $t^{chem}_{3}$ 
	(\autoref{sec:Formulation of hydrogen chemistry on the stellar surface}). 
	This characterizes the cumulation of the molecular hydrogen in the subsonic region. 
	In addition to this, when forming dust grains, the hydrogen abundance is enhanced furthermore 
	due to its formation on grain surface by adsorbing atomic hydrogen. 
	The hydrogen abundance therefore 
	is significantly increased ($x(H_{2}) \sim 0.495$, 
	corresponding to $x(H) \sim 0.01$).
	
	At such a higher radius, where the gas velocity is high enough, 
	the dynamical timescale becomes shorter than the chemical timescales.
	Hence the chemical reactions could not occur, which makes 
	the abundance of hydrogen \textit{freeze-out}.  
	
	At the radius of $\sim 9\,10^{17}\,$cm, where the detached shell appears, 
	the chemical timescale turns back to be shorter than the dynamical timescale 
	because the termination shock slows down the gas. 
	The abundance of molecular hydrogen is now reduced due to the 
	photo-dissociation by the ISRF. 
	Crossing this region, the abundance of H terminates at $\sim 0.2$, 
	with a corresponding abundance of atomic hydrogen at $x(H_{2}) \sim 0.4$. 
	
\section{Y Canum Venaticorum (Y CVn)}
	\subsection{Hydrodynamics}
	Y CVn is a spectral-type J carbon-rich star. Contrary to IRC +10216, 
	the evolution state of Y CVn is not well known. Some authors have suggested 
	that it could evolve on the red giant branch (RGB), in which the carbon 
	composition might be produced by the core He flash \citep{Dominy_1984}. However, the
	high mass-loss rate, as derived from observations of the detached dust shell 
	\citep{Izumiura_1996} conforts the belief that Y CVn is on the AGB phase. 
	Other authors suggest that it has not reached the thermal pulse yet,
	because there is no detection of technetium \citep{Little_1987} 
	and a lack of s-process elements \citep{Utsumi_1985}, 
	which are enhanced by the \textit{Third Dredge-up} process. 
	This convection process is an indication of the TP-AGB phase 
	(see \autoref{sec:ABG_introduction}).  
	Its luminosity ($L \sim 6200\,$L$_{\odot}$) 
	(\citealt{Libert_2007}), which is converted from a bolometric magnitude of $1.96$ 
	(\citealt{Lebertre_2001}) at a distance of 218 pc (\citealt{Perryman_1997}), probably 
	locates Y CVn on the early AGB. 

	Nevertheless, the distance to Y CVn is also uncertain. 
	According to the analysis of \citet{Perryman_1997} on 
	the \textit{Hipparcos} parallax 
	measurement, the distance is estimated at about 218 pc, 
	while other authors come up with different results by re-analyzing 
	this measurement. For instance, the distance is approximated as $\sim$ 272 pc 
	(\citealt{Knapp_2003}) or $\sim$ 312 (\citealt{VanLeeuwen_2007}). 
	In this work, we take the value of $\sim 218\,$pc as \citet{Libert_2007}.
	
	The effective temperature of Y CVn is $\sim 2760$ K (\citealt{Bergeat_2001}). 
	The mass-loss rate is adopted as $1.5\,10^{-7}$ M$_{\odot}$ 
	yr$^{-1}$ (\citealt{Schoier_2002}). 
	Unfortunately there is no clear constraint on its mass, 
	we thus use the arbitrary value of $1.6\,$M$_{\odot}$ 
	(mass of RS CnC star) following a suggestion by Thibaut Le Bertre. 
	The main parameters of Y CVn are listed 
	in \autoref{tab:YCVn_parameters}.

	\begin{table}
	\centering
	\begin{tabular}{l l l}
	\hline \hline 
	Parameters & \hspace{1mm} value \tabularnewline
	\hline 
                            
	Stellar radius ($R_{\ast}$)      &    $2.45\,10^{13}\,$cm     &   \tabularnewline
	Stellar effective temperature ($T_{\ast}$) &  $2760\,$K  &   \tabularnewline
	Stellar luminosity ($L_{\ast}$) &  $6200\,$L$_{\odot}$  &   \tabularnewline
	Stellar mass ($M_{\ast}$) &  $1.6\,$M$_{\odot}$  &   \tabularnewline
	Mass-loss rate ($\dot{M}_{\ast}$) &  $1.5\,10^{-7}\,$M$_{\odot}$ yr$^{-1}$ & 
	\tabularnewline
	Carbon/Oxygen ratio ($C/O$) &  $1$ & \tabularnewline
	$\bar{\Gamma}$ &  $\simeq 1.015$ & \tabularnewline
	\hline \hline 
	\end{tabular}
	\caption[Input model parameters for Y CVn]{Input model parameters for Y CVn.}
	\label{tab:YCVn_parameters}
	\end{table}
	
	\subsubsection{Freely expanding wind region}
\begin{figure}
		\centering
    	\includegraphics[width=0.9\linewidth]
      	{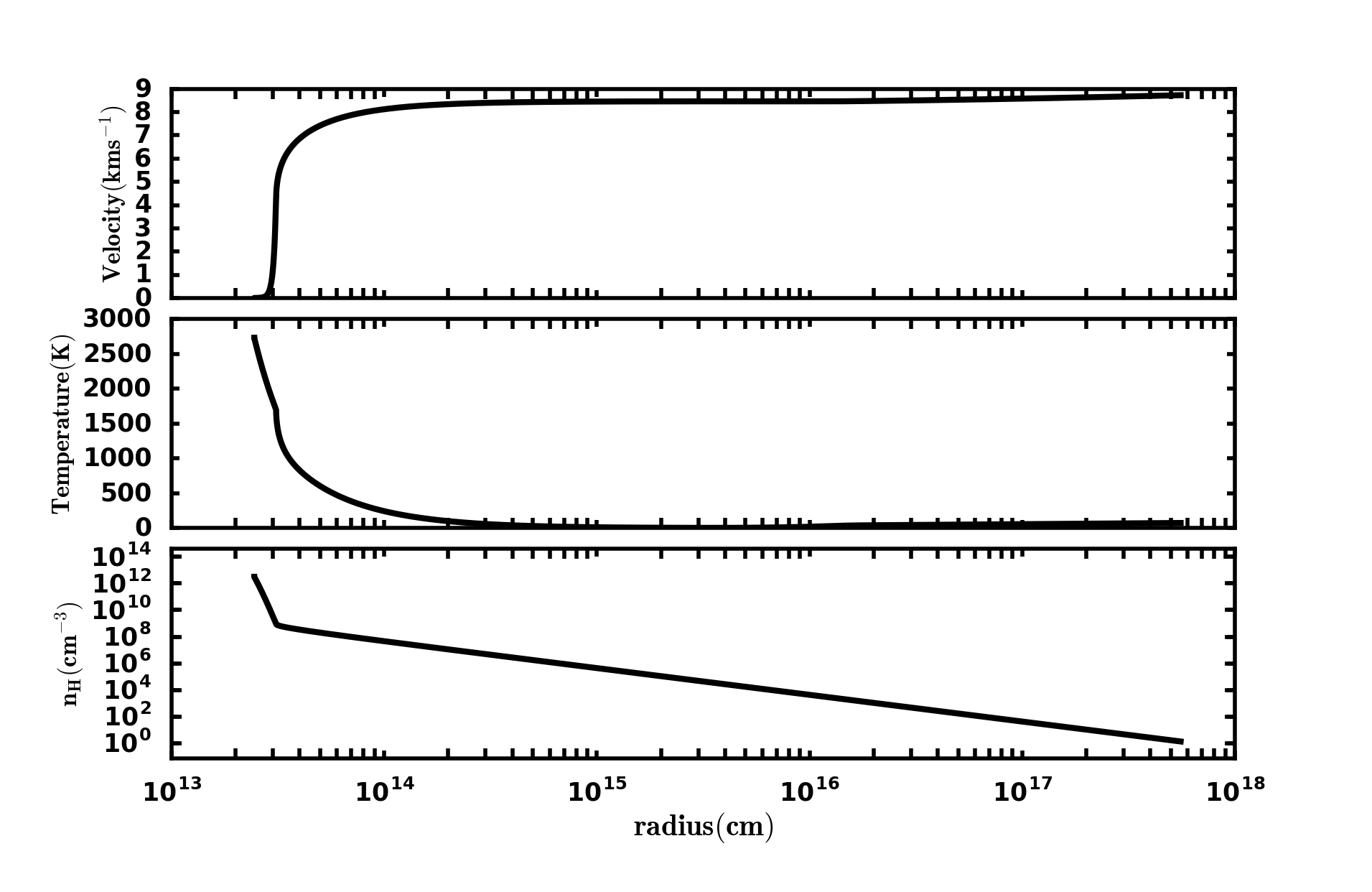}
		\caption[Physical profiles in the freely expanding wind region 
		of the Y CVn CSE]
		{Physical profiles of gas properties in the freely expanding wind region 
		of the Y CVn CSE.} 
		\label{fig:free_YCVn}
	\end{figure}

	The implementation of the stellar wind model with these parameters 
	gives us the physical properties of the gas in the region of free expansion. 
	The \textit{top panel} of \autoref{fig:free_YCVn} indicates 
	that a terminal wind velocity of about $8.7\,$km$\,$s$^{-1}$, 
	which is similar to the value obtained from the estimate standard radiative modeling 
	of CO observations (\citealt{Olofsson_1993, Schoier_2001}), 
	or from high-resolution observations of the CO spectrum (\citealt{Knapp_1998}). 
	We also fit the thermal properties of the gas to a power law in radius 
	as $r^{\alpha}$. $\alpha$ equal $-1.32$ 
	in the range $3.22\,10^{13}\,$cm $\leq r \leq$ $3\,10^{15}\,$cm. 
	Beyond this range, it turns positive due to the 
	photoelectric heating effect. 
	$\alpha = -1.92$ where $3\,10^{15}\,$cm $< r \leq 1.5\,10^{16}\,$cm, and 
	$\alpha = -0.18$ where $r > 1.5\,10^{16}\,$cm.
	
	\subsubsection{Detached shell region}
	\label{sec:YCVn_detached_shell}
	A shell in the HI 21-cm line in Y CVn was detected 
	by \citet{Lebertre_2004}, \citet{Gerard_2006} and \citet{Libert_2007} 
	with the \textit{Nan$\c{c}$ay Radio Telescope} (NRT), 
	and \citet{Matthews_2013} with the VLA telescope. 
	As discussed in \autoref{sec:IRC_detached_shell}, a detached shell 
	is likely to be the main cause for this emission. 
	Thanks to the ISO 90 $\mu$m observation, \citet{Izumiura_1996} clearly 
	showed a quasi-circular shell with an inner radius of r$_{in} \sim 5.48\,10^{17}\,$cm and 
	an outer radius of r$_{out} \sim 7.24\,10^{17}\,$cm - $\sim 9.98\,10^{17}\,$cm 
	by adopting the distance of $\sim 218\,$pc. Therefore, we turn on the J-shock, 
	whose input parameters are in \autoref{tab:YCVn_input_termination_shock}, 
	at a radius of $r_{in}$.  

	\begin{table}
	\centering
	\begin{tabular}{l c l}
	\hline \hline 
	Parameter & \hspace{10mm} Value & \hspace{1mm} Note \tabularnewline
	\hline 
	$n_{H}$ & $2.06\,$cm$^{-3}$  & Pre-shock density of H nuclei \tabularnewline
	$A_{\nu}$ & 0.0004 & Extinction shield \tabularnewline
	$N_{0}(H_2)$ & $10^{20}$cm$^{-2}$ & Buffer H$_{2}$ column density \tabularnewline
	$N_{0}(CO)$ & 0 cm$^{-2}$ & Buffer CO column density \tabularnewline
	$G_{0}$ & 1 & External radiation field \tabularnewline
	$\zeta$ & $3.10^{-17}$ s$^{-1}$ & Cosmic ray flux \tabularnewline
	$OPR$ & 3 & Pre-shock H$_{2}$ ortho/para ratio \tabularnewline
	$v_{s}$ & $8.7$ km$\,$s$^{-1}$ & Effective shock velocity \tabularnewline
	$T$ & 73 K & Initial gas temperature \tabularnewline
	$T_{d}$ & 12.8 K & Initial grain temperature \tabularnewline
	$b_{\parallel}$ & $0$   & No magnetic field \tabularnewline
	\hline \hline 
	\end{tabular}
	\caption[Main input parameters of termination shock in the CSE of Y CVn]
	{Main input parameters of termination shock in the CSE of Y CVn. 
	Note that we neglect the motion of the termination shock.}
	\label{tab:YCVn_input_termination_shock}
	\end{table}

	\begin{figure}
		\centering
    	\includegraphics[width=0.9\linewidth]
      	{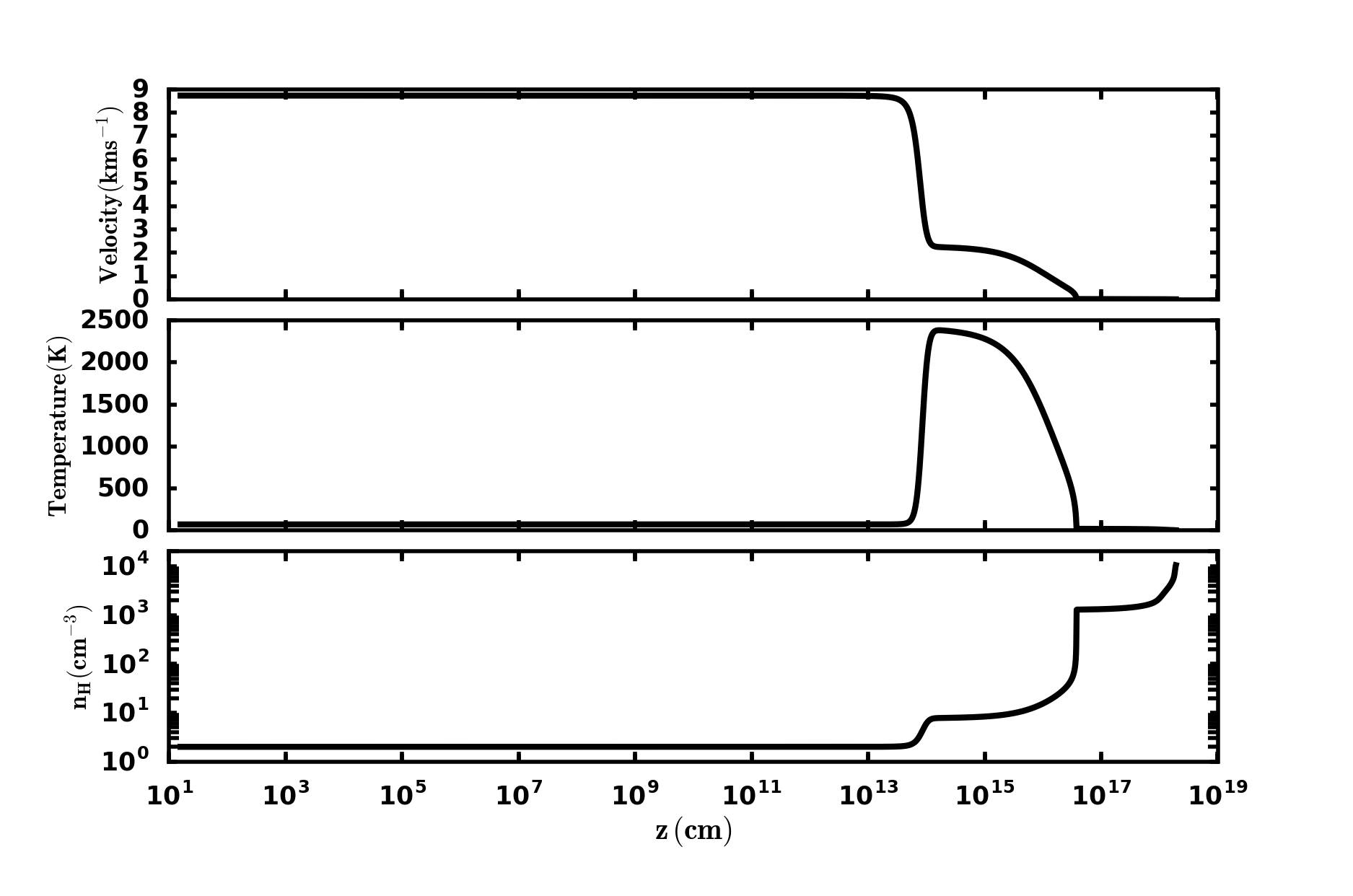}
		\caption[Physical profiles of the gas in the detached shell region 
		of the Y CVn CSE]
		{Physical profiles of the gas in the detached shell region 
		of the Y CVn CSE.} 
		\label{fig:shock_YCVn}
	\end{figure}	    	

	\begin{figure}
		\centering
    	\includegraphics[width=0.9\linewidth]
      	{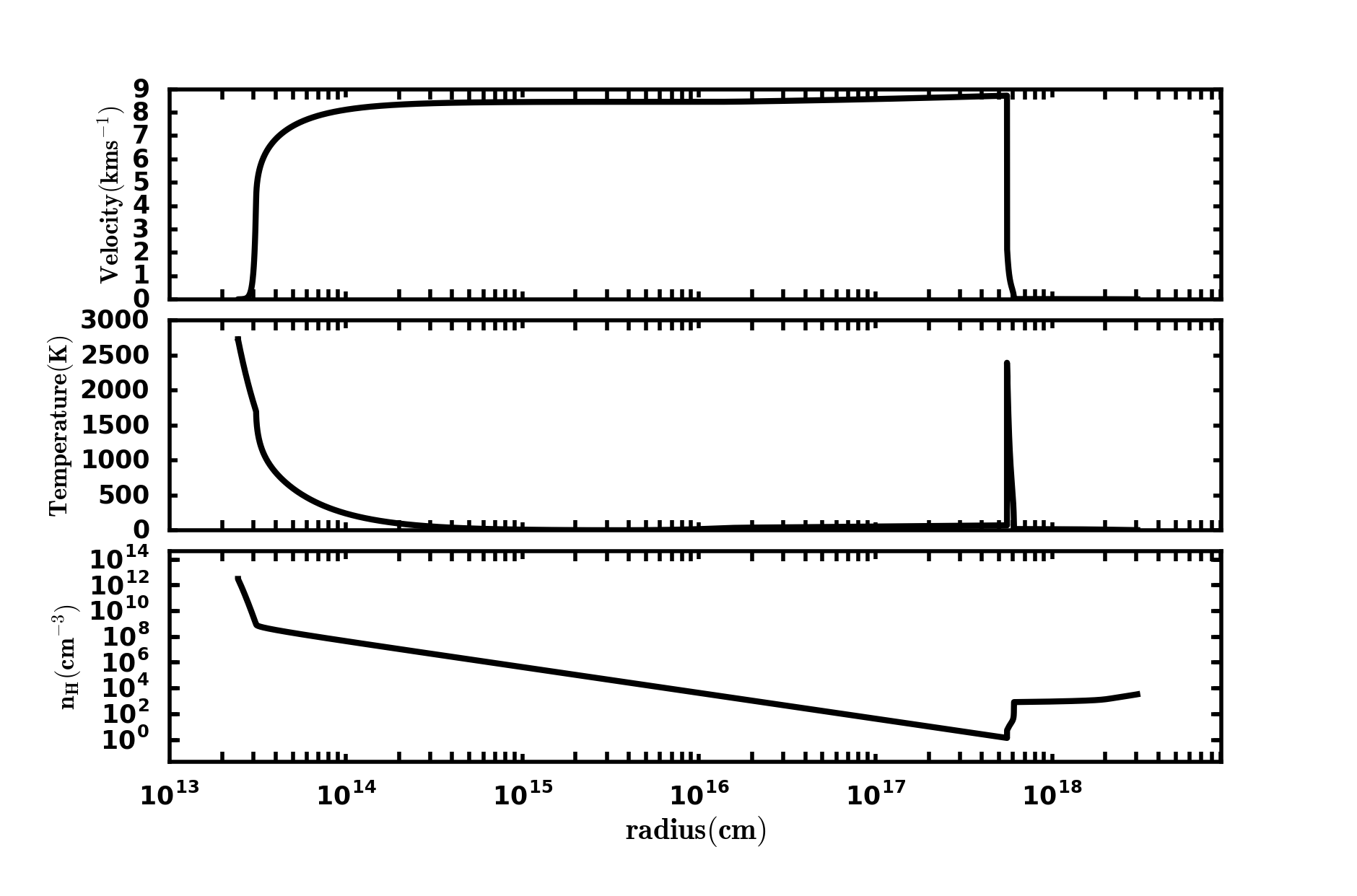}
		\caption[Physical profiles of the gas in the CSE of Y CVn]
		{Physical profiles of the gas in the CSE of Y CVn.} 
		\label{fig:full_YCVn}
	\end{figure}

	\autoref{fig:shock_YCVn} shows the gas profiles in the termination shock of Y CVn 
	in shock frame,
	Like in \autoref{sec:IRC_detached_shell}, 
	this figure indicates a factor of 4 of the acceleration of gas 
	when it crosses the termination shock and is heated. 
	The gas flow also cools down thereafter, 
	and its velocity also continues decreasing via expansion, 
	while the density is increasing. 
	Due to the J-type shock character, the maximum of the gas temperature is also large 
	$\sim 2500\,$K.     
	To sum up, \autoref{fig:full_YCVn} displays the properties of gas 
	in the CSE of YCVn, 
	starting from the stellar surface through the CSE medium 
	until its interaction with the ISM.
	
	\subsection{Hydrogen profile}
\begin{figure}
		\centering
    		\includegraphics[width=0.9\linewidth]
      		{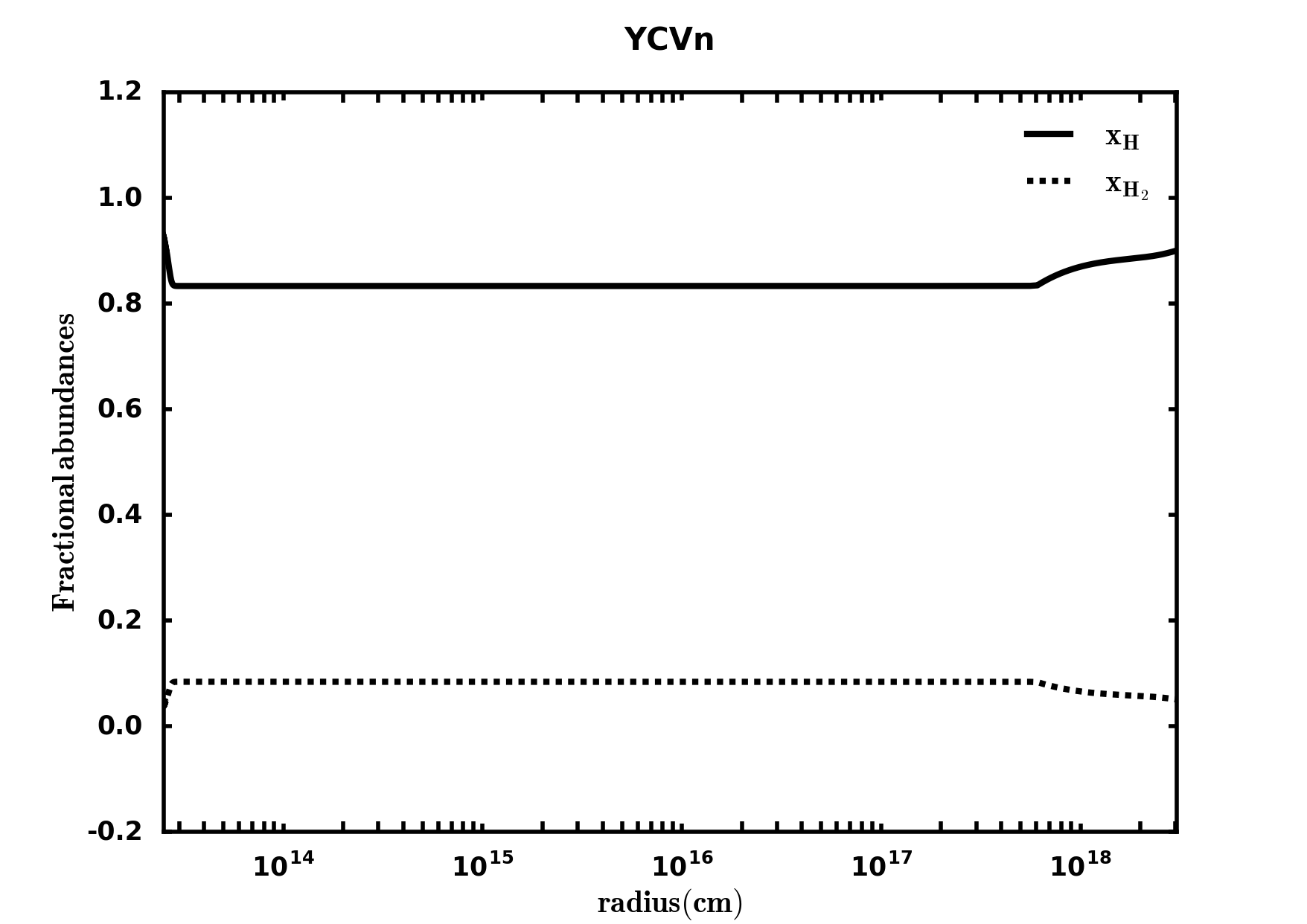}
			\caption[Abundance of hydrogen in the CSE of YCVn]
			{Abundance of hydrogen in the CSE of YCVn.} 
		\label{fig:YCVn_hydrogen}
	\end{figure}

	The profile of molecular and atomic hydrogen 
	is displayed in \autoref{fig:YCVn_hydrogen}.
	Because of its \text{high} effective temperature ($2760$ K),	
	the CSE around Y CVn is expected to contain mostly atomic hydrogen (HI). 
	Similar to \autoref{sec:IRC_Hydrogen}, 
	the variation of hydrogen in the CSE of Y CVn is explained by the interplay between 
	dynamical and chemical timescales. 
	On the stellar photosphere, the initial fractional abundance of atomic hydrogen $\sim 0.94$, 
	and that of molecular hydrogen is $\sim 0.03$ 
	due to the gas density $n_{H}\,\sim 10^{13}\,$cm$^{-3}$ (\autoref{fig:full_YCVn}). 	 
	In the innermost region, molecular hydrogen is enhanced 
	through the three-body recombination reactions. 
	Both profiles \textit{freeze-out} when the gas velocity is sufficient high 
	and atomic hydrogen is finally more enhanced 
	due to the photo-dissociation of molecular hydrogen 
	by the termination shock and the ISRF.    
	
	\newpage
	\section{HI modeling}
\begin{figure}
		\centering
    		\includegraphics[width=0.9\linewidth]
      		{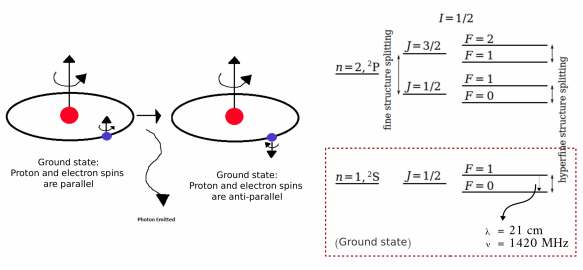}
			\caption[HI 21-cm line formation]
			{HI 21-cm line formation. 
			Hyperfine splitting at ground state emits a 21-cm wavelength radiation. 
			$I$ is the spin of electron, 
			$J$ is the nuclear angular momentum, 
			and $F$ (\textbf{F} = \textbf{I} + \textbf{J}) is the total angular momentum.} 
		\label{fig:HI_formation}
	\end{figure}

	As discussed in \autoref{sec:hydrogen}, 
	the 21-cm line ($\lambda = 21\,$cm) of HI is an excellent probe of the CSE of AGB stars.
	The radiation comes from the transition between the two split levels 
	of atomic hydrogen in its ground state, which is caused by the interaction 
	between the electron spin and the nuclear spin. 
	This splitting is known as hyperfine structure (\autoref{fig:HI_formation}). 
	The relevant frequency of the 21-cm line radiation is:
	\begin{equation}
		\nu_{10} = \frac{c_{l}}{\lambda} = 1.420\, 10^{9}\, \textrm{s}^{-1} \mbox{.}
	\end{equation}
	
	The probability of this transition is defined by: 
	\begin{equation}
		A_{10} = \frac{64\pi^{4}}{3hc^{3}_{l}}\nu^{3}_{10} \left| \frac{e \hbar}{2m_{e}c_{l}} \right|^{2}
		       = 2.8688\,10^{-15}\, \textrm{s}^{-1}	
	\end{equation}		
	where $\hbar = h/2\pi$ is the reduced Planck's constant, 
	$e$ is the charge of the electron, and $m_{e}$ is its mass. 
	This extremely small value means that, for a single atomic hydrogen, the emission of radiation in the 21-cm line pasts $\sim 10^{7}$ yr, therefore it is very difficult to observe. As the total number density of atomic hydrogen is very large in the ISM, this radiation, however, is easy to detect by radio telescopes. 

	\begin{figure}
		\centering
    		\includegraphics[width=0.8\linewidth]
      		{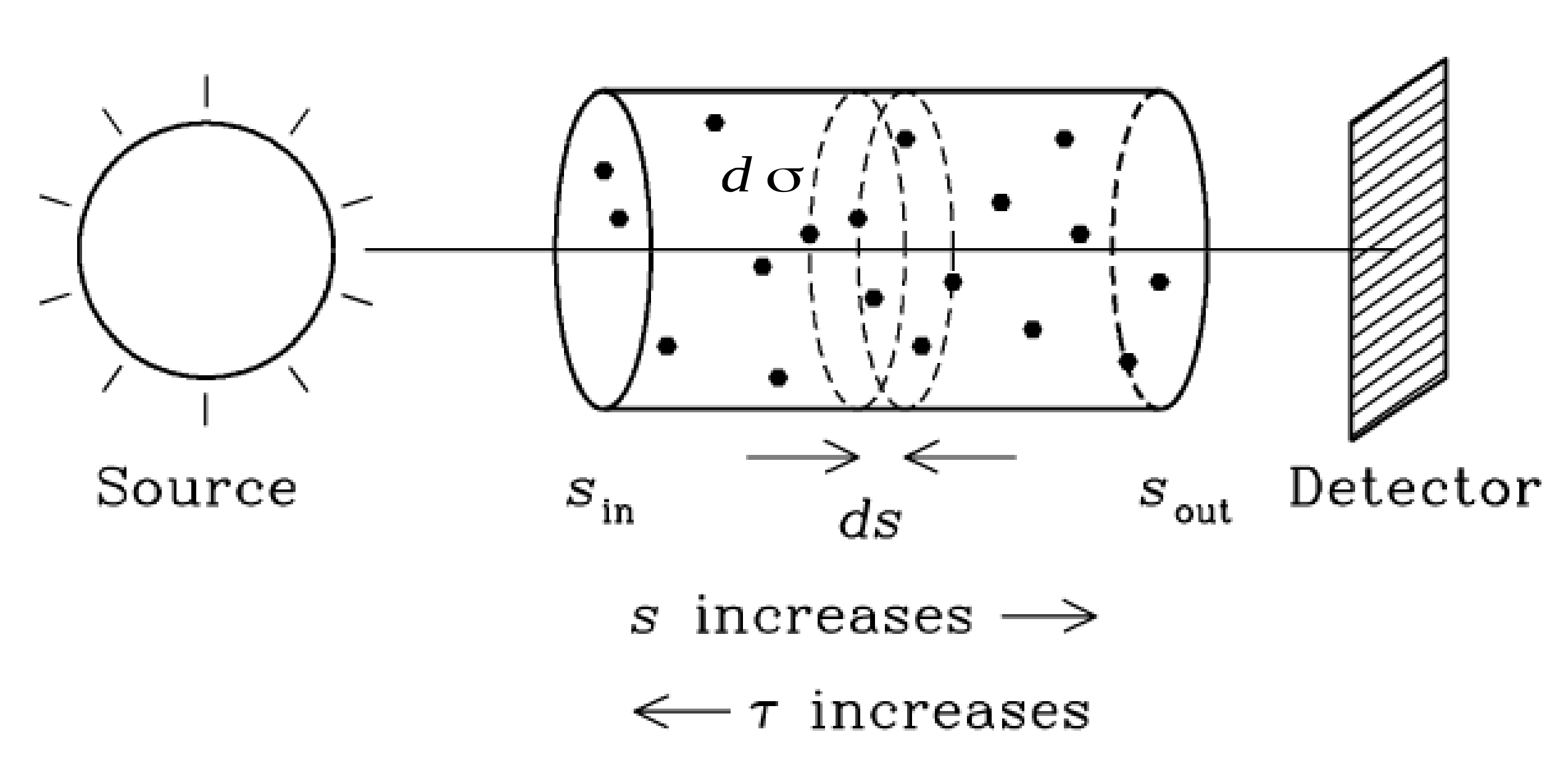}
			\caption[Sketch of the general radiative transfer problem]
			{Sketch of the general radiative transfer problem. 
			$ds$ is the differential line-of-sight, 
			$d\sigma$ is the differential cross-section.} 
		\label{fig:radiative_transfer}
	\end{figure}

	The specific intensity of the radiation $I_{\nu}$ changes according to the absorption and/or emission processes along the ray (\autoref{fig:radiative_transfer}). The \textit{radiative transfer} equation states that	
	\begin{equation} \label{eq:radiative_transfer}
		\frac{dI_{\nu}}{ds} = \epsilon_{\nu} - k_{\nu}I_{\nu}	
	\end{equation}
	where $\epsilon_{\nu}$ and $k_{\nu}$ are the emissivity and absorption coefficients 
	in the line of sight $ds$. The analytical solution of \autoref{eq:radiative_transfer} is
	\begin{equation} \label{eq:I_solution}
		I_{\nu} = e^{-\int^{s}_{0} k_{\nu}\, ds} \left(\int^{s}_{0} \epsilon_{\nu}e^{\int^{s'}_{0}k_{\nu}\,ds'}\,ds + const. \right)
	\end{equation}
	
	As in \autoref{sec:pumping}, we consider the two levels system, 
	named $0$ (lower) and $1$ (upper) 
	of the 21-cm line (\autoref{fig:HI_formation}).
	Three processes are considered to formulate $\epsilon_{\nu}$ and $k_{\nu}$: 
	(1) spontaneous emission from the upper to the lower level,	
	(2) absorption from the lower to the upper level, 
	and (3) stimulated emission from the upper to the lower level.	
	
	Differential energy emitted spontaneously:
	\begin{equation}
		dE_{e}(\nu) = h\nu_{10} A_{10} n_{1} \phi(\nu)d\nu \frac{d\omega}{4\pi} dt dV \mbox{.}	
	\end{equation}	 
	Differential energy absorbed:
	\begin{equation}
		dE_{a}(\nu) = h\nu_{10}n_{0}B_{01}\frac{4\pi}{c_{l}}I_{\nu}\phi(\nu)\frac{d\omega}{4\pi}d\nu dt dV \mbox{.}
	\end{equation}
	Differential energy of stimulated emission:
	\begin{equation}
		dE_{s}(\nu) = h\nu_{10}n_{1}B_{10}\frac{4\pi}{c_{l}}I_{\nu}\phi(\nu)\frac{d\omega}{4\pi}d\nu dt dV
	\end{equation}
	where $n_{1}$ and n$_{0}$ are the densities of the upper level, and the lower level, respectively. $dV = d\sigma ds$ is the unit volume. $A$ and $B$ are the Einstein coefficients, 
	which were introduced in \autoref{sec:pumping}. 
	As in \autoref{sec:H2_profile}, 
	we also introduce a line broadening function $\phi(\nu)$ due to the thermal Doppler effect.  
	
	The intensity of the radiation is the contribution the three terms above, which yields:
	\begin{equation} \label{eq:radiative_transfer_21cm}
		\begin{split}
		&dI_{\nu} d\omega d\nu dt d\sigma = dE_{e}(\nu) + dE_{s}(\nu) - dE_{a}(\nu) \\
		\\
		\textrm{or}\,\,\,&\frac{dI_{\nu}}{ds} = \frac{h\nu_{10}}{4\pi}A_{10}n_{1}\phi(\nu) + \frac{h\nu_{10}}{c_{l}}I_{\nu}\phi(\nu)[n_{1}B_{10} - n_{0}B_{01}] \mbox{.}
		\end{split}
	\end{equation}
	
	By comparing \autoref{eq:radiative_transfer_21cm} to \autoref{eq:radiative_transfer}, the emission and adsorption coefficients are:
	\begin{equation} \label{eq:ep_nu}
		\epsilon_{\nu} = \frac{h\nu_{0}}{4\pi}A_{10} n_{1}\phi(\nu)
	\end{equation}
	and
	\begin{equation}
		k_{\nu} = \frac{h\nu_{0}}{c_{l}}\phi(\nu) ( n_{0}B_{01} - n_{1}B_{10}) \mbox{.}
	\end{equation}
	Using the relation of the Einstein coefficients (\autoref{eq:E1} and \autoref{eq:E2}) and assuming the system is in LTE (\autoref{eq:LTE}), we have:
	\begin{equation} \label{eq:k_nu1}
		k_{\nu} = \frac{c^{2}_{l}}{8\pi \nu^{2}_{10}}\phi(\nu)n_{0}\frac{g_{1}}{g_{0}}A_{10}
		\left(1 - e^{-\frac{h\nu_{10}}{k_{B}T}}\right) \mbox{.}
	\end{equation}
	
	At small frequency ($h\nu_{10}/ k_{B} \sim 0.07\,$K), one can replace $e^{- h\nu_{10}/k_{B}T}$ 
	by $1 - h\nu_{10}/k_{B}T$ and \autoref{eq:k_nu1} can be rewritten as:
	\begin{equation} \label{eq:k_nu2}
		k_{\nu} = \frac{c^{2}_{l}}{8\pi \nu^{2}_{10}} \phi(\nu) n_{0}\frac{g_{1}}{g_{0}} A_{10}\frac{h\nu_{10}}{k_{B}T} \mbox{.}
	\end{equation}
	
	For 21-cm line, $g_{0}=1$ and $g_{1} = 3$, and the population is in statistical equilibrium $n_{0} = 1/4 n_{H}$ 
	and $n_{1} = 3/4n_{H}$, where $n_{H}$ is the total density of hydrogen in ground state.
        In practice, observers mostly express the spectra as a function of the radial velocity 
        $v_{r}$. Hence, we replace $\phi(\nu) \sim \phi(v_{r})$. 
        The line function as a function of $v_{r}$ is defined by: 
	\begin{equation}
		\phi(v_{r}) = \frac{\lambda}{\sqrt{2\pi} \sigma} e^{-\frac{\left(v_{rad} - v_{r}\right)^{2}}{2\sigma^{2}}}
	\end{equation}	 
	where $\sigma^{2} = (k_{B}/m_{H})\, T$ is the thermal velocity of atomic hydrogen. 
	The emission velocity $v_{rad}$ is illustrated in \autoref{fig:detached_shell}. 
	With these transformations, the emission coefficient (\autoref{eq:ep_nu})
        and the adsorption coefficient (\autoref{eq:k_nu2}) can be rewritten as:
	
	\begin{equation}
		\epsilon_{v_{r}} = \frac{3h\nu_{10}}{16\pi}n_{H}A_{10}\phi(v_{r}) 
	\end{equation}
	\begin{equation}
		k_{v_{r}} = \frac{3c^{2}_{l}}{32\pi \nu_{10}}n_{H} A_{10}\frac{h}{k_{B}T} \phi(v_{r})
	\end{equation}
	
	and the specific intensity (\autoref{eq:I_solution}) is become:
	\begin{equation} \label{eq:Iv}
		I_{r}(r_{0}) = e^{-\int^{s_{out}}_{-s_{out}} k_{v_{r}}\, ds} \left(\int^{s_{out}}_{-s_{out}} \epsilon_{v_{r}}e^{\int^{s}_{-s_{out}}k_{v_{r}}\,ds^{'}}\,ds + const \right) \mbox{.}
	\end{equation}
	
	As discussed earlier, we consider a spherical geometry for the CSE around an AGB star. 
	In that simple case, the computational technique of these integrations 
	in \autoref{eq:Iv} is described in \aref{app:techinique}. 
	The only needed parameter is $R_{out}$, 
	which defines the outer radius of the detached shell. 
	Finally, the observed flux density on earth is defined by:
	\begin{equation}
		F_{v_{r}} = \frac{\int^{Rmax}_{0} 2\pi r_{0}I_{r}(r_{0}) dr_{0}}{d^{2}} \mbox{.}
	\end{equation}
	where d is the distance of the star and R$_{max}$ is the size of the observation beam. 
	We also verify our method  
	by using the same physical properties as \citet{Hoai_2015} 
	and comparing the calculated flux density with their method. 
	\autoref{fig:Hoai_benchmark} is flux density calculated for the two cases of 
	$\dot{M}_{\ast} = 10^{-7}$ M$_{\odot}\,$yr$^{-1}$ 
	and $\dot{M}_{\ast} = 10^{-5}$ M$_{\odot}\,$yr$^{-1}$, 
	which are compatible to \citet{Hoai_2015}'s figure(2). 
 
        \subsection{HI modeling for IRC +10216} \label{sec:HI_IRC}
\begin{figure}
	    \begin{minipage}[c]{1\textwidth}
	    \centering
    	  	\includegraphics[width=0.9\linewidth]
      	{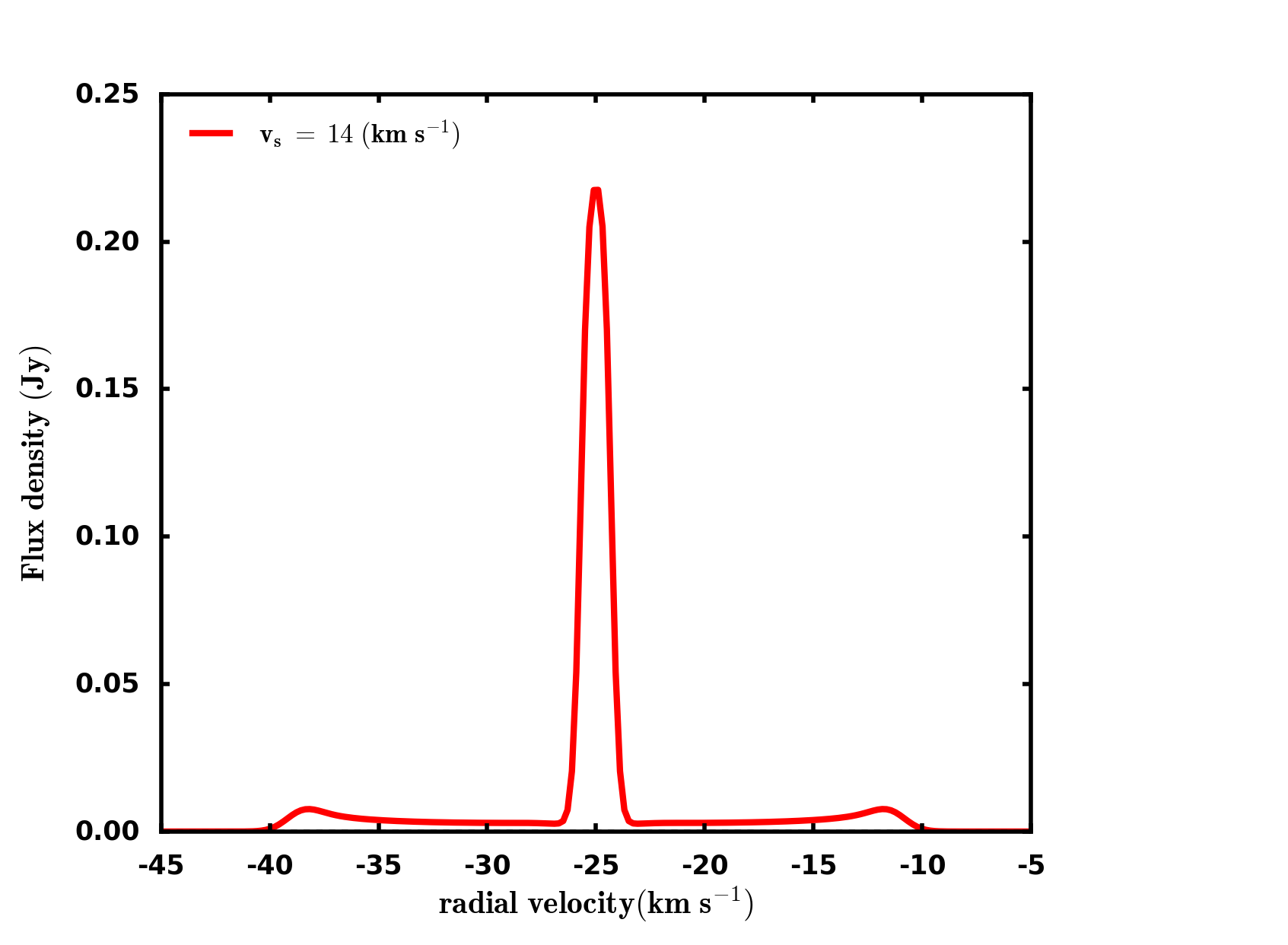}
   		\end{minipage} \\
   		\begin{minipage}[c]{1\textwidth}
   		\centering
      		\includegraphics[width=0.9\linewidth]
      	{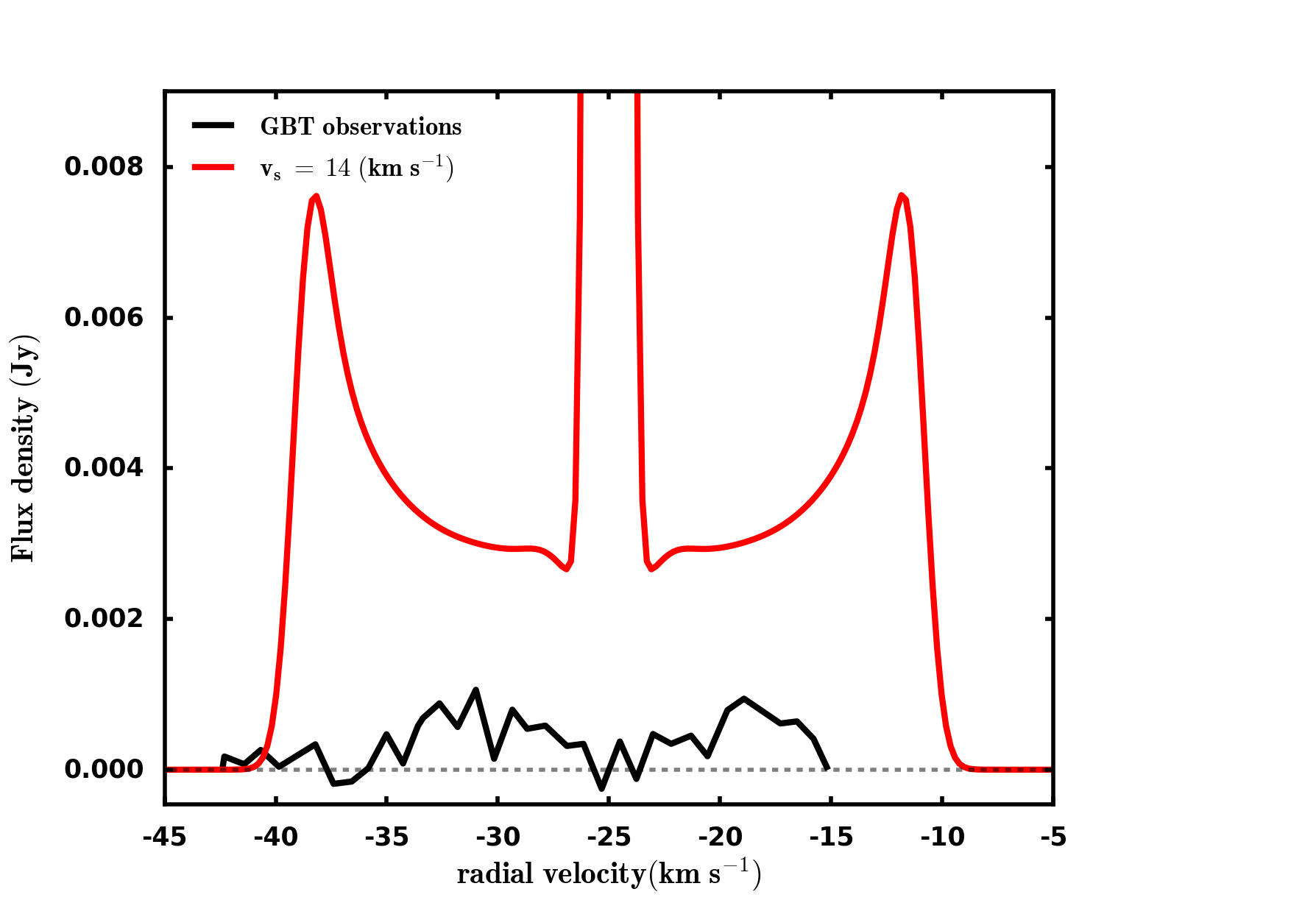}
   		\end{minipage}\\
		\caption[HI 21-cm line profile in the CSE of IRC +10216 obtained 
		by model with 14 km$\,$s$^{-1}$ J-type terminal shock and 
		compared to observations]
		{(\textit{Top}) HI 21-cm line profile in the CSE of  IRC +10216 
		obtained by model with 14 km$\,$s$^{-1}$ J-type termination shock. 
		(\textit{Bottom}) Comparison with GBT observations.} 
		\label{fig:vs14}
	\end{figure}

        \autoref{fig:vs14} shows the line profile of the HI 21-cm line profile
        in the CSE of IRC +10216 with $\propto$ 100'' beam side of radius 
        (corresponding to $R_{max} \sim 9.45\,10^{16}\,$cm). 
        To our knowledge, it is the first modeling attempt.
        \textit{Top panel} shows the flux density profile, 
		which is computed by using the physical profile of IRC +10216 (\autoref{fig:full_IRC}) 
		extended up to $R_{out} \simeq 1.008\,10^{18}\,$cm (\autoref{sec:IRC_detached_shell}).        
        The emission from the detached shell characterizes the central peak, 
        while \textit{double-horned} structure identifies 
        the freely expanding wind.
		\textit{Bottom panel} shows the zoom-in profile in comparison 
		with the \textit{Green Bank Telescope} (GBT) observations \citep{Matthews_2015}. 
        The observational data is the difference between the 
                   spectrum integrated over a 100'' aperture 
                   on the IRC +10216 position with reference spectra extracted 
                   over 1100'' east of the star (outside the HI shell). 
        Unfortunately, there are two drawbacks in our model: first,  
        the simulated flux density is about 10 times higher than
        the value inferred from observations in the freely expanding region, 
        which means 
        that the abundance of atomic hydrogen computed by our model is about 
        10 times higher than it should be in this regions. 
        Second, our model greatly overestimates 
        the emission of the detached shell. 
         
\begin{figure}
		\begin{center}
		\includegraphics[width=1\linewidth]
      	{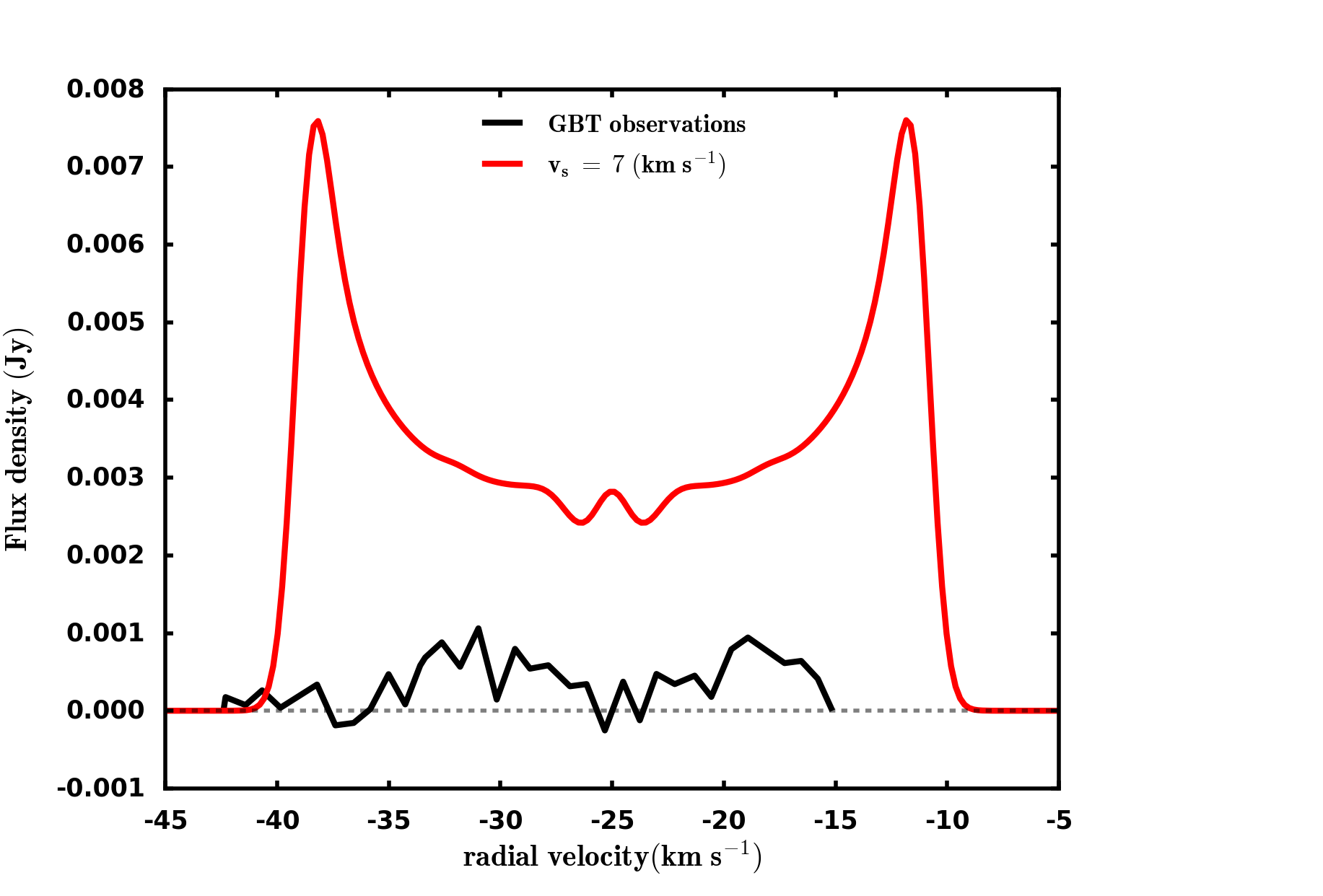}
		
		\caption[IRC +10216 HI 21-cm line profile obtained by model 
		         with 7 km$\,$s$^{-1}$ C-type terminal shock and 
		         compared to observations]
		{Same as \autoref{fig:vs14}, 
		but using 7 km$\,$s$^{-1}$ C-type terminal shock.} 
		\label{fig:vs7}
		\end{center}	
	\end{figure}

        Now, if we consider the motion of the termination shock 
        (see \aref{app:term_motion} for detail) and 
        assume that the density of the ISM is on the order of 
        the stellar wind medium, 
        the termination shock's frame will move at the velocity $u_{s} = v_{w}/2$. 
        Therefore, the input shock velocity now is $v_{w}/2$ = 7 km$\,$s$^{-1}$ 
        in the termination shock's frame. 
        Furthermore, instead of using the J-type, 
        we also consider a C-type shock for the magnetized ambient medium 
        with the typical value $b = 1$. 
        Updating the parameters in \autoref{tab:IRC_input_termination_shock} on these new values, 
        we recompute the termination shock. 
        Because of the lower compression factor, 
        the emission in the detached shell with a C-type shock 
        is now negligible 
        and we can free ourselves from the second problem above (\autoref{fig:vs7}).  
        
                   
 \newpage	  	
  \subsection{HI modeling for Y CVn}
\begin{figure}
		\begin{center}
		\begin{minipage}[c]{0.9\textwidth}
			\includegraphics[width=0.9\linewidth]
      		{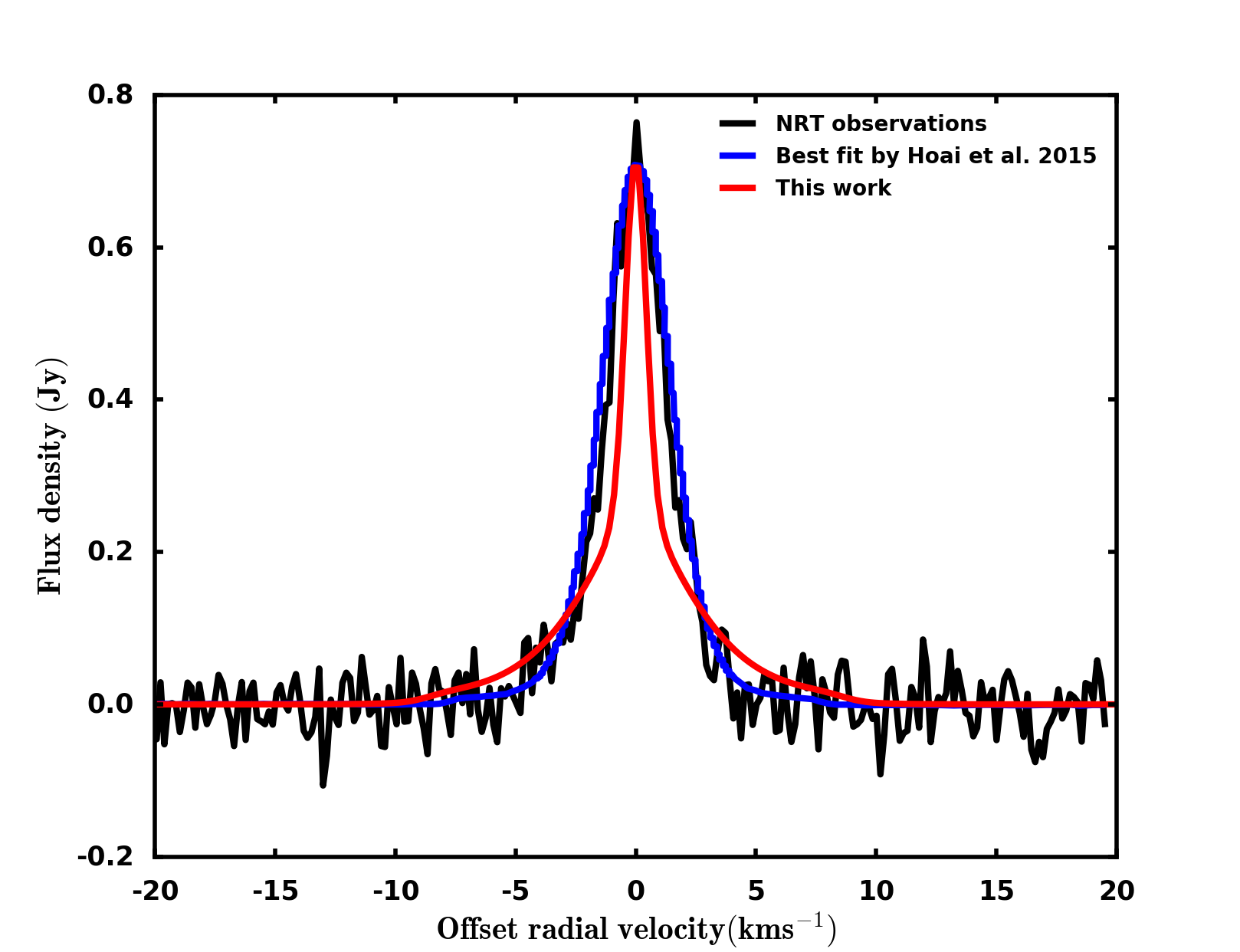}
		\end{minipage} \\
		\begin{minipage}[c]{0.9\textwidth}
			\includegraphics[width=0.9\linewidth]
      		{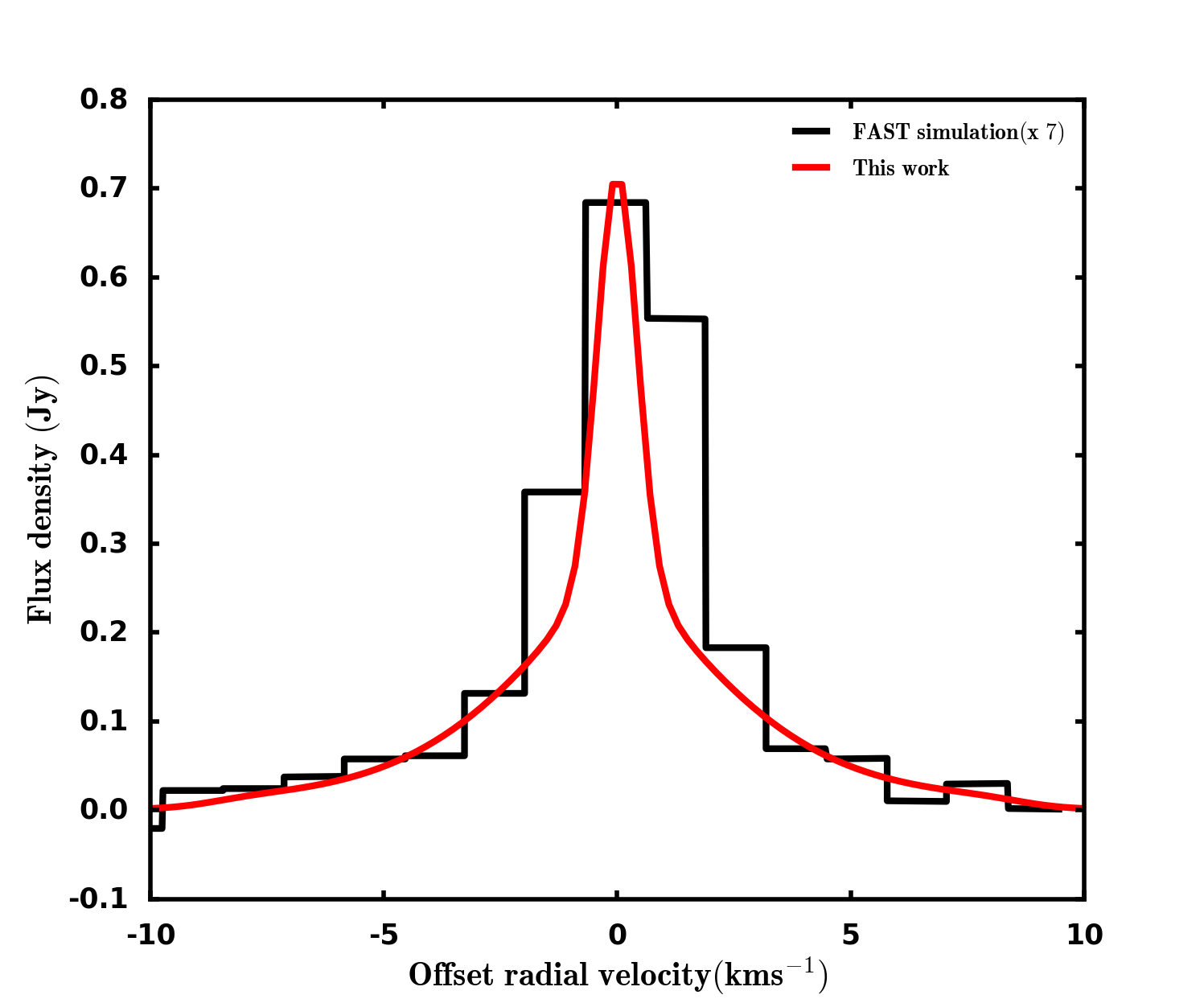}
		\end{minipage} \\
      	\end{center}
		\caption[Y CVn HI 21-cm line profile computed by the HI model in comparison 
		with the NRT observations and the FAST simulation]
			{Y CVn HI 21-cm line profile computed by the HI model in comparison 
			with the NRT observations (\textit{top panel}) 
			and the FAST simulation (\textit{bottom panel}). 
			The profile obtained with $R_{max} \sim 7.47\,10^{17}\,$cm.} 
		\label{fig:HI_YCVn}
	\end{figure}

	\autoref{fig:HI_YCVn} shows 
	the HI 21-cm line profile in the CSE of Y CVn computed by our model 
	with only one free parameter $R_{max}$ ($R_{out} = R_{max}$). 
	\textit{Top panel} shows the comparison with NRT observations \citep{Libert_2007} 
	and with the fit from a parametric model with 5 free parameters \citep{Hoai_2015}. 
	\textit{Bottom panel} shows the comparison with 
	the \textit{Five hundred-meter Aperture Spherical radio Telescope} 
	(FAST) simulation \citep{Hoai_2017}, which is based on 
	VLA observations \citep{Matthews_2013}. 
	
	Our model matches quite well the observations, however, 
	the width of the central line is narrower than observations, 
	which means that the temperature computed by our model 
	is cooler than the observational one in the detached shell. 
	On contrary, the width of the wings is broader, 
	which is synonymous with the fact that the temperature 
	computed by our model is hotter than the observational one in the freely expanding wind region. 
	Therefore, our total flux integrated over velocity is quite comparable to observations. 
	It is about $2.14\,$Jy$\,$km$\,$s$^{-1}$ 
	from our model and about 
	$2.34\,$Jy$\,$km$\,$s$^{-1}$ from NRT observations.   
	
\setstretch{1.1} 
\chapter{CONCLUSIONS AND PERSPECTIVES}
\label{Chapter10}

\lhead{Chapter 10. \emph{Conclusions and perspectives}} 
\section{Conclusions and remarks}
  In the last part of this thesis, we incorporated to the Paris-Durham shock code a few new physical ingredients to allow it to compute steady-state solutions of AGB winds. We included the geometrical dilution due to the spherical symmetry.  We provided the code with the gravitation and radiation pressure forces, with the net effect of pushing the grains outwards and launching the wind. We modified the dynamical integration to help go through the sonic point. We now compute the dust temperature subject to radiation equilibrium. We modified the computation of the extinction so that the irradiation field is external and spherical geometry is accounted in the line cooling opacity. We found an approximation to easily account for gas heating by the radiation pumping near the stellar surface. Finally, we added three-body reactions necessary to form H$_{2}$ at the high densities experienced near the stellar surface.

  With this new powerful tool, we started to examine the time-dependent conversion of atomic to molecular hydrogen along the wind trajectories, and we proceeded to model two AGB stars with surface temperatures below and above the critical temperature for hydrogen to be in atomic or molecular form. In the \textit{"low"} temperature case, we show that hydrogen quickly becomes molecular in the wind, which might explain the difficulty to detect HI with instruments such as the VLA in AGB winds. 
  
  We have tried to reproduce the HI 21-cm line profile from the CSE of the AGB stars in two example cases: IRC +10216 and Y CVn. In the case of IRC +10216, our model can produce the \textit{"double-horned"} structure but it contains about 10 higher of the atomic hydrogen than the observational expectation. In the case of Y CVn, our model reproduces quite well observations but the temperature computed from our model seems cooler than observations in the detached shell region. Hopefully, the FAST telescope will provide us with better sensitivity in the future and thus bring better statistics for a greater number of stars.

\section{Perspectives}
\label{sec:partIII_perspectives}
  Although H$_2$ is the dominant molecular species in the ISM, H$_2$ is a tracer only at the high temperatures caused by shocks and at large scales when excited by UV radiation from the ISM. Alternatively, CO is one of the major molecules used to determine the dynamics of stellar wind and also its interaction with the ISM. Therefore, we aim at updating the chemical network for CO and calculating CO emission from stellar winds. The CO chemistry will have to incorporate more three-body reactions which are not yet present in the Paris-Durham code. Fortunately, now that the framework has been developed in the code, it will be simply a matter of updating the network input file, and finding which reactions are relevant. Once we have a good description of the CO chemistry, we will be in a good position to synthesize the results of many observable data (e.g., \citealt{Truong-Bach_1991}, \citealt{Groenewegen_1998}, \citealt{Knapp_1998}, \citealt{Hoai_2014}) thanks to the post-processing tools for the emissivity of CO lines (and other molecules) developed by \citet{G08} which can be directly fed to outputs of the Paris-Durham code (assuming spherical symmetry). 
  In particular, we shall be able to control the validity of the LTE approximation made to infer temperatures from observations by \citet{Hoai_2014} and \citet{Nhung_2015b}.
     
  In these stellar wind models, the formation and evolution of circumstellar dust grains is not yet properly treated. We assumed that the grain radii follow a “standard” (MRN) grain-size distribution as soon as the gas temperature is lower than the condensation temperature. The slope of -3.5 (MRN distribution) is believed to be appropriate for typical interstellar dust  \citep{MRN_1977}, while other studies suggest that it should be steeper for circumstellar dust shells \citep{Dominik_1989}. Therefore, we will thus aim at studying in a consistent way the processes of coagulation of the circumstellar dust grains. One place to start would be to introduce Hirashita's simple coagulation model in the code, with a bin size by bin size treatment \citep{Hirashita_2009}.

\begin{figure}
	\centering
    \includegraphics[width=0.9\linewidth]
    {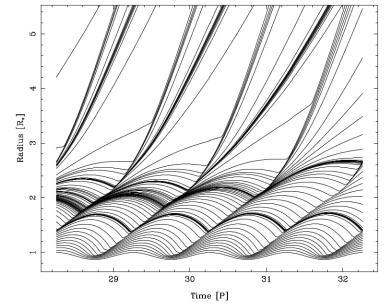}
	\caption[Affect of pulsation on the large scale motion of the gas]
	{Positions of selected mass shells in AGB 
	atmospheres showing how pulsation can affect on 
	the large scale motion of the gas \citep{Hoefner_1997}.} 
	\label{fig:pulsation}
\end{figure}

  Pulsations are the dominant theory to lift the wind from the stellar surface. Material is pushed away from the star during each period of pulsation, then the gravitational force becomes dominant and forces them backward on to the star. The falling flow meets the upward drafts from the next pulsation, and are slowed down by shocks and pushed outward again. The process is repeated until the gas flow condensates solids and gets accelerated by the newly formed dust grains (see \autoref{fig:pulsation}). With a prescribed scaling for the shock strength vs. radius, we can trigger a shock in the flowing wind every period thanks to the versatility of the Paris-Durham shock code. This way, we could model such thermally pulsing winds with our shock code.

\begin{figure}
	\centering
    \includegraphics[width=0.9\linewidth]
    {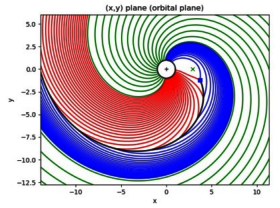}
	\caption[Effect of binary system on stellar wind]
	{Effect of binary system on stellar wind. The star 
	positions are symbolized by the signs "+" and "x".} 
	\label{fig:binary}
\end{figure}

  Thanks to ALMA high resolution data, \citet{Decin_2015} found evidence for a spiral structure in inner wind of  IRC +10216, which could be caused by a binary companion. We thus started to simulate the trajectory of a wind in a binary star system in the hypersonic regime. In the hypersonic regime, the pressure gradients are negligible, and the fluid parcels paths can be computed independently from each other. This allows to recover spiral shocks structures where two flows meet as shown in \autoref{fig:binary}. We are currently able to model only trajectories in the equatorial plane, and the full 3D structure of the wind still escapes us, but we feel this can eventually be done, and we could generate in 3D each individual fluid parcel trajectory, as well as the characteristic of the 2D manifold of the shocks generated by inter-penetrating trajectories. We then plan to post-process with our code the thermal and chemical properties of the gas along these trajectories to reproduce detailed observations and help interpret ALMA data.

  The above remark on hypersonic flows is valid for more simple winds and we expect to recover wind asymmetries such as detected by \citet{Hoai_2016} on W Aquilae star. This simple assumption can also be used to model the influence of a magnetic field, provided we neglect the back-reaction of the flow of the field (i.e., we can easily compute fluid parcel trajectories for a prescribed magnetic field configuration).  

  We thus claim that such simple models can easily provide predictions for complex dynamics and chemistry, and greatly help observers interpret the ever increasing wealth of details unveiled by the new generations of instruments such as ALMA. Thanks to their policy of quick release, ALMA observations are a gold mine for researchers who have not yet easy access to powerful facilities, as is still the case in Vietnam. We hope this work will help foster research projects in Vietnam's flowering astrophysics for quite a few years to come.



\addtocontents{toc}{\vspace{2em}} 

\begin{appendices}
\setstretch{1.1} 
\chapter{H$_{2}$ Ortho/Para ratio} 
\label{app:othor_para} 
\lhead{Appendix A}

A molecule of hydrogen consists of two atoms. Each nucleus of these atoms can have its own spin. 
Depending on the direction of the nuclei's spin, hydrogen can exist in two different configurations: \textit{ortho} in which spins of both nuclei are in the same direction, 
and \textit{para} in which spins of both nuclei are in the opposite direction. 
The difference between these two configurations manifests itself in the rotational energy of H$_{2}$. 
The rotational energy relates to the rotational quantum level $J$ as $E_{J} = BJ(J+1)$, 
where $B \sim 60.85\,$cm$^{-1}$ is the rotational constant of H$_{2}$. 
In fact, the rotational ground state ($E_{J} = 0$) of hydrogen is only occupied by \textit{para} ($J=0$), while the lowest state ($J=1$) for \textit{ortho} ($E_{rot} = 2B/k_{B} = 170.5\,$K) is the first rotational state. 
The difference in rotational energy between ortho and para is thus $\Delta E_{J} \sim 170\,$K.   

At LTE, the ratio between ortho and para populations is:
\begin{equation}
	\frac{ortho}{para} (T_{rot},LTE) = \frac{\sum_{J\,odd} 3(2J + 1)e^{-E_{J}/k_{B}T_{rot}}}{\sum_{J\,even} (2J + 1)e^{-E_{J}/k_{B}T_{rot}}}
\end{equation}

The conversion between ortho-H$_{2}$ to para-H$_{2}$ can occur via four main mechanisms:
\begin{itemize}

\item First, proton from H$_{2}$ exchanges to H$^{+}$, H$^{+}_{3}$ or other cations:
The proton exchange can be expressed as \citep{Dalgarno_1973}	
	\begin{equation}
		\ce{H_{2}(J=1) + H^{+} <=> H_{2}(J=0) + H^{+} + 170.5\,K} 
	\end{equation}
This process is dominant at low temperature ($T \leq 50\,$K) \citep{Flower_2006} 
and releases the amount of energy $170\,$K.  
Therefore, if the temperature is less than $170\,$K, this reaction destroys ortho-H$_{2}$ and 
forms para-H$_{2}$.  	

\item Second, active H and H$_{2}$ collisions \citep{Dalgarno_1973}
\begin{equation}
	\ce{H_{2}(para) + H <=> H_{2}(ortho) + H} \mbox{.}
\end{equation} 
This process mainly occurs at high temperature ($T \sim 3900\,$K) and this is therefore negligible in cold molecular clouds.

\item Third, interaction H$_{2}$ with interstellar dust grains:
\begin{equation}
	\ce{H_{2}(para) + g <=> H_{2}(ortho) + g}
\end{equation} 
This process is inefficient in \textit{low-velocity} shocks \citep{Timmermann_1998}. 

\item Fourth, the formation of H$_{2}$ onto the surface of dust grains:
\begin{equation}
	\ce{H(adsorb) + H(g) -> H_{2}}  
\end{equation}
This formation is not an important process in low-velocity shocks, 
because these shocks insufficiently produce atomic hydrogen 
and the rate coefficient for H$_{2}$
formation on dust is small \citep{Timmermann_1998}.
\end{itemize}

\setstretch{1.1} 
\chapter{Analytical expression of the distribution function of shock velocities for a parabolic bow shock} 
\label{app:pdf_accuracy} 
\lhead{Appendix B}

In \autoref{sec:1D_distribution}, we show the numerical methodology to calculate the distribution function of single shocks (defined by their velocities $u_{\bot}$) for a given arbitrary axisymmetric bow shock shape. Here we examine the accuracy of this method in the special case of the parabolic shape $z = x^{2}/R_{0} - R_{0}$, where $R_{0}$ is a curvature radius. In this case, the norm of a segment $dl$ defined in \autoref{eq:dl} becomes 
\begin{equation}
	dl = \sqrt{dx^{2} + dz^{2}} = \sqrt{1 + \frac{4\ x^{2}}{R^{2}_{0}}}dx \mbox{.}
\end{equation} 

Therefore, the element area $ds$ (\autoref{eq:ds}) can be rewritten as:
\begin{equation}
	ds = 2 \pi \sqrt{1 + \frac{4\ x^{2}}{R^{2}_{0}}}\ x\ dx \mbox{.}
\end{equation}

Following \autoref{eq:tan_alpha}, the tangent of the angle $\alpha$ is $\tan \alpha = dx/dz = 2R/x$, and noting that $\alpha = 90 - \theta, u_{\bot} = u_{0}\cos\theta$, we can find the relation between $ds$ and $u_{\bot}$ as 

\begin{equation} \label{eq:ds_parabola}
	ds = \frac{\pi\ R^{2}_{0}}{2} \frac{\cos \alpha}{\sin^{4} \alpha} d\alpha 
	= \frac{\pi R^{2}_{0}}{2} \frac{\sin \theta}{\cos^{4}\theta} d\theta 
	= \frac{\pi R^{2}_{0}}{2} u^{3}_{0}\frac{du_{\bot}}{u_{\bot}^{4}} \mbox{.}
\end{equation}

Integrating \autoref{eq:ds_parabola} over $u_{\bot}$ from $c$ to $u_{0}$, we obtain
\begin{equation}
	S_{shock} = \frac{\pi R^{2}_{0}}{2}u^{3}_{0}\int_{c}^{u{0}} \frac{du_{\bot}}{u_{\bot}^{4}} 
	= \frac{\pi R^{2}_{0}}{6} \frac{u^{3}_{0} - c^{3}}{c^{3}} \mbox{.}
\end{equation} 
Therefore, the probability density function in the case of a parabolic shape will be:
\begin{equation}
	PDF_{parabola} = \frac{ds}{\int ds} = \frac{3\ u^{3}_{0}\ c^{3}}{u_{\bot}^{4}(u^{3}_{0}-c^{3})} \mbox{.}
\end{equation}

\begin{figure}
	\begin{center}
 		\includegraphics[width=0.9\linewidth]
 		{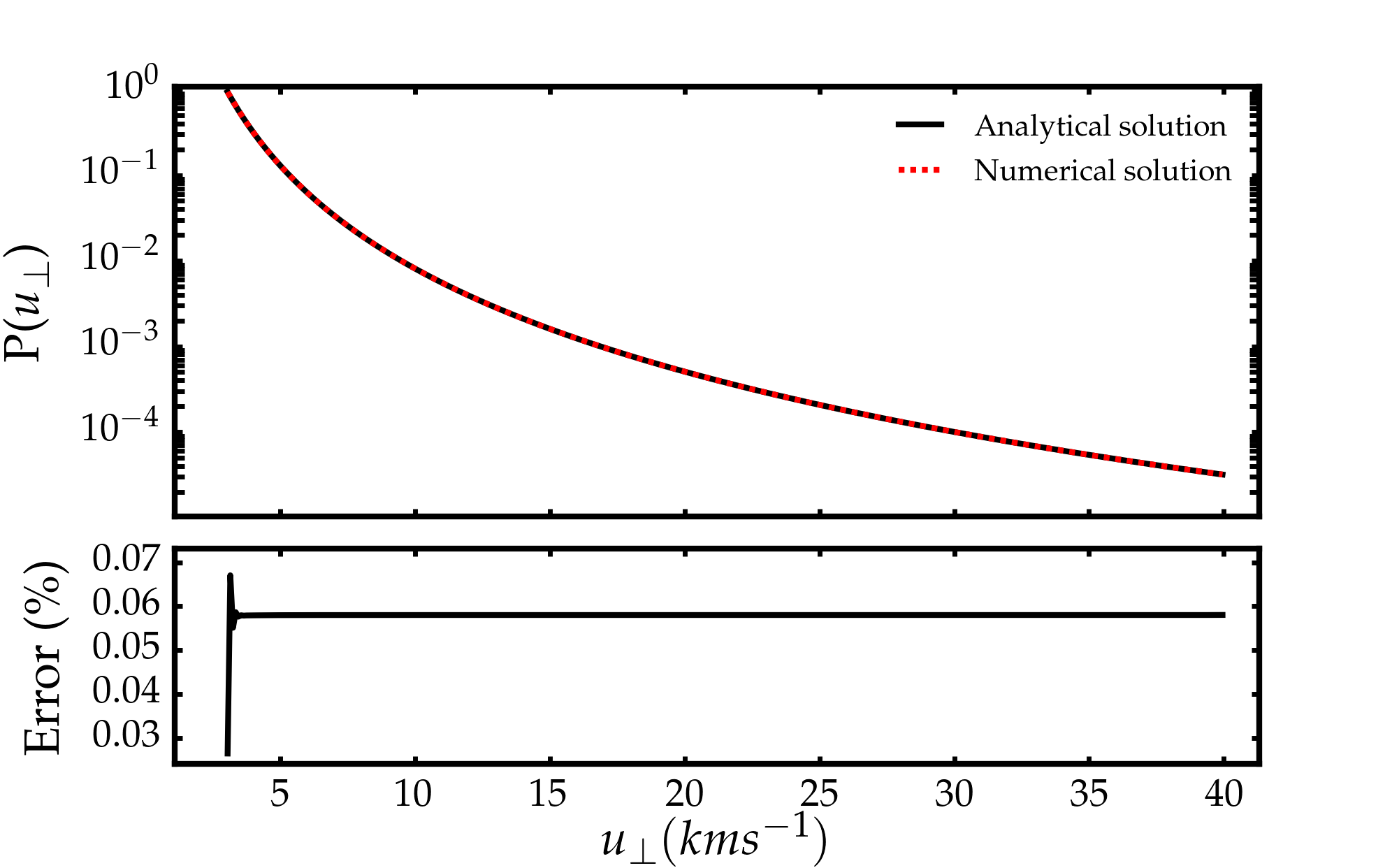}
 		\caption[An example of the accuracy of the numerical method for the PDF calculation]
 		{An example of the accuracy of the numerical method for the PDF calculation.}
 		\label{fig:pdf_accuracy}
	\end{center}
\end{figure}

\autoref{fig:pdf_accuracy} shows the comparison in PDF calculation between our numerical method for arbitrary shapes and the exact analytical method. This figure also shows that the numerical calculation has an error of $0.06 \%$ relative to the analytical one.  

\setstretch{1.1} 
\chapter{L'Hôpital's rule for stellar winds at the sonic point} 
\label{app:hopital_rulel} 
\lhead{Appendix C}
The l'Hôpital's rule gives us an expression for the first derivative of a function at a singular point.

Assume that the function $y(x)$ is described the the differential equation
\begin{equation}
	\frac{d y(x)}{dx} = \frac{f(x)}{g(x)} \mbox{.}
\end{equation}

The point $x=r_{c}$ is called a singular point of this function, when and only 
when $f(r_{c}) = g(r_{c}) = 0$. 
Then the derivative $dy(x)/dx$ can be developed by using the Taylor series expansion:
\begin{equation} \label{eq:dy/dx}
	\left. \frac{dy}{dx} \right |_{r_{c}} 
	= \frac{f(r_{c}) + \frac{x-r_{c}}{1!}f^{'}(r_{c}) 
	+ \frac{(x-r_{c})^{2}}{2!}f^{''}(r_{c}) + ...}{g(r_{c}) + \frac{x-r_{c}}{1!}g^{'}(r_{c}) + \frac{(x-r_{c})^{2}}{2!}g^{''}(r_{c}) + ...}
\end{equation}
with the function $f(x)$ expanded at $r_{c}$ as 
\begin{equation}
	f(x) = f(r_{c}) + \frac{x-r_{c}}{1!}f^{'}(r_{c}) 
	+ \frac{(x-r_{c})^{2}}{2!}f^{''}(r_{c}) + ...  \mbox{.}
\end{equation}  

Therefore, the first order differential \autoref{eq:dy/dx} becomes
\begin{equation} \label{eq:lhopital_law}
	\left. \frac{dy}{dx} \right|_{r_{c}} = \frac{f^{'}(r_{c})}{g^{'}(r_{c})} \mbox{.}
\end{equation}

If $f^{'}(r_{c})$ and $g^{'}(r_{r_{c}})$ are also equal to $0$, one goea to the next order and a  similar method gives us:
\begin{equation}
	\left. \frac{dy}{dx} \right|_{r_{c}} = \frac{f^{''}(r_{c})}{g^{''}(r_{c})} \mbox{.}
\end{equation}

This rule is very convenient to treat the velocity profile at the \textit{sonic point} of stellar winds. The momentum equation in a stellar wind is usually formulated as:
\begin{equation} \label{eq:momen_eq}
	\frac{1}{v}\frac{\partial v}{\partial r} = \left(\frac{2c^{2}_{s}}{r}-\frac{\partial c_{s}}{\partial r} - \frac{GM_{\ast}}{r^{2}} + f \right) / (v^{2} - c^{2}_{s})
\end{equation}
where the function $f$ expresses the other forces acting on the gas, such as the radiation force from dust grains. At the \textit{sonic point} r$_{c}$, the numerator and the denominator of this equation both equal to zeros, we thus apply the l'Hôpital's law \autoref{eq:lhopital_law}. Assuming constant sound speed, the application on \autoref{eq:momen_eq} gives us:
\begin{equation}
	\left.\left(\frac{1}{v}\frac{\partial v}{\partial r}\right) \right|_{r_{c}} = 
	\left[-\frac{2c^{2}_{s}}{r^{2}_{c}} - \left. \frac{\partial^{2}c^{2}_{s}}{\partial^{2} r} \right|_{r_{c}} + \frac{2GM_{\ast}}{r^{3}_{r_{c}}} + \left.\frac{\partial f}{\partial r} \right|_{r_{c}} \right] / \left[ 2\left. \left(v\frac{\partial v}{\partial r}\right) \right|_{r_{c}} \right]	
\end{equation}
with $v(r_{c}) = c_{s}$. Finally, the differential equation for the velocity of the stellar wind at $r_{c}$ is:
\begin{equation}
	\left. \frac{\partial v}{\partial r} \right|_{r_{c}} = 
	\sqrt{-\frac{c^{2}_{s}}{r_{c}} +\frac{GM_{\ast}}{r^{3}_{c}} -\frac{1}{2}\left. \frac{\partial^{2}c^{2}_{s}}{\partial^{2} r} \right|_{r_{c}} + \frac{1}{2}\left.\frac{\partial f}{\partial r} \right|_{r_{c}}} \mbox{.}
\end{equation}

  In practice, we use this rule whenever $0.99 c_{s} < v_{n} < 1.01 c_{s}$ and it allows us to go through the sonic point {\it even in the cases where the numerator of \label{eq:momen_eq} does not vanish}. We then adjust the starting velocity at the base of the wind so that this numerator is close to zero at the sonic point.

\setstretch{1.1} 
\chapter{Motion of termination shock} 
\label{app:term_motion} 
\lhead{Appendix D}

\begin{figure}
	\centering
    \includegraphics[width=0.9\linewidth]
    {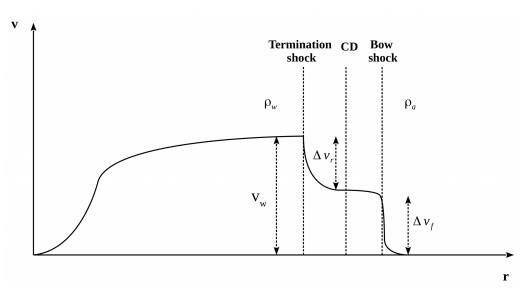}
	\caption[Schematic diagram of the interaction between the stellar wind and the surrounding ISM]
	{Schematic diagram of the interaction between the stellar wind and the surrounding ISM.} 
	\label{fig:wind_scheme}
\end{figure}

In \autoref{sec:IRC_detached_shell} and \autoref{sec:YCVn_detached_shell}, the stellar wind velocity is computed in the star frame, while the termination shock velocity is considered in the shock frame. However, since we neglect the motion of the termination shock's frame, we use the termination wind velocity as an input parameter for the termination shock. 
In this section, we will examine the relationship between the termination shock's velocity in the shock frame and the terminal wind velocity in the star frame, which can infer the effect of the motion of the termination shock's frame. The diagram of interaction between the stellar wind and the ISM is shown in \autoref{fig:wind_scheme}. 

In the termination shock's frame, we call $v_{r}$ the velocity of the termination shock and $v_{f}$ the bow shock velocity. The balance of ram-pressure between both sides of the shock region gives us:
\begin{equation} \label{eq:pressure_balance}
	\rho_{w}v^{2}_{r} = \rho_{a}v^{2}_{f}
\end{equation}
where $\rho_{w}$ and $\rho_{a}$ are the density of the inner wind region and the outer ISM ambient.

In the star's frame, the terminal wind velocity ($v_{w}$) is:
\begin{equation}
	v_{w} = \Delta v_{r} + \Delta v_{f}
\end{equation}
where $\Delta$ is the difference of the pre-shock velocity to the post-shock velocity. 
As indicated in \autoref{eq:compress_ratio_reduce}, the compression factor is $1/M^{2}$ 
in the case of isothermal ($\gamma = 1$), these differences are then:  
\begin{equation} \label{eq:deltav}
	\begin{split}
	&\Delta v_{r} = v_{r} - \frac{v_{r}}{M^{2}_{r}} \simeq v_{r}\\
	&\Delta v_{f} = v_{f} - \frac{v_{r}}{M^{2}_{f}} \simeq v_{f} \\	
	\end{split}
\end{equation}

where $M_{r}$ and $M_{f}$ are the Match numbers of the reverse and the forward shocks. Substituting $v_{f}$ from \autoref{eq:pressure_balance} and \autoref{eq:deltav}, 
we get the relationship between $v_{r}$ and $v_{w}$ as:
\begin{equation}
	v_{r} = \frac{v_{w}}{1 + \sqrt{\frac{\rho_{a}}{\rho_{w}}}}
\end{equation} 

If $\rho_{w} \ll \rho_{a}$: $v_{r} = v_{w}$. Thus, the motion of the termination shock's frame can be negligible.
If $\rho_{w} \simeq \rho_{a}$: $v_{r} = v_{w}/2$. Thus, The termination shock's frame moves 
at the velocity $u_{s} \simeq v_{w}/2$.
If $\rho_{w} \gg \rho_{a}$: $v_{r} \simeq 0$. Thus, only forward shock occurs. 

\setstretch{1.1} 
\chapter{Calculation the specific intensity of the HI radiation from the detached shell} 
\label{app:techinique} 
\lhead{Appendix E}

To calculate the specific intensity of the HI 21-cm line radiation (\autoref{eq:Iv}), we need to calculate two integrations over the whole space.

\begin{figure}
	\centering
    \includegraphics[width=0.9\linewidth]
    {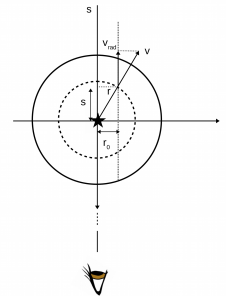}
	\caption[Spherical symmetry of the detached shell]
	{Spherical symmetry of the detached shell.} 
	\label{fig:detached_shell}
\end{figure}

Adopting the spherical symmetry of the CSE around star as in \autoref{fig:detached_shell}. At one point of a distance $r_{0}$ in the detached shell, we have:
\begin{equation}
	r^{2} = s^{2} + r^{2}_{0} \rightarrow s = \sqrt{r^{2} - r^{2}_{0}}
\end{equation}
where $s$ is the coordinate along the line-of-sight. 
Therefore, the integrated specific intensity along one line-of-sight (r$_{0}$) is:
\begin{equation}
	I_{v_{r}}(r_{0}) = e^{-\int^{b}_{a} 
	k_{v_{r}}(\sqrt{s^{2}+r^{2}_{0}})\, ds} 
	\left(\int^{b}_{a} \epsilon_{v_{r}}(\sqrt{s^{2}+r^{2}_{0}}) e^{\int^{s}_{a} 
	k_{v_{r}}(\sqrt{s'^{2}+r^{2}_{0}})\, ds'}\, ds + const. \right) \mbox{.}
\end{equation}
with $a = -\sqrt{r^{2}-r^{2}_{0}}$ and $b = \sqrt{r^{2}-r^{2}_{0}}$. 

\begin{figure}
	\centering
    \includegraphics[width=1.0\linewidth]
    {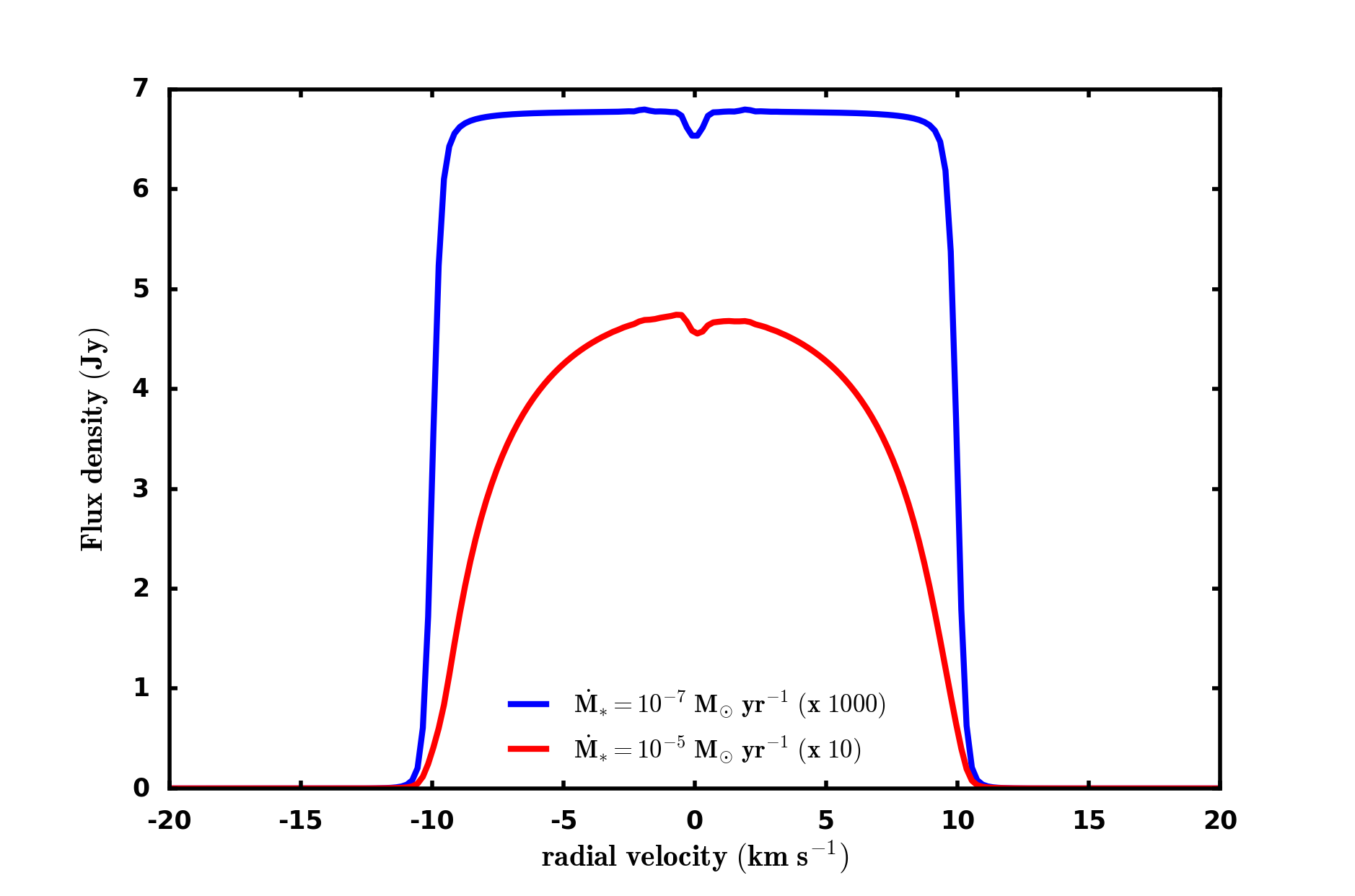}
	\caption[HI line profiles of shells in free expansion 
	for two mass loss rates with no background]
	{HI line profiles of shells in free expansion 
	for two mass loss rates with no background.} 
	\label{fig:Hoai_benchmark}
\end{figure}

To verify our method, we took the radial physical profiles of temperature, velocity and density of \citet{Hoai_2015} and computed the HI line profile in two cases of $10^{-7}$ M$_{\odot}\,$yr$^{-1}$ and $10^{-5}$ M$_{\odot}\,$yr$^{-1}$. Our results (\autoref{fig:Hoai_benchmark}) match well the figure (2) of \citet{Hoai_2015}, except at the central velocity.

\setstretch{1.1} 
\chapter{Table of H$_{2}$ rovibrational excitation levels} 
\label{app:tab_H2_excitation} 
\lhead{Appendix F}
\begin{table} [hbtp]
	\centering
    \includegraphics[width=0.8\textwidth]
    {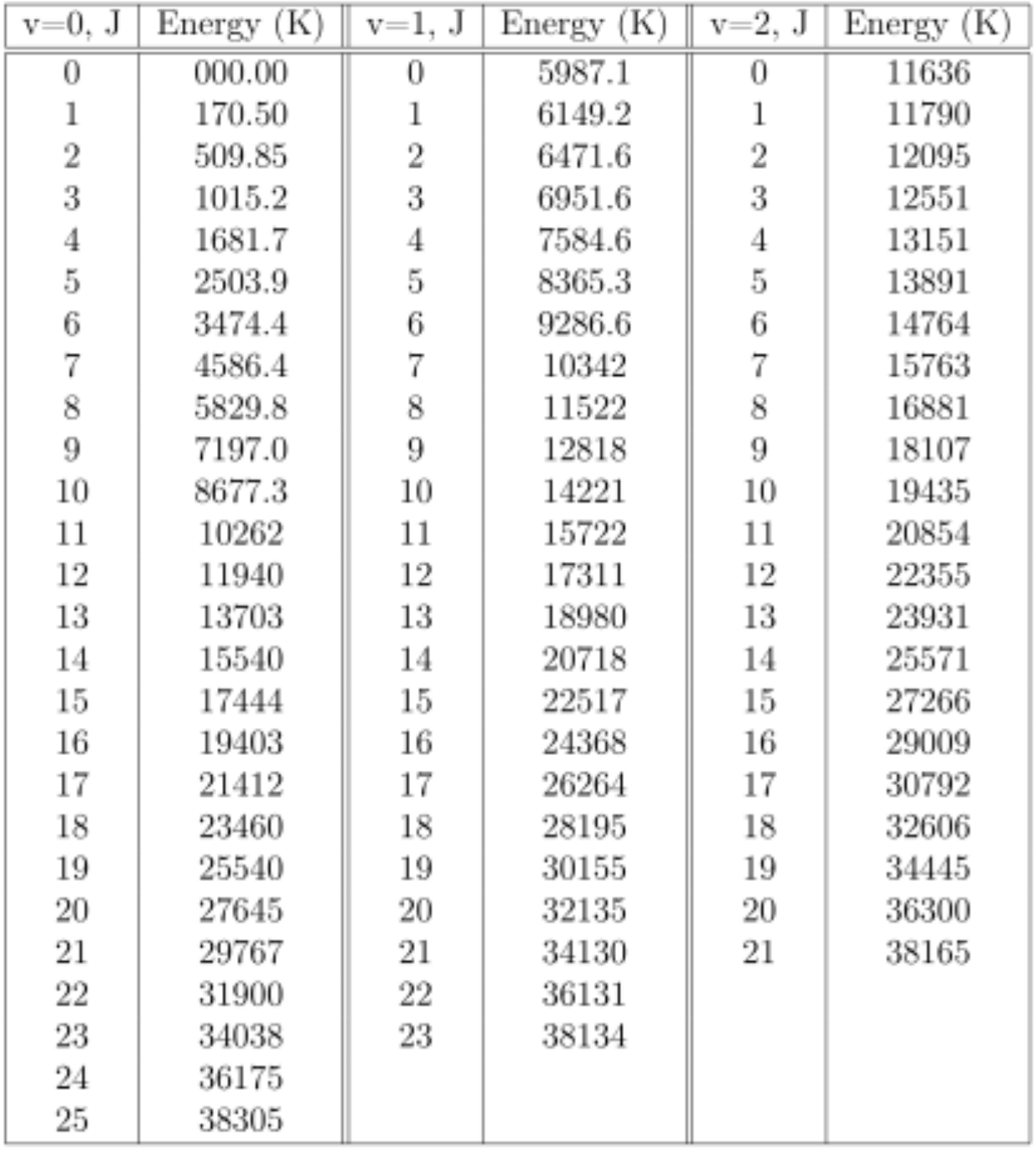}
    \caption[Table of H$_{2}$ rovibrational excitation levels with $v=\overline{0,2}$]
    {Table of H$_{2}$ rovibrational excitation levels with $v=\overline{0,2}$.}
    \label{tab:table_of_H2}
\end{table}

\begin{table} [hbtp]
	\centering
    \includegraphics[width=0.8\textwidth]
    {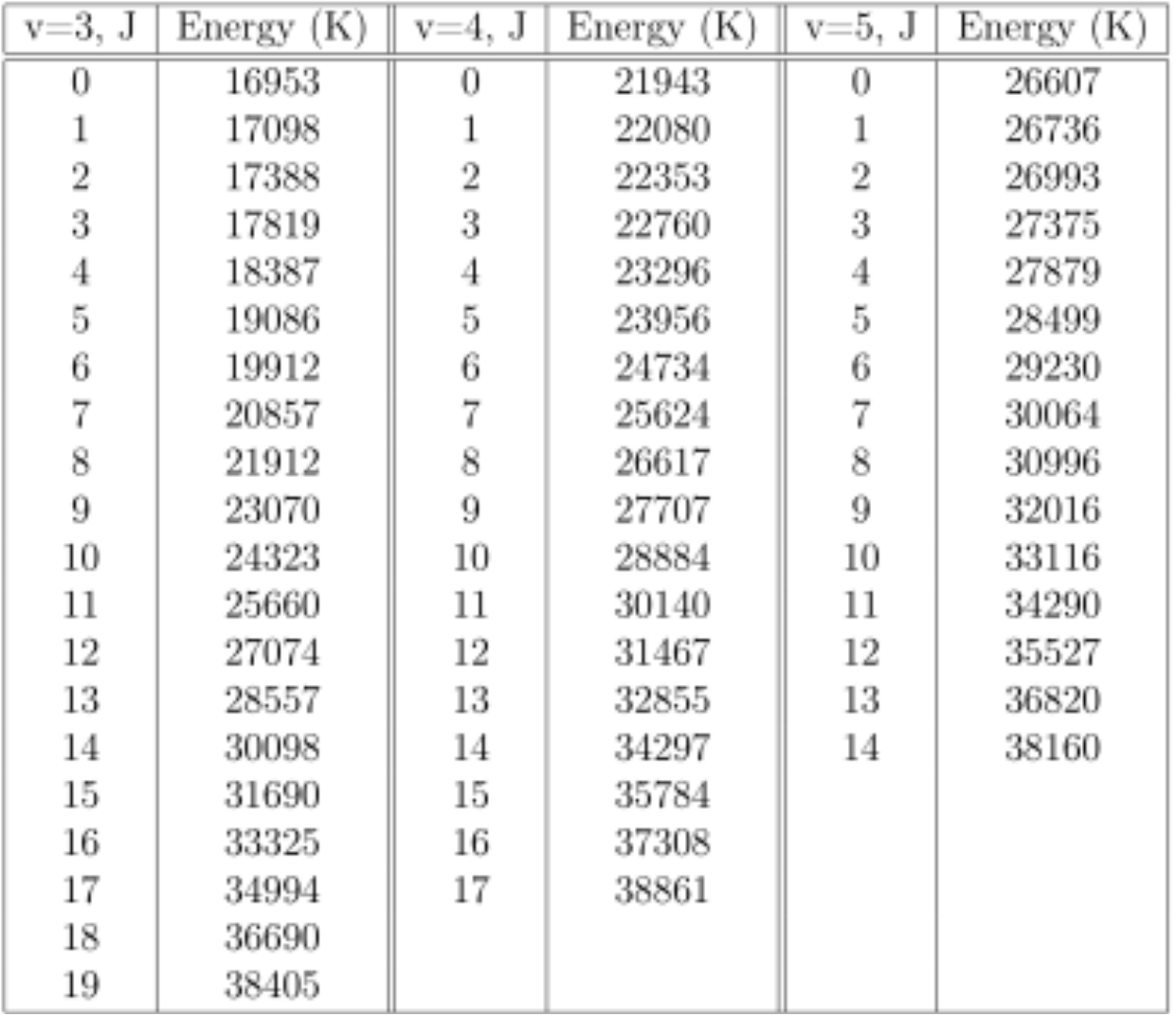}
    \caption[Table of H$_{2}$ rovibrational excitation levels with $v=\overline{3,5}$]
    {Table of H$_{2}$ rovibrational excitation levels with $v=\overline{3,5}$.}
    \label{tab:table_of_H2}
\end{table}

\begin{table} [hbtp]
	\centering
    \includegraphics[width=0.8\textwidth]
    {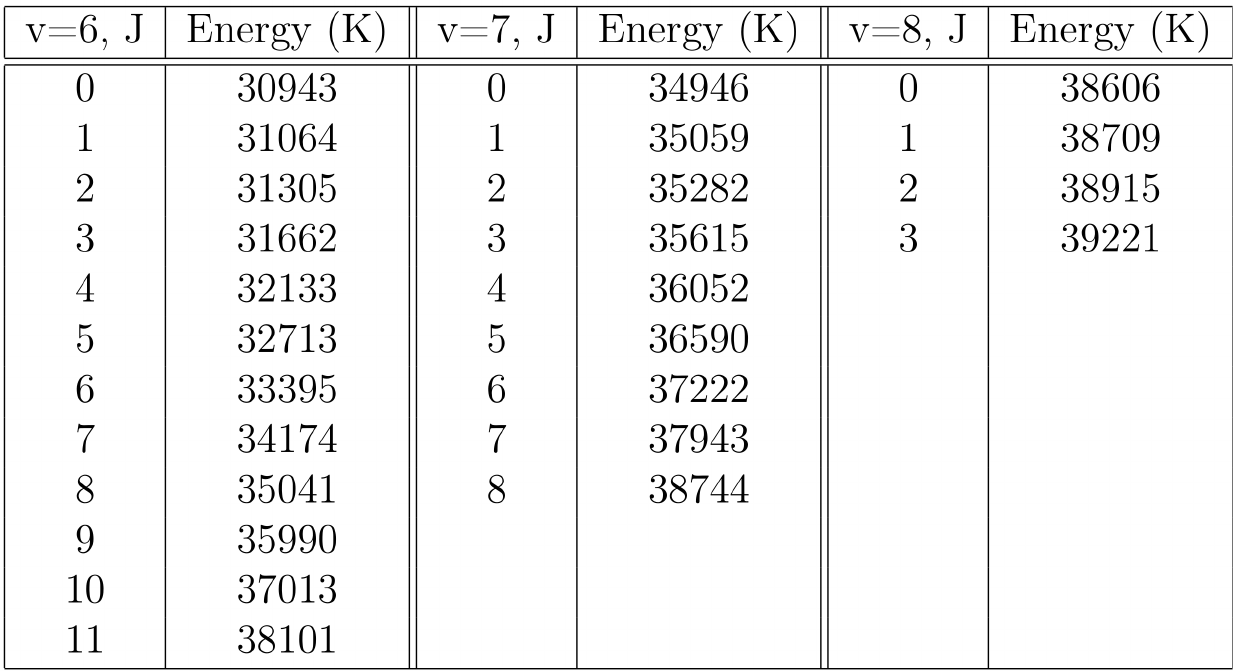}
    \caption[Table of H$_{2}$ rovibrational excitation levels with $v=\overline{6,8}$]
    {Table of H$_{2}$ rovibrational excitation levels with $v=\overline{6,8}$.}
    \label{tab:table_of_H2}
\end{table}

\setstretch{1.1} 
\chapter{Input chemical species for the Paris-Durham shock code} 
\label{app:chemical_species} 
\lhead{Appendix G}
\begin{table} [hbtp]
	\centering
    \includegraphics[width=0.9\textwidth]
    {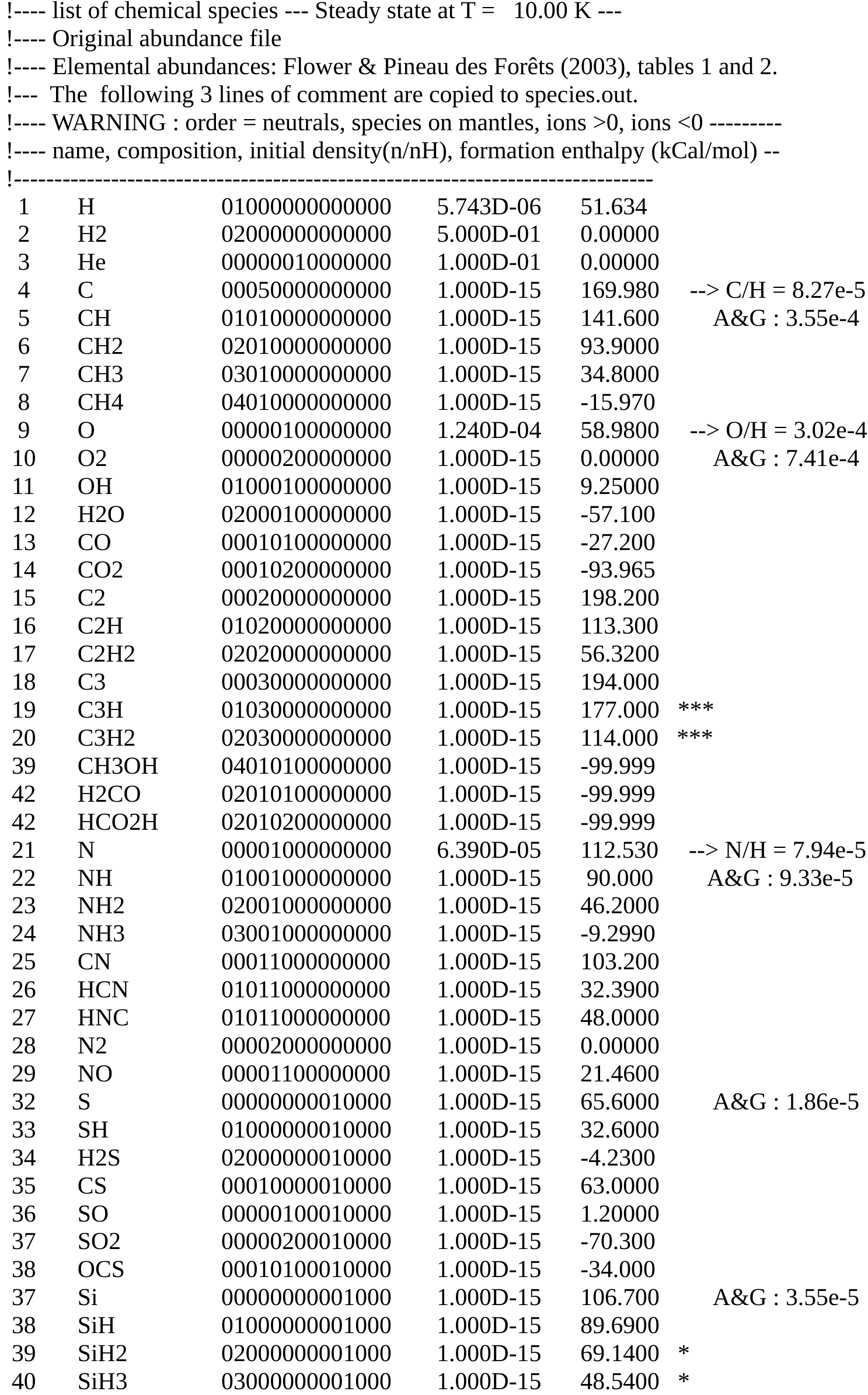}
\end{table}
\newpage
\begin{table} [hbtp]
	\centering
    \includegraphics[width=0.9\textwidth]
    {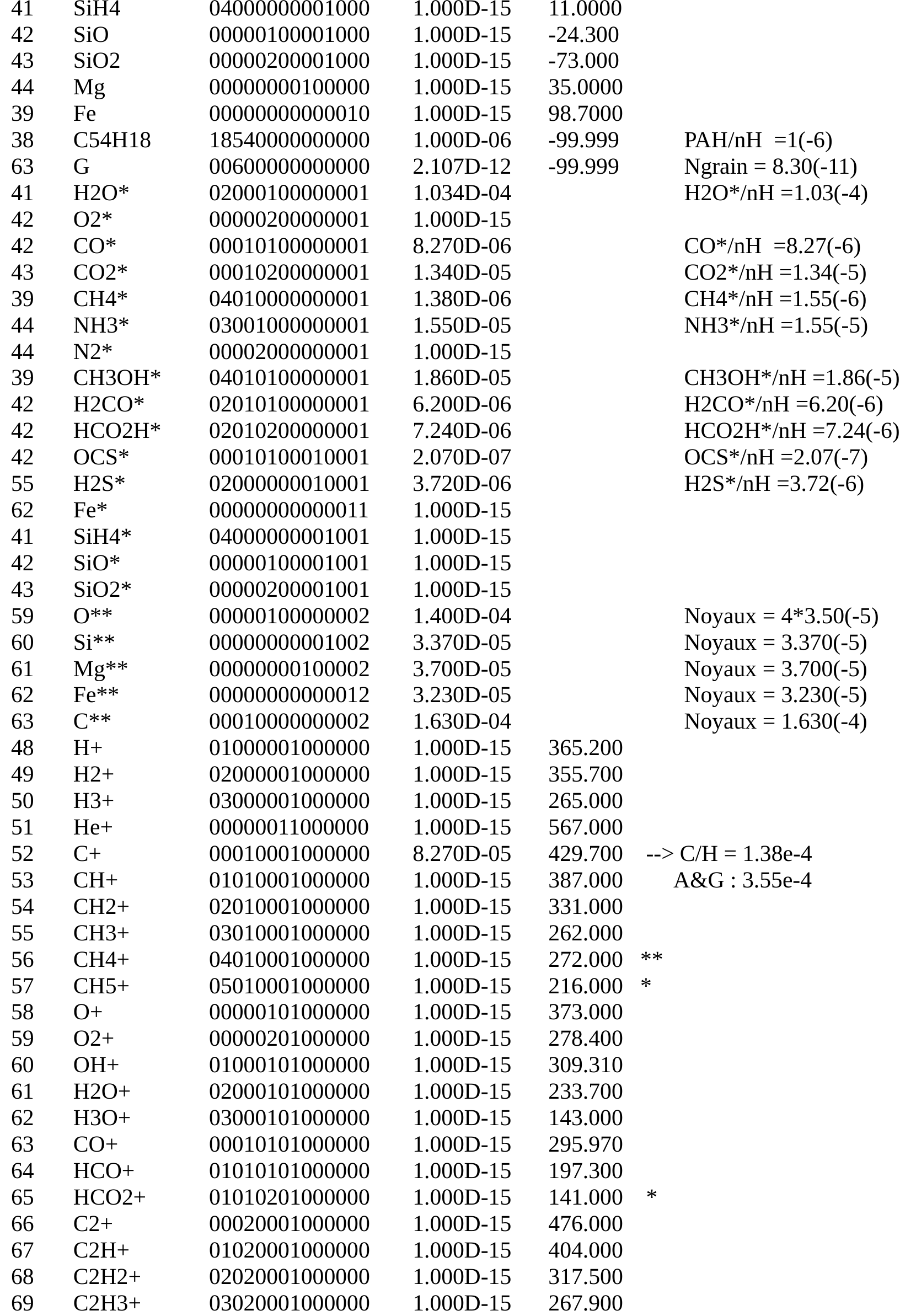}
\end{table}
\newpage
\begin{table} [hbtp]
	\centering
    \includegraphics[width=0.9\textwidth]
    {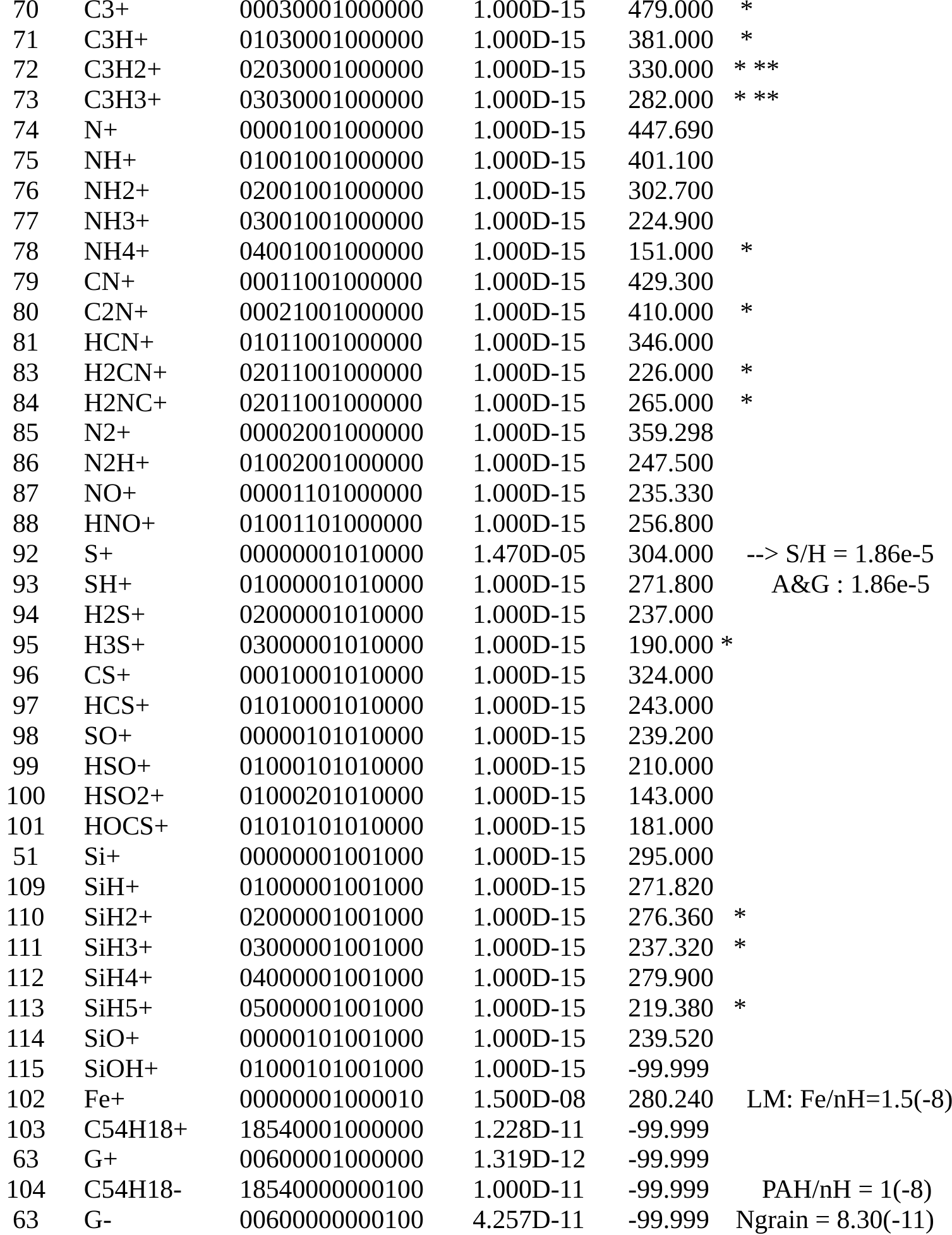}
    \caption[List of input chemical species for the Paris-Durham shock code]
    {List of input chemical species for the Paris-Durham shock code.}
\end{table}

\end{appendices}
\addtocontents{toc}{\vspace{2em}} 

\backmatter


\label{Bibliography}
\lhead{\emph{Bibliography}} 
\bibliographystyle{plainnat}
\bibliography{main} 


\end{document}